%% file: these.tex
\documentclass[twoside,12pt]{Classes/aesm_edspia}
\usepackage[latin1]{inputenc}
\usepackage[english,french]{babel}
\usepackage[T1]{fontenc}
\usepackage{amsmath}
\usepackage{lmodern}%font modern
\usepackage{pdfpages}
\rmfamily
\DeclareFontShape{T1}{lmr}{bx}{sc}{<->ssub * cmr/bx/sc}{} %manque une police en lmodern
\usepackage{lettrine}
\usepackage{tabularx}
\usepackage{epsfig, floatflt, amssymb}
\usepackage{moreverb} %% pour le verbatim en boite
\usepackage{cases}%equations en systemes num\'erot\'es - soluce possible package : CASES
\usepackage{multirow} %% pour regrouper un texte sur plusieurs lignes dans une table
\usepackage{url} %% pour citer les url par \url
\usepackage[all]{xy} %% pour la barre au dessus des symboles
\usepackage{textcomp} %% pour le symbol pour mille par \textperthousand et degr\'es par \degres
\usepackage[right]{eurosym}
\usepackage{setspace} %interligne simple, double etc...
\usepackage{eurosans} %%pour le symbole \euro
\usepackage{epic,eepic}
\usepackage{soul}
\usepackage[nottoc]{tocbibind} % tables des figures, des matieres et autres dans la TOC
\usepackage{fancybox}
\usepackage[leftcaption]{sidecap}
\usepackage[labelsep=endash, textfont={footnotesize, singlespacing}, margin=10pt, format=plain, labelfont=bf]{caption}
\usepackage[Conny]{fncychap} %en tete chapitrage

\makeatletter
\renewcommand{\thesection}{\arabic {section}.}
\renewcommand{\SC@figure@vpos}{c}% centrer verticalement le caption avec le package sidecap...
\renewcommand{\fnum@figure}{\small\textbf{Figure~\thefigure}}
\renewcommand{\fnum@table}{\small\textbf{Tableau~\thetable}}
\makeatother
\usepackage{subfig}

\def\thechapter {\arabic{chapter}}
\def\thesection {\thechapter.\arabic{section}}
\def\thesubsection {\thesection.\arabic{subsection}}
\def\thesubsubsection {\thesubsection \arabic{subsubsection}.}

\begin{document}
%%%%%%%%%%%%%%%%%%%%%%%%%%%%%%%%%%%%%%%%%%%
\renewcommand\figurename{\small\textbf{Figure}}

\addtocounter{page}{-1}%pour revenir \`a 0

%%  1ere de Couverture:
%\RapporteurA{\emph{E. H. SAIDI}}{\emph{Professeur, Facult\'e des
%Sciences, Rabat}}
%\RapporteurB{\emph{L. B.
%DRISSI}}{\emph{Professeur, Facult\'e des Sciences, Rabat}}
%\ExaminateurA{\emph{M. DAOUD}}{\emph{Professeur, Facult\'e des
%Sciences, Agadir}}
%\President{\emph{R. AHL LAAMARA}}{\emph{Dr.
%Chercheur, Facult\'e des Sciences, Rabat}}

\makethese %% cr\'ee la couverture.

\onehalfspacing
%\renewcommand\baselinestretch{1.2}
%\baselineskip=18pt plus 1pt

% une page blanche (deuxi\`eme de couverture)

%\newpage\thispagestyle{empty}\addtocounter{page}{-5}
%\null\newpage\thispagestyle{empty}

%\newpage\thispagestyle{empty}
~\newpage\thispagestyle{empty}

%profondeur dans la table des mati\`eres et de la num\'erotation des sections
\setcounter{secnumdepth}{3}
\setcounter{tocdepth}{3}

\frontmatter  \pagestyle{fancy} \fancyhf{}
%\fancyhead[LE,RO]{D\'edicace}
%\fancyhead[RO]{\bfseries\rightmark}
%\fancyhead[LE]{\bfseries\leftmark}
%\fancyfoot[RO]{\thepage} \fancyfoot[LE]{\thepage}
%\renewcommand{\headrulewidth}{0.5pt}
%\renewcommand{\footrulewidth}{0pt}
%\renewcommand{\chaptermark}[1]{\markboth{\MakeUppercase{\chaptername~\thechapter. #1 }}{}}
%\renewcommand{\sectionmark}[1]{\markright{\thechapter.\thesection~ #1}}
%\include{D\'edicace/D\'edicace}
\def\cleardoublepage{\clearpage}
\def\cleardoublepage{\clearpage}
\pagestyle{fancy} \fancyhf{} \fancyhead[LE,RO]{Remerciements}
%\fancyhead[RO]{\bfseries\rightmark}
%\fancyhead[LE]{\bfseries\leftmark}
%\fancyfoot[RO]{\thepage}
%\fancyfoot[LE]{\thepage}
\renewcommand{\headrulewidth}{0.5pt}
\renewcommand{\footrulewidth}{0pt}
\include{Remerciement/remerciement}

\def\cleardoublepage{\clearpage}
\include{Resume/Resume}
%\include{Abstract/Abstract}
%\printnomenclature
\pagestyle{fancy}
\fancyhf{}
\fancyhead[LE,RO]{Table des mati\`eres}
%\fancyhead[RO]{\bfseries\rightmark}
%\fancyhead[LE]{\bfseries\leftmark}
%\fancyfoot[RO]{\thepage}
%\fancyfoot[LE]{\thepage}
\renewcommand{\headrulewidth}{0.5pt}
\renewcommand{\footrulewidth}{0pt}
\tableofcontents
\def\cleardoublepage{\clearpage}
\pagestyle{fancy}
\fancyhf{}
\fancyhead[LE,RO]{Table des figures}
%\fancyhead[RO]{\bfseries\rightmark}
%\fancyhead[LE]{\bfseries\leftmark}
%\fancyfoot[RO]{\thepage}
%\fancyfoot[LE]{\thepage}
\renewcommand{\headrulewidth}{0.5pt}
\renewcommand{\footrulewidth}{0pt}
%\renewcommand{\chaptermark}[1]{\markboth{\MakeUppercase{\chaptername~\thechapter. #1 }}{}}
%\renewcommand{\sectionmark}[1]{\markright{\thechapter.\thesection~ #1}}
%\listoffigures
\def\cleardoublepage{\clearpage}
\mainmatter
%\listoftables

\pagestyle{fancy}
\fancyhf{}
\fancyhead[LE,RO]{Introduction g\'en\'erale}
%\fancyhead[RO]{\bfseries\rightmark}
%\fancyhead[LE]{\bfseries\leftmark}
\fancyfoot[RO]{\thepage}
\fancyfoot[LE]{\thepage}
\renewcommand{\headrulewidth}{0.5pt}
\renewcommand{\footrulewidth}{0pt}

\chapter*{Introduction g\'en\'erale}
\addcontentsline{toc}{chapter}{Introduction} \vspace{-0.5cm}
Le magn\'{e}tisme est un domaine de la physique de la mati\`{e}re condens%
\'{e}e qui ne cesse de nous surprendre par sa grande vari\'{e}t\'{e} de ph%
\'{e}nom\`{e}nes, souvent li\'{e}s \`{a} l'\'{e}mergence de nouveaux types
de mat\'{e}riaux. L'\'{e}tude concr\`{e}te du
magn\'{e}tisme a vu le jour apr\`{e}s la d\'{e}couverte de l'\'{e}lectron en $1897$. Depuis, le magn\'{e}tisme a connu un essor consid\'{e}rable, qui a conduit \`{a} deux d\'{e}couvertes majeures, durant les deux derniers si\`{e}cles \cite{1-1,1-2}. La premi\`{e}re d\'{e}couverte concerne l'\'{e}troite liaison entre le magn\'{e}tisme et l'\'{e}lectricit\'{e} que nous retrouvons dans l'onde \'{e}lectromagn\'{e}tique ou la lumi\`{e}re. La seconde d\'{e}couverte, repose sur l'\'{e}venement de la th\'{e}orie de relativit\'{e} qui a permis de mieux comprendre les m\'{e}canismes du magn\'{e}tisme d\'{e}crits, comme un effet
purement relativiste, en raison du mouvement relatif des \'{e}lectrons dans
l'atome.

Le moteur \'{e}lectrique, qui utilise le champ magn\'{e}%
tique cr\'{e}\'{e} par le courant \'{e}lectrique circulant dans une bobine \cite{1-1,1-2}, est consid\'{e}r\'{e} comme l'une des applications les plus anciennes du magn\'{e}tisme. De nos jours, les mat\'{e}riaux magn\'{e}tiques sont devenus omnipr\'{e}sents dans notre
environnement vu leur grand usage dans la technologie moderne. Ils composent
de nombreux dispositifs \'{e}lectrom\'{e}caniques et \'{e}lectroniques,
notamment les g\'{e}n\'{e}ratrices, les transformateurs, les moteurs \'{e}%
lectriques, les postes de radio ou de t\'{e}l\'{e}vision, les t\'{e}l\'{e}%
phones, les ordinateurs et les appareils audio ou vid\'{e}o. Les mat\'{e}%
riaux magn\'{e}tiques sont \'{e}galement des constituants indispensables
dans une large gamme d'\'{e}quipements industriels et m\'{e}dicaux.

Au d\'{e}but, seuls les mat\'{e}riaux dans un \'{e}tat massif \'{e}taient consid\'{e}r\'{e}s. Mais avec les nouveaux proc\'{e}d\'{e}s de fabrication, de nouvelles formes de mat\'{e}riaux sont apparues en commen\c{c}ant par les couches minces, puis ultra-mince et en arrivant aux nanofils, nanotubes, nanorubans et nanoparticules \`{a} l'\'{e}chelle nanom\'{e}trique.\newline
Une question centrale dans la physique de la mati\`{e}re condens\'{e}e concerne l'origine du magn\'{e}tisme dans les mat\'{e}riaux. L'\'{e}tude du comportement des grandeurs physiques caract\'{e}ristiques des syst\`{e}mes \'{e}tudi\'{e}s en terme de temp\'{e}rature a fourni un moyen perspicace qui a permis de r\'{e}pondre \`{a} cette question. Pour le cas de l'aimantation, elle dispara\^{\i}t au del\`{a} d'une certaine temp\'{e}rature critique,
marquant ainsi une transition magn\'{e}tique de phase \cite{1-5}.

Le magn\'{e}tisme existe en deux formes diff\'{e}rentes. Le magn\'{e}tisme intrins\`{e}%
que qui concerne les propri\'{e}t\'{e}s magn\'{e}tiques relatives \`{a} la
structure \'{e}lectronique et d'autres principes fondamentaux dans les m\'{e}%
taux ou les non-m\'{e}taux. Ainsi que, le magn\'{e}tisme extrins\`{e}que qui est relatif aux propri\'{e}t\'{e}s des domaines magn\'{e}%
tiques et des ph\'{e}nom\`{e}nes connexes. Dans les mat\'{e}riaux, l'\'{e}tude de leurs propri\'{e}t\'{e}%
s magn\'{e}tiques s'av\`{e}re extr\^{e}mement difficile. Cette difficult\'{e} est due en grande partie aux nombreux types de comportements magn%
\'{e}tiques. En effet, pas moins de quatorze comportements magn\'{e}tiques
diff\'{e}rents ont \'{e}t\'{e} identifi\'{e}s dans les solides, nous citons entre autres, le ferromagn\'{e}tisme, l'antiferromagn\'{e}tisme, le ferrimagn\'{e}tisme, l'antiferrimagn\'{e}tisme et le verre de spin. Cette diversit\'{e} des comportements
magn\'{e}tiques provient principalement de la multitude des diverses interactions spin-spin observ%
\'{e}es dans les mat\'{e}riaux \cite{1-6}.

Les propri\'{e}t\'{e}s magn\'{e}tiques des mat\'{e}riaux ont connu une progression \'{e}norme au fil du temps avec la r\'{e}duction des dimensions. En g\'{e}n\'{e}ral, la dimensionnalit\'{e} est l'une des caract\'{e}ristiques les plus
d\'{e}cisives pour un mat\'{e}riau, puisque la m\^{e}me substance pr\'{e}sente des propri\'{e}t\'{e}s compl\`{e}tement diff\'{e}rentes lors de la
formation de sa structure dans diverses dimensions ; notamment $0$, $1$, $2$,
ou $3$ \cite{1-21}. Bien que les structures de dimensions quasi nulles, telles les points
quantiques ou quasi unidimensionnelles comme les nanotubes et les nanofils ou bien
tridimensionnelle ont \'{e}t\'{e} pour longtemps bien explor\'{e}es, la
recherche sur les cristaux bidimensionnels par contre a demeur\'{e} dans l'obscurit\'{e} peu explor%
\'{e}e, jusqu'\`{a} la r\'{e}alisation exp\'{e}rimentale du graph\`{e}ne en $2004$. Ce nouveau mat\'{e}riau bidimensionnel est obtenu par l'exfoliation d'une seule couche de graphite, un mat\'{e}riau d\'{e}j\`{a} bien connu par la communaut\'{e} scientifique, depuis
les ann\'{e}es $60$. Bien que le graph\`{e}ne soit un conducteur \'{e}lectrique de gap nul non magn\'{e}tique, il a ouvert la voie \`{a} l'\'{e}tude d'une grande vari\'{e}t\'{e} de mat\'{e}riaux bidimensionnels \cite{1-6}. Les mat\'{e}riaux \`{a} base de graph\`{e}%
ne s'av\`{e}rent \'{e}galement d'un tr\`{e}s grand int\'{e}r\^{e}t pour le
magn\'{e}tisme et la spintronique.

Ce travail de th\`{e}se, comportant quatre chapitres et une conclusion g\'{e}n\'{e}rale, est une contribution \`{a} l'\'{e}tude Monte Carlo des propri\'{e}t\'{e}s magn\'{e}tiques des nanomat\'{e}riaux type graphyne et graphone. Plus pr\'{e}cis\'{e}ment, nous visons l'\'{e}tude des propri\'{e}t\'{e}s magn\'{e}tiques de deux cat\'{e}gories de syst\`{e}mes \`{a} base de graph\`{e}ne de
natures extr\^{e}mement diff\'{e}rentes. La premi\`{e}re cat\'{e}gorie est
constitu\'{e}e des mat\'{e}riaux ferromagn\'{e}tiques \`{a} base de graphone. Ce nanomat\'{e}riau est une d\'{e}riv\'{e}e magn\'{e}tique du graph\`{e}ne, dont le magn\'{e}tisme provient des \'{e}lectrons localis\'{e}s sur les atomes de
carbone qui ne sont pas hydrog\'{e}n\'{e}s. C'est un semi-conducteur ferromagn\'{e}tique, qui montre
des propri\'{e}t\'{e}s magn\'{e}tiques pertinentes, aussi bien en pratique,
qu'en th\'{e}orie. Il a \'{e}t\'{e} propos\'{e} th\'{e}oriquement en $2012$,
et synth\'{e}tis\'{e} en $2014$ \cite{O1}. Par ailleurs, la seconde cat\'{e}gorie correspond \`{a}
des syst\`{e}mes c{\oe}ur-coquille ferrimagn\'{e}tiques avec diff\'{e}rentes morphologies, tels le nanoruban et le type graphyne. Le graphyne qui est une autre d\'{e}riv\'{e}e du graph\`{e}ne est obtenu avec le m\'{e}canisme de croissance vapeur-liquide-solide. Il a \'{e}t\'{e} th\'{e}oris\'{e} que le graphyne puisse exister selon diff\'{e}rentes g\'{e}om\'{e}tries, du fait des arrangements multiples des carbones $sp$ et $sp^{2}$. Les nanoparticules type graphyne pr\'{e}sentent des comportements magn\'{e}tiques
importants et inhabituels tel que l'apparition de la temp\'{e}rature de
compensation qui est d'une tr\`{e}s grande importance dans le stockage
d'information en particulier dans l'enregistrement thermo-optique \cite{O2}.

L'\'{e}tude des propri\'{e}t\'{e}s magn\'{e}tiques des mat\'{e}riaux n\'{e}cessite l'utilisation de m\'{e}thodes de calcul et
de simulation capables de sonder la mati\`{e}re \`{a} l'\'{e}chelle atomique tout en tenant compte explicitement de la structure \'{e}lectronique des
\'{e}l\'{e}ments chimiques \cite{1-22}. Les m\'{e}thodes th\'{e}oriques se r\'{e}v\`{e}%
lent \^{e}tre des outils de choix pour mod\'{e}liser les mat\'{e}riaux \`{a}
l'\'{e}chelle atomique voire \'{e}lectronique, et acc\'{e}der de mani\`{e}re
directe \`{a} un ensemble de donn\'{e}es fondamentales comme les propri\'{e}t%
\'{e}s magn\'{e}tiques des mat\'{e}riaux \cite{1-23}. Parmi ces m\'{e}thodes th\'{e}%
oriques, nous citons la th\'{e}orie du champ effectif \cite{1-10,1-11,1-12}, qui est une m\'{e}thode
d'approximation qui utilise l'op\'{e}rateur diff\'{e}rentiel dans l'identit%
\'{e} de Callen \cite{1-13} pour calculer les diagrammes de phases des mat\'{e}riaux
consid\'{e}r\'{e}s. La matrice de transfert est un autre moyen tr\`{e}s
utilis\'{e} pour les mod\`{e}les bidimensionnels. Ce proc\'{e}d\'{e} \'{e}tudie les
comportements critiques, quand le syst\`{e}me subit une transition de phases
ordre-d\'{e}sordre \cite{1-14}. La m\'{e}thode de groupe de renormalisation, initi\'{e}e
tout d'abord par Wilson en $1971$ \cite{1-15} et g\'{e}n\'{e}ralis\'{e}e plus tard, \'{e}%
tudie les ph\'{e}nom\`{e}nes critiques dans les syst\`{e}mes d\'{e}sordonn%
\'{e}s. Cette m\'{e}thode consiste \`{a} transformer l'hamiltonien qui d\'{e}%
crit le syst\`{e}me en un autre, par une transformation qui laisse la fonction de partition invariante \cite{1-15}. Il existe d'autres m\'{e}thodes de renormalisation comme la renormalisation ph\'{e}nom\'{e}nologique, qui ressemble aux m\'{e}%
thodes du groupe de renormalisation. Cette m\'{e}thode utilise les lois d'%
\'{e}chelles de la longueur de corr\'{e}lation des syst\`{e}mes de diff\'{e}%
rentes tailles.

Les calculs ab-initio constituent un outil essentiel dans la physique
de la mati\`{e}re condens\'{e}e moderne et la chimie quantique mol\'{e}culaire. Ils permettent les calculs des propri\'{e}t\'{e}s des syst\`{e}mes \`{a} \'{e}lectrons corr\'{e}l\'{e}s. Le mouvement corr\'{e}l\'{e} des \'{e}lectrons joue un r\^{o}le crucial dans l'agr\'{e}gation des atomes dans les mol\'{e}cules et les solides, les propri\'{e}t\'{e}s de transport \'{e}lectronique et de nombreux autres ph\'{e}nom\`{e}nes physiques majeurs. Actuellement, le moyen le plus utilis\'{e} pour inclure les effets de corr\'{e}lation \'{e}lectronique dans les calculs est la th\'{e}orie
fonctionnelle de la densit\'{e} (DFT) \cite{1,2}. Cette m\'{e}thode qui est rapide et souvent tr\`{e}s pr\'{e}cise, comporte
un certain nombre de limites bien connues, telle la limite
des connaissances disponibles de la forme math\'{e}matique exacte de
la fonctionnelle d\'{e}crivant l'\'{e}change-corr\'{e}lation. La pr\'{e}%
cision de la forme approximative de la th\'{e}orie est non-uniforme et
non-universelle, et il existe des classes importantes de mat\'{e}riaux pour
lesquels, elle donne qualitativement des r\'{e}ponses inad\'{e}quates.\newline
Une alternative importante et compl\'{e}mentaire pour les situations, o\`{u} la pr%
\'{e}cision est primordiale est la m\'{e}thode Monte Carlo, qui pr\'{e}%
sente de nombreuses caract\'{e}ristiques attrayantes pour sonder la
structure \'{e}lectronique des syst\`{e}mes r\'{e}els. Cette m\'{e}thode est
consid\'{e}r\'{e}e parmi les m\'{e}thodes math\'{e}matiques les plus
importantes et techniquement sophistiqu\'{e}es dans la simulation num\'{e}%
rique des ph\'{e}nom\`{e}nes physiques. Elle est bas\'{e}e sur la th\'{e}%
orie des probabilit\'{e}s et l'utilisation de l'\'{e}chantillonnage al\'{e}%
atoire. Elle a pour but majeur de, trouver, mesurer ou v\'{e}rifier une solution
d'un mod\`{e}le quantitatif, ou de d\'{e}crire son comportement simul\'{e} et
ses \'{e}tats transitoires. Cette technique est aujourd'hui largement utilis%
\'{e}e dans toutes sortes d'algorithmes num\'{e}riques d'optimisation. De ce
fait, pour de nombreux probl\`{e}mes complexes, comme la r\'{e}solution des
probl\`{e}mes analytiques, la d\'{e}termination des probl\`{e}mes d'\'{e}%
quation diff\'{e}rentielle normale et partielle sous des conditions
complexes, l'\'{e}valuation de la fiabilit\'{e} de syst\`{e}mes techniques
et d'autres, et la r\'{e}solution des probl\`{e}mes de stockage et de
transport ou de trajets al\'{e}atoires, la m\'{e}thode Monte Carlo reste souvent la seule solution ad\'{e}quate.\cite{6}. Monte Carlo est une m\'{e}thode applicable aux syst\`{e}mes
finis et p\'{e}riodiques, prenant en compte d\`{e}s le d\'{e}part la corr\'{e}%
lation \'{e}lectronique. Elle donne des r\'{e}sultats coh\'{e}rents, tr\`{e}%
s pr\'{e}cis et en m\^{e}me temps pr\'{e}sente la possibilit\'{e} d'utiliser plusieurs mod\`{e}les ayant des structures diff\'{e}rentes. La m\'{e}thode Monte Carlo comporte trois types diff\'{e}rents, tels que : Monte Carlo statique qui est utilis\'{e}e pour simuler des ph\'{e}nom\`{e}nes physiques complexes dans plusieurs domaines scientifiques et appliqu\'{e}s \cite{1,2,7,8}, Monte Carlo cin\'{e}tique qui est employ\'{e}e pour \'{e}tudier l'\'{e}volution des syst\`{e}mes au cours du temps \cite{MC1,MC2}, et Monte Carlo quantique qui est une m\'{e}thode de simulation probabiliste de l'\'{e}quation de Schroedinger \cite{1,9,10,11}.

Compte-tenu de l'importance des \'{e}tudes sur le magn\'{e}tisme au sein de
la communaut\'{e} de physique de la mati\`{e}re condens\'{e}e, nous avons
souhait\'{e} traiter dans ce m\'{e}moire de th\`{e}se des r\'{e}cents d\'{e}%
veloppements dans ce domaine. Ce travail de recherche est organis\'{e} comme suit :

Dans le chapitre $1$, nous introduisons les concepts de base de
la m\'{e}thode Monte Carlo. Nous commen\c{c}ons par une introduction sur
quelques mod\`{e}les de spin. Ensuite, nous pr\'{e}sentons les notions de
base de la simulation Monte Carlo statique. Finalement, nous d\'{e}crivons
quelques algorithmes permettant de g\'{e}n\'{e}rer num\'{e}riquement des
configurations du mod\`{e}le d'Ising tout en mettant l'accent sur
l'algorithme de Metropolis.

Dans le second chapitre, nous consacrons la premi\`{e}re partie \`{a} la d\'{e}finition des ph\'{e}nom\`{e}nes critiques, et notamment les transitions de phase. En seconde partie, nous nous int\'{e}ressons de pr%
\`{e}s \`{a} la th\'{e}orie de champ moyen. Par la suite, nous d\'{e}crivons bri\`{e}vement la th\'{e}orie de champ effectif. \`{A} la fin du chapitre, nous pr\'{e}sentons explicitement le calcul de la fonction de partition et la fonction de corr%
\'{e}lation en utilisant la m\'{e}thode de la matrice de transfert.

Dans le chapitre $3$, nous abordons l'\'{e}tude des propri\'{e}t%
\'{e}s magn\'{e}tiques et hyst\'{e}r\'{e}tiques caract\'{e}%
ristiques des mat\'{e}riaux. Dans un premier temps, nous
fixons notre attention sur l'\'{e}tude des propri\'{e}t\'{e}s magn\'{e}%
tiques des mat\'{e}riaux. Pour ce fait, nous exposons l'origine du magn\'{e}%
tisme et la classification magn\'{e}tiques des mat\'{e}riaux. Puis, nous
donnons les diff\'{e}rents types des temp\'{e}ratures de transition ainsi
que la classification de N\'{e}el. Ensuite, nous \'{e}tudions les diff\'{e}%
rents types des interactions magn\'{e}tiques qui peuvent avoir lieu dans les
mat\'{e}riaux magn\'{e}tiques. Une description compl\`{e}te
de l'anisotropie magn\'{e}tique dans les mat\'{e}riaux magn\'{e}tiques est \'{e}galement fournit. Nous
finissons par la pr\'{e}sentation des diff\'{e}rents outils de base utiles
pour mieux comprendre les propri\'{e}t\'{e}s hyst\'{e}r\'{e}tiques des mat%
\'{e}riaux.

Le chapitre $4$ est consacr\'{e} \`{a} pr\'{e}senter les r\'{e}sum%
\'{e}s de nos diff\'{e}rentes travaux de recherche avec le contenu d\'{e}taill\'{e} de certains d'entre eux. Nous traitons dans la premi\`{e}re section des mat\'{e}riaux ferromagn\'{e}tiques type graphone. Dans la deuxi\`{e}me section, nous pr\'{e}sentons les r\'{e}sultats obtenus pour les mat\'{e}riaux c{\oe}ur-coquille type nanoruban de graph\`{e}ne et type nanoparticule de graphyne. Dans la troisi\`{e}me section, nous exposons l'effet des d\'{e}fauts et de surface sur les propri\'{e}t\'{e}s magn\'{e}tiques des nanomat\'{e}riaux.\\
Nous cl\^{o}turons ce manuscrit de th\`{e}se par une conclusion g\'{e}n\'{e}rale
rappelant les principales id\'{e}es d\'{e}velopp\'{e}es tout le long du document.
Des perspectives sont \'{e}galement pr\'{e}sent\'{e}es.

\textbf{Liste des contributions et communications}

Au cours de la pr\'{e}paration de cette th\`{e}se de doctorat, de nouvelles contributions ont vu le jour dont certaines sont d\'{e}j\`{a}
publi\'{e}es. La liste de ces travaux est la suivante :

\textit{Contributions et Publications}
\newline
$\bullet $ Magnetic phase transitions in pure zigzag graphone nanoribbons,
\newline
J. Phys. Lett. A 379 (2015) 753-760.
\newline
$\bullet $ Edge effect on magnetic phases of doped zigzag graphone
nanoribbons,
\newline
J. Magn. Magn. Mater. 374 (2015) 394-401.
\newline
$\bullet $ Monte Carlo study of magnetic behavior of core-shell nanoribbon,
\newline
J. Magn. Magn. Mater.374 (2015) 639-646.
\newline
$\bullet $ Graphyne core/shell nanoparticles : Monte Carlo study of thermal and magnetic properties,
\newline
Submitted, (2016).
\newline
$\bullet $ Stone-Wales defected molecular-based $%
AFe^{II}Fe^{III}(C_{2}O_{4})_{3}$ nanoribons : Thermal and magnetic
propertie,
\newline
In preparation.
\newline
$\bullet $ Monte Carlo study of edge effect on magnetic and hysteretic
behaviors of sulfur vacancy defected zigzag $FeS_{2}$ nanoribbon,
\newline
In preparation.
\newline
$\bullet $ Surface effect on compensation and hysteretic behavior in surface/bulk nanocube,
\newline
Submitted, (2016).

\textit{Communications}
\newline
$\bullet $ "Coh\'{e}rence quantique et effet Kondo dans les nanostructures",
pr\'{e}sent\'{e}e lors de l'Ecole "National School: Cryptography and Quantum
Information Theory", organis\'{e}e les 31 Janvier -1 F\'{e}vrier 2014 \`{a}
ENSET-Rabat, Facult\'{e} des Sciences, Rabat.
\newline
$\bullet $ "Etude Monte Carlo des propri\'{e}t\'{e}s magn\'{e}tiques et hyst%
\'{e}r\'{e}tiques des syst\`{e}mes ferrimagn\'{e}tiques", pr\'{e}sent\'{e}e
lors de Colloque Franco-Marocain sur Propri\'{e}t\'{e}s des Nouveaux Mat\'{e}%
riaux \`{a} la Facult\'{e} des Sciences de Rabat du 4 au 5 D\'{e}cembre 2014.
\newline
$\bullet $ "Etude Monte Carlo des propri\'{e}t\'{e}s magn\'{e}tiques et
thermodynamiques des syst\`{e}mes hexagonales", pr\'{e}sent\'{e}e lors des
Journ\'{e}es Doctorales nationales CPM-2015 du 11 au 13 Juin 2015 \`{a}
l'Institut Scientifique et Facult\'{e} des Sciences \`{a} Rabat.

\chapter{M\'{e}thode Monte Carlo pour les mod\`{e}les de spin}
\graphicspath{{Chapitre1/figures/}}
%==============================================================================
%\pagestyle{fancy}
%\fancyhf{}
\fancyhead[RO]{\bfseries\rightmark}
\fancyhead[LE]{\bfseries\leftmark} \fancyfoot[RO]{\thepage}
\fancyfoot[LE]{\thepage}
\renewcommand{\headrulewidth}{0.5pt}
\renewcommand{\footrulewidth}{0pt}
\renewcommand{\chaptermark}[1]{\markboth{\MakeUppercase{\chaptername~\thechapter. #1 }}{}}
\renewcommand{\sectionmark}[1]{\markright{\thechapter.\thesection~ #1}}
%==============================================================================
\vspace{-1cm}
Par principe, tout objet de la nature peut changer de phase \`{a} un instant
pr\'{e}cis sous l'influence de son environnement. Il s'agit de l'instant
critique o\`{u} se produisent des ph\'{e}nom\`{e}nes critiques dits des transitions de phase. Les transitions de phases sont des ph\'{e}nom\`{e}nes physique qui d\'{e}crivent les changements d'\'{e}tat de syst\`{e}mes physiques. Cependant, l'un des d\'{e}fis majeurs des syst\`{e}mes physique comportant un grand nombre de particules, est le calcul de la fonction de partition \`{a} la limite thermodynamique \cite{2}. Ainsi en absence des solutions exactes qui n'existent que pour le mod\`{e}le d'Ising et le mod\`{e}le de Potts \`{a} deux dimensions, l'utilisation des m\'{e}thodes d'approximations s'impose. L'usage des m\'{e}thodes d'approximation n\'{e}cessite l'\'{e}tude des mod\`{e}les de spin.

Les mod\`{e}les de spins ont initialement \'{e}t\'{e} introduits pour la description du magn\'{e}tisme dans les mat\'{e}riaux ferromagn\'{e}tiques. Les mod\`{e}les classiques, notamment le mod\`{e}le d'Ising ont jou\'{e} un r\^{o}le important depuis la naissance du magn\'{e}tisme mais, de nos jours, l'int\'{e}r\^{e}t est plus port\'{e} sur les mod\`{e}les quantiques et les ph\'{e}nom\`{e}nes qui leur sont associ\'{e}s. Les mod\`{e}les quantiques de spins ont \'{e}t\'{e} \'{e}tudi\'{e}s pour la premi\`{e}re fois par Bethe en utilisant le mod\`{e}le de Heisenberg unidimensionnel.

Par ailleurs, les mod\`{e}les pour lesquels les fonctions de partition sont
des solutions exactes sont limit\'{e}s. Ce fait a exig\'{e} le d\'{e}veloppement de diff\'{e}rentes techniques approximatives tel les d\'{e}veloppement en s\'{e}ries, les m\'{e}thodes des th\'{e}ories de champ et les m\'{e}thodes num\'{e}riques.\newline
Le calcul num\'{e}rique de la fonction de partition par la m\'{e}thode num\'{e}rique devient facile \`{a} d\'{e}terminer lorsque le mod\`{e}le consid\'{e}r\'{e} est plac\'{e} sur un r\'{e}seau de taille finie \cite{2}. Bien que cette
technique permet d'obtenir les propri\'{e}t\'{e}s critiques du syst\`{e}me,
l'exactitude des r\'{e}sultats d\'{e}pend de la taille consid\'{e}r\'{e}. Les simulations par les m\'{e}thodes num\'{e}riques apportent
des outils compl\'{e}mentaires destin\'{e}s \`{a} mieux comprendre les syst%
\`{e}mes \cite{3,4}. Elles sont essentielles pour l'\'{e}tude des syst\`{e}mes complexes, au voisinage de l'instant critique o\`{u} s'\'{e}tablit la transition.
La m\'{e}thode Monte Carlo est tr\`{e}s utilis\'{e}e et adapt\'{e}e \`{a} cet
effet.\newline
Cette m\'{e}thode est bas\'{e}e sur un jeu d'\'{e}%
chantillon d'une repr\'{e}sentation d'un syst\`{e}me physique pris dans un
\'{e}tat donn\'{e} \cite{5,6}. A ce titre, elle vise la d\'{e}termination de mani%
\`{e}re efficace et rapide des grandeurs physiques li\'{e}es au syst\`{e}me
consid\'{e}r\'{e} par des proc\'{e}d\'{e}s probabilistes \cite{6}. Il existe trois types diff\'{e}rents de Monte Carlo :\newline
$\bullet $ \textbf{Monte Carlo statique :} que nous allons \'{e}laborer au cours de ce chapitre. Cette m\'{e}thode est utilis\'{e}e pour simuler des ph\'{e}nom\`{e}nes physiques complexes dans plusieurs domaines scientifiques et appliqu\'{e}s tels que : physique de la mati\`{e}re condens\'{e}e, physique des hautes \'{e}nergies, radioactivit\'{e}, r\'{e}seaux, \'{e}conom\'{e}trie et logistique \cite{1,2,7,8},\newline
$\bullet $ \textbf{Monte Carlo cin\'{e}tique :} est utilis\'{e} pour simuler les ph\'{e}nom\`{e}nes physiques tels que la diffusion de surface, l'\'{e}pitaxie, l'\'{e}volution et la croissance de domaines ou la mobilit\'{e} des agr\'{e}gats. Cette m\'{e}thode permet d'\'{e}tudier l'\'{e}volution des syst\`{e}mes au cours du temps \cite{MC1,MC2},
\newline
$\bullet $ \textbf{Monte Carlo quantique :} est une m\'{e}thode de simulation probabiliste de l'\'{e}quation de
Schroedinger. \`{A} son tour, elle comporte diff\'{e}rents types utilis\'{e}s pour les calculs de structure \'{e}lectronique tel : le Monte Carlo variationnel \cite{29}, le Monte Carlo
diffusionnel \cite{30,31}, le Monte Carlo de la fonction de Green \cite{32,33}, et le Monte Carlo pour les int\'{e}grales de chemin \cite{34}. Bien qu'il existe de nombreuses variantes nomm\'{e}es diff\'{e}remment, l'id\'{e}e de base de la m\'{e}thode Monte Carlo quantique est toujours la m\^{e}%
me, \`{a} savoir d\'{e}finir une dynamique brownienne pour les
\'{e}lectrons et calculer les valeurs moyennes quantiques comme
valeurs moyennes le long des trajectoires stochastiques \cite{1,9,10,11}.\\
La technique Monte Carlo, d\'{e}velopp%
\'{e}e suite \`{a} de nombreux travaux \cite{1,MC1,29}, introduit divers algorithmes de
simulation, tous int\'{e}ressants et pr\'{e}sentant des sp\'{e}cificit\'{e}s
particuli\`{e}res li\'{e}es aux syst\`{e}mes \'{e}tudi\'{e}s.

Ce chapitre vise \`{a} pr\'{e}senter la m\'{e}thode Monte Carlo statique. Nous commen\c{c}ons par introduire quelques mod\`{e}les de spin. Ensuite, nous d\'{e}crirons les notions de base du Monte Carlo statique. Enfin, nous \'{e}talerons les grandes lignes de quelques algorithmes
permettant de g\'{e}n\'{e}rer num\'{e}riquement de diff\'{e}rentes configurations du mod%
\`{e}le d'Ising.
\vspace{-0.7cm}
\section{Mod\`{e}les de spin}

\subsection{Mod\`{e}le d'Ising}

Le mod\`{e}le d'Ising est l'un des mod\`{e}les les plus simples qui permet de mod\'{e}liser des syst\`{e}mes physiques trop complexes \`{a}
analyser de fa\c{c}on exacte \cite{5}. En raison de sa simplicit\'{e} et de la
richesse de son comportement, le mod\`{e}le d'Ising suscite depuis son
introduction un grand int\'{e}r\^{e}t \cite{12}. Le mod\`{e}le d'Ising est constitu%
\'{e} d'une distribution d'atomes dans un plan. Chacun de ces atomes porte
un moment magn\'{e}tique $\mu _{B}$ (magn\'{e}ton de Bohr) orient\'{e} al\'{e}%
atoirement en spin up $\left( +\mu _{B}\right) $ ou spin down $\left( -\mu
_{B}\right) $. Ces spins int\'{e}ragissent entre eux deux
\`{a} deux, uniquement entre premiers voisins avec une \'{e}nergie
d'interaction .
\begin{figure}[!ht]
  \centering
\includegraphics[scale=0.4]{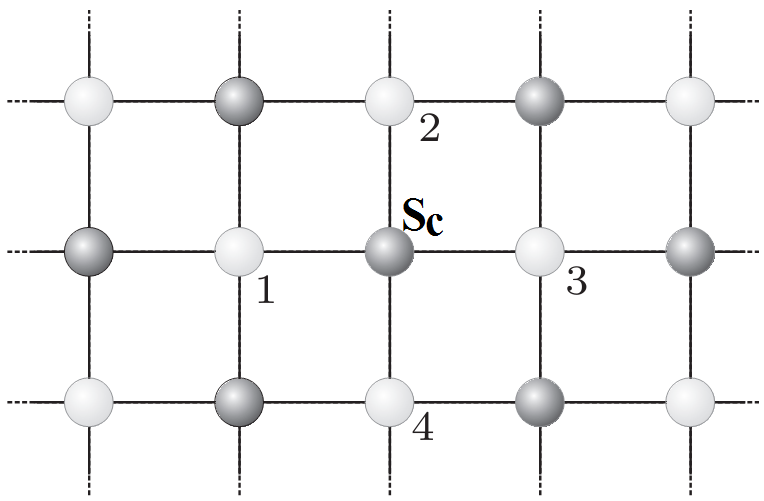}
\caption{\label{sanae1}Syst\`{e}me d'Ising \`{a} deux dimensions o\`{u} le spin central $S_{c}$ int\'{e}ragit uniquement avec les 4 spins indic\'{e}s 1, 2, 3 et 4.}
\end{figure}\newline
L'hamiltonien $H_{Ising}$ du syst\`{e}me est donn\'{e} par :
\begin{equation}
H_{Ising}=-J\sum_{\left \langle ij\right \rangle }S_{i}S_{j}-h\sum_{i}S_{i}.\label{jus1}
\end{equation}%
o\`{u} $\left \langle i,j\right \rangle $ d\'{e}signe une somme sur les sites qui sont les plus proches voisins, $h$ est le champ magn\'{e}tique ext\'{e}rieur, $S_{i}$ repr\'{e}sente le spin au site $i$ et $J$ est l'interaction d'\'{e}change. Les signes (-) dans l'\'{e}quation (\ref{jus1}) sont
classiques. Ils dictent simplement le choix du signe pour le param\`{e}tre
d'interaction $J$ et le champ externe $h$. La simulation d'un syst\`{e}me d'Ising de taille finie par la m\'{e}thode Monte Carlo permet de calculer les valeurs des grandeurs
physiques telles que l'aimantation, l'\'{e}nergie, la chaleur sp\'{e}cifique
et la susceptibilit\'{e} \`{a} une temp\'{e}rature donn\'{e}e.
\vspace{-0.6cm}
\subsection{Mod\`{e}le de Potts}

En physique statistique, le mod\`{e}le de Potts est une g\'{e}%
n\'{e}ralisation du mod\`{e}le d'Ising. C'est un mod\`{e}le d'interaction de
spins sur un r\'{e}seau cristallin \cite{12}. Ce mod\`{e}le permet de comprendre le comportement des mat\'{e}riaux ferromagn\'{e}tiques. Il est \'{e}galement utilis\'{e} pour expliquer certains ph\'{e}nom\`{e}nes relatifs \`{a} la physique des solides tels que les transitions de phases et les propri\'{e}t\'{e}s magn\'{e}tiques des structures p\'{e}riodiques en couches.\newline
Le mod\`{e}le de Potts est similaire au mod\`{e}le d'Ising, hormis le fait que le spin $Si$ sur chaque site du r\'{e}seau peut prendre plus de deux valeurs discr\`{e}%
tes diff\'{e}rentes \cite{13}. Habituellement, ces valeurs sont repr\'{e}sent\'{e}%
es par des nombres entiers positifs \`{a} partir de $1$, et le mod\`{e}le de
Potts \`{a} $q$ \'{e}tats est celui dans lequel chaque spin peut avoir des
valeurs enti\`{e}res $S_{i}=1\ldots q$. L'hamiltonien $H_{Potts}$ s'exprime comme suit :
\begin{equation}
H_{Potts}=-J\sum_{\left \langle ij\right \rangle }\delta _{S_{i}S_{j}},\label{jus2}
\end{equation}
o\`{u} $\delta _{ij}$ est le symbole de Kronecker qui satisfait :
\begin{equation}
\delta _{ij=}\left \{
\begin{array}{c}
1\qquad \text{si }i=j, \\
0\qquad \text{si }i\neq j.%
\end{array}%
\right.
\end{equation}
Le mod\`{e}le de Potts est \'{e}quivalent au mod\`{e}le d'Ising pour $q=2$. Il d\'{e}coule que l'\'{e}quation (\ref{jus2}) prend la forme suivante :
\begin{equation}
H_{Potts}=-\frac{1}{2}J\sum_{\left \langle ij\right \rangle }2\left( \delta
_{S_{i}S_{j}}-\frac{1}{2}\right) -\sum_{\left \langle ij\right \rangle }\frac{1%
}{2}J,
\end{equation}
avec
\begin{equation}
\left \{
\begin{array}{c}
\text{Si }S_{i}=S_{j}\Longrightarrow 2\left( \delta _{S_{i}S_{j}}-\frac{1}{2}%
\right) =1,\text{ \ } \\
\text{Si }S_{i}\neq S_{j}\Longrightarrow 2\left( \delta _{S_{i}S_{j}}-\frac{1%
}{2}\right) =-1.%
\end{array}%
\right.
\end{equation}
Cet hamiltonien est \'{e}quivalent \`{a} celui d'Ising plus une constante $-\sum \limits_{\left \langle i,j\right \rangle }\frac{1}{2}J$.
Le mod\`{e}le de Potts avec $q>2$ transite de l'\'{e}tat ferromagn\'{e}tique \`{a}
l'\'{e}tat paramagn\'{e}tique \cite{2}.
\vspace{-0.4cm}
\subsection{Mod\`{e}le Blume-Emery-Griffiths}

Le mod\`{e}le Blume-Emery-Griffiths (BEG) est un mod\`{e}le de spin qui pr%
\'{e}sente une grande vari\'{e}t\'{e} de ph\'{e}nom\`{e}nes critiques et
multicritiques \cite{1}. Ce mod\`{e}le a \'{e}t\'{e} introduit au d\'{e}but pour d\'{e}crire la
s\'{e}paration de phase et la superfluidit\'{e} dans les m\'{e}langes $^{3}He-^{4}He$ \cite{14}. Par la suite, il a \'{e}t\'{e} utilis\'{e} pour d\'{e}crire les syst\`{e}mes caract\'{e}ris\'{e}%
s par trois \'{e}tats de spin. Il est l'un des rares mod\`{e}les simples qui donne \`{a} la fois la transition de phase du premier ordre et du second ordre. Le mod\`{e}le Blume-Emery-Griffiths est d\'{e}crit par l'hamiltonien $H_{BEG}$ :
\begin{equation}
H_{BEG}=-J\sum_{\left \langle ij\right \rangle }S_{i}S_{j}-K\sum_{\left \langle
ij\right \rangle }S_{i}^{2}S_{j}^{2}+\Delta \sum_{i}S_{i}^{2}-h\sum_{i}S_{i},
\end{equation}%
$\left \langle ij\right \rangle $ indique que la somme est restreinte aux sites des plus proches voisins, $J$\ et $K$ sont
respectivement, l'interaction bilin\'{e}aire et l'interaction biquadratique. $\Delta $\ et $h$\ sont le champ cristallin et le champ magn\'{e}tique \cite{15}.
\vspace{-0.4cm}
\subsection{Mod\`{e}les de spin continu}

Les mod\`{e}les de spin continu constituent une autre g\'{e}n\'{e}ralisation du mod\`{e}le d'Ising. Dans ces mod\`{e}les, les spins sur le r\'{e}seau ont une gamme
continue de valeurs, plut\^{o}t qu'un spectre discret comme dans les mod%
\`{e}les cit\'{e}s auparavant \cite{2}. Les deux mod\`{e}les les plus fr\'{e}quents sont : le mod\`{e}le $XY$ et le mod\`{e}le d'Heisenberg.\newline
$\bullet $ \textbf{Mod\`{e}le $XY$}\newline
Dans le mod\`{e}le $XY$, les spins sont des vecteurs \`{a} deux composantes de
norme unit\'{e}, qui peuvent s'orienter dans n'importe quelle direction dans
un plan \`{a} deux dimensions $\left( x,y\right) $ \cite{13}. L'hamiltonien $H_{XY}$ s'\'{e}crit :
\begin{equation}
H_{XY}=-J\sum_{\left \langle ij\right \rangle }\vec{S}_{i}\vec{S}_{j}-\vec{h}%
\sum_{i}\vec{S}_{i},
\end{equation}%
o\`{u} $J$ est la constante d'\'{e}change, $\vec{h}$ est le champ magn\'{e}tique externe et $\vec{S}_{i}$ est un op\'{e}rateur de spin.\newline
Les spins peuvent \^{e}tre repr\'{e}sent\'{e}s soit par leurs composantes $%
S_{x}$ et $S_{y}$\ qui satisfont la contrainte $S^{2}=S_{x}^{2}+S_{y}^{2}=1$, soit par une variable angulaire qui indique la
direction de spin \cite{2}. Il convient donc de r\'{e}\'{e}crire l'hamiltonien $H_{XY}$ en
fonction des variables angulaires :
\begin{equation}
H_{XY}=-J\sum_{\left \langle ij\right \rangle }\cos \left( \theta _{i}-\theta
_{j}\right) -\sum_{i}\left[ h_{x}\cos \left( \theta _{i}\right) +h_{y}\sin
\left( \theta _{i}\right) \right] ,
\end{equation}%
o\`{u} les angles $\theta _{i}$ et $\theta _{j}$ sont des variables angulaires locales qui sp\'{e}cifient les orientations des spins. Le mod\`{e}le $XY$ s'\'{e}tale aussi \`{a} l'\'{e}tude des syst%
\`{e}mes tridimensionnels en d\'{e}pit du fait que les spins sont \`{a} deux dimensions.\newline
$\bullet $ \textbf{Mod\`{e}le d'Heisenberg}\newline
Dans le mod\`{e}le d'Heisenberg les
spins sont des vecteurs unitaires \`{a} trois dimensions. En effet, les spins
d'Heisenberg sont repr\'{e}sent\'{e}s soit par des vecteurs \`{a}
trois composantes $S_{x}$, $S_{y}$ et $S_{z}$\ tel que $S^{2}=S_{x}^{2}+S_{y}^{2}+S_{z}^{2}=1$, ou bien par deux angles variables $\theta $ et $\Phi $ en coordonn\'{e}es sph\'{e}riques \cite{13}. Dans ce cas, l'hamiltonien $H_{Heis}$ prend la forme :
\begin{equation}
H_{Heis}=-J\sum_{\left \langle i,j\right \rangle
}S_{xi}S_{xj}+S_{yi}S_{yj}+S_{zi}S_{zj}-\sum_{i}\left(
h_{x}S_{xi}+h_{y}S_{yi}+h_{z}S_{zi}\right) ,
\end{equation}
o\`{u} $J$ est la constante d'\'{e}change et $h_{x}$, $h_{y}$ et $h_{z}$ sont les composantes du champ magn\'{e}tique suivant l'axe $x$, $y$ et $z$ respectivement.\newline
Le mod\`{e}le d'Heisenberg est un mod\`{e}le de spin qui permet de traiter directement la d\'{e}pendance en spin d'un syst\`{e}me de plusieurs \'{e}lectrons.
\vspace{-0.8cm}
\section{Conditions aux bords}

La m\'{e}thode Monte Carlo \'{e}tudie les propri\'{e}t\'{e}s
d'un syst\`{e}me fini, alors que l'on s'int\'{e}resse g\'{e}n\'{e}ralement
aux propri\'{e}t\'{e}s d'un syst\`{e}me infini. Afin d'\^{e}tre en
mesure d'effectuer une extrapolation significative \`{a} la limite
thermodynamique, la question des conditions aux bords s'impose. Pour traiter les effets de bords il faut tenir compte de la formulation du probl\`{e}me autant bien que de la nature du syst\`{e}me. Ceci a donn\'{e} naissance \`{a} diff\'{e}rentes approches que nous \'{e}talons dans ce qui suit.
\vspace{-0.6cm}
\subsection{Conditions aux limites p\'{e}riodiques}

\begin{figure}[th]
\centering
\includegraphics[scale=0.8]{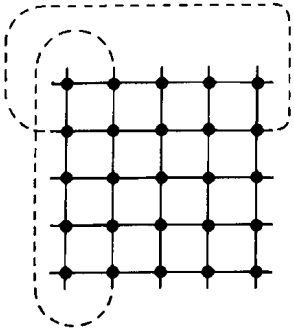}
\caption{Conditions aux bords p\'{e}riodiques pour le mod%
\`{e}le d'Ising bidimensionnel \cite{1}.}
\label{pbc}
\end{figure}

Une fa\c{c}on pour \'{e}liminer les limites au bords, connus aussi par limites du r\'{e}seau, revient \`{a} encapsuler un r\'{e}seau de dimension $d$
sur un tore de dimension $(d+1)$ \cite{4}. Cette condition aux limites p\'{e}riodiques fait que le premier spin dans une rang\'{e}e consid\`{e}re le dernier spin
dans la ligne comme un plus proche voisin et vice-versa \cite{1}. Il en est de m\^{e}me pour les spins en haut et en bas d'une colonne comme il est montr\'{e} dans la figure (\ref{pbc}) pour un r\'{e}seau carr\'{e}.
\newline
Les conditions aux limites p\'{e}riodiques est donc une proc\'{e}dure qui \'{e}limine effectivement les effets de bord pour un syst\`{e}me qui demeure caract\'{e}ris\'{e} par la taille de r\'{e}seau
fini $L$, puisque la valeur maximale de la longueur de corr\'{e}lation est
limit\'{e}e \`{a} $L/2$ et les propri\'{e}t\'{e}s qui en r\'{e}sultent du
syst\`{e}me vont diff\`{e}rer de celles du r\'{e}seau infini correspondant \cite{1}.
\vspace{-0.4cm}
\subsection{Conditions aux limites p\'{e}riodiques vis}

\begin{figure}[th]
\centering
\includegraphics[scale=0.65]{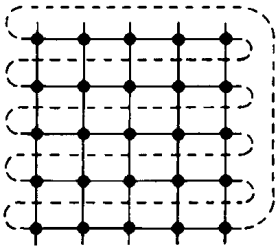}
\caption{Conditions aux limites p\'{e}riodiques vis pour le
mod\`{e}le d'Ising bidimensionnel \cite{1}.}
\label{spbc}
\end{figure}

Ce type de conditions consid\`{e}re une limite enroulable. Pour ce fait, les spins sur le r\'{e}seau sont repr\'{e}sent\'{e}s en tant qu'entr\'{e}es dans
un vecteur unidimensionnel enroul\'{e} autour du syst\`{e}me \cite{1}. Dans cette approche, le dernier spin dans une ligne et le premier spin dans la
ligne suivante sont proches voisins comme illustr\'{e} sur la figure (\ref{spbc}) pour le mod\`{e}le d'Ising bidimensionnel.
\newline
Outre la limitation de la longueur de corr\'{e}lation maximale possible, une
couture est introduite en raison de la forme de bord p\'{e}riodique. Cela
signifie que les propri\'{e}t\'{e}s du syst\`{e}me ne seront pas compl\`{e}%
tement homog\`{e}nes. Dans la limite de r\'{e}seau de taille infinie, cet
effet devient n\'{e}gligeable. Cependant, pour des syst\`{e}mes finis, il y
a une diff\'{e}rence syst\'{e}matique en ce qui concerne les conditions aux
limites compl\`{e}tement p\'{e}riodiques qui peuvent ne pas \^{e}tre n\'{e}%
gligeables \cite{1}.
\vspace{-0.4cm}
\subsection{Conditions aux limites antip\'{e}riodiques}

Si les conditions aux limites p\'{e}riodiques sont appliqu\'{e}es avec un changement de signe du couplage aux bords, alors une interface sera introduite dans le syst\`{e}me. Cette op\'{e}ration m\`{e}ne aux conditions aux limites antip\'{e}riodiques \cite{16}. Les conditions aux limites antip\'{e}riodiques s'effectuent selon la direction normale \`{a} l'interface que l'on souhaite \'{e}tudier, tandis que les conditions aux limites p\'{e}riodiques seront appliqu\'{e}es dans l'autre direction.
\vspace{-0.4cm}
\subsection{Conditions aux limites de bord libre}

\begin{figure}[th]
\centering
\includegraphics[scale=0.65]{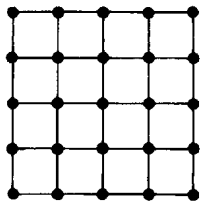}
\caption{Conditions aux limites de bord libre pour le mod%
\`{e}le d'Ising bidimensionnel \cite{1}.}
\label{fpb}
\end{figure}

C'est un autre type de limite qui ne comporte aucun type de connexion entre
la fin d'une ligne et de n'importe quelle autre ligne dans un r\'{e}seau. Par cons\'{e}quent, les spins \`{a} la fin d'une ligne ne poss\`{e}dent aucun proche voisin comme pr\'{e}sent\'{e} sur la figure (\ref{fpb}) pour le mod\`{e}le d'Ising bidimensionnel.
\newline
De plus, la limite de bord libre n'introduit pas seulement la bavure de taille
finie, mais elle tient compte \'{e}galement des effets de surfaces et de coins d\^{u}s aux liaisons
pendantes sur les bords puisque d'importants changements peuvent se
produire pr\`{e}s des surfaces o\`{u} le comportement du syst\`{e}me n'est pas
homog\`{e}ne \cite{1}. Notons aussi que les limites de bord libres constitue l'approche la plus r\'{e}aliste pour certaines situations tel que la mod\'{e}lisation du comportement des particules ou des nanoparticules superparamagn\'{e}tiques. En g\'{e}n\'{e}ral, les propri\'{e}%
t\'{e}s des syst\`{e}mes avec des limites de bord libre diff\`{e}rent des propri\'{e}t\'{e}s du syst\`{e}me infini avec des conditions aux limites p\'{e}riodiques \cite{16}.
\vspace{-0.6cm}
\section{Monte Carlo statique}

Dans cette section, nous pr\'{e}senterons les \'{e}l\'{e}ments de base de la
simulation Monte Carlo, qui est crucial dans la compr\'{e}hension des simulations Monte Carlo thermiques r\'{e}alis%
\'{e}es au cours des trente derni\`{e}res ann\'{e}es.
\vspace{-0.4cm}
\subsection{\'{E}quation Ma\^{\i}tresse et estimateur}

L'id\'{e}e de base de la m\'{e}thode Monte Carlo est de simuler la
fluctuation thermique al\'{e}atoire du syst\`{e}me d'un \'{e}tat \`{a} un
autre. La probabilit\'{e} de trouver le syst\`{e}me dans l'\'{e}tat $a$ au
cours de la simulation est \'{e}gale au poids de cet \'{e}tat dans le syst%
\`{e}me r\'{e}el. Ceci n\'{e}cessite le choix d'une r\`{e}gle qui r\'{e}git le passage d'un \'{e}tat \`{a} un
autre au cours de la simulation \cite{2}. C'est ce que nous allons discuter dans ce qui suit. Mais tout d'abord, nous allons commencer par introduire les processus physiques qui donnent l'\'{e}quation ma\^{\i}tresse :
\begin{equation}
\frac{dp_{a}}{dt}=\sum_{b}\left( p_{b}\left( t\right) w_{ba}-p_{a}\left(
t\right) w_{ab}\right) ,\label{1-2}
\end{equation}%
o\`{u} $p_{a}\left( t\right) $ est la probabilit\'{e}
d'observer le syst\`{e}me dans la configuration $a$ \`{a} l'instant $t$
connaissant la distribution des probabilit\'{e}s \`{a} l'instant initial alors que $%
w_{ab}$ pr\'{e}sente la probabilit\'{e} de passer de la configuration $a$ \`{a} $b$
par unit\'{e} de temps. Les $w_{ab}$ sont choisis de fa\c{c}on que la
solution d'\'{e}quilibre de l'\'{e}quation ma\^{\i}tresse rel\`{e}ve de la
distribution de Boltzmann $p_{a}$ est donn\'{e}e par :
\begin{equation}
p_{a}=\frac{e^{\frac{-E_{a}}{kT}}}{Z}.\label{1-8}
\end{equation}%
o\`{u} $E_{a}$ est l'\'{e}nergie de l'\'{e}tat $a$, $k$ est la constante de Boltzmann, $T$ est la temp\'{e}rature et $Z$ est la fonction de partition du syst\`{e}me.\newline
L'\'{e}quation ma\^{\i}tresse (\ref{1-2}) est une \'{e}quation qui donne la
variation de la probabilit\'{e} de chaque \'{e}tat ; c'est-\`{a}-dire l'%
\'{e}volution de la probabilit\'{e}. L'avantage de cette \'{e}quation est
d'avoir une bonne estimation des grandeurs physiques \`{a} partir de petits
\'{e}chantillons. N\'{e}anmoins, cette \'{e}quation induit des erreurs
statistiques, parmi lesquelles on a un bruit dans $Z$ qui peut induire des d\'{e}riv\'{e}es mal
calcul\'{e}es. Il est donc pr\'{e}f\'{e}rable de calculer plusieurs
moyennes. Par exemple au lieu de consid\'{e}rer la chaleur sp\'{e}cifique $C$ comme \'{e}tant :
\begin{equation}
C=\frac{dE}{dT},
\end{equation}%
il est pr\'{e}f\'{e}rable de travailler avec la forme suivante :
\begin{equation}
C=\frac{k\beta ^{2}}{N}\left( \left \langle E^{2}\right \rangle -\left \langle
E\right \rangle ^{2}\right) ,
\end{equation}%
o\`{u} $\beta =\frac{1}{kT}$ est le beta thermodynamique et $N$ est le nombre total de particules.\newline
Il s'ensuit que la recherche et le calcul des valeurs moyennes pr\'{e}sentent les principaux objectifs
des simulations Monte Carlo \cite{5,8}. Mais pour acc\'{e}l\'{e}rer les calculs, il est pr\'{e}f\'{e}rable d'utiliser les probabilit\'{e}s.\newline
La valeur moyenne d'une quantit\'{e} $Q$, par sommation sur tous les \'{e}%
tats $a$ du syst\`{e}me et sur leurs probabilit\'{e}s respectives, est donn%
\'{e}e par :
\begin{equation}
\left \langle Q\right \rangle =\frac{\sum_{a}Q_{\alpha }e^{-\beta E_{a}}}{%
\sum_{a}e^{-\beta E_{a}}}.
\end{equation}%
Cette moyenne ne peut \^{e}tre calcul\'{e}e que pour de syst\`{e}mes tr\`{e}s petits. Pour des syst\`{e}mes de grande taille, la somme sur un sous-ensemble d'\'{e}tats induit des impr\'{e}cisions. La m\'{e}thode Monte Carlo consiste \`{a}
choisir au hasard un sous-ensemble d'\'{e}tats \`{a} partir d'une
distribution $p_{a}$ \`{a} sp\'{e}cifier \cite{17}.\newline
Supposons que le choix porte sur le sous-ensemble $M=\left \{
a_{1},....,a_{M}\right \} $. L'estimation $Q_{M}$ de la quantit\'{e} $Q$ s'\'{e}crit comme suit :
\begin{equation}
Q_{M}=\frac{\sum_{i=1}^{M}Q_{ai}p_{ai}^{-1}e^{-\beta E_{ai}}}{%
\sum_{j=1}^{M}p_{aj}^{-1}e^{-\beta E_{aj}}}.\label{1-11}
\end{equation}%
L'\'{e}quation (\ref{1-11}) donne une estimation de $Q$ sur un
mod\`{e}le r\'{e}duit o\`{u} plus le nombre $M$ d'\'{e}tats dans l'\'{e}%
chantillon augmente, plus on se rapproche de la vraie valeur de $\left \langle
Q\right \rangle $. Ceci peut \^{e}tre exprim\'{e} comme suit :
\begin{equation}
\left \langle Q\right \rangle =\lim_{M\rightarrow \infty }Q_{M}=Q_{M}.
\end{equation}%
Il reste alors \`{a} d\'{e}terminer $M$ pour une meilleure expression de $Q$. Pour
ce faire, il suffit de consid\'{e}rer une \'{e}quiprobabilit\'{e} entre les
\'{e}tats du syst\`{e}me. C'est-\`{a}-dire tous les $p_{a}$ sont \'{e}gaux \cite{2}, d'o\`{u}%
\begin{equation}
\begin{array}{ccc}
Q_{M} & = & \frac{\sum_{i=1}^{M}p_{a}^{-1}Q_{ai}e^{-\beta E_{ai}}}{%
\sum_{j=1}^{M}p_{a}^{-1}e^{-\beta E_{aj}}} \\
& = & \frac{\sum_{i=1}^{M}Q_{ai}e^{-\beta E_{ai}}}{\sum_{j=1}^{M}e^{-\beta
E_{aj}}}.\text{ \ }%
\end{array}\label{1-15}
\end{equation}
En fait, c'est le plus mauvais choix.
\vspace{-0.5cm}
\subsection{Principes de la simulation Monte Carlo}

La simulation Monte Carlo repose sur trois param\`{e}tres importants :

$\bullet $ \emph{l'\'{e}chantillon important,}

$\bullet $ \emph{la balance d\'{e}taill\'{e}e,}

$\bullet $ \emph{le taux d'acceptation.}\vspace{0.3cm}\newline
$\bullet $ \textbf{\textit{\'{E}chantillon important}}\vspace{0.3cm}\newline
La m\'{e}thode Monte Carlo consiste \`{a} choisir un \'{e}chantillon qui
contient les \'{e}tats dominants. Cette op\'{e}ration s'appelle
l'\'{e}chantillon important \cite{18}. Les \'{e}%
tats de l'\'{e}chantillon ne sont pas \'{e}quiprobables, mais distribu\'{e}s
selon la distribution de probabilit\'{e} de Boltzmann donn\'{e}e dans l'\'{e}quation (\ref{1-8}) qui permet d'am\'{e}liorer l'estimation. Cette distribution est la forme la plus connue de l'\'{e}chantillon important \cite{2}. Dans ce cas, l'estimation $Q_{M}$ devient :
\begin{equation}
Q_{M}=\frac{1}{M}\sum_{i=1}^{M}Q_{ai}.
\end{equation}%
Ce r\'{e}sultat est mieux que l'estimation obtenue en \'{e}quation (\ref{1-15}) lorsque tous les $p_{a}$ sont \'{e}gaux. Avant de pr\'{e}senter la balance d\'{e}taill\'{e}e et le taux d'acceptation, il sera plus fondamental d'exposer tout d'abord le processus de Markov et l'ergodicit\'{e}.\vspace{0.3cm}\newline
$\bullet $ \textit{Processus de Markov}\newline
La partie la plus complexe dans la simulation Monte Carlo est la g\'{e}n\'{e}%
ration d'un ensemble al\'{e}atoire appropri\'{e}e des \'{e}tats en fonction
de la distribution de probabilit\'{e} de Boltzmann \cite{16}. En d'autres termes, on ne peut pas
choisir au hasard certains \'{e}tats puis les accepter ou les rejeter avec une
probabilit\'{e} proportionnelle \`{a} $%
e^{-\beta E_{a}}$ puisque le r\'{e}sultat n'en sera pas meilleur que celui issu d'un \'{e}chantillonnage hasardeux. Dans ce cas, nous risquons de r\'{e}p\'{e}ter virtuellement certains \'{e}tats autant que leurs probabilit\'{e}s sont exponentiellement petites \cite{19}. Pour \'{e}viter cette contrainte,
presque tous les algorithmes des m\'{e}thodes Monte Carlo utilisent le
processus de Markov pour choisir les \'{e}tats utilis\'{e}s \cite{13}.\newline
Le processus de Markov \cite{12} n'est autre que le m\'{e}canisme qui g\'{e}n\`{e}re un \'{e}%
tat $b$ du syst\`{e}me \`{a} partir d'un autre $a$ connu \cite{8,10}. L'\'{e}tat g\'{e}n%
\'{e}r\'{e} n'est pas toujours le m\^{e}me. Ainsi, il parcourt le syst\`{e}%
me \`{a} la recherche d'un nouvel \'{e}tat avec une probabilit\'{e} de
transition $w\left( a\rightarrow b\right) $ \`{a} laquelle il impose deux conditions \cite{2} :

$\bullet $ ne pas varier avec le temps,

$\bullet $ d\'{e}pendre uniquement des propri\'{e}t\'{e}s du syst\`{e}me sur les
\'{e}tats $a$ et $b$.\newline
Ceci traduit le fait que la probabilit\'{e} de transition $w\left(
a\rightarrow b\right) $ d'un \'{e}tat $a$ \`{a}
un autre $b$ du processus de Markov est toujours constante et devra satisfaire
\`{a} la relation de fermeture \cite{2} :%
\begin{equation}
\sum_{b}w\left( a\rightarrow b\right) =1.  \label{ty}
\end{equation}
$w\left( a\rightarrow a\right) $ n'est pas obligatoirement nul.\newline
Dans la simulation Monte Carlo, le processus de Markov est utilis\'{e} \`{a} plusieurs reprises pour g\'{e}n\'{e}rer une cha\^{\i}ne de Markov des
\'{e}tats \cite{5,9}. Cette cha\^{\i}ne est g\'{e}n\'{e}ralement utilis\'{e}s lorsqu'on
veut partir de n'importe quel \'{e}tat du syst\`{e}me et g\'{e}n\'{e}rer,
par exemple, une suite de configurations de certains \'{e}tats pr\'{e}cis
(finaux). Pour parachever cet objectif, il est utile d'imposer deux nouvelles conditions au processus de
Markov, notamment l'ergodicit\'{e} et la balance d\'{e}taill\'{e}e \cite{2}.
\vspace{0.3cm}\newline
$\bullet $ \textit{Ergodicit\'{e}}\vspace{0.3cm}\newline
La condition d'ergodicit\'{e} est le fait que le syst\`{e}me peut \`{a}
partir d'un \'{e}tat donn\'{e} passer lors du processus de Markov par
n'importe quel \'{e}tat initial pour un nouvel \'{e}tat.\newline
Ceci est n\'{e}cessaire pour g\'{e}n\'{e}rer des
\'{e}tats avec leurs probabilit\'{e}s de Boltzmann. Chaque \'{e}tat $b$ appara%
\^{\i}t avec une certaine probabilit\'{e} non nulle $p_{b}$ dans la distribution de
Boltzmann. L'acc\`{e}s de cet \'{e}tat \`{a} partir d'un autre \'{e}tat
distinct de $a$ ne provoquera aucun probl\`{e}me. Le processus de l'\'{e}tat
initial est repris pour le nouvel \'{e}tat \cite{8}.\newline
La condition d'ergodicit\'{e} nous montre que nous pouvons prendre certaines
probabilit\'{e}s de transition nulles dans le processus de Markov. Ce ne
sera pas le cas pour deux \'{e}tats distincts que nous prenons dans un
espace restreint \cite{18}. En pratique, la plupart des algorithmes de Monte Carlo
configure toutes les probabilit\'{e}s de transition \`{a} z\'{e}ro. Il nous
faut, dans ce cas, faire attention \`{a} ne pas cr\'{e}er un algorithme qui ne satisfait pas la condition d'ergodicit\'{e}.\vspace{0.3cm}\newline
$\bullet $ \textbf{\textit{Balance d\'{e}taill\'{e}e}}\vspace{0.3cm}\newline
La condition de la balance d\'{e}taill%
\'{e}e assure que la distribution de probabilit\'{e} de
Boltzmann g\'{e}n\'{e}r\'{e}e apr\`{e}s que le syst\`{e}me consid\'{e}r\'{e} atteint l'\'{e}quilibre, est la plus grande de toutes les autres
distributions \cite{1}.%
\newline
Si le syst\`{e}me est en \'{e}quilibre, les taux de transition \`{a} partir d'un \'{e}tat et vers le m\^{e}me \'{e}tat sont \'{e}gaux \cite{2}. On \'{e}crit donc :
\begin{equation}
\sum_{b}p_{a}w\left( a\rightarrow b\right) =\sum_{b}p_{b}w\left(
b\rightarrow a\right) .  \label{nb}
\end{equation}%
\`{A} partir de
\begin{equation}
\sum_{b}w\left( a\rightarrow b\right) =1,
\end{equation}%
on a
\begin{equation}
p_{a}=\sum_{b}p_{b}w\left( b\rightarrow a\right) .  \label{saaa}
\end{equation}%
Pour tout ensemble de probabilit\'{e}s de transitions qui satisfait cette
\'{e}quation, la distribution $p_{a}$ sera un \'{e}tat d'\'{e}quilibre \`{a} partir
de la dynamique du processus de Markov. Malheureusement, r\'{e}pondre \`{a}
cette \'{e}quation ne nous garantit pas d'atteindre $p_{a}$ \`{a} partir de n'importe
quel \'{e}tat du syst\`{e}me \cite{16}.\newline
En effet, la probabilit\'{e} de transition $w\left( a\rightarrow b\right) $ peut \^{e}tre d\'{e}termin\'{e}e
comme un \'{e}l\'{e}ment de la matrice $W$. Cette matrice est appel\'{e}e matrice de Markov ou la matrice stochastique pour le processus de Markov.\newline Consid\'{e}rant $q_{a}\left( t\right) $, si nous mesurons le temps mis dans chaque \'{e}tat le long de la cha\^{\i}%
ne de Markov, la probabilit\'{e} d'\^{e}tre dans l'\'{e}tat $b$ \`{a} un
instant $t+1$ est donn\'{e}e par \cite{2} :%
\begin{equation}
q_{b}\left( t+1\right) =\sum_{a}w\left( a\rightarrow b\right) q_{a}\left(
t\right) .
\end{equation}%
Sous forme matricielle, on obtient :
\begin{equation}
Q\left( t+1\right) =W.Q\left( t\right) ,
\end{equation}%
$Q\left( t\right) $ est le vecteur dont les coordonn\'{e}es sont les diff\'{e}rents poids
statistiques $q_{a}\left( t\right) $.\newline
\`{A} l'\'{e}quilibre, le processus de Markov satisfera \`{a} :
\begin{equation}
Q\left( \infty \right) =W.Q\left( \infty \right) .
\end{equation}%
Cependant, il est \'{e}galement possible au processus d'atteindre l'\'{e}%
quilibre dynamique par rotation de $Q$ sur toute la cha\^{\i}ne. Une telle
rotation est appel\'{e}e cycle limite \cite{2}.\newline
Dans ce cas $Q\left( \infty \right)
$, est :
\begin{equation}
Q\left( \infty \right) =W^{n}.Q\left( \infty \right) ,  \label{reee}
\end{equation}%
o\`{u} $n$ est la longueur du cycle limite.\newline
Si nous choisissons une probabilit\'{e} de transition (ou de mani\`{e}re
\'{e}quivalente une matrice de Markov) pour satisfaire \`{a} la relation (\ref{saaa}), nous garantirons que la cha\^{\i}ne de Markov aura une simple probabilit%
\'{e} d'\'{e}quilibre de distribution $p_{a}$. Mais, elle peut aussi avoir un
nombre quelconque de cycles limites de la forme (\ref{reee}). Cela signifie que rien
ne garantit que l'\'{e}tat d'\'{e}quilibre g\'{e}n\'{e}r\'{e} aura la
probabilit\'{e} de distribution d\'{e}sir\'{e}e \cite{16}.\newline
Pour contourner ce probl\`{e}me, nous appliquerons une condition suppl\'{e}%
mentaire \`{a} nos probabilit\'{e}s de transition \cite{2} :
\begin{equation}
p_{a}w\left( a\rightarrow b\right) =p_{b}w\left( b\rightarrow a\right) .
\label{az}
\end{equation}%
C'est la condition de la balance d\'{e}taill\'{e}e. Puisque l'\'{e}quation
(\ref{nb}) est une sommation de (\ref{az}), sur les diff\'{e}rents \'{e}tats, alors
chaque \'{e}tat satifaisant (\ref{az}) va satisfaire (\ref{nb}). Nous pouvons \'{e}%
galement montrer que cette condition \'{e}limine les cycles limites \cite{18}. En
effet, la balance d\'{e}taill\'{e}e nous enseigne qu'en moyenne, le syst\`{e}%
me peut quitter un \'{e}tat $a$ vers un autre $b$ indiff\'{e}remment du chemin
choisi et apr\`{e}s un temps infini. Une fois qu'on enl\`{e}ve les cycles
limites de cette fa\c{c}on, il est facile de v\'{e}rifier que le syst\`{e}me
aura toujours une probabilit\'{e} de distribution $p_{a}$ lorsque $t\rightarrow \infty $. \`{A} $t\rightarrow \infty $, $Q\left( t\right) $, tendra
exponentiellement vers le vecteur propre correspondant \`{a} la plus grande
valeur propre de $W$ (propri\'{e}t\'{e} des matrices stochastiques) \cite{2}.\newline
On remarque que les grandes valeurs propres des matrices de Markov $W$ pourront
\^{e}tre \'{e}quivalentes \`{a} partir de l'\'{e}quation (\ref{reee}%
). Si les cycles
limites de la forme (\ref{az}) \'{e}taient pr\'{e}sents, nous pourrions aussi
avoir des valeurs propres qui seraient des racines complexes. Mais la
condition de balance d\'{e}taill\'{e}e nous pr\'{e}vient de cette possibilit%
\'{e}.\vspace{0.3cm}\newline
$\bullet $ \textbf{\textit{Taux d'acceptation}}\vspace{0.3cm}\newline
Apr\`{e}s l'avoir d\'{e}crit pr\'{e}c\'{e}demment comme \'{e}l\'{e}ment
important pour l'obtention rapide et efficace d'un syst\`{e}me \`{a} l'\'{e}%
tat d'\'{e}quilibre, nous avons aussi d\'{e}montr\'{e} que nous pouvons g%
\'{e}n\'{e}rer un processus de Markov avec lequel nous pouvons retrouver de
nouveaux \'{e}tats avec une probabilit\'{e} qui puisse r\'{e}pondre \`{a} l'%
\'{e}quation (\ref{az}). Cependant, il est difficile de pr\'{e}voir le processus
de Markov appropri\'{e} pour g\'{e}n\'{e}rer un nouvel \'{e}tat \`{a} partir
d'un autre \'{e}tat pr\'{e}c\'{e}dent avec un bon ensemble de probabilit\'{e}%
s de transition \cite{2}. D'apr\`{e}s l'\'{e}quation de la balance d\'{e}taill%
\'{e}e, les probabilit\'{e}s de transition doivent satisfaire :
\begin{equation}
\frac{w\left( a\rightarrow b\right) }{w\left( b\rightarrow a\right) }=\frac{%
p_{b}}{p_{a}}=e^{-\beta \left( E_{b}-E_{a}\right) }.  \label{hu}
\end{equation}%
Cela implique que les inconnues $w\left( a\rightarrow b\right) $ \`{a} d\'{e}terminer ne d\'{e}pendent pas de
la fonction $Z$, mais uniquement du facteur de Boltzmann reli\'{e} \`{a} l'\'{e}%
nergie de chaque \'{e}tat qui peut \^{e}tre calcul\'{e}e \cite{17}.\newline
Les m\'{e}thodes standards ne s'appliquent pas toujours aux nouveaux probl%
\`{e}mes. On construit de nouveaux algorithmes pour des besoins sp\'{e}%
cifiques. M\^{e}me si on peut proposer plusieurs processus de Markov, on
peut ne pas trouver celui qui donne le bon ensemble de probabilit\'{e}s de
transition. De ce fait, il est n\'{e}cessaire d'avoir une probabilit\'{e} de
transition souhait\'{e}e par l'introduction d'une condition d'acceptation du
taux de transition qui va permettre de trouver les bonnes probabilit\'{e}s
de transition \`{a} partir d'un processus de Markov quelconque \cite{18}. L'id\'{e}%
e sous-jacente de cette astuce est la suivante : nous avons mentionn\'{e} pr%
\'{e}c\'{e}demment que nous pouvons introduire une probabilit\'{e} de
transition de base $w\left( a\rightarrow b\right) \neq 0$. Si nous fixons $a=b$ dans l'\'{e}quation (\ref{hu}), nous obtenons
la tautologie simple $(1=1)$. Ce qui signifie que la condition de la balance d\'{e}%
taill\'{e}e est toujours satisfaite pour $w\left( a\rightarrow
b\right) $, peu importe la valeur de cette probabilit\'{e}. De fait, nous disposons d'une certaine flexibilit\'{e} sur la
fa\c{c}on dont nous choisissons les autres probabilit\'{e}s de transition
avec $a=b$. Nous pouvons donc ajuster la valeur de n'importe quelle $w\left( a\rightarrow b\right) $, telle que
la r\`{e}gle de fermeture (\ref{ty}) soit v\'{e}rifi\'{e}e par une simple
compensation de cet ajustement avec un autre ajustement \'{e}quivalent, mais
oppos\'{e}e \`{a} $w\left( a\rightarrow a\right) $. Le seul point que nous devons examiner est que $w\left( a\rightarrow a\right) $ ne d\'{e}%
passe pas ses limites (soit $0<w\left( a\rightarrow a\right) <1$). Si nous faisons un ajustement de ce genre
dans $w\left( a\rightarrow b\right) $, nous pouvons \'{e}galement nous organiser afin que l'\'{e}quation
(\ref{hu}) soit satisfaite, en faisant simultan\'{e}ment un changement en $w\left( a\rightarrow b\right)
$, afin
de pr\'{e}server le rapport entre les deux \cite{2}.\newline Il s'av\`{e}re que ces consid\'{e}rations nous donnent effectivement assez
de libert\'{e} sur la possibilit\'{e} de donner aux probabilit\'{e}s de
transition n'importe quel ensemble de valeurs que nous souhaitons en
ajustant les valeurs des probabilit\'{e}s $w\left( a\rightarrow a\right) $ \cite{17}. Pour voir cela, d\'{e}%
composons la probabilit\'{e} de transition en deux parties :
\begin{equation}
w\left( a\rightarrow b\right) =\alpha \left( a\rightarrow b\right) \rho
\left( a\rightarrow b\right) ,
\end{equation}%
o\`{u} $\alpha \left( a\rightarrow b\right) $ est la probabilit\'{e} de s\'{e}lection ; probabilit\'{e} pour que
notre algorithme g\'{e}n\`{e}re l'\'{e}tat final $b$ \`{a} partir de l'\'{e}tat
initial $a$ et $\rho \left(
a\rightarrow b\right) $ \'{e}tant le taux d'acceptation (appel\'{e} aussi probabilit\'{e}
d'acceptation). Le rapport d'acceptation indique que si on commence \`{a} $a$ et
que notre algorithme g\'{e}n\`{e}re $b$, on devrait accepter $b$ et on change l'%
\'{e}tat du syst\`{e}me \`{a} $b$, pendant une fraction de temps $\rho \left( a\rightarrow b\right) $. Le reste du
temps, il reste dans $a$. L'acceptation est arbitraire entre z\'{e}ro et un ($0<\rho \left( a\rightarrow b\right) <1$). Opter $\rho \left(
a\rightarrow b\right) =0$ pour toutes les transitions, est \'{e}quivalent \`{a} choisir $w\left( a\rightarrow a\right) =1$, qui est la plus grande valeur qu'elle puisse prendre et signifie que nous
ne pourrons jamais quitter l'\'{e}tat $b$ \cite{2}. Ceci nous donne une libert\'{e}
totale du choix de la probabilit\'{e} de s\'{e}lection $\alpha \left( a\rightarrow b\right) $, puisque la
contrainte (\ref{hu}) ne fixe que le rapport :
\begin{equation}
\frac{w\left( a\rightarrow b\right) }{w\left( b\rightarrow a\right) }=\frac{%
\alpha \left( a\rightarrow b\right) \rho \left( a\rightarrow b\right) }{%
\alpha \left( b\rightarrow a\right) \rho \left( b\rightarrow a\right) },
\label{aa}
\end{equation}%
o\`{u} $\frac{\rho \left( a\rightarrow b\right)}{\rho \left( b\rightarrow a\right)}\in \left[ 0,\infty \right[ $, donc $\alpha \left( a\rightarrow
b\right) $ et $\alpha \left( b\rightarrow a\right) $ peuvent prendre n'importe quelle valeur souhait\'{e}e.\newline
La relation de fermeture (\ref{ty}) est toujours satisfaite, car le syst\`{e}me
doit se retrouver dans un \'{e}tat apr\`{e}s chaque \'{e}tape de la cha\^{\i}%
ne de Markov, m\^{e}me si cet \'{e}tat est celui avec lequel nous avons
commenc\'{e} \cite{17}.\newline
Donc, pour cr\'{e}er un algorithme de Monte Carlo, nous devons \'{e}laborer
un algorithme qui g\'{e}n\`{e}re de nouveaux \'{e}tats al\'{e}atoires $a$ \`{a}
partir des anciens \'{e}tats $b$, avec un ensemble des probabilit\'{e}s $%
\alpha \left( a\rightarrow b\right) $. Puis,
nous acceptons ou rejetons ces \'{e}tats avec le rapport d'acceptation $\rho \left( a\rightarrow b\right) $ choisi pour satisfaire l'\'{e}quation (\ref{aa}). Cela satisfait, alors, toutes
les exigences pour les probabilit\'{e}s de transition et produira ainsi une
cha\^{\i}ne des \'{e}tats qui appara\^{\i}tront avec leurs probabilit\'{e}s
de Boltzmann lorsque le syst\`{e}me atteint l'\'{e}quilibre \cite{18}.\newline
Il est, toutefois un point que nous devons toujours garder \`{a} l'esprit
qui reste l'un des aspects les plus importants dans la conception des
algorithmes de Monte Carlo. Si le rapport d'acceptation est faible,
l'algorithme para\^{\i}tra immobile, ce qui bloque naturellement l'\'{e}%
volution du syst\`{e}me. Il nous faut donc trouver un algorithme qui puisse
\'{e}voluer entre les \'{e}tats pour un large \'{e}chantillonnage. En effet,
pour \'{e}viter que l'algorithme soit lent, on choisit un rapport
d'acceptation proche de 1. Une fa\c{c}on de le faire est de noter que l'\'{e}%
quation (\ref{aa}) fixe seulement le rapport $\frac{\rho
\left( a\rightarrow b\right) }{\rho \left( b\rightarrow a\right) }$ de taux d'acceptation entre deux \'{e}%
tats distincts dans n'importe quelle direction. Avec comme contrainte que ce
taux soit compris entre z\'{e}ro et un, bien que math\'{e}matiquement, on
puisse le multiplier proportionnellement par un coefficient r\'{e}el \cite{17}.\newline
Cependant, la meilleure option \`{a} prendre pour garder des rapports
d'acceptation \'{e}lev\'{e}s est d'essayer d'incarner dans la probabilit\'{e}
de s\'{e}lection $\alpha \left( a\rightarrow b\right) $ autant que nous le pouvons de d\'{e}pendance de $w\left( a\rightarrow b\right) $ des caract%
\'{e}ristiques des \'{e}tats $a$ et $b$ et d'en mettre le moins possible dans le
taux d'acceptation. Un bon algorithme est celui dans lequel la probabilit%
\'{e} d'acceptation est g\'{e}n\'{e}ralement proche de 1 \cite%
{2}.\vspace{-0.5cm}
\subsection{Temps d'\'{e}quilibre}

La m\'{e}thode Monte Carlo ne prend pas en compte toutes les configurations
possibles. Mais, les configurations n\'{e}glig\'{e}es en majorit\'{e} sont
les moins probables et donc influencent peu les r\'{e}sultats.
Effectivement, les configurations peu probables sont naturellement \'{e}limin%
\'{e}es par la fonction de probabilit\'{e} de transition. Par exemple, il
est quasi impossible que la configuration d'un spin orient\'{e} vers le bas
dans un domaine de spins orient\'{e}s vers le haut se r\'{e}alise pour une
temp\'{e}rature assez faible \cite{1,2}.\newline
On fait tourner le programme suffisamment longtemps pour que le r\'{e}sultat
ne d\'{e}pende pas de la configuration initiale afin d'atteindre l'\'{e}%
quilibre \cite{2,5,18}. Ce temps s'appelle temps d'\'{e}quilibre \cite{2,5,18}. Apr\`{e}s l'\'{e}quilibre, on calcule sur une
nouvelle p\'{e}riode l'estimation de la grandeur physique qui nous int\'{e}%
resse par la relation :
\begin{equation}
Q_{M}=\frac{1}{M}\sum_{i=1}^{M}Q_{ai}.
\end{equation}%
Pour visualiser l'\'{e}quilibre, on trace une quantit\'{e} physique, comme
l'aimantation ou l'\'{e}nergie. Apr\`{e}s l'\'{e}quilibre, la quantit\'{e} physique se
stabilise et seules les fluctuations restent. Dans certains cas, le syst\`{e}%
me reste pi\'{e}g\'{e} dans un minimum local d'\'{e}nergie. Pour \'{e}viter
cette situation, on part de diff\'{e}rentes configurations initiales \cite{1,2}.
\vspace{-0.5cm}
\subsection{Mesures}

D\`{e}s que nous sommes s\^{u}rs que le syst\`{e}me a atteint l'\'{e}%
quilibre, nous pouvons mesurer la moyenne des grandeurs physiques qui nous
int\'{e}resse \cite{2}. Pour calculer l'\'{e}nergie, on utilise $\Delta
E=E_{b}-E_{a}$ qui a \'{e}t\'{e}
calcul\'{e}e au cours de la simulation :
\begin{equation}
E_{b}=E_{a}+\Delta E,
\end{equation}%
et pour calculer l'aimantation, on utilise :
\begin{equation}
\Delta
M=M_{b}-M_{a}=\sum_{i}S_{i}^{b}-%
\sum_{i}S_{i}^{a}=S_{k}^{b}-S_{k}^{a}=2S_{k}^{b},
\end{equation}%
avec :
\begin{equation}
M_{a}=\sum_{i}S_{i}^{a},
\end{equation}%
d'o\`{u} :%
\begin{equation}
M_{b}=M_{a}+\Delta M=M_{a}+2S_{k}^{b}.
\end{equation}%
Nous pouvons \'{e}galement calculer la moyenne des carr\'{e}s de l'\'{e}%
nergie et de l'aimantation pour d\'{e}finir les quantit\'{e}s de chaleur sp%
\'{e}cifique et de susceptibilit\'{e} magn\'{e}tique. \`{A} ce titre, la chaleur
sp\'{e}cifique est donn\'{e}e par :
\begin{equation}
C=\frac{\beta ^{2}}{N}\left( \left \langle E^{2}\right \rangle -\left
\langle E\right \rangle ^{2}\right) ,
\end{equation}%
et la susceptibilit\'{e} magn\'{e}tique est donn\'{e}e par :
\begin{equation}
\chi =\frac{\beta }{N}\left( \left \langle M^{2}\right \rangle -\left
\langle M\right \rangle ^{2}\right) .
\end{equation}
\vspace{-1.5cm}
\section{Algorithmes de simulation}
\vspace{-0.2cm}
\subsection{Algorithme de M\'{e}tropolis}

L'algorithme de M\'{e}tropolis est l'une des plus
efficaces et simples solutions en ce qui concerne les probl\`{e}mes de
simulation en transition de phase \cite{2}. Nous utiliserons cet algorithme pour
illustrer les nombreux concepts g\'{e}n\'{e}raux impliqu\'{e}s dans un
calcul de Monte Carlo, y compris l'\'{e}quilibre, la mesure de valeurs
moyennes et le calcul des erreurs.\newline
L'algorithme Metropolis suit les \'{e}tapes de la m\'{e}thode Monte Carlo d%
\'{e}finies dans le taux d'acceptation \cite{12}. Nous choisissons un ensemble de
probabilit\'{e}s de s\'{e}lection $\alpha \left( a\rightarrow b\right) $ pour chaque transition $(%
a\rightarrow b)$. Puis on opte pour
un ensemble de probabilit\'{e}s d'acceptation $\rho \left( a\rightarrow b\right) $ tel que l'\'{e}quation (%
\ref{aa}) r\'{e}ponde \`{a} la condition de la balance d\'{e}taill\'{e}e. Cet
algorithme consiste \`{a} g\'{e}n\'{e}rer, \`{a} partir d'une configuration
de d\'{e}part, de nouvelles configurations par la modification al\'{e}atoire
des coordonn\'{e}es d'une ou de plusieurs particules \cite{1,3,12}. La probabilit%
\'{e} de s\'{e}lection $\alpha \left( a\rightarrow b\right) $ doit \^{e}tre d\'{e}finie pour satisfaire \`{a} la
condition d'ergodicit\'{e}. On consid\`{e}re la dynamique de retourner un
seul spin \`{a} chaque \'{e}tape \cite{3,5}. En effet, l'utilisation de cette
dynamique garantit que la diff\'{e}rence d'\'{e}nergie entre deux \'{e}tats
successifs ne d\'{e}passe pas $%
2zJ$, o\`{u} $z$ est le nombre des sites voisins d'un
site donn\'{e} et $J$ l'interaction d'\'{e}change \cite{2}. Avec la dynamique de
retourner un seul spin \`{a} chaque \'{e}tape, nous poss\'{e}dons $N$ spins diff%
\'{e}rents et donc $N$ \'{e}tats possibles $b$ \`{a} atteindre \`{a} partir d'un
\'{e}tat donn\'{e} $a$. Ainsi, il y a $N$ probabilit\'{e}s de s\'{e}lection $\alpha \left( a\rightarrow b\right) $ qui ne
sont pas nulles \cite{1}. Chacune d'entre elles prend alors la valeur suivante :
\begin{equation}
\alpha \left( a\rightarrow b\right) =\frac{1}{N}.
\end{equation}%
Avec ces probabilit\'{e}s de s\'{e}lection, la condition de la balance d\'{e}%
taill\'{e}e \cite{18}, se r\'{e}v\`{e}le sous la forme suivante :
\begin{equation}
\frac{w\left( a\rightarrow b\right) }{w\left( b\rightarrow a\right) }=\frac{%
\alpha \left( a\rightarrow b\right) \rho \left( a\rightarrow b\right) }{%
\alpha \left( b\rightarrow a\right) \rho \left( b\rightarrow a\right) }=%
\frac{\rho \left( a\rightarrow b\right) }{\rho \left( b\rightarrow a\right) }%
=e^{-\beta \left( E_{b}-E_{a}\right) },  \label{3-3}
\end{equation}%
o\`{u} nous choisissons :
\begin{equation}
\rho \left( a\rightarrow b\right) =\rho _{0}e^{-\frac{\beta }{2}\left(
E_{b}-E_{a}\right) },  \label{3-4}
\end{equation}%
La constante de proportionnalit\'{e} $\rho _{0}$ est arbitraire et la plus grande valeur
de $e^{-\frac{\beta }{2}\left( E_{b}-E_{a}\right) }$ est $%
e^{\beta zJ}$. Ainsi, pour assurer $\rho \left( a\rightarrow b\right)
\leqslant 1$ on d\'{e}termine
\begin{equation}
\rho _{0}=e^{-\beta zJ}.
\end{equation}%
L'algorithme est d'autant plus efficace que $\rho \left( a\rightarrow
b\right) $ est grand. Cela nous donne :
\begin{equation}
\rho \left( a\rightarrow b\right) =e^{-\frac{\beta }{2}\left(
E_{b}-E_{a}+2zJ\right) },
\end{equation}%
qui n'est pas Metropolis, mais en utilisant cette probabilit\'{e}
d'acceptation, nous pouvons effectuer une simulation Monte Carlo du mod%
\`{e}le d'Ising \cite{2}. Toutefois, la simulation sera inefficace car elle
rejette la majorit\'{e} des transitions. La solution \`{a} ce probl\`{e}me
est la suivante : dans l'\'{e}quation (\ref{3-4}), nous avons pris une forme
fonctionnelle particuli\`{e}re pour le taux d'acceptation, mais la condition
de la balance d\'{e}taill\'{e}e (\ref{3-3}), n'avait pas besoin de prendre cette
forme \cite{1,2}. De ce fait, comme nous avons indiqu\'{e} dans le taux
d'acceptation, pour que l'algorithme soit encore plus efficace, il convient
que le rapport d'acceptation soit maximal. C'est-\`{a}-dire proche de 1.
Dans ce cas, pour voir comment cela fonctionne, nous supposerons que $E_{a}<E_{b}$ et $\rho
\left( b\rightarrow a\right) =1$.
Afin de satisfaire l'\'{e}quation (\ref%
{3-3}), $\rho \left( a\rightarrow b\right) $ doit alors prendre la valeur $%
e^{-\beta \left( E_{b}-E_{a}\right) }$.
Ainsi, l'algorithme optimal est celui o\`{u} :
\begin{equation}
\rho \left( a\rightarrow b\right) =\left \{
\begin{array}{c}
e^{-\beta \left( E_{b}-E_{a}\right) }\qquad \text{si }E_{b}-E_{a}>0, \\
1\qquad \qquad \qquad \text{ailleurs,}\qquad \  \  \  \
\end{array}%
\right.   \label{3-7}
\end{equation}%
qui est l'algorithme de Metropolis lanc\'{e} pour le mod\`{e}le d'Ising avec
la dynamique de retournement d'un seul spin \cite{1,2}.\vspace{0.1cm}\newline
$\bullet $ \textbf{\textit{Impl\'{e}mentation de l'algorithme Metropolis}}\vspace{0.1cm}\newline
La premi\`{e}re \'{e}tape de l'algorithme Metropolis consiste \`{a} g\'{e}n%
\'{e}rer un nouvel \'{e}tat $b$ \`{a} partir de l'\'{e}tat $a$. Les deux \'{e}tats
diff\`{e}rent par le retournement d'un seul spin. Il importe de choisir un
spin $k$ au hasard sur le r\'{e}seau et de consid\`{e}rer son renversement $S_{k}\rightarrow
-S_{k}$. Ensuite, on calcule \`{a} partir de l'expression de l'hamiltonien \cite{2}. La
variation d'\'{e}nergie entre les deux \'{e}tats est donc :
\begin{equation}
E_{b}-E_{a}=-J\sum_{\left \langle ij\right \rangle
}S_{i}^{b}S_{j}^{b}+J\sum_{\left \langle ij\right \rangle
}S_{i}^{a}S_{j}^{a}=-J\sum_{n.p.v}S_{i}^{a}\left( S_{k}^{b}-S_{k}^{a}\right)
.
\end{equation}%
\begin{eqnarray}
\text{Si \  \ }S_{k}^{a} &=&1\text{ \  \  \  \ alors \  \ }S_{k}^{b}=-1\text{ \  \
et \  \ }S_{k}^{b}-S_{k}^{a}=-2, \\
\text{si \  \ }S_{k}^{a} &=&-1\text{ \  \ alors \  \ }S_{k}^{b}=1\text{ \  \ et
\  \ }S_{k}^{b}-S_{k}^{a}=2.  \nonumber
\end{eqnarray}%
\`{A} ce titre, nous pouvons \'{e}crire :
\begin{equation}
S_{k}^{b}-S_{k}^{a}=-2S_{k}^{a},
\end{equation}%
et donc%
\begin{equation}
E_{b}-E_{a}=2J\sum_{n.p.v}S_{i}^{a}S_{k}^{a}=2JS_{k}^{a}%
\sum_{n.p.v}S_{i}^{a}.  \label{3-10}
\end{equation}%
L'algorithme consiste donc \`{a} calculer $E_{b}-E_{a}$ \`{a} partir de l'\'{e}%
quation (\ref{3-10}), puis en suivant la r\`{e}gle donn\'{e}e dans l'\'{e}quation
(\ref{3-7}) \cite{1} :\newline
$\bullet $ Si $E_{b}-E_{a}\leq 0$ on accepte le retournement $%
S_{k}\rightarrow -S_{k}$.\newline
$\bullet $ Si $E_{b}-E_{a}>0$ on admet le retournement $S_{k}\rightarrow
-S_{k}$ avec la probabilit\'{e} $\rho \left( a\rightarrow b\right)
=e^{-\beta \left( E_{b}-E_{a}\right) }$.\newline
$\bullet $ On g\'{e}n\`{e}re alors un nombre al\'{e}atoire $\xi $, tel que $%
0\leqslant \xi <1$. Si $\xi \leq \rho \left( a\rightarrow b\right) $, on consent
au retournement $S_{k}\rightarrow -S_{k}$, sinon on ne fait rien.
\newline
$\bullet $ On tire ensuite un nouveau site au hasard et on recommence la proc\'{e}dure.\newline
Dans la pratique, l'algorithme de Metropolis se pr\'{e}sente de la mani\`{e}%
re suivante (voir figure (\ref{p})) :
\begin{figure}[th]
\centering
\includegraphics[scale=0.35]{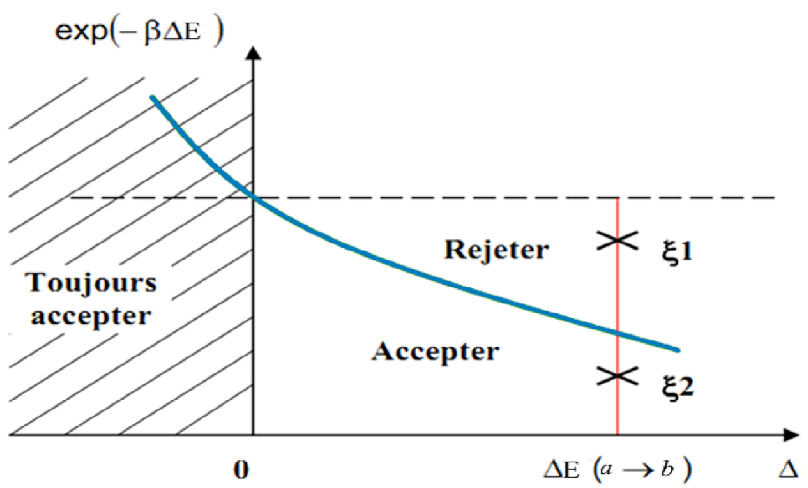}
\caption{Crit\`{e}re d'acceptation de la m\'{e}thode de Metropolis.}\label{p}
\end{figure}\newline
Le calcul d'une moyenne thermique ne peut commencer que lorsque le syst\`{e}%
me atteint l'\'{e}quilibre. Ainsi, dans une simulation Monte Carlo, il y a g%
\'{e}n\'{e}ralement deux p\'{e}riodes : la premi\`{e}re, o\`{u} partant
d'une configuration initiale, on r\'{e}alise une dynamique afin d'amener le
syst\`{e}me pr\`{e}s de l'\'{e}quilibre ; la seconde, o\`{u} le syst\`{e}me
\'{e}volue au voisinage de l'\'{e}quilibre et le calcul des moyennes est r%
\'{e}alis\'{e} \cite{2}. En l'absence de crit\`{e}re pr\'{e}cis, la dur\'{e}e de
la premi\`{e}re p\'{e}riode n'est pas facilement pr\'{e}visible. Une premi%
\`{e}re technique a \'{e}t\'{e} pendant longtemps, de suivre l'\'{e}nergie
instantan\'{e}e du syst\`{e}me en consid\'{e}rant que l'\'{e}quilibre est
atteint lorsque l'\'{e}nergie se stabilise autour d'une valeur
quasi-stationnaire \cite{1}.\vspace{-0.55cm}
\subsection{Algorithme du bain thermique (heat-bath algorithm)}

L'algorithme du bain thermique est \'{e}galement un algorithme de retournement d'un seul
spin \cite{2,16,18}, mais cet algorithme s'av\`{e}re plus efficace pour trouver les \'{e}%
tats \'{e}nergiquement d\'{e}sirables des spins dans le cas o\`{u} nous
avons un seul spin avec plusieurs \'{e}tats \cite{16}. Le principe de cet
algorithme est \cite{2} :\newline
$\bullet $ On choisit un spin $k$ au hasard sur le r\'{e}seau.\newline
$\bullet $ On d\'{e}finit une nouvelle valeur pour $s_{k}$, c'est-\`{a}-dire une valeur $n$ entre $1$ et $q$ avec la probabilit\'{e} :
\begin{equation}
p_{n}=\frac{e^{-\beta E_{n}}}{\sum_{m=1}^{q}e^{-\beta E_{m}}},
\end{equation}%
$E_{n}$ est l'\'{e}nergie du syst\`{e}me lorsque $s_{k}=n$. Il est clair que cet algorithme
satisfait l'ergodicit\'{e} et la balance d\'{e}taill\'{e}e. La probabilit%
\'{e} de transition de $s_{k}=n$ \`{a} $s_{k}=n^{\prime
}$ est $p_{n^{\prime }}$ et la probabilit\'{e} de retour est $p_{n}$, donc :
\begin{equation}
\frac{p\left( n\rightarrow n^{\prime }\right) }{p\left( n^{\prime
}\rightarrow n\right) }=\frac{p_{n^{\prime }}}{p_{n}}=\frac{e^{-\beta
E_{n^{\prime }}}}{\sum_{m=1}^{q}e^{-\beta E_{m}}}\times \frac{%
\sum_{m=1}^{q}e^{-\beta E_{m}}}{e^{-\beta E_{n}}}=e^{-\beta \left(
E_{n^{\prime }}-E_{n}\right) },
\end{equation}%
ainsi, la balance d\'{e}taill\'{e}e est satisfaite. Cet algorithme est plus
efficace que celui de Metropolis pour un grand $q$, car il choisit l'\'{e}tat
qui a un plus grand poids que celui de Boltzmann \cite{1}.
\vspace{-0.6cm}
\subsection{Algorithme de BKL (Bortz, Kalos et Lebowitz)}

L'algorithme de BKL ou la m\'{e}thode Monte Carlo \`{a} temps continu est une autre technique \`{a}
adjoindre \`{a} notre processus de Markov \cite{20}. Cet algorithme n'est pas
suffisamment utilis\'{e} comme il devrait \^{e}tre. N\'{e}anmoins, il reste
une technique importante et puissante pour de nombreux calculs \cite{18}.\newline
Consid\'{e}rons un syst\`{e}me \`{a} basse temp\'{e}rature. Ces syst\`{e}mes
ont toujours un probl\`{e}me o\`{u} le passage d'un \'{e}tat \`{a} un autre
est tr\`{e}s lent et le taux d'acceptation faible. En effet, le syst\`{e}me
passe la majorit\'{e} du temps dans l'\'{e}tat fondamental. Dans cette
technique, l'\'{e}tape de temps d\'{e}pend de l'estimation du temps pendant
lequel le syst\`{e}me reste dans son \'{e}tat avant de passer \`{a} un
nouvel \'{e}tat. On donne un grand poids dans $\left \langle Q\right \rangle $, aux \'{e}tats qui sont occup%
\'{e}s le plus longtemps. Soit $\Delta t$\ le temps que le syst\`{e}me passe dans un
\'{e}tat $a$ avant de passer \`{a} un autre. Le temps est obtenu \`{a} partir de $w\left( a\rightarrow a\right) $ \cite{2}. La probabilit\'{e} de rester dans $a$ apr\`{e}s $t$ \'{e}tapes est :
\begin{equation}
\left[ w\left( a\rightarrow a\right) \right] ^{t}=e^{t\log w\left(
a\rightarrow a\right) },
\end{equation}%
et le temps $\Delta t$\ est donn\'{e} par :
\begin{equation}
\Delta t=\frac{1}{\log w\left( a\rightarrow a\right) }=-\frac{1}{\log \left[
1-\sum \limits_{b\neq a}w\left( a\rightarrow b\right) \right] }\simeq \frac{%
1}{\sum \limits_{b\neq a}w\left( a\rightarrow b\right) }.
\end{equation}%
Si on peut calculer $\Delta t$, alors au lieu d'attendre plusieurs \'{e}tapes
Monte Carlo pour que le changement soit accept\'{e}, on suppose cela et on
passe directement \`{a} un autre \'{e}tat. Ainsi, l'algorithme Monte Carlo
\`{a} temps continu se compose des \'{e}tapes suivantes \cite{2,18,20} :\newline
$\bullet $ on \'{e}value les probabilit\'{e}s $w\left( a\rightarrow b\right) $ \`{a} partir de $a$. On choisit un nouvel
\'{e}tat $b$ avec une probabilit\'{e} proportionnelle \`{a} $w\left( a\rightarrow b\right) $ et on change l'\'{e}%
tat de syst\`{e}me \`{a} $b$,\newline
$\bullet $ on calcule $\Delta t$ et on le recalcule \`{a} chaque \'{e}tape,\newline
$\bullet $ on incr\'{e}mente le temps $t$ de $\Delta t$ pour tenir compte du temps que le syst\`{e}me
passe dans le m\^{e}me \'{e}tat.\newline
Cette technique tr\`{e}s \'{e}l\'{e}gante pour les probl\`{e}mes de la
simulation d'un syst\`{e}me \`{a} basse temp\'{e}rature. Elle poss\`{e}de quelques
inconv\'{e}nients qu'il faut prendre en compte :\newline
$\bullet $ dans la premi\`{e}re \'{e}tape, le nombre d'\'{e}tats $b$ cro\^{\i}t
exponentiellement avec la taille du syst\`{e}me et le temps pour calculer
les probabilit\'{e}s $w\left( a\rightarrow b\right) $ est long,\newline
$\bullet $ souvent l'ensemble des probabilit\'{e}s de transition ne change pas beaucoup
d'une \'{e}tape \`{a} une autre. Il est possible de ne stocker \`{a} chaque
\'{e}tape que les nouvelles transitions.
\vspace{-0.5cm}
\subsection{Algorithmes d'amas}

$\bullet $ \textbf{\textit{Algorithme de Wolff}}

Lorsque la longueur de corr\'{e}lation devient importante au voisinage de la
temp\'{e}rature critique $T_{c}$, des domaines se forment, o\`{u} tous les spins
pointent dans la m\^{e}me direction. Il est alors tr\`{e}s difficile pour
l'algorithme de Metropolis de retourner ce domaine, car il doit le faire
spin par spin avec une grande probabilit\'{e} de rejet du retournement \cite{2}.
En deux dimensions retourner un spin co\^{u}te $8J$ et en utilisant la valeur de $T_{c}=2.269$, la probabilit\'{e} d'accepter un tel retournement est :
\begin{equation}
\rho \left( a\rightarrow b\right) \simeq e^{-8J/T_{c}}=0.0294.
\end{equation}%
La probabilit\'{e} de retourner un spin du bord est plus grande. La solution
\`{a} ce probl\`{e}me a \'{e}t\'{e} propos\'{e}e par l'algorithme de Wolff.
L'id\'{e}e de base est de retourner tout le domaine en un seul coup. Ces
algorithmes sont appel\'{e}s aussi algorithmes de retournement d'amas ou algorithmes de l'amas. Au cours des derni\`{e}res ann\'{e}es, ils sont tr%
\`{e}s appr\'{e}ci\'{e}s pour toutes sortes de probl\`{e}mes, car il s'av%
\`{e}re qu'au moins dans le cas du mod\`{e}le d'Ising, ils enl\`{e}vent
presque enti\`{e}rement la difficult\'{e} du ralentissement critique \cite{1,2}.\newline
Pour trouver les amas \`{a} retourner, on tire un spin au hasard et on
cherche si ses voisins sont dans le m\^{e}me \'{e}tat. Puis, les voisins des
voisins et ainsi de suite. Le nombre d'amas retourn\'{e}s d\'{e}pend de la
temp\'{e}rature et les tailles des amas cro\^{\i}ssent lorsque la temp\'{e}%
rature d\'{e}cro\^{\i}t. Par cons\'{e}quent, nous avons $P_{add}$ qui est la probabilit%
\'{e} d'ajouter un spin \`{a} un amas, lorsque la temp\'{e}rature baisse. Apr%
\`{e}s l'ajout des spins, on retourne l'amas avec un rapport d'acceptation
qui d\'{e}pend du co\^{u}t d'\'{e}nergie pour le retournement \cite{2}.\newline
Consid\'{e}rons le passage de l'\'{e}tat $a$ \`{a} l'\'{e}tat $b$ par le
retournement d'un amas. Le point important est l'\'{e}tat des spins au bord
de l'amas, o\`{u} certains de ces spins sont dans le m\^{e}me \'{e}tat que l'amas. Les
liaisons entre ces spins et l'amas doivent \^{e}tre bris\'{e}es lorsque
l'amas est retourn\'{e}. De ce fait, les liaisons qui ne sont pas bris\'{e}%
es lors du passage de $a$ \`{a} $b$ le seront lors du passage de $b$ \`{a} $a$ \cite{1}. Consid%
\'{e}rons maintenant le passage de $a$ \`{a} $b$. Supposons qu'il y a $%
m$ liaisons bris%
\'{e}es. Ces liaisons correspondent aux spins sur le bord, qui sont dans le m%
\^{e}me \'{e}tat que l'amas, mais qui n'appartiennent pas \`{a} l'amas. La
probabilit\'{e} de ne pas les ajouter \`{a} l'amas est $%
\left( 1-P_{add}\right) ^{m}$. Si $n$ est le nombre
de liaisons bris\'{e}es dans le flip inverse, alors la probabilit\'{e} cette
fois est : $\left(
1-P_{add}\right) ^{n}$ \cite{1}. La condition de la balance d\'{e}taill\'{e}e \cite{2} est
donc :%
\begin{equation}
\frac{\alpha \left( a\rightarrow b\right) \rho \left( a\rightarrow b\right)
}{\alpha \left( b\rightarrow a\right) \rho \left( b\rightarrow a\right) }%
=\left( 1-P_{add}\right) ^{m-n}\frac{\rho \left( a\rightarrow b\right) }{%
\rho \left( b\rightarrow a\right) }=e^{-\beta \left( E_{b}-E_{a}\right) },
\end{equation}%
o\`{u} $\rho \left( a\rightarrow b\right) $ et $\rho \left( b\rightarrow
a\right) $ sont les taux d'acceptation dans les deux directions et $%
E_{b}-E_{a}=2J\left( m-n\right) $, d'o\`{u}%
\begin{equation}
\frac{\rho \left( a\rightarrow b\right) }{\rho \left( b\rightarrow a\right) }%
=\left[ e^{2\beta J}\left( 1-P_{add}\right) \right] ^{n-m},
\end{equation}%
pour :
\begin{equation}
P_{add}=1-e^{-2\beta J},
\end{equation}%
Les deux taux d'acceptation sont \'{e}gaux et on les pose \'{e}gaux \`{a} 1
quel que soit les \'{e}tats $a$ et $b$. Ce choix d\'{e}finit l'algorithme de l'amas
de Wolff pour le mod\`{e}le d'Ising, dont les d\'{e}tails de cet algorithme
sont les suivants \cite{21} :\newline
$\bullet $ on choisit au hasard sur le r\'{e}seau le spin de d\'{e}part,\newline
$\bullet $ on visite les sites voisins du site choisi. S'ils sont dans le m\^{e}me \'{e}%
tat que le spin choisi, on les ajoute \`{a} l'amas avec la probabilit\'{e} $P_{add}=1-e^{-2\beta J}$.\newline
$\bullet $ pour chaque spin ajout\'{e} dans l'\'{e}tape pr\'{e}c\'{e}dente, on cherche,
parmi ses voisins, ceux qui pointent dans la m\^{e}me direction. On les
cumule \`{a} l'amas avec la probabilit\'{e} $P_{add}$.\newline
$\bullet $ on retourne l'amas.\newline
Cet algorithme satisfait aux conditions de l'ergodicit\'{e} et de la balance
d\'{e}taill\'{e}e. Il convient de noter que la probabilit\'{e} $P_{add}$ cro\^{\i}t
lorsque la temp\'{e}rature d\'{e}cro\^{\i}t . Elle est nulle \`{a} $T$ infini et
\'{e}gale \`{a} 1 \`{a} $T=0$ et donc les amas sont grands \`{a} basse temp\'{e}%
rature \cite{1,2}. L'algorithme de Wolf est plus rapide que celui de Metropolis
au voisinage de $T_{c}$ mais un peu plus lent \`{a} haute et basse temp\'{e}rature. A basse temp\'{e}rature, l'algorithme de Wolff, engendre une
nouvelle configuration \`{a} chaque \'{e}tape Monte Carlo, alors que
Metropolis g\'{e}n\`{e}re une nouvelle configuration ind\'{e}pendante apr%
\`{e}s chaque \'{e}tape Monte Carlo par site \cite{21}.\newline
Nous avons examin\'{e} jusqu'\`{a} maintenant deux algorithmes de simulation
du mod\`{e}le d'Ising, \`{a} savoir l'algorithme de Metropolis et
l'algorithme de Wolff destin\'{e}es \`{a} simuler le mod\`{e}le en \'{e}%
quilibre. Pour \'{e}tudier le comportement du mod\`{e}le loin de la temp\'{e}%
rature critique, l'algorithme de Metropolis offre un moyen simple et
efficace d'obtenir des r\'{e}sultats qui ne peuvent \^{e}tre am\'{e}lior\'{e}%
s par autres algorithmes. Pr\`{e}s de $T_{c}$, l'algorithme de Wolff est meilleur
par rapport \`{a} l'algorithme de Metropolis, m\^{e}me si il s'av\`{e}re
plus complexe que ce dernier.

$\bullet $ \textbf{\textit{Algorithme de Swendsen-Wang}}

Apr\`{e}s les algorithmes de Metropolis et Wolff quoique similaires, l'algorithme de Swendsen-Wang \cite{22} reste tr\`{e}s important. Dans cet algorithme, on
construit les amas comme l'algorithme de Wolff et on lie les spins voisins,
dans le m\^{e}me \'{e}tat, avec la probabilit\'{e} $P_{add}=1-e^{-2\beta J}$. Mais au lieu de
retourner un seul amas, on retourne tous les amas, chacun avec la probabilit%
\'{e} $1/2$ \cite{1,2,16,18,22}. Cet algorithme satisfait \`{a} l'ergodicit\'{e} ainsi
qu'\`{a} la balance d\'{e}taill\'{e}e. La preuve de ce fait est exactement
la m\^{e}me que celle de l'algorithme de Wolff \cite{16,18,22}. Le choix de la
probabilit\'{e} $P_{add}=1-e^{-2\beta J}$ assure que la probabilit\'{e} d'acceptation ne d\'{e}pend ni
de $m$ ni de $n$. La mise \`{a} jour, de tous les sites du r\'{e}seau, se fait apr%
\`{e}s chaque mouvement. Pour mesurer le temps de corr\'{e}lation, on mesure
le nombre d'\'{e}tapes Monte Carlo et non pas le nombre d'\'{e}tapes par
site comme dans Metropolis \cite{1,2}. Il est l\'{e}g\`{e}rement plus lent que
Metropolis \`{a} haute temp\'{e}rature. Par contre, \`{a} basse temp\'{e}%
rature, il reste identique \`{a} l'algorithme de Wolff. Mais l'algorithme de
Wolff est deux fois plus rapide que celui de Swendsen-Wang \cite{1,2,16,18}.

$\bullet $ \textbf{\textit{Algorithme de Niedermayer}}

L'algorithme de Niedermayer est une extension de l'algorithme de Wolff et de Swendsen-Wang et
s'applique \`{a} tous les types de mod\`{e}le \cite{23}. Dans cet algorithme,
Niedermayer utilise deux probabilit\'{e}s : la premi\`{e}re pour lier deux
spins voisins dans le m\^{e}me \'{e}tat et la deuxi\`{e}me pour lier deux
spins voisins dans deux \'{e}tats oppos\'{e}s \cite{2}. Dans le cas du mod\`{e}le
Ising, l'\'{e}nergie de liaison d'une paire est donn\'{e}e par :
\begin{equation}
E_{ij}=-JS_{i}S_{j}.
\end{equation}%
La probabilit\'{e} de faire un lien entre deux spins voisins est fonction de
cette \'{e}nergie $P_{add}\left( E_{ij}\right) $. Dans le mod\`{e}le d'Ising, l'\'{e}nergie $E_{ij}$ ne peut prendre que deux valeurs $\pm J$. Nous avons alors deux valeurs de $P_{add}\left( E_{ij}\right) $, soit $P_{add}\left( -J\right) =1-e^{-2\beta J}$ et $%
P_{add}\left( J\right) =0$. D'autre part, l'algorithme de Neidermayer ob\'{e}it \`{a} l'ergodicit%
\'{e} sauf pour $P_{add}\left( E\right) =1$ quel que
soit $E$. Il r\'{e}pond aussi \`{a} la balance d\'{e}%
taill\'{e}e \cite{2}. Par d\'{e}finition, la probabilit\'{e} d'avoir $m$ spins parall%
\`{e}les $n$ sur le bord et antiparall\`{e}les \`{a} l'amas est $\left[ 1-P_{add}\left( -J\right) %
\right] ^{m}\left[ 1-P_{add}\left( J\right) \right] ^{n}$. Dans la
direction oppos\'{e}e, la probabilit\'{e} s'\'{e}crit $\left[ 1-P_{add}\left(
-J\right) \right] ^{n}\left[ 1-P_{add}\left( J\right) \right] ^{m}$ \cite{23}. Comme
l'algorithme de Wolf, le passage de $a$ \`{a} $b$ co\^{u}te :
\begin{equation}
E_{b}-E_{a}=2J\left( m-n\right) .
\end{equation}%
La balance d\'{e}taill\'{e}e est v\'{e}rifi\'{e}e par les taux d'acceptation
:
\begin{equation}
\frac{\rho \left( a\rightarrow b\right) }{\rho \left( b\rightarrow a\right) }%
=\left[ e^{2\beta J}\frac{1-P_{add}\left( -J\right) }{1-P_{add}\left(
J\right) }\right] ^{n-m},  \label{4-31}
\end{equation}%
Niedermayer a choisi les $P_{add}$, qui satisfont \`{a} :
\begin{equation}
\frac{1-P_{add}\left( -E\right) }{1-P_{add}\left( E\right) }=e^{-2\beta E},
\end{equation}%
La solution de Niedermayer de cette \'{e}quation \'{e}tait $P_{add}\left(
E\right) =1-\exp \left[ \beta \left( E-E_{0}\right) \right] $ avec $E_{0}$ qui est un
param\`{e}tre arbitraire. Toutefois, la probabilit\'{e} $%
P_{add}\left( E\right) $ est positive \cite{2}. En
outre, nous pouvons \'{e}crire :
\begin{equation}
P_{add}\left( E_{ij}\right) =\left \{
\begin{array}{c}
1-e^{\beta \left( E_{ij}-E_{0}\right) }\qquad \text{si }E_{ij}\leqslant
E_{0}, \\
0\qquad \qquad \qquad \  \  \  \text{ailleurs, \  \  \  \ }%
\end{array}%
\right.
\end{equation}%
Ce qui d\'{e}finit l'algorithme de Niedermayer qui peut \^{e}tre r\'{e}sum%
\'{e} ainsi \cite{23} :\newline
$\bullet $ si $E_{ij}\leqslant E_{0}$ quel que soit $ij$, on peut poser que les taux d'acceptation sont \'{e}gaux
\`{a} 1 pour n'importe quel mouvement. Ceci revient \`{a} chercher la plus
grande valeur de $E_{ij}$. Pour le mod\`{e}le d'Ising la plus grande valeur est $J$,\newline
$\bullet $ si on augmente $E_{0}$ au-dessus de $J$, $P_{add}\left(
E_{ij}\right) $ s'approche de 1 et l'amas devient de plus en
plus grand. Ceci permet de contr\^{o}ler la taille des amas form\'{e}s par
l'algorithme,\newline
$\bullet $ si $E_{0}$ est inf\'{e}rieure \`{a} la valeur maximale de $E_{ij}$, alors le membre droit de
l'\'{e}quation (\ref{4-31}) n'est pas \'{e}gal \`{a} 1 et on ne peut pas choisir le
taux d'acceptation \'{e}gal \`{a} 1. Pour le mod\`{e}le d'Ising, si nous
choisissons $-J\leqslant E_{0}<J$, nous avons :
\begin{equation}
\frac{\rho \left( a\rightarrow b\right) }{\rho \left( b\rightarrow a\right) }%
=\left[ e^{2\beta J}e^{-\beta \left( J+E_{0}\right) }\right] ^{n-m}=\left[
e^{\beta \left( J-E_{0}\right) }\right] ^{n-m},
\end{equation}%
on choisit l'un des taux \'{e}gal \`{a} 1 et l'autre ob\'{e}it \`{a} cette
\'{e}quation. Ainsi, l'algorithme ob\'{e}it \`{a} la balance d\'{e}taill\'{e}%
e,\newline
$\bullet $ si $E_{0}$ est inf\'{e}rieure \`{a} la valeur minimale de $E_{ij}$ qui est $-J$ pour le mod\`{e}le
d'Ising, alors $P_{add}\left( E_{ij}\right)
=0$ et les taux d'acceptation sont donn\'{e}s par :
\begin{equation}
\frac{\rho \left( a\rightarrow b\right) }{\rho \left( b\rightarrow a\right) }%
=\left[ e^{2\beta J}\right] ^{n-m},
\end{equation}%
Ce qui est l'\'{e}quivalent de l'algorithme de Metropolis. Ainsi, en faisant
varier $E_{0}$ on passe de l'algorithme Metropolis \`{a} l'algorithme de Wolff.
\vspace{-0.5cm}
\subsection{Algorithme de Kawasaki}

L'algorithme le plus simple qui conserve la valeur de l'aimantation apr\`{e}%
s le flip est l'algorithme de Kawasaki \cite{1,2}. Il se d\'{e}%
finit ainsi :\newline
$\bullet $ on choisit au hasard une paire de spins adjacents $k$ et $%
k\prime $,\newline
$\bullet $ on \'{e}change leurs valeurs, pour pr\'{e}server l'aimantation totale. Puis
on calcule la diff\'{e}rence d'\'{e}nergie $\Delta E=E_{b}-E_{a}$ entre les \'{e}tats $a$ et $b$ du syst%
\`{e}me avant et apr\`{e}s le changement,\newline
$\bullet $ la probabilit\'{e} d'acceptation du mouvement est donn\'{e}e par :
\begin{equation}
\rho \left( a\rightarrow b\right) =\left \{
\begin{array}{c}
e^{-\beta \Delta E}\qquad \text{si }\Delta E>0, \\
1\qquad \qquad \text{ailleurs. \  \ }%
\end{array}%
\right.
\end{equation}%
Cet algorithme est ergodique et v\'{e}rifie la balance d\'{e}taill\'{e}e. En
effet, le nombre de paires est $zN/2$ et la probabilit\'{e} de choisir une paire
\`{a} n'importe quelle \'{e}tape de l'algorithme devient :
\begin{equation}
\alpha \left( a\rightarrow b\right) =\frac{2}{zN}.
\end{equation}%
La probabilit\'{e} de choisir la m\^{e}me paire pour un mouvement inverse
est la m\^{e}me :
\begin{equation}
\frac{w\left( a\rightarrow b\right) }{w\left( b\rightarrow a\right) }=\frac{%
\alpha \left( a\rightarrow b\right) \rho \left( a\rightarrow b\right) }{%
\alpha \left( b\rightarrow a\right) \rho \left( b\rightarrow a\right) }=%
\frac{\rho \left( a\rightarrow b\right) }{\rho \left( b\rightarrow a\right) }%
=e^{-\beta \Delta E}.
\end{equation}
Dans ce chapitre, nous avons d\'{e}taill\'{e} la m\'{e}thode Mont Carlo
statique utilis\'{e}e ayant pour finalit\'{e} d'\'{e}tudier
les propri\'{e}t\'{e}s magn\'{e}tiques des syst\`{e}mes physiques.\newline
Dans une premi\`{e}re partie nous avons pr\'{e}sent\'{e} quelques mod\`{e}%
les de spins, \'{e}tudi\'{e}s en m\'{e}canique statistique. Ces mod\`{e}les nous ont
aid\'{e} \`{a} mod\'{e}liser tous les ph\'{e}nom\`{e}nes au cours desquels
des effets collectifs se sont produits du fait des interactions locales
entre les particules. \newline
Puis, nous avons abord\'{e} les principes de la simulation
Monte Carlo \`{a} l'\'{e}quilibre thermique afin d'\'{e}tudier les ph\'{e}nom%
\`{e}nes critiques. Nous avons d\'{e}velopp\'{e} les trois id\'{e}es de base
de la simulation Monte Carlo, \`{a} savoir : l'\'{e}chantillon important, la
balance d\'{e}taill\'{e}e et le taux d'acceptation.\newline
Enfin, nous avons expos\'{e} quelques algorithmes de
la m\'{e}thode Monte Carlo pour d\'{e}terminer les propri\'{e}t\'{e}s d'une
vari\'{e}t\'{e} de diff\'{e}rents mod\`{e}les d'\'{e}quilibre. Nous avons
\'{e}tudi\'{e} aussi certaines astuces utilis\'{e}es pour mettre en {\oe}uvre
les algorithmes de Monte Carlo en mati\`{e}re de programmes informatiques.%
\newline
Ce chapitre porte sur la m\'{e}thode Monte Carlo statique et afin d'examiner les propri\'{e}t\'{e}s magn\'{e}tiques des mat\'{e}riaux.
Dans le prochain chapitre, nous pr\'{e}sentons d'autres m\'{e}thodes
sophistiqu\'{e}es ; telles que celles de l'approximation de champ moyen, la
th\'{e}orie du champ effectif et la matrice de transfert. Ces derni\`{e}res
ont jou\'{e} un r\^{o}le important dans la description des ph\'{e}nom\`{e}%
nes critiques de transitions de phase.
\def\cleardoublepage{\clearpage}
\setcounter{chapter}{1}
\chapter{\'{E}tude des ph\'{e}nom\`{e}nes critiques}
       \graphicspath{{Chapitre2/figures/}}
\vspace{-0.8cm}
Les ph\'{e}nom\`{e}nes critiques ont fait l'objet de plusieurs \'{e}tudes approfondies. La plupart d'entre eux r\'{e}sultent de la divergence de la longueur de corr\'{e}lation. Provenant de la divergence, ces ph\'{e}nom\`{e}nes incluent les divergences en loi de puissance de
certaines quantit\'{e}s d\'{e}crites par les exposants critiques tels que : l'universalit\'{e}, le comportement fractal, et la rupture d'ergodicit\'{e} \cite{50,52}.
De nombreuses et nouvelles id\'{e}es ont \'{e}t\'{e} d\'{e}velopp\'{e}es
afin de mieux comprendre le comportement critique de syst\`{e}mes ayant des structures cristallographiques de plus en
plus complexes \cite{50}.\newline
Au cours d'une transition de phases les interactions entre champs et particules brisent spontan\'{e}ment la sym\'{e}trie du syst\`{e}me \cite{crit1}. La notion de sym\'{e}trie et de sa brisure, permettent de classer les diff\'{e}rentes phases, tout en expliquant les transitions de phase qui les s\'{e}parent \cite{51}.\newline
La compr\'{e}hension des ph\'{e}nom\`{e}nes critiques n\'{e}cessite l'usage de th\'{e}ories tels que : la th\'{e}orie du champ moyen,
la th\'{e}orie du champ effectif et la m\'{e}thode de la matrice de
transfert. Les grandes lignes de ces th\'{e}ories seront \'{e}tal\'{e}es le long de ce chapitre tout en mettant l'accent sur la description th\'{e}orique des transitions de phases. Notons au passage que pour mieux illustrer l'origine physiques des ph\'{e}nom\`{e}nes critiques nous utiliserons le mod\`{e}le simple d'Ising.\vspace{-0.6cm}
\section{Ph\'{e}nom\`{e}nes critiques}

En physique, les ph\'{e}nom\`{e}nes critiques sont les ph\'{e}nom\`{e}nes
qui se d\'{e}roulent aux points critiques o\`{u} les grandeurs physiques
varient tr\`{e}s rapidement et pr\'{e}sentent ainsi des discontinuit\'{e}s. Le point critique est une singularit\'{e} du diagramme
de phase. En ce point, la divergence des grandeurs thermodynamiques caract%
\'{e}ristiques du syst\`{e}me est donn\'{e}e par des lois de puissances caract%
\'{e}ris\'{e}es par des exposants critiques. Les ph\'{e}nom\`{e}nes
critiques incluent \'{e}galement les relations d'\'{e}chelle entre les diff%
\'{e}rentes quantit\'{e}s, l'universalit\'{e}, le comportement fractal et la
brisure d'ergodicit\'{e} \cite{51}. Cette section est consacr\'{e}e \`{a} une br\`{e}ve
description des ph\'{e}nom\`{e}nes critiques et des transitions de phase.
\vspace{-0.6cm}
\subsection{Singularit\'{e}}

Le comportement critique est caract\'{e}ris\'{e} par des singularit\'{e}s
de certaines fonctions thermodynamiques. Cette singularit\'{e} peut \^{e}tre
une discontinuit\'{e} ou une divergence \cite{53}. Les
singularit\'{e}s sont interpr\'{e}t\'{e}es comme des changements dans la
structure de phase du syst\`{e}me, o\`{u} plus pr\'{e}cis\'{e}ment comme des transitions de
phase \cite{51}. Les transitions de phase sont class%
\'{e}es selon la nature de la singularit\'{e} typique qui s'y produit. Les singularit\'{e}s des propri\'{e}t\'{e}s thermodynamiques, telles
que la chaleur sp\'{e}cifique et la susceptibilit\'{e}, sont tr\`{e}s faibles et tr\`{e}s difficiles
\`{a} d\'{e}tecter exp\'{e}rimentalement. De plus, elles n'apparaissent qu'\`{a} la limite thermodynamique pour des syst\`{e}mes infiniment grands \cite{53}. Cependant, un syst\`{e}me fini ne peut pas pr\'{e}senter une vraie singularit\'{e} \`{a} une temp\'{e}rature non-nulle, mais une temp\'{e}rature
pseudo-critique qui peut \^{e}tre li\'{e}e au pic pointu de la chaleur sp%
\'{e}cifique et de la susceptibilit\'{e}. Dans le d\'{e}veloppement des
fluctuations critiques, les singularit\'{e}s des grandeurs thermodynamiques et des fonctions de corr%
\'{e}lation ont la m\^{e}me origine physique. La connexion est assur\'{e}e par le th\'{e}or\`{e}me
de fluctuation-dissipation \cite{51}.
\vspace{-0.5cm}
\subsection{Transition de phase}

Une transition de phase est un ph\'{e}nom\`{e}ne spectaculaire, qui
correspond \`{a} une transformation qualitative et quantitative des propri%
\'{e}t\'{e}s macroscopiques d'un syst\`{e}me thermodynamique. Elle est
provoqu\'{e}e par la variation typique d'un param\`{e}tre intensif de contr%
\^{o}le de syst\`{e}me tel que la temp\'{e}rature, la pression, le champ \'{e}lectrique ou
magn\'{e}tique. Les \'{e}tats de la mati\`{e}re ont des propri\'{e}t%
\'{e}s physiques uniformes. Au cours d'une transition de phase, certaines
propri\'{e}t\'{e}s d'un syst\`{e}me thermodynamique donn\'{e}, changent de
mani\`{e}re discontinue \cite{53}. L'exemple fondamental le plus connu de transition
de phase est celui de l'eau, qui passe de l'\'{e}tat solide \`{a} l'\'{e}tat
liquide et de l'\'{e}tat liquide \`{a} l'\'{e}tat gazeux. Mais, il existe
d'autres transitions de phases. Par exemple : les transitions magn\'{e}%
tiques, superfluides, supraconductrices, d'ordre-d\'{e}sordre dans les
alliages m\'{e}talliques et les cristaux liquides. Les transitions de
phase se produisent lorsque l'\'{e}nergie libre d'un syst\`{e}me devient
non-analytique pour certaines variables thermodynamiques \cite{crit1}. G\'{e}n\'{e}%
ralement, cette condition r\'{e}sulte de l'interaction d'un grand nombre de
particules dans un syst\`{e}me et n'appara\^{\i}t pas dans les syst\`{e}mes de petites tailles. Il est important de noter que ces transitions peuvent se
manifester dans des syst\`{e}mes non-thermodynamiques \`{a} temp\'{e}rature
nulle. Les exemples incluent les transitions de phase quantiques, dynamiques
et topologiques. Dans ces syst\`{e}mes, d'autres param\`{e}%
tres prennent la place de la temp\'{e}rature. \`{A} titre d'exemple, la
probabilit\'{e} de connexion qui remplace la temp\'{e}rature pour les r\'{e}%
seaux de percolation \cite{56}.
\vspace{-0.5cm}
\subsection{Brisure de sym\'{e}trie et param\`{e}tre d'ordre}

Les transitions de phase impliquent souvent un processus de brisure de sym%
\'{e}trie \cite{51}. G\'{e}n\'{e}ralement, dans une transition de phase, les phases
\`{a} haute temp\'{e}rature sont plus sym\'{e}triques que les phases \`{a}
basse temp\'{e}rature dues \`{a} la rupture spontan\'{e}e de sym\'{e}trie.
En physique de mati\`{e}re condens\'{e}e, il existe plusieurs types de
brisure de sym\'{e}trie \cite{crit1} dont la brisure par formation de r\'{e}seau qui a lieu lors de la
transition entre un fluide et un solide cristallin, la brisure par inversion qui se produit lors de la transition
ferromagn\'{e}tique, et la brisure de la sym\'{e}trie de jauge qui appara\^{\i}t dans les
supraconducteurs.\newline
Pour d\'{e}crire les transitions de phase avec changement de sym\'{e}trie,
Landau a introduit la notion de param\`{e}tre d'ordre qui est consid\'{e}r%
\'{e} comme \'{e}tant une mesure du degr\'{e} d'ordre de l'\'{e}tat d'un syst%
\`{e}me physique lors d'une transition de phase \cite{50}. En effet, le param\`{e}tre d'ordre est uniforme
et nul au-dessus de la temp\'{e}rature critique dans la phase sym\'{e}trique
(d\'{e}sordonn\'{e}e). Cependant, Il est non-uniforme et non-nul au dessous de la temp%
\'{e}rature critique dans la phase moins sym\'{e}trique (ordonn\'{e}e). Du
point de vue th\'{e}orique, les param\`{e}tres d'ordre proviennent de la
brisure de sym\'{e}trie. Lorsque cela se produit, il est n\'{e}cessaire
d'introduire une ou plusieurs variables suppl\'{e}mentaires pour d\'{e}crire
l'\'{e}tat du syst\`{e}me. Le param\`{e}tre d'ordre est d\'{e}fini diff\'{e}%
remment dans les diff\'{e}rents types de syst\`{e}mes physiques.
\begin{table}[!ht]
  \centering
{\renewcommand{\arraystretch}{2} {\setlength{\tabcolsep}{1.5cm}}}
\begin{tabular}{ll}
\hline
Nature de la transition & Param\`{e}tre d'ordre \\ \hline
gaz-liquide & masse volumique \\
para-ferromagn\'{e}tique & aimantation \\
para-antiferromagn\'{e}tique & aimantation des sous r\'{e}seaux \\
para-ferro\'{e}lectrique & polarisation \\
ordre-d\'{e}sordre dans un alliage binaire & probabilit\'{e} d'occupation des deus sites \\
d\'{e}mixion d'un binaire AB & fractions molaires \\
supraconductivit\'{e} & gap supraconducteur \\
superfluidit\'{e} & fonction d'onde superfluide \\ \hline
\end{tabular}
\caption{\label{tab1} Quelques exemples des param\`{e}tres d'ordre utilis\'{e}s selon la nature de transition.}
\end{table}\newline
Dans le tableau (\ref{tab1}), nous donnons quelques exemples de param\`{e}tres
d'ordre. Le choix du param\`{e}tre d'ordre est une question ph\'{e}nom\'{e}%
nologique pas toujours \'{e}vidente. Cependant, les param\`{e}tres d'ordre
peuvent \'{e}galement \^{e}tre d\'{e}finis pour des transitions qui ne brisent pas la sym\'{e}trie. Certaines transitions de phase, comme celles
dans les supraconducteurs, peuvent avoir des param\`{e}tres d'ordre \`{a}
plusieurs degr\'{e}s de libert\'{e}. Dans ce cas par exemple, le param\`{e}tre
d'ordre peut \^{e}tre un nombre complexe, un vecteur ou m\^{e}me un tenseur,
dont la magnitude tend vers z\'{e}ro \`{a} la transition de phase. Par
ailleurs, on associe un champ conjugu\'{e} au param\`{e}tre d'ordre qui traduit l'\'{e}tat du syst%
\`{e}me. Pour les syst\`{e}mes magn\'{e}tiques, le champ conjugu\'{e} n'est autre que le champ magn%
\'{e}tique \cite{56}.
\vspace{-0.4cm}
\subsection{Classification des transitions de phase}

On distingue deux classes de transitions de phase qui sont les transitions d'Ehrenfest et les transitions de Landau.%
\newline
$\bullet $ \textbf{\textit{Classification d'Ehrenfest}}\newline
Ehrenfest est le premier \`{a} avoir classifier les transitions de
phase selon le degr\'{e} de non-analyticit\'{e} de l'\'{e}nergie ou la
continuit\'{e} des d\'{e}riv\'{e}es d'ordre de l'\'{e}nergie libre. Dans la
classification d'Ehrenfest, on distingue deux types de transition de phase :
transition du premier ordre et transition du second ordre \cite{crit2}.\newline
$\bullet $ \textit{Transition du premier ordre}\newline
Les transitions de phase du premier ordre pr\'{e}sentent une discontinuit%
\'{e} dans la d\'{e}riv\'{e}e premi\`{e}re de l'\'{e}nergie libre par
rapport \`{a} une variable thermodynamique. Les transitions entre les trois
\'{e}tats standards (solide, liquide, gaz), sont des transitions de phase du
premier ordre, car la d\'{e}riv\'{e}e premi\`{e}re de l'\'{e}nergie libre
par rapport au potentiel chimique est discontinue \cite{51}.\newline
$\bullet $ \textit{Transition du second ordre}\newline
Une transition de phase est du second ordre, si la d\'{e}riv\'{e}e premi\`{e}%
re de l'\'{e}nergie libre est continue, mais la d\'{e}riv\'{e}e seconde est
en revanche discontinue. Au cours de cette transition, le passage d'une
phase \`{a} une autre se fait de fa\c{c}on continue. Par exemple dans le cas
de la transition de phase entre l'\'{e}tat ferromagn\'{e}tique et l'\'{e}tat
paramagn\'{e}tique. La classification d'Ehrenfest peut \^{e}tre g\'{e}n\'{e}%
ralis\'{e}e pour d\'{e}finir des transitions multicritiques d'ordre sup\'{e}%
rieur \cite{51}.\newline
La classification d'Ehrenfest permet de mettre en \'{e}vidence
les diff\'{e}rences et les similitudes entre diverses transitions. Cependant, elle
reste inexacte au voisinage d'une transition de phase puisqu'elle ne pr%
\'{e}voit pas la possibilit\'{e} de divergence \`{a} la limite
thermodynamique.\newline
$\bullet $ \textbf{\textit{Classification de Landau}}\newline
En $1937$, Landau a remarqu\'{e} que la transition de phase sans chaleur
latente peut s'accompagner d'un changement de la sym\'{e}trie du syst\`{e}me
comme dans le cas de la classification d'Ehrenfest. Alors, il a propos\'{e} deux
types de transition de phase :\newline
$\bullet $ \textit{Transition sans param\`{e}tre d'ordre}\newline
Lors de ces transitions, le param\`{e}tre d'ordre
est discontinu et les groupes de sym\'{e}trie des deux phases ne sont pas
inclus l'un dans l'autre \cite{crit1}. Ces transitions sont toujours des transitions du premier ordre au sens d'Ehrenfest. \newline
$\bullet $ \textit{Transition avec param\`{e}tre d'ordre}\newline
Dans ce type de transition, le groupe de sym\'{e}trie de la phase ordonn\'{e}e est inclus
dans le groupe de la phase la plus sym\'{e}trique (d\'{e}sordonn\'{e}e). Au
sens de Landau, cette transition est du premier ordre si le param\`{e}tre
d'ordre est discontinu au point de transition et elle est du second ordre si le param%
\`{e}tre d'ordre est continu \`{a} la transition \cite{crit1}.\newline
Pour les transitions sans changement de sym\'{e}trie, le formalisme de Landau d\'{e}finit un pseudo-param\`{e}tre d'ordre vu la difficult\'{e} de d\'{e}terminer le param\`{e}tre d'ordre.\newline
Dans ce qui suit, nous allons pr\'{e}senter quelques m\'{e}thodes d'approximation de la physique statistique qui sont n\'{e}cessaires pour \'{e}tudier la singularit\'{e} des grandeurs physiques au voisinage d'une transition de phase.
\vspace{-0.6cm}
\section{Approximation de champ moyen}

L'approximation de champ moyen a \'{e}t\'{e} d\'{e}velopp\'{e}e pour \'{e}%
tudier les syst\`{e}mes \`{a} corps en interaction \cite{50}. Elle consiste \`{a}
remplacer le probl\`{e}me des spins en interaction par un probl\`{e}me de
spins ind\'{e}pendants plac\'{e}s dans un champ moyen produit par l'ensemble
des autres spins \cite{51}. Dans cette section, nous allons d\'{e}velopper l'approximation du champ moyen pour le mod\`{e}le d'Ising.\vspace{-0.5cm}
\subsection{\'{E}quation du champ moyen}

L'approximation du champ moyen consiste \`{a} prendre un seul spin $S_{i}$ pour
calculer son \'{e}nergie en rempla\c{c}ant tous les autres spins par leur
valeur moyenne $\left \langle S_{j}\right \rangle $ \cite{52}. L'hamiltonien englobant le champ magn\'{e}tique ext\'{e}%
rieur pour le mod\`{e}le d'Ising est :
\begin{equation}
H=-J\sum \limits_{\left \langle i,j\right \rangle }S_{i}S_{j}-\mu
h\sum \limits_{i}S_{i},
\end{equation}%
o\`{u} $J$ est la constante de couplage, $h$ le champ magn\'{e}tique ext\'{e}rieur et $\mu $ le
moment magn\'{e}tique des spins. Dans l'approximation du champ moyen, l'\'{e}%
nergie du spin $S_{i}$ est donn\'{e}e par :
\begin{equation}
E_{i}=-JS_{i}\sum \limits_{j}\left \langle S_{j}\right \rangle -\mu hS_{i}.
\end{equation}%
Selon que le spin est dirig\'{e} vers le haut $\left( +1\right) $ ou vers le bas $\left( -1\right) $, l'\'{e}nergie
du site a deux valeurs $E_{i+}$ et
$E_{i-}$,%
\begin{equation}
\begin{array}{c}
E_{i+}=-J\sum \limits_{j}\left \langle S_{j}\right \rangle -\mu h=-zJM-\mu h,
\\
E_{i-}=J\sum \limits_{j}\left \langle S_{j}\right \rangle +\mu h=zJM+\mu h,%
\text{ \  \  \ }%
\end{array}%
\end{equation}%
$M$ est la valeur moyenne de $S_{j}$ et $z$ est le nombre des plus proches voisins.\newline
La valeur moyenne de $S_{i}$ est donn\'{e}e en terme de la constante de Boltzmann $k_{B}$ et la temp\'{e}rature $T$ par l'expression suivante \cite{52} :%
\begin{equation}
\left \langle S_{i}\right \rangle =\tanh \left( \frac{zJM+\mu h}{k_{B}T}%
\right) .
\end{equation}%
L'\'{e}quation du
champ moyen est attribu\'{e}e par :
\begin{equation}
M=\tanh \left( \frac{zJM+\mu h}{k_{B}T}\right) ,
\end{equation}%
qui s'\'{e}crit sous la forme :
\begin{equation}
\tanh ^{-1}M=\frac{1}{2}\ln \frac{1+M}{1-M}=\frac{zJ}{k_{B}T}M+\frac{\mu h}{%
k_{B}T}.  \label{M1}
\end{equation}%
Cette \'{e}quation est la base de l'approximation du champ moyen que l'on peut r\'{e}soudre graphiquement. Dans ce qui suit, nous pr\'{e}sentons les solutions graphiques de l'\'{e}quation (\ref{M1}) qui font appara\^{\i}tre l'existence d'une transition ferromagn\'{e}tique.
\vspace{-0.6cm}
\subsection{Transition ferromagn\'{e}tique}
\begin{figure}[th]
\centering
\includegraphics[scale=0.35]{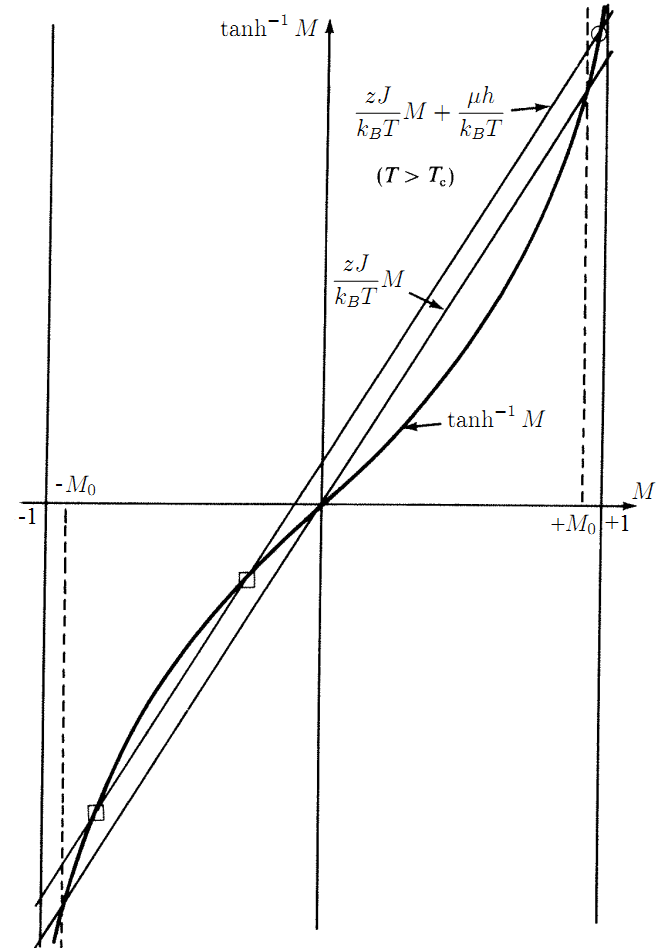}
\caption{La solution graphique de l'\'{e}quation (\ref{M1}) \cite{52}.}
\label{tanh}
\end{figure}

Les solutions de l'\'{e}quation (\ref{M1}) r\'{e}sultent des intersections de la ligne droite $\left( zJ/k_{B}T\right) M+\mu h/k_{B}T$ avec la
courbe $\tanh ^{-1}M$ pr\'{e}sent\'{e}es dans la figure (\ref{tanh}).
\newline
Il est utile de rappeler que la courbe $\tanh ^{-1}M$ a deux asymptotes verticales
\`{a} $M=\pm 1$, et que sa tangente \`{a} l'origine poss\`{e}de une pente unitaire
\cite{52}.\newline
Ainsi, si $h>0$, l'\'{e}quation d'\'{e}tat (\ref{M1}) admet trois solutions, deux solutions m\'{e}tastables ou instables ayant $M<0$ et une solution physique pour $M>0$.\newline
L'approximation du champ moyen pr\'{e}dit une aimantation
spontan\'{e}e pour le cas $h=0$ \cite{54}. Cette aimantation $M_{0}$ satisfait :
\begin{equation}
\left \{
\begin{array}{c}
M_{0}\neq 0\qquad \text{si\qquad }T<T_{c} \\
M_{0}=0\qquad \text{si\qquad }T>T_{c}.%
\end{array}%
\right.
\end{equation}
Ainsi, la temp\'{e}rature de transition est :
\begin{equation}
T_{c}=zJ/k_{B}.\label{M0}
\end{equation}
\vspace{-1.5cm}
\subsection{Comportement au voisinage de la transition de phase}

Au voisinage de la temp\'{e}rature critique, l'aimantation est faible ce qui permet d'utiliser le d\'{e}veloppement en s\'{e}rie suivant \cite{52},%
\begin{equation}
\tanh ^{-1}M=M+\frac{1}{3}M^{3}+O\left( M^{5}\right) .  \label{M2}
\end{equation}
Afin d'\'{e}tudier le comportement du syst\`{e}me au voisinage de la temp%
\'{e}rature critique, nous pr\'{e}sentons dans ce qui suit les grandeurs
physiques telles que l'aimantation, la suceptibilit\'{e} et la chaleur sp%
\'{e}cifique. Pour simplifier nous supposons que le champ magn\'{e}tique ext%
\'{e}rieur est nul $h=0$.\newline
$\bullet $ \textbf{\emph{Aimantation}}
\newline
En utilisant l'\'{e}quation (\ref{M0}) avec $k_{B}=1$, nous trouvons que :
\begin{equation}
\frac{T}{zJ}=1+t,
\end{equation}%
o\`{u} $t$ est la temp\'{e}rature r\'{e}duite d\'{e}finie par :
\begin{equation}
t=\frac{T-T_{c}}{T_{c}}.
\end{equation}%
L'\'{e}quation du champ moyen s'\'{e}crit :
\begin{equation}
M\simeq \left( 1+t\right) \left( M+\frac{1}{3}M^{3}\right) ,
\end{equation}%
d'o\`{u}%
\begin{equation}
M_{0}\simeq \sqrt{-3t}.
\end{equation}%
Au voisinage de $T_{c}$, l'aimantation spontan\'{e}e varie comme suit :%
\begin{equation}
M_{0}\sim \left( T_{c}-T\right) ^{1/2}.
\end{equation}
Pour examiner le comportement singulier de la susceptibilit\'{e} et la
chaleur sp\'{e}cifique au voisinage de la temp\'{e}rature critique, nous les
repr\'{e}sentons par la suite comme des lois de puissance.\newline
$\bullet $ \textbf{\emph{Susceptibilit\'{e}}}
\newline
Nous distinguons deux cas en absence du champ magn\'{e}tique $h$ :\newline
\textbf{\emph{(a) $T>T_{c}$}}
\newline
Au voisinage de $T_{c}$, le terme avec $M^{3}$ dans l'\'{e}quation (\ref{M2}) peut \^{e}tre n%
\'{e}glig\'{e} :
\begin{equation}
M\simeq \left( 1+t\right) M-\frac{\mu h}{k_{B}T_{c}},
\end{equation}%
d'o\`{u}%
\begin{equation}
M\simeq \frac{\mu h}{k_{B}\left( T-T_{c}\right) }.
\end{equation}%
Pour une configuration du syst\`{e}me de $N$ spins, l'aimantation totale $M_{t}$ est donn\'{e}e par \cite{51} :%
\begin{equation}
M_{t}=N\mu M=\frac{\mu ^{2}Nh}{k_{B}\left( T-T_{c}\right) }.
\end{equation}%
Dans ce cas la susceptibilit\'{e} magn\'{e}tique $\chi$ varie comme :
\begin{equation}
\chi \sim \left( T-T_{c}\right) ^{-1},\label{a1}
\end{equation}
car elle s'\'{e}crit sous la forme \cite{50} :%
\begin{equation}
\chi =\left. \frac{\partial M_{t}}{\partial h}\right \vert _{h=0}=\frac{\mu
^{2}N}{k_{B}\left( T-T_{c}\right) }.
\end{equation}\newline
\textbf{\emph{(b) $T<T_{c}$}}
\newline
Dans ce cas, $M$ s'\'{e}crit en terme de l'aimantation spontan\'{e}e $M_{0}$ sous la forme :
\begin{equation}
M=M_{0}+\varepsilon .
\end{equation}%
En n\'{e}gligeant les termes d'ordre sup\'{e}rieur, l'\'{e}quation (\ref%
{M2}) devient alors :
\begin{equation}
M_{0}+\varepsilon =\left( 1+t\right) \left( M_{0}+\varepsilon \right) +\frac{%
1}{3}\left( M_{0}+\varepsilon \right) ^{3}-\frac{\mu h}{k_{B}T_{c}}.
\end{equation}%
$\varepsilon $ est de l'ordre de $h$ et comme $h\longrightarrow 0$, les termes en $\varepsilon ^{2}$ et $\varepsilon ^{3}$ sont n\'{e}gligeables. Il s'\'{e}coule que :
\begin{equation}
\varepsilon =\frac{\mu h}{2k_{B}\left( T-T_{c}\right) },\qquad \chi =\frac{%
\mu ^{2}N}{2k_{B}\left( T_{c}-T\right) };
\end{equation}%
d'o\`{u}
\begin{equation}
\chi \sim \left( T_{c}-T\right) ^{-1}.  \label{b1}
\end{equation}%
Nous d\'{e}duisons des \'{e}quations (\ref{a1}) et (\ref{b1}) que la susceptibilit\'{e} $\chi$ a la m\^{e}me loi de puissance pour les deux cas $T>T_{c}$ et $T<T_{c}$ mais avec une diff\'{e}rence des coefficients num\'{e}riques.\newline
$\bullet $ \textbf{\emph{Chaleur sp\'{e}cifique}}
\newline
Dans l'approximation du champ moyen, l'\'{e}nergie interne en absence du champ magn\'{e}tique \cite{50} est donn\'{e}e par :
\begin{equation}
\begin{array}{cc}
E=-\frac{1}{2}zJNM_{0}^{2} & T<T_{c}, \\
E=0\qquad \qquad \qquad & T>T_{c},%
\end{array}%
\end{equation}%
o\`{u} $N$ est le nombre total de spins dans le syst\`{e}me.\newline
Lorsque $T<T_{c}$ mais toujours pr\`{e}s de $T_{c}$, nous trouvons :
\begin{equation}
E=-\frac{1}{2}zJN\frac{3\left( T_{c}-T\right) }{T_{c}}=\frac{3}{2}%
k_{B}N\left( T-T_{c}\right) .
\end{equation}%
La chaleur sp\'{e}cifique $C$ en champ nul \cite{50} d\'{e}finit par :
\begin{equation}
C=\left. \frac{dE}{dT}\right \vert _{h=0},
\end{equation}%
est donn\'{e}e alors par :
\begin{equation}
C=\frac{3}{2}k_{B}N.
\end{equation}%
Lorsque $T>T_{c}$, la chaleur sp\'{e}cifique $C$ dispara\^{\i}t. Nous d\'{e}duisons alors que la chaleur sp\'{e}cifique est
discontinue, avec une discontinuit\'{e} \`{a} $3k_{B}N/2$ pour $%
T=T_{c}$.
\vspace{-0.6cm}
\subsection{Exposants critiques $\protect \alpha ,$ $\protect \beta ,$ $%
\protect \gamma ,$ $\protect \delta $}

Il est constat\'{e} exp\'{e}rimentalement que l'aimantation spontan\'{e}e,
la susceptibilit\'{e}, l'isotherme critique et la chaleur sp\'{e}cifique ob%
\'{e}issent aux lois de puissance pr\`{e}s de $T_{c}$. Les exposants critiques sont d\'{e}finis comme suit \cite{50,52} :
\[
\begin{array}{ccccc}
C\sim \left \vert T-T_{c}\right \vert ^{-\alpha }, & \qquad  & M_{0}\sim
\left( T_{c}-T\right) ^{\beta } & \qquad  & \left( T<T_{c}\right) , \\
\chi \sim \left \vert T-T_{c}\right \vert ^{-\gamma }, & \qquad  & h\sim
M_{t}^{\delta }\qquad \  \  \  \  \  \  & \qquad  & \left( T=T_{c}\right) .%
\end{array}%
\]
Les valeurs pr\'{e}vues par l'approximation du champ moyen sont $\alpha =0$,
$\beta =\frac{1}{2}$, $\gamma =1$ et $\delta =3$ pour une dimension $D\geq4$.\newline
L'approximation du champ moyen ne tient pas compte des fluctuations de tous
les spins, puisque chaque spin est remplac\'{e} par sa moyenne. Il n'existe
pas de solution pour \'{e}valuer cette approximation \`{a} l'avance. Tout ce
que nous pouvons faire est une \'{e}valuation \`{a} posteriori en comparant
ses r\'{e}sultats avec les r\'{e}sultats exactes obtenus pour le cas unidimensionnel et le cas bidimensionnel, ou avec des calculs
num\'{e}riques effectu\'{e}s pour le cas tridimensionnel. Il est int\'{e}ressant de comparer les exposants du champ moyen avec ceux du
calcul analytique pour le cas bidimensionnel ou ceux obtenus \`{a} partir d'un calcul num\'{e}rique
utilisant une expansion \`{a} haute temp\'{e}rature pour le cas tridimensionnel \cite{55}.
\begin{table}[th]
\centering{\renewcommand{\arraystretch}{1.5} {\setlength{\tabcolsep}{0.3cm}
\begin{tabular}{c|c|c|c|}
\cline{2-4}
& champ moyen & $D=2$ & $D=3$ \\ \hline
\multicolumn{1}{|c|}{$\alpha $} & $0$ & $ln\left \vert T-T_{c}\right \vert $ &
$0.110\pm 0.005$ \\ \hline
\multicolumn{1}{|c|}{$\beta $} & $0.5$ & $0.125$ & $0.312\pm 0.003$ \\ \hline
\multicolumn{1}{|c|}{$\gamma $} & $1$ & $1.75$ & $1.238\pm 0.002$ \\ \hline
\multicolumn{1}{|c|}{$\delta $} & $3$ & $15$ & $5.0\pm 0.05$ \\ \hline
\end{tabular}%
}}
\caption{Les exposants critiques $\protect \alpha ,$\ $\protect \beta ,$\ $%
\protect \gamma ,$\ $\protect \delta $ pour la th\'{e}orie de champ moyen, mod%
\`{e}le d'Ising bidimensionnel ($D=2$) et mod\`{e}le d'Ising tridimensionnel $D=3$.}
\label{tabCM}
\end{table}
\newline
Le tableau (\ref{tabCM}) montre que l'augmentation de la dimension conduit \`{a} une am%
\'{e}lioration des r\'{e}sultats du champ moyen. En effet, l'approximation
du champ moyen devient conforme lorsque $D\longrightarrow
\infty $ \cite{56}. Elle est exacte \'{e}galement
dans le cas des interactions \`{a} tr\`{e}s longue port\'{e}e. \`{A} titre d'exemple, la solution exacte d'un mod\`{e}le d'Ising \`{a} port\'{e}e infinie co\"{\i}ncide avec le r\'{e}sultat du
champ moyen si chaque spin int\'{e}ragit avec tout les autres spins du r\'{e}%
seau par une constante de couplage $J/N$ \cite{52}.
\vspace{-0.5cm}
\section{Th\'{e}orie du champ effectif}

Cette m\'{e}thode a \'{e}t\'{e} initi\'{e} en premier temps par Honmura et Kaneyoshi \cite{57,58} et d\'{e}%
velopp\'{e}e par Boccara \cite{59}. Elle a la simplicit\'{e} du champ moyen et
donne de meilleurs r\'{e}sultats qualitatifs et quantitatifs. Elle ne
s'applique qu'\`{a} des syst\`{e}mes dans lesquels le d\'{e}sordre est d\'{e}%
crit par des variables discr\`{e}tes. Elle consiste \`{a} prendre un spin
central et \`{a} calculer sa valeur moyenne en fixant tous les autres spins
du r\'{e}seau. La valeur moyenne sur toutes les configurations, donne l'\'{e}%
quation d'\'{e}tat du syst\`{e}me qui va permettre de d\'{e}terminer la temp%
\'{e}rature de transition ainsi que d'autres propri\'{e}t\'{e}s. Dans cette section, nous \'{e}tudions quelques aspects de la th\'{e}orie du
champ effectif. Pour cela, nous commen\c{c}ons par introduire la technique
d'op\'{e}rateur diff\'{e}rentiel. Ensuite, nous examinons avec quelques d%
\'{e}tails l'approximation de d\'{e}couplage (ou Zernike). Nous exposons
\'{e}galement les grandes lignes de l'approximation du champ effectif corr%
\'{e}l\'{e} (ou Bethe-Peierls). Finalement, nous discutons la m\'{e}thode du
groupe de renormalisation du champ effectif.
\vspace{-0.8cm}
\subsection{Technique d'op\'{e}rateur diff\'{e}rentiel}

En pr\'{e}sence d'un champ magn\'{e}tique ext\'{e}rieur $h$, l'hamiltonien du
mod\`{e}le d'Ising est donn\'{e} par :
\begin{equation}
H=-\frac{1}{2}\sum_{i,j}J_{ij}\mu _{i}\mu _{j}-h\sum_{i}\mu _{i},
\label{4-1}
\end{equation}
o\`{u} $\mu _{i}$ est une variable dynamique qui peut prendre deux valeurs $\pm1$ et $J_{ij}$ est l'interaction d'%
\'{e}change entre le site $i$ et le site $j$. Notons que $\mu _{i}$ est la
composante $z$ d'un op\'{e}rateur de spin $\left( S_{i}^{z}=1/2\mu _{i}\right) $ associ\'{e} \`{a} l'ion localis\'{e}e
sur le site qui peut prendre un spin up $\left( \mu
_{i}=+1\right) $ ou un spin down $\left( \mu
_{i}=-1\right) $. Le r\^{o}le du champ magn\'{e}%
tique dans cette partie est important. En effet, il permet de briser la sym\'{e}trie et de favoriser la phase
ordonn\'{e}e \cite{61}. Le param\`{e}tre d'ordre est d\'{e}termin\'{e} par $%
m=\left
\langle \mu _{i}\right \rangle $. Dans
la phase ordonn\'{e}e $m\neq
0$, tandis que dans la phase d\'{e}sordonn\'{e}e $m=0$. La
valeur moyenne du spin est donn\'{e}e par :
\begin{equation}
\left \langle m_{i}\right \rangle =\frac{1}{Z}Tr\mu _{i}e^{-\beta H},
\label{4-2}
\end{equation}%
avec la fonction de partition $Z$ :
\begin{equation}
Z=Tre^{-\beta H},
\end{equation}%
$\beta =1/k_{B}T$, $k_{B}$ est la constante de Boltzmann et $T$ la temp\'{e}rature absolue. L'hamiltonien de l'\'{e}quation (\ref{4-1}) s'exprime comme la somme suivante :
\begin{equation}
H=H_{i}+H^{\prime },
\end{equation}%
o\`{u} la premi\`{e}re partie $H_{i}$ comprend toutes les contributions associ\'{e}es au site, alors que la second partie $H^{\prime }$ ne d\'{e}pend pas de l'emplacement $i$.\newline
Alors,
\begin{equation}
H_{i}=-\mu _{i}E_{i},
\end{equation}%
tel que $E_{i}$ est l'op\'{e}rateur exprimant le champ local sur le site $i$, il prend la forme suivante :
\begin{equation}
E_{i}=\sum_{i,j}J_{ij}\mu _{j}+h.
\end{equation}%
Sachant que les
variables de spin commutent $\left[ \mu _{i},\mu _{j}\right]
=0 $, il s'ensuit que :
\begin{equation}
\left[ H_{i},H^{\prime }\right] =\left[ H_{i},H\right] =0.
\end{equation}%
En raison de cette relation commutative, la valeur moyenne (\ref{4-2}) s'exprime sous la forme :
\begin{equation}
\left \langle \mu _{i}\right \rangle =\frac{1}{Z}\left \{ Tre^{-\beta H}%
\left[ \frac{tr_{\left( i\right) }\mu _{i}\exp \left( -\beta H_{i}\right) }{%
tr_{\left( i\right) }\exp \left( -\beta H_{i}\right) }\right] \right \} ,
\end{equation}%
o\`{u} $tr_{\left( i\right) }=\sum_{\mu _{i}=-1}^{+1}$.\newline
Par cons\'{e}quent, nous obtenons l'identit\'{e} premi\`{e}re qui fut d\'{e}riv\'{e}e par Callen en $1963$ \cite{60} :
\begin{equation}
\left \langle \mu _{i}\right \rangle =\left \langle \tanh \left( \beta
E_{i}\right) \right \rangle .  \label{4-9}
\end{equation}%
En \'{e}tendant la proc\'{e}dure ci-dessus, l'identit\'{e} peut \^{e}tre
facilement g\'{e}n\'{e}ralis\'{e}e \`{a} :
\begin{equation}
\left \langle \left \{ f_{i}\right \} \mu _{i}\right \rangle =\left \langle
\left \{ f_{i}\right \} \tanh \left( \beta E_{i}\right) \right \rangle ,
\label{4-10}
\end{equation}%
o\`{u} $\left \{ f_{i}\right \} $ peut \^{e}tre n'importe quelle fonction des variables d'Ising tant qu'il ne
s'agit pas d' une fonction du site $i$.\newline
Pour traiter l'identit\'{e} de Callen (\ref{4-9}), il convient
de remarquer les relations exactes suivantes valables pour $\mu _{i}=\pm 1$ et $K=\beta J$ :
\begin{equation}
\begin{array}{c}
\tanh \left( K\mu _{1}\right) =A\mu _{1},\qquad A=\tanh K,\qquad \tanh \left[
K\left( \mu _{1}+\mu _{2}\right) \right] =B\left( \mu _{1}+\mu _{2}\right)
, \  \  \\
B=\frac{1}{2}\tanh 2K,\qquad \tanh \left[ K\left( \mu _{1}+\mu _{2}+\mu
_{3}\right) \right] =C_{1}\left( \mu _{1}+\mu _{2}+\mu _{3}\right) +C_{2}\mu
_{1}\mu _{2}\mu _{3}, \\
C_{1}=\frac{1}{4}\left( \tanh 3K+\tanh K\right) ,\qquad C_{2}=\frac{1}{4}%
\left( \tanh 3K-\tanh K\right) , \qquad \qquad \qquad \  \
\end{array}
\label{4-15}
\end{equation}%
Dans le cas du r\'{e}seau en nid d'abeilles o\`{u} le nombre de coordination $z=3$, l'identit\'{e} (\ref{4-9}) peut \^{e}tre r%
\'{e}\'{e}crite sous la forme suivante :
\begin{equation}
\left \langle \mu _{i}\right \rangle =C_{1}\left( \left \langle \mu
_{i+1}\right \rangle +\left \langle \mu _{i+2}\right \rangle +\left \langle
\mu _{i+3}\right \rangle \right) +C_{2}\left \langle \mu _{i+1}\mu _{i+2}\mu
_{i+3}\right \rangle ,  \label{4-16}
\end{equation}%
o\`{u} $i+\delta $ $\left( \delta =1,2,3\right) $ d\'{e}signent les plus proches voisins du site $i$. Le calcul de l'\'{e}%
quation d'\'{e}tat se fait par l'utilisation de la technique d'op\'{e}rateur
diff\'{e}rentiel introduit par Honmura et Kaneyoshi est d\'{e}%
finie par \cite{61}:
\begin{equation}
\tanh \left( \beta E_{i}\right) =\exp \left( E_{i}\nabla \right) \left.
\tanh x\right \vert _{x=0},  \label{4-17}
\end{equation}%
o\`{u} $\nabla =\partial /\partial x$ est un op\'{e}rateur diff\'{e}rentiel.\newline
En faisant appel aux relations math%
\'{e}matiques suivantes :
\begin{equation}
\exp \left( a\nabla \right) \varphi \left( x\right) =\varphi \left(
x+a\right) ,\qquad \text{et\qquad }e^{a\mu _{i}}=\cosh a+\mu _{i}\sinh a,
\end{equation}
l'\'{e}quation (\ref{4-17}) s'\'{e}crit pour $h=0$ comme :
\begin{equation}
\tanh \left( \beta J\sum_{\delta }\mu _{i+\delta }\right) =\prod
\limits_{\delta =1}^{z}\left[ \cosh \left( J\nabla \right) +\mu _{i+\delta
}\sinh \left( J\nabla \right) \right] \left. \tanh x\right \vert _{x=0}.
\end{equation}%
Lorsque $z=1,2$ ou $3$, les m\^{e}mes relations exactes de (\ref%
{4-15}) peuvent
\^{e}tre facilement d\'{e}duites. Par exemple, lorsque $z=2$,
\begin{equation}
\begin{array}{ccc}
\tanh \left[ K\left( \mu _{i+1}+\mu _{i+2}\right) \right]  & = & \left( \mu
_{i+1}+\mu _{i+2}\right) \sinh \left( J\nabla \right) \cosh \left( J\nabla
\right) \left. \tanh x\right \vert _{x=0} \\
& = & \frac{1}{4}\left( \mu _{i+1}+\mu _{i+2}\right) \left( e^{2J\nabla
}-e^{-2J\nabla }\right) \left. \tanh x\right \vert _{x=0}\text{ \  \  \ } \\
& = & B\left( \mu _{i+1}+\mu _{i+2}\right) .\qquad \qquad \qquad \qquad \qquad
\qquad \qquad
\end{array}%
\end{equation}
La forme g\'{e}n\'{e}rale de la relation exacte (\ref{4-10}) est alors :
\begin{equation}
\begin{array}{c}
\left \langle \left \{ f_{i}\right \} \mu _{i}\right \rangle =\left \langle
\left \{ f_{i}\right \} e^{E_{i}\nabla }\right \rangle \left. \tanh \left(
\beta x\right) \right \vert _{x=0}\qquad \qquad \qquad \qquad \qquad \qquad \qquad \qquad \ \ \ \ \ \ \ \
\\
=\left \langle \left \{ f_{i}\right \} \prod \limits_{j}\left[ \cosh \left(
J_{ij}\nabla \right) +\mu _{j}\sinh \left( J_{ij}\nabla \right) \right]
\right \rangle \left. \tanh \left( \beta x+h^{\prime }\right) \right \vert
_{x=0},%
\end{array}
\label{4-23}
\end{equation}%
o\`{u} $h^{\prime }=\beta h$.\newline
Pour $\left \{ f_{i}\right \} =\mu
_{k}\left( k\neq i\right) $ et $h=0$, l'\'{e}quation (\ref{4-23}) devient :
\begin{equation}
\left \langle \mu _{k}\mu _{i}\right \rangle =\frac{1}{2}\tanh \left( 2\beta
J\right) \left( \left \langle \mu _{k}\mu _{i-1}\right \rangle +\left
\langle \mu _{k}\mu _{i+1}\right \rangle \right) .  \label{4-24}
\end{equation}%
En raison de l'invariance par translation, la fonction de corr\'{e}lation $%
\left \langle \mu _{k}\mu _{i}\right \rangle $ ne
d\'{e}pend que de la distance entre $i$ et $k$ :
\begin{equation}
\left \langle \mu _{k}\mu _{i}\right \rangle =\left \langle \mu _{0}\mu
_{i-k}\right \rangle =\left \langle \mu _{0}\mu _{r}\right \rangle =g\left(
r\right) ,  \label{4-25}
\end{equation}%
o\`{u} $r=i-k$ est une mesure de la distance entre les spins.\newline
L'\'{e}quation (\ref{4-25}) permet d'exprimer l'\'{e}quation (\ref{4-24}) comme suit :
\begin{equation}
2\coth \left( 2\beta J\right) =\frac{g\left( r+1\right) }{g\left( r\right) }+%
\left[ \frac{g\left( r\right) }{g\left( r-1\right) }\right] ^{-1},
\label{4-26}
\end{equation}%
En supposant que%
\begin{equation}
\frac{g\left( r+1\right) }{g\left( r\right) }=\frac{g\left( r\right) }{%
g\left( r-1\right) }=\gamma ,
\end{equation}%
et en ne retenant que la solution physiquement acceptable, la solution de l'%
\'{e}quation (\ref{4-26}) s'\'{e}crit :
\begin{equation}
\gamma =\tanh \left( \beta J\right),
\end{equation}%
d'o\`{u}%
\begin{equation}
g\left( r\right) =g_{i-k}=\left[ \tanh \left( \beta J\right) \right] ^{r},
\end{equation}%
qui est un r\'{e}sultat exact bien connu de la cha\^{\i}ne d'Ising \cite{62}.
\vspace{-0.8cm}
\subsection{Approximation de d\'{e}couplage (ou Zernike)}

Pour d\'{e}coupler l'\'{e}quation (\ref{4-16}) et (\ref{4-23}), on utilise :
\begin{equation}
\left \langle \mu _{j}\mu _{k}\ldots \mu _{l}\right \rangle \approx \left
\langle \mu _{j}\right \rangle \left \langle \mu _{k}\right \rangle \ldots
\left \langle \mu _{l}\right \rangle ,  \label{4-30}
\end{equation}%
pour $j\neq k\neq \cdots \neq l$.\newline Introduisant l'approximation (\ref{4-30}), la valeur moyenne des $\mu _{i}$ (l'\'{e}quation (\ref{4-23}) avec $\left \{ f_{i}\right\}=1 $) peut \^{e}tre \'{e}crite sous une forme compacte \cite{63}:
\begin{equation}
\left \langle \mu _{i}\right \rangle =\prod \limits_{j}\left[ \cosh \left(
J_{ij}\nabla \right) +\mu _{j}\sinh \left( J_{ij}\nabla \right) \right]
\left. \tanh \left( \beta x+h^{\prime }\right) \right \vert _{x=0}.
\label{4-31}
\end{equation}%
Pour un syst\`{e}me ferromagn\'{e}tique avec un nombre
de coordination $z$ et en absence du champ magn\'{e}tique, l'\'{e}quation (\ref{4-31}) se r\'{e}duit alors \`{a} :
\begin{equation}
m=\left \langle \mu _{i}\right \rangle =\left[ \cosh \left( J\nabla \right)
+m\sinh \left( J\nabla \right) \right] ^{z}\left. \tanh \left( \beta
x\right) \right \vert _{x=0}.  \label{4-32}
\end{equation}%
La temp\'{e}rature de transition $T_{c}$ peut \^{e}tre obtenue par lin\'{e}arisation de l'\'{e}quation (\ref{4-32}) ; en d\'{e}veloppant le terme de droite de l'\'{e}quation (\ref{4-32}), on obtient :
\begin{equation}
z\sinh \left( J\nabla \right) \cosh ^{z-1}\left( J\nabla \right) \left.
\tanh \left( \beta _{c}x\right) \right \vert _{x=0}=1,  \label{4-33}
\end{equation}%
o\`{u} $\beta _{c}=1/k_{B}T_{c}$. Pour un cubique simple o\`{u} $z=6$, l'\'{e}quation (\ref{4-33}) se r%
\'{e}duit \`{a} :
\begin{equation}
\tanh \left( 6J\beta _{c}\right) +4\tanh \left( 4J\beta _{c}\right) +5\tanh
\left( 2J\beta _{c}\right) =\frac{16}{3},
\end{equation}%
et la temp\'{e}rature de transition $T_{c}$ est donn\'{e}e par :
\begin{equation}
\frac{k_{B}T_{c}}{J}=5.073.
\end{equation}%
L'\'{e}quation (\ref{4-31}) peut \^{e}tre \'{e}galement r\'{e}\'{e}crite
comme suit \cite{64} :
\begin{equation}
\left \langle \mu _{i}\right \rangle =\prod \limits_{j}\left[ \frac{1}{2}%
\left( 1+\left \langle \mu _{j}\right \rangle \right) e^{J_{ij}\nabla }+%
\frac{1}{2}\left( 1-\left \langle \mu _{j}\right \rangle \right)
e^{-J_{ij}\nabla }\right] \left. \tanh \left( \beta x\right) \right \vert
_{x=0}.
\end{equation}%
Les facteurs $\frac{1}{2}\left( 1+\left \langle \mu _{j}\right \rangle
\right) $ et $\frac{1}{2}\left( 1-\left \langle \mu _{j}\right \rangle
\right) $ correspondent aux probabilit\'{e}s du spin voisin respectivement up et down. Au cas o\`{u} l'interaction d'\'{e}change de l'\'{e}quation (\ref{4-31}) prend la forme :
\begin{equation}
J_{ij}=\frac{j^{\prime }}{N},
\end{equation}%
o\`{u} $j^{\prime }$ est une constante finie et $N$ est le nombre total de sites de r\'{e}seau, alors, l'\'{e}%
quation (\ref{4-31}) se r\'{e}duit \`{a} :
\begin{equation}
m=\left \langle \mu _{i}\right \rangle =\left[ \cosh \left( \frac{j^{\prime }%
}{N}\nabla \right) +m\sinh \left( \frac{j^{\prime }}{N}\nabla \right) \right]
^{N-1}\left. \tanh \left( \beta x+h^{\prime }\right) \right \vert _{x=0}.
\label{4-39}
\end{equation}%
Lorsque $N$ est tr\`{e}s grand, on a alors :
\begin{equation}
\cosh \left( \frac{j^{\prime }}{N}\nabla \right) \approx 1\qquad \text{%
et\qquad }\sinh \left( \frac{j^{\prime }}{N}\nabla \right) \approx \frac{%
j^{\prime }}{N}\nabla ,
\end{equation}%
d'o\`{u} l'\'{e}quation (\ref{4-39}) se r\'{e}duit \`{a}%
\begin{equation}
m=\left[ 1+m\frac{j^{\prime }}{N}\nabla \right] ^{N-1}\left. \tanh \left(
\beta x+h^{\prime }\right) \right \vert _{x=0}.  \label{4-40}
\end{equation}%
Pour $N\longrightarrow \infty $, l'\'{e}quation (\ref{4-40}) est donn\'{e}e
par :%
\begin{equation}
m=e^{N\left[ m\left( j^{\prime }/N\right) \nabla \right] }\left. \tanh
\left( \beta x+h^{\prime }\right) \right \vert _{x=0}=\tanh \left( h^{\prime
}+\beta mj^{\prime }\right) .
\end{equation}%
La technique de l'op\'{e}rateur diff\'{e}rentiel est g\'{e}n\'{e}ralement
plus favorable \`{a} cette approximation en raison de la facilit\'{e} relative de la formulation
d'autres propri\'{e}t\'{e}s thermodynamiques, de l'extension \`{a} des probl%
\`{e}mes de spins plus \'{e}lev\'{e}s ainsi que des syst\`{e}mes de spins d%
\'{e}sordonn\'{e}s.
\vspace{-0.8cm}
\subsection{Approximation du champ effectif corr\'{e}l\'{e} (ou
Bethe-Peierls)}
Contrairement \`{a} l'approximation de d\'{e}couplage, cette approximation du champ effectif corr\'{e}l\'{e} \cite{63} consid\`{e}re que le spin central $\mu _{i}$ est li\'{e} aux plus proches voisins $\mu _{i+\delta }$ via :
\begin{equation}
\mu _{i+\delta }=\left \langle \mu _{i+\delta }\right \rangle +\lambda
\left( \mu _{i}-\left \langle \mu _{i}\right \rangle \right) ,  \label{4-42}
\end{equation}%
o\`{u} $\lambda $ est un param\`{e}tre d\'{e}pendant de la temp\'{e}rature.
Lorsque l'\'{e}quation (\ref{4-42}) est remplac\'{e} dans l'hamiltonien (\ref%
{4-1}) avec $h=0$, nous trouvons que :%
\begin{equation}
H=-\sum_{i}H_{i}^{eff}\mu _{i}+\text{constante,}
\end{equation}%
avec%
\begin{equation}
H_{i}^{eff}=J\sum_{j}\left \langle \mu _{j}\right \rangle -\lambda Jz\left
\langle \mu _{i}\right \rangle .
\end{equation}%
En substituant l'\'{e}quation (\ref{4-42}) dans (\ref{4-23}) avec $\left \{
f_{i}\right \} =1$ et en supposant que $m=\left \langle \mu _{i}\right
\rangle =\left \langle \mu _{i+\delta }\right \rangle $ et $h=0$, on obtient
:%
\begin{equation}
\begin{array}{ccc}
m & = & \left \langle \left \{ P\left( m;J\nabla \right) +\lambda \left[ \cosh
\left( J\nabla \right) +\mu _{i}\sinh \left( J\nabla \right) \right]
\right \} ^{z}\right \rangle \left. \tanh \left( \beta x\right) \right \vert
_{x=0}\qquad \qquad \  \  \\
& = & \sum \limits_{\nu =0}^{z}\frac{z!}{\nu !\left( z-\nu \right) !}\lambda
^{\nu }\left[ P\left( m;J\nabla \right) \right] ^{z-\nu }\left[ \cosh \left(
\nu J\nabla \right) +m\sinh \left( \nu J\nabla \right) \right] \left. \tanh
\left( \beta x\right) \right \vert _{x=0}%
\end{array}
\label{4-46}
\end{equation}
avec%
\begin{equation}
P\left( m;J\nabla \right) =\left( 1-\lambda \right) \left[ \cosh \left(
J\nabla \right) +m\sinh \left( J\nabla \right) \right] .
\end{equation}%
En posant le param\`{e}tre corr\'{e}l\'{e} $\lambda =0$, l'\'{e}quation (\ref{4-46}) co\"{i}ncide exactement avec (\ref%
{4-32}).\newline
Pour l'\'{e}valuation de $\lambda $, on pose $\left \{ f_{i}\right
\} =\mu _{i+\delta }$ dans l'\'{e}quation (\ref{4-23}),%
\begin{equation}
\begin{array}{c}
\left \langle \mu _{i+\delta }\mu _{i}\right \rangle =\left \langle \left[
\sinh \left( J\nabla \right) +\mu _{i+\delta }\cosh \left( J\nabla \right) %
\right] \right. \qquad \qquad \qquad \qquad \qquad \qquad \ \ \ \ \\
\times \left. \prod \limits_{\delta ^{\prime }\left( \neq \delta \right) }
\left[ \cosh \left( J\nabla \right) +\mu _{i+\delta ^{\prime }}\sinh \left(
J\nabla \right) \right] \right \rangle \left. \tanh \left( \beta x\right)
\right \vert _{x=0}.%
\end{array}
\label{4-48}
\end{equation}%
En substituant l'\'{e}quation (\ref{4-42}) dans l'\'{e}quation (\ref{4-48}),
on obtient :%
\begin{equation}
\begin{array}{ccc}
\left \langle \mu _{i+\delta }\mu _{i}\right \rangle  & = & \sum \limits_{\nu
=0}^{z-1}\frac{\left( z-1\right) !}{\nu !\left( z-1-\nu \right) !}\lambda
^{\nu }P\left( m;J\nabla \right) \left[ P\left( m;J\nabla \right) \right]
^{z-1-\nu }\qquad \qquad \qquad \qquad \qquad \qquad \  \  \  \  \\
& \times  & \left[ \cosh \left( \nu J\nabla \right) +m\sinh \left( \nu
J\nabla \right) \right] \left. \tanh \left( \beta x\right) \right \vert
_{x=0}+\sum \limits_{\nu =0}^{z-1}\frac{\left( z-1\right) !}{\nu !\left(
z-1-\nu \right) !}\lambda ^{\nu +1}\qquad \qquad  \\
& \times  & \left[ P\left( m;J\nabla \right) \right] ^{z-1-\nu }\left[
m\cosh \left( \left( \nu +1\right) J\nabla \right) +\sinh \left( \left( \nu
+1\right) J\nabla \right) \right] \left. \tanh \left( \beta x\right)
\right \vert _{x=0}%
\end{array}
\label{4-49}
\end{equation}
avec%
\begin{equation}
\bar{P}\left( m;J\nabla \right) =\left( 1-\lambda \right) \left[ m\cosh
\left( J\nabla \right) +\sinh \left( J\nabla \right) \right] .
\end{equation}%
Ainsi, l'aimantation $m$ et le param\`{e}tre corr\'{e}l\'{e} $\lambda $
peuvent \^{e}tre \'{e}valu\'{e}s \`{a} partir des \'{e}quations coupl\'{e}es
(\ref{4-46}) et (\ref{4-49}) \cite{64}. Pour mieux illustrer ceci, consid\'{e}rons un r\'{e}seau carr\'{e} avec $z=4$. Dans ce cas, nous avons
\begin{equation}
m=4\left( K_{1}+3K_{2}\lambda ^{2}-2K_{2}\lambda ^{3}\right) m+4K_{2}\left(
1-3\lambda ^{2}+2\lambda ^{3}\right) m^{3},
\end{equation}%
et%
\begin{equation}
\begin{array}{c}
m^{2}+\lambda \left( 1-m^{2}\right) =K_{1}\left( 1+3\lambda ^{2}\right)
+K_{2}\lambda ^{2}\left( 3+\lambda ^{2}\right) \qquad \  \  \\
+m^{2}\left[ 3K_{1}\left( 1-\lambda ^{2}\right) +K_{2}\left( 3+3\lambda
^{2}-8\lambda ^{3}+2\lambda ^{4}\right) \right] \\
+m^{4}\left[ 1-6\lambda ^{2}+8\lambda ^{3}-3\lambda ^{4}\right] ,\qquad
\qquad \qquad \qquad \  \
\end{array}%
\end{equation}%
o\`{u} les coefficients $K_{1}$ et $K_{2}$ prennent les formes suivantes :%
\begin{equation}
\begin{array}{c}
K_{1}=\frac{1}{8}\left[ \tanh \left( 4\beta J\right) +2\tanh \left( 2\beta
J\right) \right] , \\
K_{2}=\frac{1}{8}\left[ \tanh \left( 4\beta J\right) -2\tanh \left( 2\beta
J\right) \right] .%
\end{array}%
\end{equation}%
La temp\'{e}rature de transition $T_{c}$ peut \^{e}tre d\'{e}termin%
\'{e}e apr\`{e}s la r\'{e}solution analytique des \'{e}quations coupl\'{e}es \cite{64},%
\begin{equation}
\begin{array}{c}
1=4\left( K_{1}+3K_{2}\lambda ^{2}-2K_{2}\lambda ^{3}\right) ,\text{ \  \  \  \
} \\
\lambda =K_{1}+3\lambda ^{2}\left( K_{1}+K_{2}\right) +K_{2}\lambda ^{4}.%
\end{array}%
\end{equation}%
Il s'ensuit alors que :
\begin{equation}
\frac{k_{B}T_{c}}{J}=\frac{2}{\ln 2}\qquad \text{et\qquad }\lambda \left(
T=T_{c}\right) =\frac{1}{3}.
\end{equation}%
En g\'{e}n\'{e}ral, la temp\'{e}rature de transition et le param\`{e}tre corr%
\'{e}l\'{e}e $\lambda $\  \`{a} $T=T_{c}$\ sont donn\'{e}s par :%
\begin{equation}
\frac{k_{B}T_{c}}{J}=\frac{2}{\ln \left[ z/\left( z-2\right) \right] },
\end{equation}%
et%
\begin{equation}
\lambda \left( T=T_{c}\right) =\frac{1}{z-1}.
\end{equation}
\vspace{-1.2cm}
\subsection{Groupe de renormalisation du champ effectif}

Nous avons d\'{e}montr\'{e} dans la sous-section pr\'{e}c\'{e}dente comment les corr\'{e}lations
de spin peuvent \^{e}tre d\'{e}coupl\'{e}es pour transformer la fonction
transcendantale en une forme polynomiale. En outre, nous avons trouv\'{e}
que les r\'{e}sultats d\'{e}pendent du nombre de coordination et non de la
dimension du syst\`{e}me. Afin de tenir compte de la dimension du r\'{e}%
seau, ainsi que du nombre de coordination, il convient de traiter les
fonctions de corr\'{e}lation multispin en fonction de ces param\`{e}tres.
Pour y parvenir, la moyenne thermique de la fonction
transcendantale doit \^{e}tre exprim\'{e} comme une moyenne sur un polyn\^{o}me fini de l'op\'{e}%
ration de spin dans un amas de n-site $\left(
n>1\right) $ \cite{61,65}.\newline
Dans cette partie, nous montrerons comment la proc\'{e}dure traditionnelle
de l'obtention des \'{e}quations d'\'{e}tat par la th\'{e}orie du champ
effectif, peut \^{e}tre convertie en un outil moderne pour la construction
d'une cartographie r\'{e}guli\`{e}re du groupe de renormalisation selon les
id\'{e}es de Wilson. En raison de sa relation avec la m\'{e}thode du champ
moyen standard, la d\'{e}nomination du groupe de renormalisation du champ
moyen a \'{e}t\'{e} utilis\'{e}e dans la litt\'{e}rature \cite{66}. Il a \'{e}t\'{e} utilis\'{e}
avec succ\`{e}s pour fournir des indications qualitatives et quantitatives
sur le comportement critique des syst\`{e}mes de spins.\newline
Le principe du groupe de renormalisation ph\'{e}nom\'{e}nologique est bas%
\'{e} sur la comparaison de deux amas de diff\'{e}rentes tailles $N$ et $%
N^{\prime }$ $\left( N>N^{\prime }\right) $. Pour les deux amas, on calcule l'aimantation par site, \`{a} savoir $m_{N}$ et $m_{N^{\prime }}$.
Dans le groupe de renormalisation du champ moyen, cela se fait dans le sch%
\'{e}ma traditionnel du champ moyen, dans lequel les effets des spins autour
de chaque amas sont remplac\'{e}s par de tr\`{e}s petits champs de
rupture de sym\'{e}trie $b$ et $b^{\prime }$, agissants sur les bords de chaque amas \cite{64}.
Les aimantations des amas sont donn\'{e}es par :
\begin{equation}
\left. \frac{\partial m_{N}\left( K,b\right) }{\partial b}\right \vert
_{b=0}=\left. \frac{\partial m_{N^{\prime }}\left( K^{\prime },b^{\prime
}\right) }{\partial b^{\prime }}\right \vert _{b^{\prime }=0},  \label{4-58}
\end{equation}%
Ce rapport donne une relation de r\'{e}currence entre les constantes de
couplage $K$ et $K^{\prime }$ dans les syst\`{e}mes. \`{A} partir de la relation $%
K^{\prime }=K^{\prime }\left( K\right) $, le couplage critique $K_{c}$ peut
\^{e}tre extrait par la r\'{e}solution de l'\'{e}quation du point fixe $%
K^{\ast }=K^{\prime }\left( K^{\ast }\right) $ qui est invariant par
changement d'\'{e}chelle \cite{64}. De surcro%
\^{\i}t, l'exposant critique $\nu $ de la
longueur de corr\'{e}lation $\xi $ est d\'{e}fini par :%
\begin{equation}
\xi \propto \left \vert T-T_{c}\right \vert ^{-\nu },
\end{equation}%
qui peut \^{e}tre \'{e}galement obtenue en lin\'{e}arisant la relation de r%
\'{e}currence au voisinage du point fixe $K^{\ast }$\ :%
\begin{equation}
\left( \frac{\partial K^{\prime }}{\partial K}\right) _{K=K^{\prime
}}=l^{1/\nu },  \label{4-60}
\end{equation}%
o\`{u} $l=\left( N/N^{\prime }\right) ^{1/d}$ est le facteur d'\'{e}chelle
et $d$ est la dimension du syst\`{e}me.\newline Supposons deux amas, l'un avec
un spin $\left( N^{\prime }=1\right) $ et l'autre avec deux spins $\left(
N=2\right) $. Dans l'amas \`{a} un spin, le spin $\mu _{1}$ int\'{e}ragit
avec les sites de plus proches voisins $z_{1}$ via les constantes de
couplage $K_{ij}^{\prime }$. Dans l'amas \`{a} deux-spins,
les spins $\mu _{1}$ et $\mu _{2}$ int\'{e}ragissent directement entre eux via le couplage
$K_{12}$, et int\'{e}ragissent avec leurs sites voisins via les constantes de couplage $K_{1i}$ et $K_{2j}^{\prime }$. Les aimantations
moyennes $m_{N^{\prime }}$\ et $m_{N}$ associ\'{e}s aux amas $N^{\prime
}=1$ et $N=2$\ sont donn\'{e}es par :%
\begin{equation}
m_{N^{\prime }}=\left \langle \mu _{1}\right \rangle =\left \langle \tanh
\left( \sum_{j}K_{1j}^{\prime }\mu _{j}^{\prime }\right) \right \rangle ,
\label{4-61}
\end{equation}%
et%
\begin{equation}
m_{N}=\left \langle \frac{1}{2}\left( \mu _{1}+\mu _{2}\right) \right
\rangle =\left \langle \frac{\sinh \left( u+\upsilon \right) }{\cosh \left(
u+\upsilon \right) +\exp \left( -2K_{12}\right) \cosh \left( u-\upsilon
\right) }\right \rangle ,  \label{4-62}
\end{equation}%
o\`{u} $u=\sum_{j}K_{1j}\mu _{j}$ et $\upsilon =\sum_{j^{\prime
}}K_{2j^{\prime }}\mu _{j^{\prime }}$. Les \'{e}quations (\ref{4-61}) et (%
\ref{4-62}) peuvent \^{e}tre reformul\'{e}es comme :%
\begin{equation}
m_{N^{\prime }}=\left \langle \prod \limits_{j}\exp \left( K_{1j}^{\prime
}\mu _{j}^{\prime }\nabla _{x}\right) \right \rangle \left. f\left( x\right)
\right \vert _{x=0},  \label{4-63}
\end{equation}%
et%
\begin{equation}
\begin{array}{c}
m_{N}=\left \langle \prod \limits_{j}{}^{\prime }\exp \left( K_{1j}\mu
_{j}\nabla \right) \prod \limits_{j^{\prime }}{}^{\prime }\exp \left(
K_{2j^{\prime }}\mu _{j^{\prime }}\nabla _{y}\right) \right. \qquad \qquad \qquad \qquad
\\
\times \left. \prod \limits_{k}{}^{\prime }\exp \left[ \mu _{k}\left(
K_{1k}\nabla _{x}+K_{2k}\nabla _{y}\right) \right] \right \rangle \left.
f\left( x,y\right) \right \vert _{x=0,y=0},%
\end{array}
\label{4-64}
\end{equation}%
o\`{u} $\nabla _{\mu }=\partial /\partial \mu $ $\left( \mu =x\text{ ou }%
y\right) $ est l'op\'{e}rateur diff\'{e}rentiel et les fonctions $f\left(
x\right) $ et $f\left( x,y\right) $ sont d\'{e}finis par \cite{64} :%
\begin{equation}
f\left( x\right) =\tanh x,
\end{equation}%
et%
\begin{equation}
f\left( x,y\right) =\frac{\sinh \left( x+y\right) }{\cosh \left( x+y\right)
+\exp \left( -2K_{12}\right) \cosh \left( x-y\right) }.
\end{equation}%
En introduisant l'approximation de d\'{e}couplage (\ref{4-30}) dans les relations
exactes (\ref{4-63}) et (\ref{4-64}), nous obtenons :
\begin{equation}
m_{N^{\prime }}\left( K^{\prime },b^{\prime }\right) =A_{N^{\prime
}}^{\left( z\right) }\left( K^{\prime }\right) b^{\prime }+O\left( b^{\prime
3}\right) ,
\end{equation}%
et%
\begin{equation}
m_{N}\left( K,b\right) =A_{N}^{\left( z\right) }\left( K\right) b+O\left(
b^{3}\right) ,
\end{equation}%
o\`{u} les coefficients $A_{N^{\prime }}^{\left( z\right) }\left( K^{\prime
}\right) $ et $A_{N}^{\left( z\right) }\left( K\right) $\ sont donn\'{e}s
par :%
\begin{equation}
A_{N^{\prime }}^{\left( z\right) }\left( K^{\prime }\right) =z_{1}\cosh
^{z-1}\left( K\nabla _{x}\right) \left. f\left( x\right) \right \vert _{x=0},
\label{4-69}
\end{equation}%
et%
\begin{equation}
\begin{array}{c}
A_{N}^{\left( z\right) }\left( K\right) =\left \{ 2z^{\prime }\sinh \left(
K\nabla _{x}\right) \cosh ^{z^{\prime }-1}\left( K\nabla _{x}\right) \cosh
^{z^{\prime }}\left( K\nabla _{y}\right) \cosh ^{z^{"}}\left[ K\left( \nabla
_{x}+\nabla _{y}\right) \right] \right. \\
+z"\sinh \left[ K\left( \nabla _{x}+\nabla _{y}\right) \right] \cosh
^{z^{\prime }}\left( K\nabla _{x}\right) \cosh ^{z^{\prime }}\left( K\nabla
_{y}\right) \qquad \qquad \  \  \\
\left. \times \cosh ^{z^{"}-1}\left[ K\left( \nabla _{x}+\nabla _{y}\right) %
\right] \right \} \left. f\left( x,y\right) \right \vert _{x=0,y=0},\qquad
\qquad \qquad \qquad \
\end{array}
\label{4-70}
\end{equation}%
$z^{\prime }$ est le nombre de sites plus proches voisins de $\mu _{1}$ (ou $%
\mu _{2}$) mais qui ne sont pas proches voisins de $\mu _{2}$ (ou $%
\mu _{1}$) et $z"$ repr\'{e}sente le nombre de plus proches voisins de $\mu
_{1}$ et $\mu _{2}$. En combinant l'\'{e}quation (\ref{4-69}) et (\ref{4-70}%
) avec l'hypoth\`{e}se d'\'{e}chelle \cite{64}, on obtient \`{a} partir de l'\'{e}%
quation (\ref{4-58})%
\begin{equation}
A_{N^{\prime }}^{\left( z\right) }\left( K^{\prime }\right) =A_{N}^{\left(
z\right) }\left( K\right) ,
\end{equation}%
qui est la relation de r\'{e}currence entre les constantes de couplage $K$
et $K^{\prime }$ pour les deux syst\`{e}mes $N^{\prime }=1$ et $N=2$.
\vspace{-0.6cm}
\section{M\'{e}thode de la matrice de transfert}

Cette derni\`{e}re section met le point sur les matrices de
transfert qui sont tr\`{e}s utilis\'{e}es pour r\'{e}soudre les mod\`{e}les
de spins classiques. L'id\'{e}e de base est d'\'{e}crire la
fonction de partition en terme d'une matrice appel\'{e}e matrice de transfert. Les propri\'{e}t\'{e}s
thermodynamiques du mod\`{e}le sont ensuite enti\`{e}rement d\'{e}crites par
les valeurs propres de la matrice. En particulier, \`{a} la limite thermodynamique l'\'{e}nergie libre par
spin d\'{e}pend exclusivement de la plus
grande valeur propre et de la longueur de corr\'{e}lation. La m\'{e}thode
de la matrice de transfert s'av\`{e}re tr\`{e}s simple pour obtenir la
solution exacte du mod\`{e}le de spin unidimensionnel poss\'{e}dant un nombre
fini de sites voisins et un nombre fini d'\'{e}tats de spin. Cette m\'{e}thode s'utilise \'{e}galement dans l'\'{e}tude des mod\`{e}les bidimensionnels exactement solubles. Cependant, pour les dimensions sup\'{e}rieures, l'analyse des matrices infinies-dimensionnelles n\'{e}cessite des
techniques math\'{e}matiques extr\'{e}mement sophistiqu\'{e}es \cite{1-14}.
\vspace{-0.6cm}
\subsection{Matrice de transfert}

Le mod\`{e}le d'Ising unidimensionnel dans un champ magn%
\'{e}tique, est l'exemple explicite qui permet de configurer la
matrice de transfert. L'hamiltonien de ce mod\`{e}le s'\'{e}crit :
\begin{equation}
H=-J\sum_{i=0}^{N-1}S_{i}S_{i+1}-h\sum_{i=0}^{N-1}S_{i}.
\end{equation}%
Pour des raisons de commodit\'{e}, nous consid\'{e}rons des conditions aux
limites p\'{e}riodiques \cite{53} qui s'identifient comme $S_{N}\equiv S_{0}$. Le choix des conditions
aux limites devient inutile dans la limite thermodynamique, $N\longrightarrow \infty $.\newline
La fonction de partition est donn\'{e}e par :
\begin{equation}
Z=\sum_{\left \{ S\right \} }e^{\beta J\left( S_{0}S_{1}+S_{1}S_{2}+\ldots
+S_{N-1}S_{0}\right) +\beta h\left( S_{0}+S_{1}+\ldots +S_{N-1}\right) },
\label{5--2}
\end{equation}%
o\`{u} $\left \{ S\right \} $ repr\'{e}sente la trace sur tous les \'{e}tats possibles du syst\`{e}%
me, qui est la somme sur $S_{i}=\pm 1$ pour tous les spins $S_{i}$. Une propri\'{e}t\'{e}
importante de l'\'{e}quation (\ref{5--2}) r\'{e}side dans le fait qu'elle peut \^{e}%
tre repr\'{e}sent\'{e}e par un produit de matrices et r\'{e}organis\'{e}e en
produits de termes d\'{e}pendants seulement des paires de plus proches voisins :
\begin{equation}
Z=\sum \limits_{\left \{ S\right \} }e^{\beta JS_{0}S_{1}+\beta h\left(
S_{0}+S_{1}\right) /2}e^{\beta JS_{1}S_{2}+\beta h\left( S_{1}+S_{2}\right)
/2}\ldots e^{\beta JS_{N-1}S_{0}+\beta h\left( S_{N-1}+S_{0}\right) /2}.
\label{5--3}
\end{equation}%
En terme matricielle, nous avons :%
\begin{equation}
Z=\sum \limits_{\left \{ S\right \} }T_{0,1}T_{1,2}\ldots T_{N-1,0},
\label{5--4}
\end{equation}%
o\`{u}%
\begin{equation}
T_{i,i+1}=e^{\beta JS_{i}S_{i+1}+\beta h\left( S_{i}+S_{i+1}\right) /2},
\end{equation}%
sont les \'{e}l\'{e}ments d'une matrice $T$ dont les lignes sont marqu\'{e}es par les
valeurs de $S_{i}$ alors que les colonnes sont marqu\'{e}es par $S_{i+1}$. Pour le mod\`{e}le consid\'{e}r\'{e}, $T$ prend la forme explicite :
\begin{equation}
\begin{array}{cc}
&
\begin{array}{cc}
{\small S}_{i+1}{\small =1} & {\small S}_{i+1}{\small =-1}%
\end{array}
\\
\begin{array}{c}
{\small S}_{i}{\small =1} \\
{\small S}_{i}{\small =-1}%
\end{array}
& \left(
\begin{array}{cc}
e^{\beta \left( J+h\right) } & e^{-\beta J} \\
e^{-\beta J} & e^{\beta \left( J-h\right) }%
\end{array}%
\right)%
\end{array}%
.  \label{5--6}
\end{equation}
L'\'{e}quation (\ref{5--4}) est facilement simplifi\'{e}e, en notant qu'elle peut s'%
\'{e}noncer comme le produit matriciel s'\'{e}crivant en termes des
composantes de la matrice $T$ \cite{54}. En prenant la trace sur les spins $%
i=1,2,\ldots ,N-1$ qui correspond au produit suivant :
\begin{equation}
Z=\sum_{S_{0}=\pm 1}\left( T^{N}\right) _{0,0},
\end{equation}
de sorte que seule la sommation sur $S_{0}$ des \'{e}l\'{e}ments diagonaux de $T^{N}$ reste. La trace de $T^{N}$ en termes de
valeurs propres $\lambda _{i}$ de $T$ donne :
\begin{equation}
Z=\sum_{i}\lambda _{i}^{N}.  \label{5--8}
\end{equation}%
L'\'{e}quation (\ref{5--8}) est un r\'{e}sultat g\'{e}n\'{e}ral malgr\'{e} le fait qu'elle est obtenue \`{a} partir du mod\`{e}le d'Ising unidimensionnel. La m\'{e}thode de la matrice de transfert est utile lorsque la fonction de
partition prend la forme de l'\'{e}quation (\ref{5--3}) et s'exprime alors comme un produit de matrices.\newline
Les syst\`{e}mes de spins classiques unidimensionnels avec des interactions de port\'{e}e finie est l'une des applications les plus courantes de la m\'{e}thode de la matrice de transfert \cite{53}. Dans ce cas, la taille de la matrice de transfert d\'{e}%
pend du nombre d'\'{e}tats de spin par site et de la port\'{e}e des
interactions. Par cons\'{e}quent, le mod\`{e}le devient plus compliqu\'{e}, l'utilit\'{e}
du formalisme d\'{e}pend du fait que la matrice de transfert puisse \^{e}tre
diagonalis\'{e}e analytiquement ou num\'{e}riquement.
\vspace{-0.7cm}
\subsection{\'{E}nergie libre}
\vspace{-0.1cm}
La puissance du formalisme de la matrice de transfert devient apparente dans
la formule de l'\'{e}nergie libre \cite{53}. En effet, pour une matrice de
transfert g\'{e}n\'{e}rale de taille $n\times n$, les valeurs propres rang\'{e}es
par module d\'{e}croissant sont \'{e}tiquet\'{e}es $\lambda
_{0},\lambda _{1},\lambda _{2}\ldots \lambda _{n-1}$. \`{A} la limite thermodynamique, l'\'{e}nergie libre par
spin est donn\'{e}e par :
\begin{equation}
\begin{array}{ccc}
f & = & -k_{B}T\lim_{N\rightarrow \infty }\frac{1}{N}\ln Z_{N}\qquad \qquad
\qquad \qquad \ \\
& = & -k_{B}T\lim_{N\rightarrow \infty }\frac{1}{N}\ln \left \{ \lambda
_{0}^{N}\left( 1+\sum \limits_{i}\frac{\lambda _{i}^{N}}{\lambda _{0}^{N}}%
\right) \right \}. \  \  \
\end{array}%
\end{equation}
Pour $N\rightarrow \infty $ et $\left( \lambda _{i}/\lambda _{0}\right)
^{N}\rightarrow 0$, l'\'{e}nergie $f$ devient :
\begin{equation}
f=-k_{B}T\ln \lambda _{0}.  \label{5--11}
\end{equation}%
Ce r\'{e}sultat est d'une grande importance, car il est souvent beaucoup plus facile de
calculer $\lambda _{0}$ que de calculer l'ensemble des valeurs propres d'une matrice.\newline
De plus, le fait que les matrices de transfert appartiennent \`{a} une classe de matrices non d\'{e}g\'{e}n\'{e}r\'{e}e, fait tomber la d\'{e}g\'{e}n\'{e}rescence de $\lambda_{0}$. La plus grande
valeur propre positive $\lambda _{0}$, donne ainsi une \'{e}nergie libre physiquement
sensible \cite{crit3}. Nous avons suppos\'{e} que les $\lambda
_{i}$ sont r\'{e}els ; ce qui n'est pas n%
\'{e}cessairement le cas pour $i\neq 0$, mais la formule (\ref{5--11}) reste toujours valable.
\vspace{-0.8cm}
\subsection{Fonction de corr\'{e}lation}

La longueur de corr\'{e}lation est une autre quantit\'{e} importante qui est aussi li\'{e}e aux valeurs propres
de la matrice de transfert. Pour calculer
cette quantit\'{e}, nous avons besoin de la fonction de corr\'{e}lation
spin/spin servant \`{a} obtenir des moyennes de produits de spins en
utilisant des matrices de transfert \cite{66}. La fonction de corr\'{e}lation \`{a}
deux spins $\Gamma _{kl}$, et la longueur de corr\'{e}lation $\xi $ sont donn\'{e}es par :
\begin{eqnarray}
\Gamma _{kl} &=&\left( \left \langle S_{k}S_{l}\right \rangle -\left \langle
S_{k}\right \rangle \left \langle S_{l}\right \rangle \right) ,
\label{5--12} \\
\xi ^{-1} &=&\lim_{r_{kl}\longrightarrow \infty }\left \{ -\frac{1}{r_{kl}}\ln \left
\vert \left \langle S_{k}S_{l}\right \rangle -\left \langle S_{k}\right
\rangle \left \langle S_{l}\right \rangle \right \vert \right \} ,
\label{5--13}
\end{eqnarray}%
o\`{u} $S_{k}$ et $S_{l}$ repr\'{e}sentent respectivement les spins du site $k$ et du site $l$, et $r_{kl}$ d\'{e}note la distance entre spin $S_{k}$ et spin $S_{l}$.\newline
Pour un anneau de $N$ spins, les spins satisfont :
\begin{equation}
\left \langle S_{k}S_{l}\right \rangle _{N}=\frac{\sum \limits_{\left \{
S\right \} }S_{k}S_{l}e^{-\beta H}}{\sum \limits_{\left \{ S\right \}
}e^{-\beta H}}\equiv \frac{1}{Z}\sum_{\left \{ S\right \}
}S_{k}S_{l}e^{-\beta H},
\end{equation}%
o\`{u}%
\begin{equation}
\begin{array}{ccc}
\sum \limits_{\left \{ S\right \} }S_{k}S_{l}e^{-\beta H} & = & \sum
\limits_{\left \{ S\right \} }S_{k}T_{S_{k-1}S_{k}}T_{S_{k}S_{k+1}}\ldots
T_{S_{l}S_{l-1}}S_{l}T_{S_{l}S_{l+1}}\ldots T_{N-1,0} \\
& = & \sum \limits_{S_{k}S_{l}}S_{k}\left( T^{l-k}\right) _{S_{k}S_{l}}S_{l}\left(
T^{N-l+k}\right) _{S_{l}S_{k}}.\qquad \qquad \qquad \qquad%
\end{array}
\label{5--15}
\end{equation}%
Soit $\left \vert \vec{u}_{i}\right \rangle $ les vecteurs propres qui correspondent aux valeurs propres $\lambda _{i}$ de la
matrice $T$, o\`{u} $%
i=0,1,2,\ldots n-1$. La matrice de transfert $T$ est d\'{e}finit par :
\begin{equation}
T=\sum_{i}\left \vert \vec{u}_{i}\right \rangle \lambda _{i}\left \langle \vec{u%
}_{i}\right \vert ,
\end{equation}
et%
\begin{equation}
\left( T^{l-k}\right) _{S_{k}S_{l}}=\sum_{i}\left \langle \vec{S}_{k}{\Large |}\vec{u}%
_{i}\right \rangle \lambda _{i}^{l-k}\left \langle \vec{u}_{i}{\Large |}\vec{S}%
_{l}\right \rangle ,
\end{equation}%
La matrice diagonale $s_{l}$ dont les valeurs propres sont celles de $S_{l}$, correspondant aux
vecteurs propres $\left \langle \vec{S}_{l}\right \vert $ \cite{53} est donn\'{e}e par :
\begin{equation}
s_{l}=\sum_{S_{l}}\left \vert \vec{S}_{l}\right \rangle S_{l}\left \langle \vec{%
S}_{l}\right \vert .  \label{5--16}
\end{equation}
En termes de ces param\`{e}tres, l'\'{e}quation (\ref{5--15}) devient :%
\begin{equation}
\begin{array}{ccc}
\sum \limits_{\left \{ S\right \} }S_{k}S_{l}e^{-\beta H} & = &
\sum \limits_{S_{k}S_{l}}\sum \limits_{i,j}S_{k}\left \langle \vec{S}_{k}{\Large |}%
\vec{u}_{i}\right \rangle \lambda _{i}^{l-k}\left \langle \vec{u}_{i}{\Large |}%
\vec{S}_{l}\right \rangle S_{l}\left \langle \vec{S}_{l}{\Large |}\vec{u}%
_{j}\right \rangle \lambda _{j}^{N-l+k}\left \langle \vec{u}_{j}{\Large |}\vec{%
S}_{k}\right \rangle  \\
& = & \sum \limits_{i,j}\left \langle \vec{u}_{j}{\Large |}s_{k}{\Large |}\vec{u}%
_{i}\right \rangle \lambda _{i}^{l-k}\left \langle \vec{u}_{i}{\Large |}s_{l}%
{\Large |}\vec{u}_{j}\right \rangle \lambda _{j}^{N-l+k}.\qquad \qquad \qquad
\  \  \  \  \  \  \  \
\end{array}%
\end{equation}
En utilisant l'\'{e}quation (\ref{5--8}), nous obtenons :%
\begin{equation}
\left \langle S_{k}S_{l}\right \rangle _{N}=\frac{\sum \limits_{i,j}\left
\langle \vec{u}_{j}{\Large |}s_{k}{\Large |}\vec{u}_{i}\right \rangle \left(
\frac{\lambda _{i}}{\lambda _{0}}\right) ^{l-k}\left \langle \vec{u}_{i}%
{\Large |}s_{l}{\Large |}\vec{u}_{j}\right \rangle \left( \frac{\lambda _{j}%
}{\lambda _{0}}\right) ^{N-l+k}}{\sum \limits_{k'}\left( \frac{\lambda _{k'}}{%
\lambda _{0}}\right) ^{N}},
\end{equation}%
o\`{u} nous avons divis\'{e} par $\lambda _{0}$. Il est alors \'{e}vident qu'\`{a} la
limite thermodynamique, seuls les termes en $j=0$ et $k'=0$ subsistent, ainsi :
\begin{equation}
\begin{array}{ccc}
\left \langle S_{k}S_{l}\right \rangle & = & \lim_{N\longrightarrow \infty }\left \langle S_{k}S_{l}\right \rangle _{N}\qquad \qquad \qquad
\qquad \qquad \  \  \  \\
& = & \left \langle S_{k}\right \rangle \left \langle S_{l}\right \rangle
+\sum \limits_{i\neq 0}\left( \frac{\lambda _{i}}{\lambda _{0}}\right)
^{l-k}\left \langle \vec{u}_{0}{\Large |}s_{k}{\Large |}\vec{u}_{i}\right
\rangle \left \langle \vec{u}_{i}{\Large |}s_{l}{\Large |}\vec{u}_{0}\right
\rangle ,%
\end{array}%
\end{equation}%
o\`{u}, nous avons employ\'{e} :
\begin{equation}
\left \langle S_{l}\right \rangle =\left \langle \vec{u}_{0}{\Large |}s_{l}%
{\Large |}\vec{u}_{0}\right \rangle . \label{5--24}
\end{equation}%
Finalement, la fonction de corr\'{e}lation de l'\'{e}quation (\ref{5--12}) est de la forme :
\begin{equation}
\Gamma _{kl}=\sum \limits_{i\neq 0}\left( \frac{\lambda _{i}}{\lambda _{0}}%
\right) ^{l-k}\left \langle \vec{u}_{0}{\Large |}s_{k}{\Large |}\vec{u}%
_{i}\right \rangle \left \langle \vec{u}_{i}{\Large |}s_{l}{\Large |}\vec{u}%
_{0}\right \rangle .  \label{5--25}
\end{equation}%
L'\'{e}quation (\ref{5--25}) montre que la fonction de corr\'{e}lation d\'{e}pend de toutes les valeurs et les vecteurs propres de
la matrice de transfert. Une formule plus simple est obtenue pour la
longueur de corr\'{e}lation de l'\'{e}quation (\ref{5--13}) en consid\'{e}rant la limite $%
r_{kl}\longrightarrow \infty $, o\`{u} $i=1$ domine
la somme dans l'\'{e}quation (\ref{5--25}). Dans ce cas :
\begin{equation}
\begin{array}{ccc}
\xi ^{-1} & = & \lim_{r_{kl}\longrightarrow \infty }-\frac{1}{r_{kl}}\ln \left \{
\left( \frac{\lambda _{1}}{\lambda _{0}}\right) ^{l-k}\left \langle \vec{u}_{0}%
{\Large |}s_{k}{\Large |}\vec{u}_{1}\right \rangle \left \langle \vec{u}_{1}%
{\Large |}s_{l}{\Large |}\vec{u}_{0}\right \rangle \right \} \\
& = & -\ln \left( \lambda _{1}/\lambda _{0}\right) .\qquad \qquad \qquad
\qquad \qquad \qquad \qquad \qquad%
\end{array}%
\end{equation}%
Remarquons que l'hamiltonien consid\'{e}r\'{e} est invariant par translation. Par cons\'{e}%
quent, le produit des \'{e}l\'{e}ments de la matrice dans l'\'{e}quation (\ref%
{5--25}) peut \^{e}tre r\'{e}\'{e}crit sous la forme suivante :
\begin{equation}
\left \langle \vec{u}_{0}{\Large |}s_{k}{\Large |}\vec{u}_{i}\right \rangle
\left \langle \vec{u}_{i}{\Large |}s_{l}{\Large |}\vec{u}_{0}\right \rangle
=\left \vert \left \langle \vec{u}_{i}{\Large |}s_{k}{\Large |}\vec{u}%
_{0}\right \rangle \right \vert ^{2}.
\end{equation}%
\vspace{-1.8cm}
\subsection{Cas du mod\`{e}le d'Ising}

\begin{figure}[!ht]
  \centering
\includegraphics[scale=0.25]{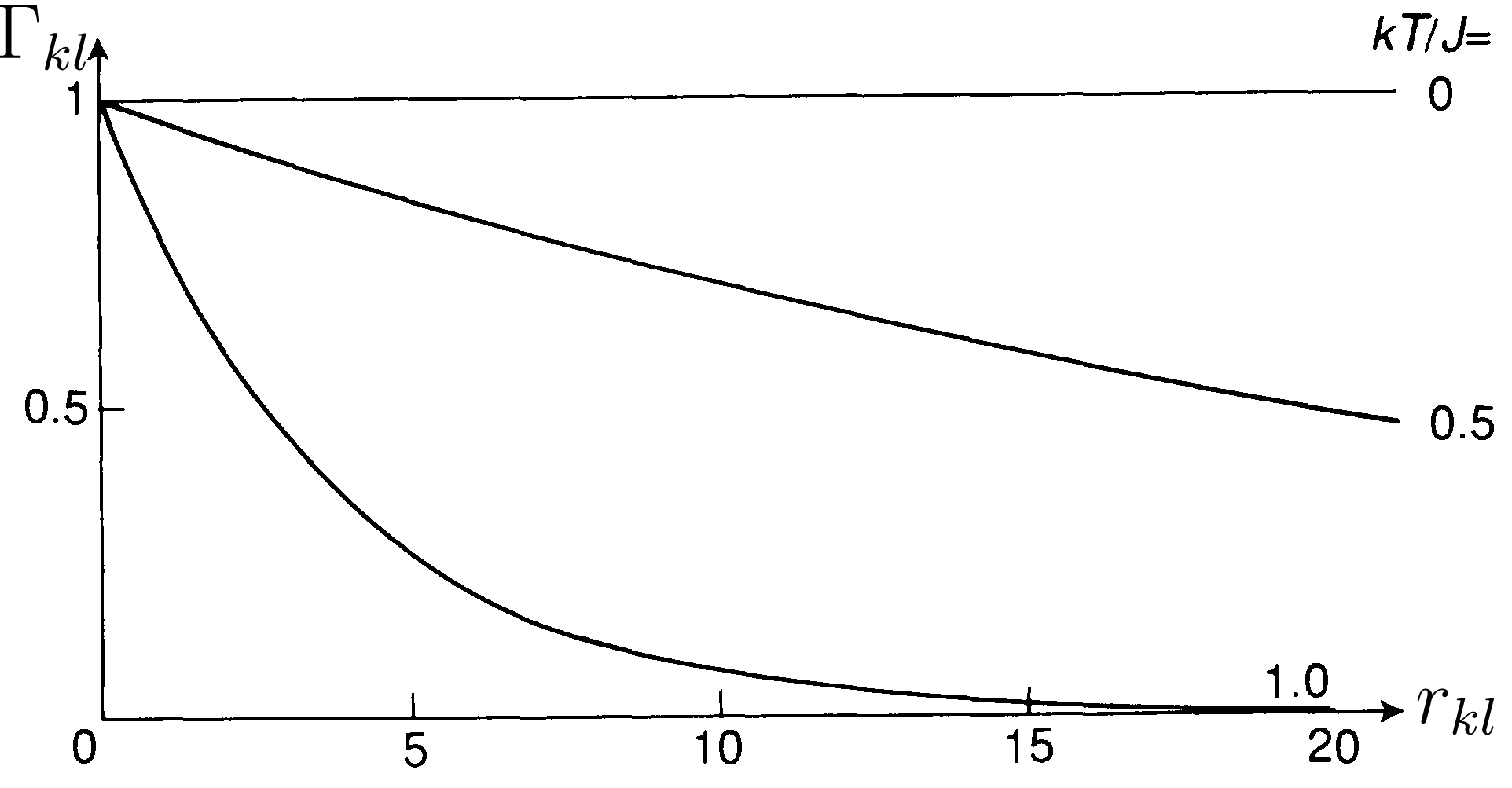}
\caption{\label{corre}La d\'{e}pendance de la fonction de corr\'{e}lation spin/spin du
mod\`{e}le d'Ising unidimensionnel en fonction de la distance et de la temp%
\'{e}rature en absence du champ magn\'{e}tique \cite{54}.}
\end{figure}

Pour le cas du mod\`{e}le d'Ising dans un champ magn\'{e}tique, la diagonalisation de la matrice de l'\'{e}quation (\ref{5--6}) donne les vecteurs propres :
\begin{equation}
\left \langle \vec{u}_{0}\right \vert =\left( \alpha _{+},\alpha _{-}\right)
,\qquad \left \langle \vec{u}_{1}\right \vert =\left( \alpha _{-},-\alpha
_{+}\right) ,
\end{equation}
et les valeurs propres :
\begin{equation}
\lambda _{0,1}=e^{\beta J}\cosh \beta h\pm \sqrt{e^{2\beta J}\sinh ^{2}\beta
h+e^{-2\beta J}},  \label{5--29}
\end{equation}
o\`{u}%
\begin{equation}
\alpha _{\pm }^{2}=\frac{1}{2}\left( 1\pm \frac{e^{\beta J}\sinh \beta h}{%
\sqrt{e^{2\beta J}\sinh ^{2}\beta h+e^{-2\beta J}}}\right) .  \label{5--31}
\end{equation}%
Ces \'{e}quations (\ref{5--29}) et (\ref{5--31}) permettent de donner les formes explicites des quantit\'{e}s physiques suivantes : l'\'{e}nergie libre par spin $f$, l'aimantation par spin $%
\left \langle S\right \rangle $, la fonction de
corr\'{e}lation $\Gamma $, et la longueur de corr\'{e}lation $\xi $ \cite{54}.\newline
$\bullet$ \textbf{\emph{\'{E}nergie libre}}
\newline
\`{A} partir des deux \'{e}quations (\ref{5--11}) et (\ref{5--29}), l'\'{e}nergie libre prend la forme :
\begin{equation}
f=-k_{B}T\ln \left \{ e^{\beta J}\cosh \beta h+\sqrt{e^{2\beta J}\sinh
^{2}\beta h+e^{-2\beta J}}\right \} .
\end{equation}%
Puisque $\beta \longrightarrow \infty $, alors l'\'{e}nergie par spin s'\'{e}crit :
\begin{equation}
f\longrightarrow -k_{B}T\ln \left \{ e^{\beta J}\left( \cosh \beta h+\sinh
\beta h\right) \right \} =-J-h,
\end{equation}%
$\bullet$ \textbf{\emph{Aimantation}}
\newline
L'aimantation peut \^{e}tre obtenue, soit en diff\'{e}renciant la forme n%
\'{e}gative de l'\'{e}nergie libre par rapport au champ magn\'{e}tique $h$,
soit \`{a} partir de l'\'{e}quation (\ref{5--24}). Rappelons que :
\begin{equation}
\begin{array}{ccc}
\left \langle S\right \rangle & = & \left( \alpha _{+},\alpha _{-}\right)
\left(
\begin{array}{cc}
1 & 0 \\
0 & -1%
\end{array}%
\right) \left(
\begin{array}{c}
\alpha _{+} \\
\alpha _{-}%
\end{array}%
\right) \\
& = & \frac{e^{\beta J}\sinh \beta h}{\sqrt{e^{2\beta J}\sinh ^{2}\beta
h+e^{-2\beta J}}}.\qquad \qquad \qquad \  \  \
\end{array}%
\end{equation}%
Pour les spins sans interaction $J=0$ nous avons $%
T=\infty $, d'o\`{u} :
\begin{equation}
\left \langle S\right \rangle =\tanh \beta h,
\end{equation}%
comme pr\'{e}vu pour un mat\'{e}riau paramagn\'{e}tique.\newline
En absence du champ magn\'{e}tique ext\'{e}rieur et \`{a} une temp\'{e}rature finie
quelconque $\left \langle S\right \rangle =0$, comme attendu \`{a} partir de la sym\'{e}trie du mod\`{e}le, en
excluant que la temp\'{e}rature soit nulle et que le champ magn\'{e}tique ext%
\'{e}rieur que l'on applique soit fini ou nul \cite{54}, nous avons :
\begin{equation}
\lim_{h\longrightarrow 0^{\pm }}\lim_{T\longrightarrow 0}\left \langle
S\right \rangle =\pm 1.
\end{equation}%
Ce r\'{e}sultat confirme l'existence d'une transition de phase \`{a} temp\'{e}%
rature nulle vers un \'{e}tat fondamental totalement ordonn\'{e}.
\newline
$\bullet$ \textbf{\emph{Fonction de corr\'{e}lation}}
\newline
L'\'{e}quation (\ref{5--25}) nous permet d'\'{e}crire la fonction de corr\'{e}lation comme suit :%
\begin{equation}
\Gamma _{kl}=\left( \frac{\lambda _{1}}{\lambda _{0}}\right) ^{l-k}\frac{%
e^{-2\beta J}}{e^{2\beta J}\sinh ^{2}\beta h+e^{-2\beta J}}.
\end{equation}%
En absence du champ magn\'{e}tique, cette \'{e}quation devient :
\begin{equation}
\Gamma _{kl}\left( h=0\right) =\tanh ^{l-k}\beta J.
\end{equation}%
La figure (\ref{corre}) repr\'{e}sente la fonction de corr\'{e}lation en fonction de $r_{kl}$ pour diff\'{e}rentes valeurs de temp\'{e}rature.
\newline
Il est important de noter que pour tout $T\neq 0$, la fonction $\Gamma _{kl}$ d\'{e}cro\^{\i}t avec $r_{kl}$. De plus, si le couplage est
antiferromagn\'{e}tique $\left( J<0\right) $, la fonction de corr\'{e}lation change de signe
pour les valeurs impaires de $r_{kl}$.
\newline
$\bullet$ \textbf{\emph{Longueur de corr\'{e}lation}}
\newline
\`{A} partir de l'\'{e}quation (\ref{5--29}), nous d\'{e}duisons la longueur de corr\'{e}lation du syst\`{e}me :
\begin{equation}
\xi ^{-1}=-\ln \left \{ \frac{e^{\beta J}\cosh \beta h-\sqrt{e^{2\beta
J}\sinh ^{2}\beta h+e^{-2\beta J}}}{e^{\beta J}\cosh \beta h+\sqrt{e^{2\beta
J}\sinh ^{2}\beta h+e^{-2\beta J}}}\right \} ,
\end{equation}
lorsque $T\rightarrow0$, la longueur de corr\'{e}lation $\xi ^{-1}\rightarrow0$ c'est-\`{a}-dire la longueur de corr\'{e}lation diverge ce qui signale
l'apparition d'un point critique \cite{66}.\newline
Pour conclure ce chapitre, rappelons que les m\'{e}thodes de la physique
statistique nous ont permis d'expliquer et de d\'{e}crire quantitativement
les ph\'{e}nom\`{e}nes collectifs dans les mat\'{e}riaux. Elles nous ont
\'{e}galement permis de comprendre les transitions de phase et les ph\'{e}nom%
\`{e}nes critiques qui appara\^{\i}ssent dans les mat\'{e}riaux.\newline
Les objectifs les plus fondamentaux du d\'{e}veloppement des m\'{e}thodes de
la physique statistique comme la th\'{e}orie du champ moyen, la th\'{e}orie
du champ effectif et la m\'{e}thode de la matrice de transfert, ainsi que la m%
\'{e}thode Monte Carlo sont l'\'{e}tude et le contr\^{o}le des propri\'{e}t%
\'{e}s magn\'{e}tiques dans les mat\'{e}riaux. \`{A} ce tire, il nous a sembl%
\'{e} int\'{e}ressant de consacrer le prochain chapitre \`{a} la d\'{e}%
finition des notions de base en magn\'{e}tisme avant de pr\'{e}senter nos
contributions dans ce domaine.

\def\cleardoublepage{\clearpage}
\newpage
\strut
\newpage
\setcounter{chapter}{2}
\chapter{Formalisme des propri\'{e}t\'{e}s magn\'{e}tiques et hyst\'{e}r\'{e}tiques des mat%
\'{e}riaux}
       \graphicspath{{Chapitre2/figures/}}
\setcounter{chapter}{3}
Dans les mat\'{e}riaux, les propri\'{e}t\'{e}s magn\'{e}tiques sont dues aux moments magn\'{e}tiques des
particules les constituant. En effet, les atomes, qui forment la mati\`{e}re, sont compos\'{e}s de noyaux autour desquel gravitent des \'{e}lectrons. Le mouvement suppl\'{e}mentaire des \'{e}lectrons, d\^{u} \`{a} l'application d'un champ magn\'{e}tique ext\'{e}rieur, cr\'{e}e les moments magn\'{e}tiques.\newline
Il est vrai que tous les mat\'{e}riaux sont influenc\'{e}s, de mani\`{e}re plus ou moins complexe, par la pr\'{e}sence d'un champ magn\'{e}tique \cite{1-1}, cependant, ils peuvent r\'{e}agir diff\'{e}remment \`{a} sa pr%
\'{e}sence. Ainsi, l'\'{e}tat magn\'{e}tique d'un mat\'{e}riau ne d%
\'{e}pend pas seulement du champ magn\'{e}tique, mais d\'{e}pend \'{e}galement de plusieurs autres param\`{e}tres tel la temp\'{e}rature, la pression ainsi que les interactions entre les atomes. Par ailleurs, l'ordre magn\'{e}tique dans les mat\'{e}riaux r\'{e}sulte de l'existence des interactions magn\'{e}tiques entre les atomes dans le syst\`{e}me. Ces interactions se traduisent par la naissance d'une force magn\'{e}tique qui a tendance \`{a} aligner les moments magn\'{e}tiques. Il existe diff\'{e}rents types d'interactions. Elles sont directes si elles s'exercent sans interm\'{e}diaire entre porteurs des moments magn\'{e}tiques et indirectes si elles sont m\'{e}di\'{e}es soit par les \'{e}lectrons de conduction du cristal dans lequel les impuret\'{e}s sont introduites tel est le cas pour l'interaction RKKY, soit par l'interm\'{e}diaire d'un ion non-magn\'{e}tique situ\'{e} entre les ions magn\'{e}tiques comme c'est le cas dans l'interaction super\'{e}change.\newline
Ce chapitre vise l'\'{e}tude des propri\'{e}t\'{e}s magn%
\'{e}tiques et hyst\'{e}r\'{e}tiques des mat\'{e}riaux. Dans la premi\`{e}re section, nous allons pr\'{e}senter bri\`{e}vement l'origine du magn\'{e}tisme puis nous allons citer \`{a} ce titre les diff\'{e}rents types des mat\'{e}riaux class\'{e}s selon leur comportement magn\'{e}tique. Nous d\'{e}crirons \'{e}galement les diff%
\'{e}rents types de temp\'{e}rature de transition ainsi que la
classification de N\'{e}el. La seconde section d\'{e}taillera les
interactions magn\'{e}tiques tandis que la troisi\`{e}me section \'{e}voquera quelques types d'anisotropie magn\'{e}tique. Enfin, nous cl\^{o}turons ce chapitre par une pr\'{e}sentation d\'{e}taill\'{e}e des propri\'{e}t\'{e}s hyst%
\'{e}r\'{e}tiques.
\vspace{-0.5cm}
\section{Propri\'{e}t\'{e}s magn\'{e}tiques}

\subsection{Origine du magn\'{e}tisme}

\begin{figure}[!ht]
  \centering
\includegraphics[scale=0.4]{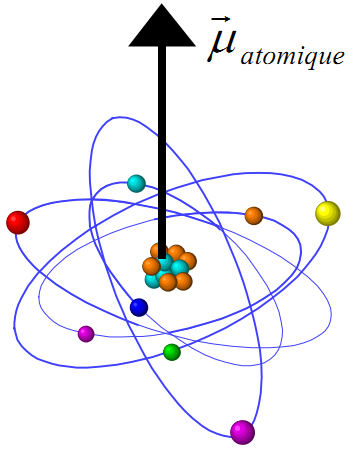}
\caption{\label{pic1}Trajectoire d'un \'{e}lectron autour du noyau \cite{I1}.}
\end{figure}

\begin{figure}[!ht]
  \centering
\includegraphics[scale=0.25]{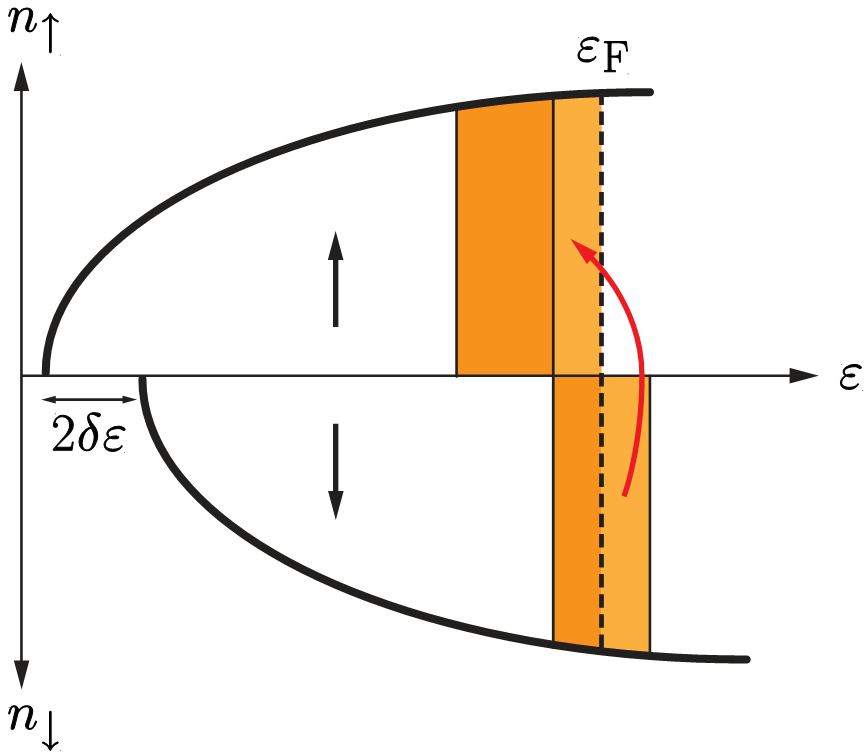}
\caption{\label{pic2}\'{E}volution des populations de la bande $3d$ dans le mod\`{e}le de
Stoner. $\left( \uparrow \right) $ et $%
\left( \downarrow \right) $ repr\'{e}sentent respectivement le spin des \'{e}lectrons up et
down \cite{c4}.}
\end{figure}

L'explication et l'interpr\'{e}tation de l'\'{e}volution de la structure magn\'{e}tique d'un mat\'{e}riau, imposent la n\'{e}cessit\'{e} de remonter \`{a} l'\'{e}chelle atomique puis progressivement de passer \`{a} l'\'{e}chelle de l'arrangement d'atomes et enfin du cristal. Pour comprendre l'origine du magn\'{e}tisme de l'atome, il faut \'{e}tudier le moment
magn\'{e}tique atomique r\'{e}sultant des \'{e}lectrons non appari\'{e}s occupant les orbitales localis\'{e}es des couches incompl\`{e}tes. Ce moment
magn\'{e}tique est d\'{e}termin\'{e} par le moment cin\'{e}tique \cite{I4,S1}. Plus explicitement, chaque \'{e}lectron individuel dans un atome est caract\'{e}ris\'{e} par un moment cin\'{e}tique $L$ associ\'{e} \`{a} son mouvement orbital et un moment
cin\'{e}tique intrins\`{e}que ou de spin $S$. Par cons\'{e}quent, il y a deux
sources de moment magn\'{e}tique atomique. Le premier est le courant associ%
\'{e} au mouvement orbital des \'{e}lectrons et le deuxi\`{e}me le spin de l'%
\'{e}lectron (voir la figure (\ref{pic1})).
\newline
Dans un atome \`{a} $n-$\'{e}lectrons, le moment cin\'{e}tique orbital est la
somme des moments orbitaux de tous les \'{e}lectrons. Le m\^{e}me
raisonnement est valable pour les moments cin\'{e}tiques de spin \cite{c1,c2,c3,c4}. De ce
fait, les moments magn\'{e}tiques orbitaux et de spin peuvent \^{e}tre exprim%
\'{e}s par :%
\begin{equation}
\begin{array}{c}
\vec{\mu}_{L}=-\mu _{B}\vec{L},\text{ \ } \\
\vec{\mu}_{S}=-2\mu _{B}\vec{S},%
\end{array}%
\end{equation}%
o\`{u} $\mu _{B}$ est le magn\'{e}ton de Bohr. Le moment magn\'{e}tique total $\mu $ est alors
donn\'{e} par :
\begin{equation}
\vec{\mu}=-\mu _{B}\left( \vec{L}+2\vec{S}\right) .
\end{equation}%
Pour un atome \`{a} $n-$\'{e}lectrons, les moments r\'{e}sultants $\vec{L}$
et $\vec{S}$ se combinent pour donner le moment cin\'{e}tique total $\vec{J}$ de l'atome \`{a} travers
le couplage spin-orbit. En utilisant les r\`{e}gles de Hund, nous pouvons
calculer le moment magn\'{e}tique d'un atome isol\'{e}. Lorsque les atomes
forment un solide ou une mol\'{e}cule, la situation est diff\'{e}rente de
celle du magn\'{e}tisme dans un atome isol\'{e}. Dans ce cas, la formation
de liaisons chimiques modifie en g\'{e}n\'{e}ral la structure \'{e}%
lectronique du syst\`{e}me de sorte que son magn\'{e}tisme dispara\^{\i}t.
Par cons\'{e}quent, seuls les solides constitu\'{e}s d'atomes avec des
couches incompl\`{e}tes, faiblement affect\'{e}s par les liaisons chimiques
sont magn\'{e}tiques. \`{A} titre d'exemple ; les compos\'{e}s des \'{e}l%
\'{e}ments de terres rares d\'{e}crites par le mod\`{e}le de Bohr. Dans ce mod\`{e}le, les sous couches $4f$ ne sont pas (ou peu) influenc\'{e}es par la pr\'{e}sence des atomes voisins puisqu'elles sont incompl\`{e}tes et leurs fonctions d'onde sont tr\`{e}s localis\'{e}es. Au contraire, les propri\'{e}t\'{e}s magn\'{e}%
tiques des m\'{e}taux $3d$ ne peuvent pas \^{e}tre d\'{e}crits par le mod\`{e}le de Bohr,
parce que les sous-couches $3d$ incompl\`{e}tes est impliqu\'{e}es dans les
liaisons chimiques. N\'{e}anmoins, ils peuvent \^{e}tre formul\'{e}s par le
mod\`{e}le du magn\'{e}tisme itin\'{e}rant ( mod\`{e}le de Stoner) \cite{c2,c3}.\newline
Dans le cas des m\'{e}taux $3d$, les fonctions d'onde $3d$ responsables des propri%
\'{e}t\'{e}s magn\'{e}tiques sont relativement plus \'{e}tendues et
fortement influenc\'{e}es par le champ cristallin. Il en r\'{e}sulte que les
moments orbitaux bloqu\'{e}s par la pr\'{e}sence des atomes voisins
n'interviennent pas dans les propri\'{e}t\'{e}s magn\'{e}tiques. Par cons%
\'{e}quent, leur magn\'{e}tisme r\'{e}sulte d'un d\'{e}s\'{e}quilibre entre
les populations de spin $1/2$ et de spin $-1/2$. Ces moments orbitaux cr\'{e}ent une aimantation d\'{e}%
finie par $M=\frac{1}{V}\sum_{i}M_{i}$ o\`{u} la somme porte sur tous les moments magn\'{e}tiques
atomiques $M_{i}$ du syst\`{e}me de volume $V$. Le mod\`{e}le de Stoner donne une description de l'origine de ce d%
\'{e}s\'{e}quilibre \cite{c5}. En effet, dans le mod\`{e}le de Stoner, les \'{e}lectrons $3d$ sont consid\'{e}r\'{e}s comme des \'{e}%
lectrons libres avec une densit\'{e} d'\'{e}tat parabolique divis\'{e}e en
deux branches : une pour les spins-up et l'autre pour les spins-down comme illustr\'{e} sur la figure (\ref{pic2}).
\newline
L'explication de Stoner s'appuie sur le principe d'exclusion de Pauli. \`{A} savoir, deux \'{e}lectrons avec le m\^{e}me spin ne peuvent pas exister dans
le m\^{e}me \'{e}tat quantique et donc dans la m\^{e}me r\'{e}gion de
l'espace. En cons\'{e}quence, la r\'{e}pulsion coulombienne entre deux \'{e}%
lectrons de spins oppos\'{e}s, qui peuvent se rapprocher les uns des autres, est sup\'{e}rieure \`{a} celle entre deux \'{e}%
lectrons de m\^{e}me spin. Dans le mod\`{e}le de Stoner, cet effet est pris en compte par une
\'{e}nergie potentielle d'interaction entre $N$ \'{e}lectrons donn\'{e}e par $\eta N_{\uparrow }N_{\downarrow }$ o\`{u} $\eta $ exprime la diff\'{e}rence de r\'{e}pulsion entre deux \'{e}lectrons de
spins identiques et oppos\'{e}s et $N_{\uparrow }\left( N_{\downarrow }\right) $ repr\'{e}sente le nombre d'\'{e}lectrons
de spin up (down). Dans ce cas, nous pouvons calculer la diff\'{e}rence d'\'{e}nergie entre un \'{e}tat
non-magn\'{e}tique, o\`{u} les deux branches de la densit\'{e} d'\'{e}tats
sont sym\'{e}triques, et un \'{e}tat magn\'{e}tique o\`{u} les deux branches
sont l\'{e}g\`{e}rement d\'{e}cal\'{e}es par une \'{e}nergie $2\delta \varepsilon $ \cite{S1}, comme le montre la figure (\ref{pic2}). En raison de ce d\'{e}calage, la quantit\'{e} $n\left( \varepsilon _{F}\right) \delta\varepsilon $ de spins-down devient
up. Dans cette expression, $n\left( \varepsilon _{F}\right)
$ repr\'{e}sente la densit\'{e} d'\'{e}tat au niveau de Fermi. Par cons%
\'{e}quent, la variation d'\'{e}nergie cin\'{e}tique est donn\'{e}e par :
\begin{equation}
\Delta E_{c}=n\left( \varepsilon _{F}\right) \left( \delta \varepsilon
\right) ^{2},
\end{equation}%
et la variation de l'\'{e}nergie potentielle se d\'{e}finit par :
\begin{equation}
\Delta E_{p}=\eta \left[ \frac{N}{2}+n\left( \varepsilon _{F}\right) \delta
\varepsilon \right] \left[ \frac{N}{2}-n\left( \varepsilon _{F}\right)
\delta \varepsilon \right] -\eta \left( \frac{N}{2}\right) ^{2}=-\eta \left[
n\left( \varepsilon _{F}\right) \delta \varepsilon \right] ^{2}.
\end{equation}%
Finalement, la variation de l'\'{e}nergie totale du syst\`{e}me est de la forme :
\begin{equation}
\Delta E=n\left( \varepsilon _{F}\right) \left( \delta \varepsilon \right)
^{2}\left[ 1-\eta n\left( \varepsilon _{F}\right) \right] .
\end{equation}%
Au cas o\`{u} $\eta n\left( \varepsilon_{F}\right) >1$ nous trouvons $\Delta E<0$ qui est la condition pour que l'\'{e}tat magn\'{e}tique soit stable selon le crit\`{e}re de stabilit\'{e} de Stoner.
\vspace{-0.4cm}
\subsection{Classification magn\'{e}tiques des mat\'{e}riaux}

\begin{figure}[!ht]
  \centering
\includegraphics[scale=0.3]{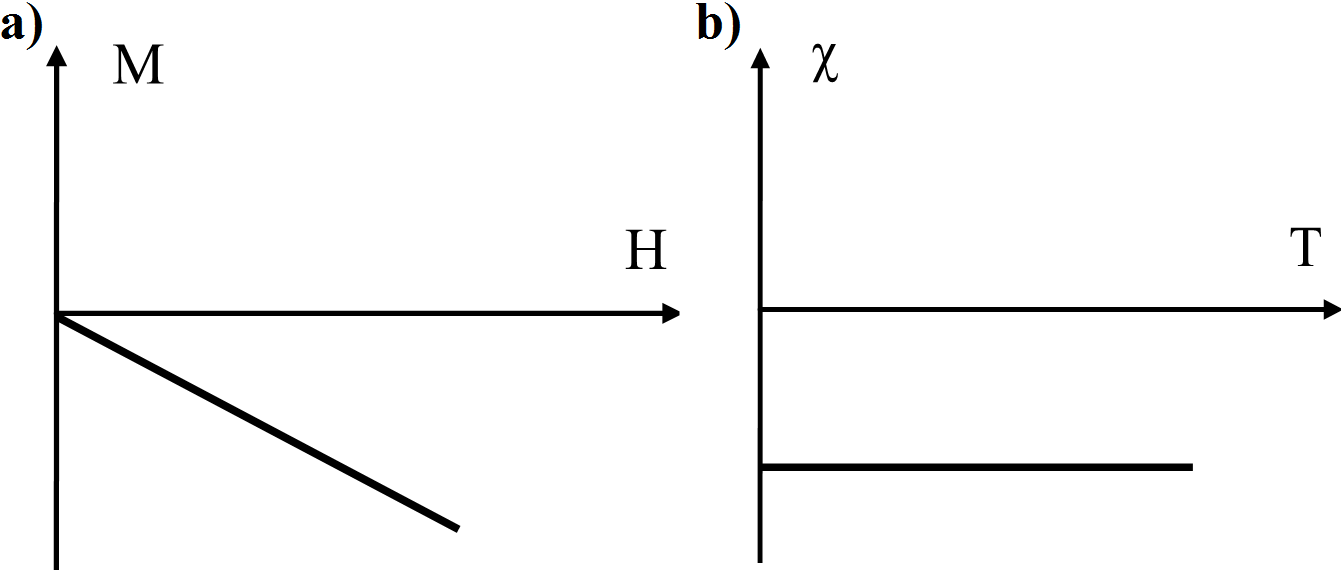}
\caption{\label{pic3}Variation de (a) l'aimantation en fonction du champ appliqu\'{e}
et (b) la susceptibilit\'{e} en fonction de la temp\'{e}rature pour un mat%
\'{e}riau diamagn\'{e}tique.}
\end{figure}

\begin{figure}[!ht]
  \centering
\includegraphics[scale=0.8]{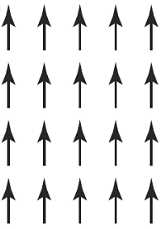}
\caption{\label{pic5}Arrangement des moments magn\'{e}tiques dans les mat\'{e}riaux
ferromagn\'{e}tiques.}
\end{figure}

\begin{figure}[!ht]
  \centering
\includegraphics[scale=0.8]{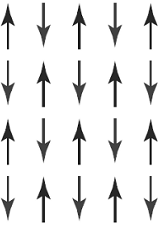}
\caption{\label{pic6}Alignements de spins dans un mat\'{e}riau antiferromagn\'{e}tique.}
\end{figure}

\begin{figure}[!ht]
  \centering
\includegraphics[scale=0.45]{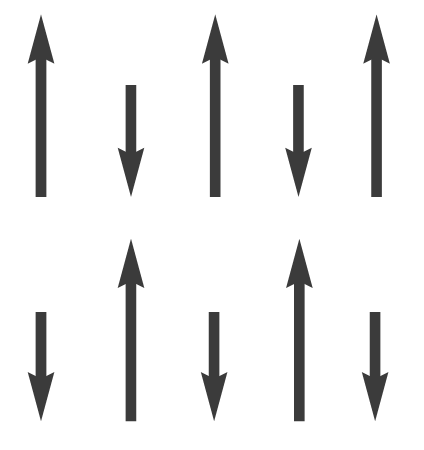}
\caption{\label{pic7}Alignements de spins dans un mat\'{e}riau ferrimagn\'{e}tique.}
\end{figure}

Les propri\'{e}t\'{e}s magn\'{e}tiques des mat\'{e}riaux proviennent du leurs moments magn\'{e}tiques atomiques produits par le spin et le
moment cin\'{e}tique de leurs \'{e}lectrons. Par
cons\'{e}quent, les mat\'{e}riaux sont class\'{e}s en fonction de leur
comportement dans un champ magn\'{e}tique ext\'{e}rieur. En effet, certains
mat\'{e}riaux sont beaucoup plus magn\'{e}tiques que d'autres \cite{I3}. Cela peut
s'expliquer par la nature des interactions entre les moments magn\'{e}tiques
atomiques qui s'\'{e}tablissent au sein de la structure. Les mat\'{e}%
riaux sont class\'{e}s en fonction de leur comportement magn\'{e}%
tique en deux cat\'{e}gories :\newline
$\bullet $ les mat\'{e}riaux magn\'{e}tiques non-ordonn\'{e}s (magn\'{e}%
tisme non-coop\'{e}ratif), tel que les diamagn\'{e}tiques, les
paramagn\'{e}tiques et les superparamagn\'{e}tiques \cite{I3,I4},\newline
$\bullet $ les mat\'{e}riaux magn\'{e}tiques ordonn\'{e}s et qui pr%
\'{e}sentent des \'{e}lectrons non appari\'{e}s (magn\'{e}tisme coop\'{e}%
ratif), comme les ferromagn\'{e}tiques, les antiferromagn%
\'{e}tiques, les ferrimagn\'{e}tiques et les antiferrimagn\'{e}tiques \cite{I3,I4}.\newline
En outre, plus la science se d\'{e}veloppe et plus de nouveaux groupes de mat%
\'{e}riaux appara\^{\i}ssent dont le comportement magn\'{e}tique ne
correspond \`{a} aucune de ces deux classifications. Par exemple, les mat\'{e}%
riaux h\'{e}limagn\'{e}tiques, m\'{e}tamagn\'{e}tiques, sp\'{e}romagn\'{e}%
tiques, asp\'{e}romagn\'{e}tiques et les verres de spins. Il est \'{e}galement
important de remarquer que le comportement magn\'{e}tique des mat\'{e}riaux
peut varier en fonction des changements associ\'{e}s \`{a} d'autres facteurs,
notamment la temp\'{e}rature.\newline
$\bullet $ \textbf{\emph{Diamagn\'{e}tisme}}\newline
Le diamagn\'{e}tisme est une propri\'{e}t\'{e} magn\'{e}tique fondamentale.
Il est extr\^{e}mement faible par rapport aux autres effets magn\'{e}tiques.
Par ailleurs, il tend \`{a} \^{e}tre submerg\'{e} par tous les autres types
de comportement magn\'{e}tique. Les mat\'{e}riaux diamagn\'{e}tiques sont
des mat\'{e}riaux ne comportant que des atomes non-magn\'{e}tiques dont tous
les \'{e}lectrons sont appari\'{e}s \cite{I4,I5}. Sous l'influence d'un champ magn\'{e}%
tique ext\'{e}rieur, le mouvement orbital des \'{e}lectrons est l\'{e}g\`{e}%
rement modifi\'{e}. Ainsi, le courant induit g\'{e}n\`{e}re une aimantation
dans la direction oppos\'{e}e de celle du champ magn\'{e}tique appliqu\'{e}
selon la loi de Lenz. Par cons\'{e}quent, l'aimantation d'un mat\'{e}riau
diamagn\'{e}tique est proportionnelle au champ magn\'{e}tique appliqu\'{e} comme illustr\'{e} sur la figure (\ref{pic3}). Ce type de mat\'{e}riaux ont une susceptibilit\'{e} relative n\'{e}gative et tr\`{e}s faible d'environ $10^{-5}$ \cite{I6}.
\newline
Les mat\'{e}riaux diamagn\'{e}tiques s'aimantent
faiblement et leur aimantation est perdue d\`{e}s que le champ magn\'{e}%
tique est supprim\'{e}. Pratiquement tous les mat\'{e}riaux ont une
contribution diamagn\'{e}tique en raison de leur r\'{e}ponse totale \`{a} un
champ magn\'{e}tique \cite{I3}. Cependant, dans les mat\'{e}riaux contenant des
moments magn\'{e}tiques permanents, la contribution diamagn\'{e}tique est g%
\'{e}n\'{e}ralement \'{e}clips\'{e}e par la r\'{e}ponse de ces moments.
Parmi les substances qui pr\'{e}sentent un comportement diamagn\'{e}tique, il existe : le quartz, la calcite, les feldspaths, le bismuth m\'{e}tallique
et certaines autres mol\'{e}cules organiques comme le benz\`{e}ne \cite{I5}.\newline
$\bullet $ \textbf{\emph{Paramagn\'{e}tisme}}\newline
Dans ces mat\'{e}riaux, les atomes ou les ions poss\`{e}dent des \'{e}%
lectrons non-appari\'{e}s dans des orbitales partiellement remplies. Cela
signifie que dans une substance paramagn\'{e}tique, chaque atome a un petit
moment magn\'{e}tique net \cite{I5}. Il convient de noter qu'il n'y a pas
d'interaction entre ces moments magn\'{e}tiques. Par cons\'{e}quent, en pr%
\'{e}sence d'un champ magn\'{e}tique ext\'{e}rieur, l'agitation thermique emp%
\^{e}che l'alignement partiel de ces moments magn\'{e}tiques atomiques dans
la direction du champ magn\'{e}tique appliqu\'{e}. Cela entra\^{\i}ne une
aimantation nette positive et une susceptibilit\'{e} positive de l'ordre $10^{-4}$ \`{a} $10^{-5}$. Cependant, l'aimantation d'un mat\'{e}riau paramagn\'{e}tique est
perdue quand le champ est supprim\'{e} en raison des effets thermiques \cite{I3,I6}. Si
la temp\'{e}rature de la substance paramagn\'{e}tique augmente, alors
l'alignement des moments magn\'{e}tique sera perturb\'{e}. Par cons\'{e}quent, la susceptibilit\'{e} magn\'{e}tique d\'{e}pend de la temp\'{e}%
rature plus pr\'{e}cis\'{e}ment, la susceptibilit\'{e} paramagn\'{e}tique est inversement
proportionnelle \`{a} la temp\'{e}rature absolue. Cette loi est la loi de Curie. \`{A} temp\'{e}rature ambiante,
la plupart des min\'{e}raux contenant du fer sont paramagn\'{e}tiques \cite{I4,I5}.\newline
$\bullet $ \textbf{\emph{Ferromagn\'{e}tisme}}\newline
Quand nous pensons \`{a} des mat\'{e}riaux magn\'{e}tiques, les premiers
\'{e}l\'{e}ments qui se pr\'{e}sentent \`{a} notre esprit sont le fer, le
nickel, le cobalt et la magn\'{e}tite. Ces derniers sont g\'{e}n\'{e}%
ralement appel\'{e}s mat\'{e}riaux ferromagn\'{e}tiques. Dans ces mat\'{e}riaux, il existe une forte interaction
entre les moments magn\'{e}tiques qui entra\^{\i}ne les forces
d'\'{e}change et qui est d\^{u}e \`{a} l'\'{e}%
change des \'{e}lectrons \cite{I4}. Sous l'influence des forces d'\'{e}change, les
moments magn\'{e}tiques sont align\'{e}s parall\`{e}lement comme montr\'{e} dans la figure (\ref{pic5}) \cite{I3,I7}.
\newline
Dans les mat\'{e}riaux ferromagn\'{e}tiques, les spins de deux \'{e}%
lectrons voisins sont orient\'{e}s de telle fa\c{c}on qu'une forte
interaction se d\'{e}veloppe entre les atomes contenant ces \'{e}lectrons.
Il s'agit d'un effet quantique. C'est la raison pour laquelle ces moments
magn\'{e}tiques sont align\'{e}s parall\`{e}lement les uns aux autres, m\^{e}%
me en absence du champ externe. Par ailleurs, la temp\'{e}rature influence
fortement les propri\'{e}t\'{e}s magn\'{e}tiques des mat\'{e}riaux. En
effet, l'aimantation d'un tel mat\'{e}riau d\'{e}cro\^{\i}t lorsque la temp%
\'{e}rature augmente pour s'annuler \`{a} une temp\'{e}rature de transition
ordre/d\'{e}sordre caract\'{e}ristique appel\'{e}e temp\'{e}rature de Curie $T_{C}$ \cite{I3,I6}. \`{A} des temp\'{e}ratures sup\'{e}rieures \`{a} $T_{C}$, ces mat\'{e}riaux
deviennent paramagn\'{e}tiques. Le comportement ferromagn\'{e}tique existe
dans les m\'{e}taux de transition et les terres rares, mais aussi dans les
alliages, les oxydes et les complexes de ces \'{e}l\'{e}ments \cite{I4,I7}.\newline
$\bullet $ \textbf{\emph{Antiferromagn\'{e}tisme}}\newline
L'ordre antiferromagn\'{e}tique se produit dans les m\'{e}taux de transition $3d$ lorsque les distances interatomiques sont suffisamment petites pour que le
couplage entre spin voisin devient n\'{e}gatif. Dans les mat\'{e}riaux
antiferromagn\'{e}tiques les atomes s'organisent de fa\c{c}on que deux
atomes voisins puissent avoir des moments oppos\'{e}s \cite{I3,I5}, comme le montre la
figure (\ref{pic6}).
\newline
En effet, les mat\'{e}riaux antiferromagn\'{e}tiques se composent de deux
sous-r\'{e}seaux magn\'{e}tiques aimant\'{e}s en sens inverse. Par cons\'{e}%
quent, l'aimantation de l'ensemble est nulle. Ainsi la susceptibilit\'{e}
est faiblement positive \`{a} cause des valeurs n\'{e}gatives des interactions d'%
\'{e}change entre les atomes voisins. La susceptibilit\'{e} d'un mat\'{e}%
riau antiferromagn\'{e}tique atteint son maximum \`{a} la temp\'{e}rature de N%
\'{e}el \cite{I5,I7}. Au del\`{a} de cette temp\'{e}rature, l'ordre antiferromagn\'{e}%
tique dispara\^{\i}t et le mat\'{e}riau se comporte comme un paramagn\'{e}%
tique. De nombreux compos\'{e}s des m\'{e}taux de transition connus tels que
les oxydes de cobalt, nickel, chrome et mangan\`{e}se sont antiferromagn\'{e}%
tiques \cite{I4}.\newline
$\bullet $ \textbf{\emph{Ferrimagn\'{e}tisme}}\newline
Le ferrimagn\'{e}tisme est observ\'{e} uniquement dans les compos\'{e}s qui
ont des structures cristallines plus complexes que celles des m\'{e}taux
purs. Le ferrimagn\'{e}tisme peut \^{e}tre consid\'{e}r\'{e} comme un
comportement interm\'{e}diaire entre le ferromagn\'{e}tisme et
l'antiferromagn\'{e}tisme. Les mat\'{e}riaux ferrimagn\'{e}tiques se d\'{e}%
composent en deux (ou plusieurs) sous-r\'{e}seaux magn\'{e}tiques avec des
spins in\'{e}gaux \cite{I4,I6,I7}. Au sein de ces mat\'{e}riaux, les interactions d'\'{e}%
change m\`{e}nent \`{a} un alignement parall\`{e}le des spins du m\^{e}me
sous-r\'{e}seau et \`{a} un alignement antiparall\`{e}le des spins des deux
sous-r\'{e}seaux diff\'{e}rents comme illustr\'{e} sur la figure (\ref{pic7}).
\newline
Cependant, comme les spins ne se compensent pas totalement, le mat\'{e}riau
ferrimagn\'{e}tique pr\'{e}sente une aimantation plus faible que celle des
mat\'{e}riaux ferromagn\'{e}tiques en absence d'un champ magn\'{e}tique
appliqu\'{e}. Cependant, Sous l'action d'un champ magn\'{e}tique ext\'{e}rieur, les
spins ont tendance \`{a} s'aligner parall\`{e}lement au champ \cite{I5}. Les mat\'{e}%
riaux ferrimagn\'{e}tiques ont des conductivit\'{e}s \'{e}lectriques faibles
et trouvent de nombreuses applications industrielles tel que les applications qui n\'{e}cessitent un mat\'{e}riau avec une aimantation spontan\'{e}e pour fonctionner \`{a} des fr\'{e}quences
\'{e}lev\'{e}es (transformateurs) \cite{I3,I6}.\newline
$\bullet $ \textbf{\emph{Antiferrimagn\'{e}tisme}}\newline
Les mat\'{e}riaux antiferrimagn\'{e}tiques ont une apparence similaire \`{a}
celle des mat\'{e}riaux antiferromagn\'{e}tiques. Effectivement dans les mat%
\'{e}riaux antiferrimagn\'{e}tiques, les spins sont r\'{e}partis en deux (ou
plusieurs) sous-r\'{e}seaux magn\'{e}tiques in\'{e}gaux, \`{a} l'int\'{e}%
rieur desquels les interactions sont antiferromagn\'{e}tiques. Les spins
sont donc antiparall\`{e}les entre eux. C'est le cas notamment dans certains
compos\'{e}s comportant deux types d'atomes de moments magn\'{e}tiques de
spins diff\'{e}rents \cite{I5}.\newline
$\bullet $ \textbf{\emph{Superparamagn\'{e}tisme}}\newline
Le superparamagn\'{e}tisme est un type de magn\'{e}tisme, qui appara\^{\i}t
dans des particules ferromagn\'{e}tiques ou ferrimagn\'{e}tiques nanom\'{e}%
triques. Les mat\'{e}riaux superparamagn\'{e}tiques pr\'{e}sentent un
comportement paramagn\'{e}tique en dessous de leur temp\'{e}rature critique o\`{u} les agitations
thermiques ne sont pas assez fortes. Les forces d'interaction entre les
atomes individuels dominent ces agitations \cite{I5}. Mais, ces forces parviennent \`{a}
changer la direction de l'aimantation de la particule enti\`{e}re. Par cons%
\'{e}quent, les directions des moments magn\'{e}tiques des particules dans
le cristal sont dispos\'{e}es de fa\c{c}on al\'{e}atoire. Ainsi, le moment
magn\'{e}tique net est nul. Les particules superparamagn\'{e}tiques sont donc souvent utilis\'{e}es dans
de nombreux syst\`{e}mes magn\'{e}tiques dans le domaine biom\'{e}dical.
Leurs avantages c'est qu'elles sont petites et ne conservent pas l'aimantation remanente \cite{I3}.\newline
$\bullet $ \textbf{Autres structures magn\'{e}tiques}\newline
Il existe \'{e}galement d'autres types de structures magn\'{e}tiques plus
complexes :\newline
$\bullet $ \textit{Antiferromagn\'{e}tisme non colin\'{e}aire}\newline
L'ordre antiferromagn\'{e}tique non-colin\'{e}aire se produit dans les mat%
\'{e}riaux o\`{u} les spins sont inclin\'{e}s les uns par rapport aux
autres. En d'autres termes, il se produit lorsque les moments magn\'{e}tiques des deux sous-r%
\'{e}seaux ne sont pas rigoureusement antiparall\`{e}les \cite{I5}.\newline
$\bullet $ \textit{H\'{e}limagn\'{e}tisme}\newline
L'h\'{e}limagn\'{e}tisme est une forme d'ordre magn\'{e}tique qui r\'{e}%
sulte de la comp\'{e}tition entre les interactions ferromagn\'{e}tiques et
antiferromagn\'{e}tiques. C'est la propri\'{e}t\'{e} \`{a} basse temp\'{e}rature de certains m\'{e}taux et de sels de m\'{e}taux de transition, dans lesquels les moments magn\'{e}tiques atomiques sont dispos\'{e}s en h%
\'{e}lice avec un angle de rotation pouvant varier entre $0%
%TCIMACRO{\U{b0}}%
%BeginExpansion
{{}^\circ}%
%EndExpansion
$ et $180%
%TCIMACRO{\U{b0}}%
%BeginExpansion
{{}^\circ}%
%EndExpansion
$ \cite{I4}.
L'antiferromagn\'{e}tisme et le ferromagn\'{e}tisme peuvent \^{e}tre consid%
\'{e}r\'{e}s comme des cas limites d'h\'{e}limagn\'{e}tisme, avec un angle
de rotation respectivement de $180%
%TCIMACRO{\U{b0}}%
%BeginExpansion
{{}^\circ}%
%EndExpansion
$ et $0%
%TCIMACRO{\U{b0}}%
%BeginExpansion
{{}^\circ}%
%EndExpansion
$. L'ordre h\'{e}limagn\'{e}tique brise la sym\'{e}trie d'inversion spatiale des moments magn\'{e}tiques comme il peut pr\'{e}senter une
rotation horaire ou antihoraire dans la nature \cite{I3,S1}.\newline
$\bullet $ \textit{M\'{e}tamagn\'{e}tisme}\newline
Le m\'{e}tamagn\'{e}tisme est une augmentation soudaine de l'aimantation
d'un mat\'{e}riau avec une faible variation du champ magn\'{e}tique appliqu%
\'{e}. Le comportement m\'{e}tamagn\'{e}tique peut avoir diff\'{e}rentes
causes physiques entra\^{\i}nant ainsi diff\'{e}rents types de m\'{e}tamagn\'{e}tisme. Selon le mat\'{e}%
riau et les conditions exp\'{e}rimentales, le m\'{e}tamagn\'{e}tisme peut
\^{e}tre associ\'{e} \`{a} une transition de phase du premier ordre, une
transition de phase continue \`{a} un point critique, ou crossovers au-del%
\`{a} d'un point critique ne comportant pas de transition de phase \cite{I5}.\newline
$\bullet $ \textit{Sp\'{e}romagn\'{e}tisme}\newline
Les mat\'{e}riaux sp\'{e}romagn\'{e}tiques se caract\'{e}risent par une
distribution al\'{e}atoire de moments magn\'{e}tiques induite par la
topologie du r\'{e}seau cationique \cite{I5}.\newline
$\bullet $ \textit{Asp\'{e}romagn\'{e}tisme}\newline
Les mat\'{e}riaux asp\'{e}romagn\'{e}tiques sont d\'{e}termin\'{e}s par une
configuration magn\'{e}tique gel\'{e}e dont les orientations des moments magn%
\'{e}tiques sont distribu\'{e}es dans un demi-espace \cite{I3,I5}.\newline
$\bullet $ \textit{Sp\'{e}rimagn\'{e}tisme}\newline
Les mat\'{e}riaux sp\'{e}rimagn\'{e}tiques sont constitu\'{e}s de deux
configurations de moments in\'{e}gaux correspondant aux deux sous-r\'{e}%
seaux. Ils ressemblent aux mat\'{e}riaux asp\'{e}romagn\'{e}tiques, du fait
que les orientations de leurs moments magn\'{e}tiques sont distribu\'{e}es
dans un demi-espace \cite{I5}.
\vspace{-0.4cm}
\subsection{Temp\'{e}rature de transition}

L'ordre magn\'{e}tique des mat\'{e}riaux est fortement influenc\'{e} par le
changement de la temp\'{e}rature. Ainsi, un mat\'{e}riau pr\'{e}sente diff%
\'{e}rents comportements magn\'{e}tiques en fonction de sa temp\'{e}rature.\newline
$\bullet $ \textbf{\emph{Temp\'{e}rature de Curie}}\newline
Les mat\'{e}riaux ferromagn\'{e}tiques perdent leurs propri%
\'{e}t\'{e}s particuli\`{e}res au-dessus d'une temp\'{e}rature critique $%
T_{C}$ appel\'{e}e temp\'{e}rature de Curie. La temp%
\'{e}rature de Curie est la temp\'{e}rature \`{a} laquelle un mat\'{e}riau
ferromagn\'{e}tique devient paramagn\'{e}tique sous l'influence d'une hausse de temp%
\'{e}rature. Au dessous de la temp\'{e}rature de Curie,
les interactions ferromagn\'{e}tiques tendent \`{a} aligner parall\`{e}%
lement les moments magn\'{e}tiques voisins dans le mat\'{e}riau \cite{I8}. Cependant, en augmentant la temp\'{e}rature, les spins fluctuent rapidement. En
effet, la transition se produit \`{a} la temp\'{e}rature critique quand l'%
\'{e}nergie d'agitation thermique des spins domine l'\'{e}nergie
d'interaction magn\'{e}tique. Ce processus est r\'{e}versible car l'ordre
ferromagn\'{e}tique r\'{e}appara\^{\i}t dans le syst\`{e}me quand sa temp%
\'{e}rature redescend en dessous de la temp\'{e}rature de Curie. La valeur
de la temp\'{e}rature de Curie varie d'un mat\'{e}riau \`{a} l'autre. \`{A} titre d'exemple, les
temp\'{e}ratures de Curie, sont respectivement pour le fer, le cobalt, le
nickel et la magn\'{e}tite : $1044K$, $1388K$, $628K$ et $856K$ \cite{I9}.\newline
$\bullet $ \textbf{\emph{Temp\'{e}rature de N\'{e}el}}\newline
Dans un mat\'{e}riau antiferromagn\'{e}tique, la temp\'{e}rature de N\'{e}el $T_{N}$, est la temp\'{e}rature \`{a} laquelle le mat\'{e}riau pr\'{e}sente une
transition de phase vers un comportement paramagn\'{e}tique. Cette temp\'{e}%
rature est similaire \`{a} la temp\'{e}rature de Curie pour les mat\'{e}%
riaux ferromagn\'{e}tiques. Comme les mat\'{e}riaux antiferromagn\'{e}tiques
ne pr\'{e}sentent pas une aimantation spontan\'{e}e, la transition de phase
se manifeste par l'apparition de pic dans le graphe de la susceptibilit\'{e}. Au dessus
de la temp\'{e}rature de N\'{e}el, la susceptibilit\'{e} ob\'{e}it \`{a} la
loi de Curie-Weiss \cite{I9,I10}.\newline
$\bullet $ \textbf{\emph{Temp\'{e}rature de compensation}}\newline
Contrairement aux mat\'{e}riaux ferromagn\'{e}tiques et antiferromagn\'{e}%
tiques, les mat\'{e}riaux ferrimagn\'{e}tiques peuvent pr\'{e}senter, sous
certaines conditions, une temp\'{e}rature de compensation $T_{comp}$ proche de la temp\'{e}rature ambiante. La temp\'{e}rature de compensation $T_{comp}$ est la temp\'{e}rature \`{a} laquelle, l'aimantation totale du syst\`{e}me dispara\^{\i}t en dessous de la temp\'{e}rature critique. Elle appara\^{\i}t en raison de la nature d'int\'{e}raction d'\'{e}change entre les deux sous-r\'{e}seaux in\'{e}gaux $A$ et $B$ qui forment le
mat\'{e}riau ferrimagn\'{e}tique \cite{I11,I12}. \`{A} la temp\'{e}rature de compensation $%
T_{comp}$, les moments magn\'{e}tiques des deux sous-r\'{e}seaux sont align\'{e}s
antiparall\`{e}lement et ont la m\^{e}me valeur absolue. Ainsi, la temp%
\'{e}rature de compensation $T_{comp}$ peut \^{e}tre d\'{e}termin\'{e}e par le point
d'intersection des valeurs absolues des aimantations des deux sous-r\'{e}%
seaux $A$ et $B$. Par cons\'{e}quent, au point de compensation, nous devons avoir :
\begin{equation}
\left \vert m_{A}\left( T_{comp}\right) \right \vert =\left \vert
m_{B}\left( T_{comp}\right) \right \vert ,
\end{equation}%
et%
\begin{equation}
sign\left \vert m_{A}\left( T_{comp}\right) \right \vert =-sign\left \vert
m_{B}\left( T_{comp}\right) \right \vert .
\end{equation}%
L'apparition du point de compensation rend le mat\'{e}riau prometteur pour des
applications technologiques importantes, notamment dans le domaine de l'enregistrement
thermomagn\'{e}tique. En outre, certaines nouvelles propri\'{e}t\'{e}s
physiques ont \'{e}t\'{e} observ\'{e}es au point de compensation. Par
exemple, il a \'{e}t\'{e} constat\'{e} que le champ c{\oe}rcitif pr\'{e}sente
un pic au point de compensation, favorisant ainsi la cr\'{e}ation de petits
domaines magn\'{e}tiques stables \cite{I11}. Cette d\'{e}pendance en temp\'{e}rature de ce dernier au point de compensation peut \^{e}tre appliqu\'{e}e \`{a} l'%
\'{e}criture et \`{a} l'effacement dans les m\'{e}dias d'enregistrement magn%
\'{e}to-optiques \`{a} haute densit\'{e}, o\`{u} les changements de temp\'{e}%
rature sont atteints par un \'{e}chauffement local des films par un faisceau
laser focalis\'{e}. Il a notamment \'{e}t\'{e} d\'{e}montr\'{e} que
l'utilisation des films amorphes ferrimagn\'{e}tiques avec des temp\'{e}%
ratures de compensation plus \'{e}lev\'{e}es que la temp\'{e}rature ambiante, permet d'atteindre une capacit\'{e} d'\'{e}crasement direct dans des supports
d'enregistrement magn\'{e}to-optique. L'importance technologique de ce genre
de temp\'{e}rature est \'{e}vidente puisqu'un petit champ est suffisant pour
changer le signe de l'aimantation totale du syst\`{e}me \cite{I10}.\newline
$\bullet $ \textbf{Temp\'{e}rature de blocage}\newline
\`{A} l'\'{e}chelle nanom\'{e}trique, la temp\'{e}rature et le temps ont un
effet crucial sur les moments magn\'{e}tiques des nanoparticules. De nombreuses
notions tr\`{e}s importantes en d\'{e}coulent, comme le superparamagn\'{e}%
tisme et la temp\'{e}rature de blocage. La temp\'{e}ratre de blocage $T_{B}$ est une
grandeur physique qui d\'{e}pend du mat\'{e}riau lui-m\^{e}me. Elle peut
\^{e}tre obtenue \`{a} partir du pic de la susceptibilit\'{e} magn\'{e}tique
en fonction de la temp\'{e}rature. La phase magn\'{e}tique des
nanoparticules est d\'{e}termin\'{e}e par la temp\'{e}rature de blocage. En
effet, au-dessus de cette temp\'{e}rature, un mat\'{e}riau ferromagn\'{e}%
tique, antiferromagn\'{e}tique ou ferrimagn\'{e}tique devient superparamagn%
\'{e}tique. La temp\'{e}rature de blocage peut \'{e}galement \^{e}%
tre d\'{e}finie comme celle \`{a} laquelle le temps de relaxation devient
\'{e}gal au temps de mesure exp\'{e}rimentale \cite{I9,I13}. Dans ce cas, la temp\'{e}rature de blocage $T_{B}$ s'exprime comme :
\begin{equation}
T_{B}=\frac{KV}{k_{B}\ln \left( \frac{\tau }{\tau _{0}}\right) },
\end{equation}
o\`{u} $K$ et $V$ sont respectivement, l'anisotropie magn\'{e}tique de la nanoparticule et son
volume, $k_{B}$ est la constante de Boltzmann, $\tau $ est le temps de mesure et $\tau _{0}$ est le temps
d'essai d'une valeur comprise entre $%
10^{-9}$ et $10^{-10}$ seconde \cite{I13}.\newline
$\bullet $ \textbf{Temp\'{e}rature de d\'{e}blocage}\newline
La temp\'{e}rature de d\'{e}blocage est la temp\'{e}rature \`{a} laquelle une composante
de l'aimantation d'un mat\'{e}riau devient thermiquement d\'{e}magn\'{e}%
tis\'{e}e dans une exp\'{e}rience de laboratoire. Le d\'{e}blocage se
produit au cours du chauffage en laboratoire lorsque le temps de relaxation
des nanoparticules portant l'aimantation devient \'{e}quivalent \`{a} celui auquel le mat\'{e}riau est maintenu \`{a} une temp\'{e}rature \'{e}lev\'{e}e \cite{I9,I13}.
\vspace{-0.45cm}
\subsection{Classification de N\'{e}el}

\begin{figure}[!ht]
  \centering
\includegraphics[scale=0.9]{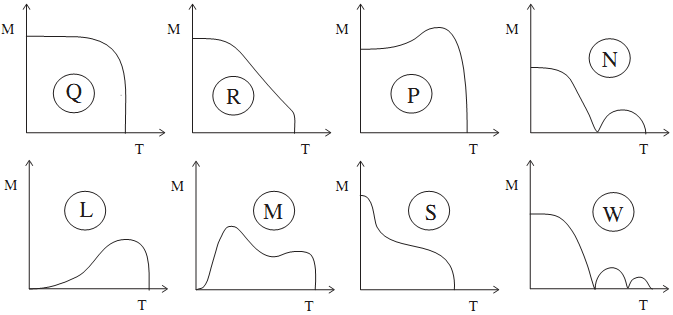}
\caption{\label{pic13}Repr\'{e}sentation sch\'{e}matique de la variation thermique de
diff\'{e}rents types d'aimantation \cite{I2}.}
\end{figure}

Dans la th\'{e}orie du ferrimagn\'{e}tisme de N\'{e}el, il est possible de
classer la variation thermique de l'aimantation spontan\'{e}e totale en cinq
cat\'{e}gories principales \cite{I14}. Ces cat\'{e}gories sont : type $Q$, type $P$, type $N$, type $L$ et type $M$. En outre, Stre\v{c}ka a montr\'{e} l'existence d'autres types d'aimantation \cite{I2}, tels que le type $R$, le type $S$ et le type $W$. La variation thermique de diff\'{e}rents types
d'aimantation est montr\'{e}e dans la figure (\ref{pic13}).
\newline
Selon la figure (\ref{pic13}), les d\'{e}pendances de type $Q$ et de type $R$ pr\'{e}sentent une
diminution monotone de l'aimantation lorsque la temp\'{e}rature augmente. De plus, la variation thermique de l'aimantation de type $Q$ est presque analogue
\`{a} celle de type $R$. La seule diff\'{e}rence est la diminution rapide de l'aimantation type $Q$ au voisinage de la temp%
\'{e}rature critique, tandis que l'aimantation type $R$ se caract\'{e}rise d'une
baisse relativement rapide dans la gamme de temp\'{e}%
ratures interm\'{e}diaires avant de s'annuler brusquement au point critique. L'aimantation de type $P$ augmente et
atteint une valeur maximale lorsque la temp\'{e}rature augmente ; alors que
la courbe de type $N$ est caract\'{e}ris\'{e}e par un point de compensation o%
\`{u} l'aimantation r\'{e}sultante dispara\^{\i}t. La courbe de type $L$ est tr\`{e}s
analogue \`{a} la d\'{e}pendance de type $P$. Cependant, l'aimantation r\'{e}%
sultante commence \`{a} partir de z\'{e}ro dans ce cas particulier. De m\^{e}%
me, la d\'{e}pendance de type $M$ commence \'{e}galement \`{a} partir de z\'{e}%
ro tout en ayant deux maxima distincts avant la temp%
\'{e}rature critique. L'aimantation de type $S$ pr\'{e}%
sente trois phase principales. Dans un premier temps, nous constatons une d\'{e}croissance rapide, suivie d'une diminution presque compl\`{e}te dans la
gamme de temp\'{e}ratures interm\'{e}diaires, et enfin, une troisi\`{e}me baisse rapide au voisinage de la temp\'{e}rature critique. La courbe de type $W$ pr\'{e}sente deux points de compensation avant la temp\'{e}rature critique.
\vspace{-0.6cm}
\section{Interactions magn\'{e}tiques}

L'origine du champ effectif est l'interaction d'\'{e}change qui refl\`{e}te
la r\'{e}pulsion de Coulomb de deux \'{e}lectrons voisins. Selon le principe
d'exclusion de Pauli, les \'{e}lectrons ne peuvent pas \^{e}tre au m\^{e}me
endroit dans le m\^{e}me \'{e}tat quantique s'ils ont le m\^{e}me spin. Il y
a une diff\'{e}rence d'\'{e}nergie entre les configurations \`{a} spins
parall\`{e}les $\uparrow _{i}\uparrow _{j}$ et antiparall\`{e}les $\uparrow
_{i}\downarrow _{j}$ des atomes voisins $i$ et $j$. L'\'{e}change
interatomique dans les isolants est g\'{e}n\'{e}ralement plus faible que l'%
\'{e}change intra-atomique ferromagn\'{e}tique entre les \'{e}lectrons d'un m%
\^{e}me atome, ce qui conduit \`{a} la premi\`{e}re r\`{e}gle de Hund \cite{I1}.%
\newline
Les \'{e}lectrons sont indiscernables et donc l'\'{e}%
change de deux \'{e}lectrons doit donner la m\^{e}me densit\'{e} \'{e}%
lectronique $\left \vert \Psi \left( 1,2\right) \right \vert
^{2}=\left
\vert \Psi \left( 2,1\right) \right \vert ^{2}$. Comme les \'{e}lectrons sont des fermions, la seule solution
pour la fonction d'onde totale des deux \'{e}lectrons est d'\^{e}tre antisym%
\'{e}trique :
\begin{equation}
\Psi \left( 1,2\right) =-\Psi \left( 2,1\right) .  \label{5-19}
\end{equation}
La fonction d'onde totale $\Psi $ est le produit des fonctions d'onde d'espace et de
spin $\phi \left( r_{1},r_{2}\right) $ et $\chi \left(
s_{1},s_{2}\right) $.\newline
Pour mieux comprendre l'interaction d'\'{e}change, nous consid\'{e}rons l'exemple le plus simple : la mol\'{e}cule d'hydrog\`{e}ne avec deux atomes
d'hydrog\`{e}ne ayant chacun un \'{e}lectron de valence dans une orbitale $1s$
$\psi _{i}\left( r_{i}\right) $. L'\'{e}quation de Schr\"{o}dinger est $H\left( r_{1},r_{2}\right)
\Psi \left( r_{1},r_{2}\right) =\varepsilon \Psi \left( r_{1},r_{2}\right) $ o\`{u} on n\'{e}glige les interactions entre les \'{e}%
lectrons :
\begin{equation}
\left[ -\frac{\hbar ^{2}}{2m}\left( \frac{\partial ^{2}}{\partial r_{1}^{2}}+%
\frac{\partial ^{2}}{\partial r_{2}^{2}}\right) -\frac{e^{2}}{4\pi \epsilon
_{0}}\left( \frac{1}{r_{1}}+\frac{1}{r_{2}}\right) \right] \Psi \left(
r_{1},r_{2}\right) =\varepsilon \Psi \left( r_{1},r_{2}\right) .
\label{5-20}
\end{equation}
Il y a deux types d'orbites mol\'{e}culaires, les orbitales sym\'{e}triques
liantes $%
\phi _{s}$ et les orbitales antisym\'{e}triques antiliantes $\phi _{a}$. Elles s'\'{e}%
crivent sous la forme :
\begin{equation}
\phi _{s}=\left( 1/\sqrt{2}\right) \left( \psi _{1}+\psi _{2}\right) \qquad
\text{et}\qquad \phi _{a}=\left( 1/\sqrt{2}\right) \left( \psi _{1}-\psi
_{2}\right) .
\end{equation}
$\psi _{1}$ et $\psi _{2}$ sont les composantes spatiales des fonctions d'onde des \'{e}lectrons
individuels $1$ et $2$ respectivement. Les fonctions d'onde $\psi _{1}\left( r_{1}\right) $ et $\psi _{2}\left(
r_{2}\right) $ sont les solutions de l'\'{e}quation de Schr\"{o}dinger pour chaque atome individuel \cite{I4}.\newline
Les fonctions de spins sym\'{e}triques et antisym\'{e}triques sont les \'{e}%
tats de spin triplet et singulet :
\begin{equation}
\begin{array}{c}
S=1;\qquad M_{s}=1,0,-1,\qquad \qquad \qquad \qquad \qquad \qquad \qquad
\qquad \qquad \\
\chi _{s}=\left \vert \uparrow _{1},\uparrow _{2}\right \rangle ;\qquad
\left( 1/\sqrt{2}\right) \left[ \left \vert \uparrow _{1},\downarrow
_{2}\right \rangle +\left \vert \downarrow _{1},\uparrow _{2}\right \rangle %
\right] ;\qquad \left \vert \downarrow _{1},\downarrow _{2}\right \rangle ,\qquad \qquad
\\
S=0;\qquad M_{s}=0,\qquad \qquad \qquad \qquad \qquad \qquad \qquad
\qquad \qquad \ \ \  \\
\chi _{a}=\left( 1/\sqrt{2}\right) \left[ \left \vert \uparrow
_{1},\downarrow _{2}\right \rangle -\left \vert \downarrow _{1},\uparrow
_{2}\right \rangle \right] .\qquad \qquad \qquad \qquad \qquad \qquad  \ \ \ \
\end{array}%
\end{equation}
Selon l'\'{e}quation (\ref{5-19}), multiplier la fonction d'espace sym\'{e}trique par la
fonction de spin antisym\'{e}trique, et vice-versa nous m\`{e}ne \`{a} des fonctions d'ondes totales antisym\'{e}triques :
\begin{equation}
\begin{array}{c}
\Psi _{I}=\phi _{s}\left( 1,2\right) \chi _{a}\left( 1,2\right) ,\text{ } \\
\Psi _{II}=\phi _{a}\left( 1,2\right) \chi _{s}\left( 1,2\right) .%
\end{array}%
\end{equation}
Lorsque les deux \'{e}lectrons sont dans l'\'{e}tat de spin triplet, il n'y
a aucune probabilit\'{e} de les retrouver au m\^{e}me point de l'espace.
Mais si les \'{e}lectrons sont dans l'\'{e}tat singulet de spin, avec des
spins antiparall\`{e}les, il est possible de les trouver au m\^{e}me
endroit, parce que la partie spatiale de la fonction d'onde est sym\'{e}%
trique par rapport \`{a} l'\'{e}change des \'{e}lectrons \cite{I1,I4}.\newline
Les \'{e}nergies des deux \'{e}tats peuvent \^{e}tre \'{e}valu\'{e}es \`{a}
partir de l'hamiltonien $H\left( r_{1},r_{2}\right) $ dans l'\'{e}quation (\ref%
{5-20}) :%
\begin{equation}
\varepsilon _{I,II}=\int \phi _{s,a}^{\ast }\left( r_{1},r_{2}\right)
H\left( r_{1},r_{2}\right) \phi _{s,a}\left( r_{1},r_{2}\right)
dr_{1}^{3}dr_{2}^{3}.
\end{equation}
Pour la mol\'{e}cule d'hydrog\`{e}ne, $\varepsilon _{I}$ est inf\'{e}rieur \`{a} $\varepsilon _{II}$. En d'autres
termes, les orbitales mol\'{e}culaires liantes sont au dessous des orbitales
mol\'{e}culaires antiliantes. Pour l'int\'{e}grale d'\'{e}change $J=\left( \varepsilon _{I}-\varepsilon
_{II}\right) /2$, nous pouvons d\'{e}crire l'\'{e}nergie par :
\begin{equation}
\varepsilon =-2\left( J/\hbar ^{2}\right) s_{1}.s_{2},
\end{equation}%
o\`{u}%
\begin{equation}
s_{1}.s_{2}=\frac{1}{2}\left[ \left( s_{1}+s_{2}\right)
^{2}-s_{1}^{2}-s_{2}^{2}\right] .
\end{equation}
Selon la valeur du nombre quantique de spin $S=s_{1}+s_{2}=0$ ou $1$, les valeurs propres sont $-\frac{3}{4}\hbar ^{2}$ ou $+\frac{1}{4}\hbar ^{2}$. La diff\'{e}rence d'\'{e}nergie entre l'\'{e}tat singulet et l'\'{e}tat
triplet est $2J$. L'int\'{e}grale d'\'{e}change $J$ s'\'{e}crit sous la forme :
\begin{equation}
J\mathbb{=}\int \psi _{1}^{\ast }\left( r^{\prime }\right) \psi _{2}^{\ast
}\left( r\right) H\left( r,r^{\prime }\right) \psi _{1}\left( r\right) \psi
_{2}\left( r^{\prime }\right) dr^{3}d^{3}r^{\prime }.
\end{equation}
Dans la mol\'{e}cule de $H_{2}$, l'\'{e}tat de spin singulet \'{e}tant plus faible,
l'int\'{e}grale est n\'{e}gative. Cependant, dans un atome, les orbitales
sont orthogonales et $J$ est positif \cite{I4}. L'hamiltonien de Heisenberg est donn\'{e}
par :
\begin{equation}
H=-2J\hat{S}_{1}.\hat{S}_{2},
\end{equation}
o\`{u} $\hat{S}_{1}$ et $\hat{S}_{2}$ sont les op\'{e}rateurs de spin. $%
\hbar ^{2}$ a \'{e}t\'{e} absorb\'{e} dans la
constante d'\'{e}change $J$. Pour la suite, nous adopterons cette convention,
afin d'\'{e}viter d'avoir \`{a} \'{e}crire $\hbar $ partout. Nous enleverons \'{e}%
galement le chapeau sur les op\'{e}rateurs de spin $\hat{S}_{i}$. $J>0$ indique une interaction
ferromagn\'{e}tique, tendant \`{a} aligner les deux spins parall\`{e}%
les ; $J<0$ indique une interaction antiferromagn\'{e}tique, qui tend \`{a}
aligner les deux spins antiparall\`{e}les \cite{I1}.\newline
Dans un r\'{e}seau, l'hamiltonien est g\'{e}n\'{e}ralis\'{e} \`{a}
une somme sur toutes les paires d'atomes sur les sites $i$ et $j$ du r\'{e}seau :
\begin{equation}
H=-2\sum_{i>j}J_{ij}S_{i}.S_{j}.
\end{equation}
Le couplage d'\'{e}change interatomique d\'{e}crit par l'hamiltonien de
Heisenberg ne peut \^{e}tre que ferromagn\'{e}tique ou antiferromagn\'{e}%
tique \cite{I2}.\newline
La constante d'\'{e}change de Heisenberg $J$ peut \^{e}tre li\'{e}e \`{a} la
constante de Weiss $n_{W}$ de la th\'{e}orie de champ mol\'{e}culaire \cite{I4}. Si le moment $g\mu _{B}S_{i}$ int\'{e}ragit avec un champ effectif $%
H^{i}=n_{W}M=n_{W}ng\mu _{B}S$, l'hamiltonien s'\'{e}crit alors comme :
\begin{equation}
H_{i}=-2\left[ \sum_{j}JS_{j}\right] .S_{i}\approx -\mu _{0}H^{i}g\mu
_{B}S_{i}.
\end{equation}
L'approximation du champ mol\'{e}culaire s'\'{e}l\`{e}ve \`{a} la moyenne
sur les corr\'{e}lations locales entre $S_{i}$\ et $S_{j}$. Si $z$ est le nombre de plus proches
voisins, alors $J=\mu _{0}n_{W}ng^{2}\mu _{B}^{2}/2z$. Par cons\'{e}quent, nous pouvons \'{e}crire :
\begin{equation}
T_{C}=\frac{2zJS\left( S+1\right) }{3k_{B}}.
\end{equation}
En g\'{e}n\'{e}ral, lorsque de nombreux \'{e}lectrons plus ou moins d\'{e}%
localis\'{e}s sont pr\'{e}sents dans les diff\'{e}rentes orbitales, le
calcul de l'\'{e}change est une question d\'{e}licate.\newline
Dans cette section, nous allons pr\'{e}senter divers mod\`{e}les en pr\'{e}sence d'une impuret\'{e} dans les syst\`{e}mes m\'{e}%
talliques. Ensuite, nous discuterons les diff\'{e}rents types d'interaction
magn\'{e}tique responsables des propri\'{e}t\'{e}s magn\'{e}tiques.\newline
\vspace{-1.4cm}
\subsection{Mod\`{e}les d'impuret\'{e}s}\vspace{-0.15cm}
\hspace{-0.6cm}$\bullet $ \textbf{\emph{Mod\`{e}le de diffusion pour une impuret\'{e} non magn\'{e}%
tique}}\newline
Ce mod\`{e}le d\'{e}crit l'effet d'une impuret\'{e} sans spin (ou d'un d\'{e}%
faut quelconque non magn\'{e}tique) sur les \'{e}lectrons de conduction
consid\'{e}r\'{e}s comme libres \cite{t33}. L'hamiltonien prend la forme:
\begin{equation}
H=\sum \limits_{k\sigma }\varepsilon _{k}a_{k\sigma }^{\dag }a_{k\sigma
}+\sum \limits_{kk^{\prime }}V_{kk^{\prime }}a_{k\sigma }^{\dag
}a_{k^{\prime }\sigma }.
\end{equation}
o\`{u} $a_{k\sigma }$ et $a_{k\sigma }^{\dag }$ sont les op\'{e}rateurs de cr\'{e}ation et d'annihilation d'un \'{e}lectron de conduction de spin $\sigma $ et de moment $k$, $a_{k^{\prime }\sigma}$ est l'op\'{e}rateur de cr\'{e}ation d'un \'{e}lectron de conduction de spin $\sigma$ et de moment $k^{\prime }$, $\varepsilon _{k}$ est l'\'{e}nergie d'un \'{e}lectron de conduction de moment $k$ et $V_{kk^{\prime }}$ repr\'{e}sente le potentiel de l'impuret\'{e} dans la base des ondes planes.\newline
Dans ce cas, nous n'avons pas d'orbitale localis\'{e}e associ\'{e}e \`{a}
l'impuret\'{e}, car elle n'intervient que sous la forme d'un potentiel ext%
\'{e}rieur. Cet hamiltonien permet de calculer les modifications de la densit%
\'{e} d'\'{e}tats induites par le potentiel, lesquelles sont donn\'{e}es par la r\`{e}gle de somme de Friedel.\newline
$\bullet $ \textbf{\emph{Mod\`{e}le d'Anderson}}\newline
Ce mod\`{e}le d\'{e}crit l'effet des impuret\'{e}s sur les propri\'{e}t\'{e}%
s de transport \'{e}lectronique dans les cristaux imparfaits.\newline
$\bullet $ \textit{Mod\`{e}le d'Anderson sans interaction de Coulomb}\newline
Le plus souvent, les atomes d'impuret\'{e}s pr\'{e}sents dans un m\'{e}tal
portent des orbitales atomiques dont les \'{e}nergies sont voisines de
celles des \'{e}lectrons de conduction du m\'{e}tal. L'orbitale de l'impuret%
\'{e} va donc se m\'{e}langer \`{a} celles des \'{e}lectrons de conduction
donnant lieu \`{a} une redistribution des niveaux d'\'{e}nergie \cite{I4,t33}. Dans ce cas, l'hamiltonien prend la forme :
\begin{equation}
H_{A}^{\left( 1\right) }=\sum \limits_{k\sigma }\varepsilon _{k}a_{k\sigma
}^{\dag }a_{k\sigma }+\varepsilon _{d}\sum \limits_{\sigma }a_{d\sigma
}^{\dag }a_{d\sigma }+\sum \limits_{k\sigma }\left( V_{k}a_{k\sigma }^{\dag
}a_{d\sigma }+h.c.\right) ,  \label{sa}
\end{equation}
o\`{u} $a_{k\sigma }$ et $a_{k\sigma }^{\dag }$ sont les op\'{e}rateurs de cr\'{e}ation et d'annihilation d'un \'{e}lectron de conduction de spin $\sigma $ et de moment $k$, $a_{d\sigma }$ et $a_{d\sigma }^{\dag }$ sont les op\'{e}rateurs de cr\'{e}ation et d'annihilation d'un \'{e}lectron $d$ de spin $\sigma $, $\varepsilon _{k}$ et $\varepsilon _{d}$ sont respectivement les \'{e}nergies d'un \'{e}lectron de conduction de moment $k$ et d'un \'{e}lectron $d$ et $V_{k}$ est le potentiel d'hybridation pour un moment $k$.\newline
$\bullet $ \textit{Mod\`{e}le d'Anderson avec interaction de Coulomb}\newline
Afin de tenir compte de la r\'{e}pulsion coulombienne lorsqu'il y a deux
\'{e}lectrons sur l'orbitale $d$ \cite{I4,t33}, l'hamiltonien consid\'{e}r\'{e} s'%
\'{e}crit donc :
\begin{equation}
H_{A}^{\left( 2\right) }=H_{A}^{\left( 1\right) }+Un_{d\uparrow
}n_{d\downarrow }.  \label{ko}
\end{equation}
o\`{u} $n_{d\uparrow }$ et $n_{d\downarrow }$ sont les nombres d'\'{e}lectrons $d$ de spin up et down et $U$ est l'interaction coulombienne. Lorsque $U=0$, l'hybridation conduit \`{a} un \'{e}tat final de largeur \'{e}gale \`{a} :
\begin{equation}
\Delta =\pi \rho \left( E_{F}\right) \left \vert V_{k}\right \vert ^{2},
\end{equation}
o\`{u} $\rho \left( E_{F}\right) $ est la densit\'{e} des \'{e}lectrons de conduction d'\'{e}nergie de Fermi $E_{F}$ de la bande de conduction. Pour le cas d'une impuret\'{e} de transition, le niveau $d$ a une \'{e}nergie $\varepsilon _{d}$ l\'{e}g\`{e}rement inf\'{e}rieure \`{a} l'\'{e}nergie de Fermi $E_{F}$ de la bande de conduction. Du fait de l'interaction coulombienne $U$, le nombre $n_{d}$ d'\'{e}lectrons port\'{e}s par les orbitales $3d$ doit \^{e}tre inf\'{e}rieur \`{a} $1$. Lorsque l'hybridation $V$ est importante, c'est \`{a} dire lorsque $\Delta >E_{F}-\epsilon _{d}$, un \'{e}tat de valence interm\'{e}diaire est obtenu pour lequel on a alors $n_{d}$ nettement inf\'{e}rieur \`{a} $1$. Par contre, lorsque l'hybridation est suffisamment faible pour que $\Delta <E_{F}-\epsilon _{d}$, on a alors $n_{d}<1$, ce qui correspond \`{a} la limite Kondo du hamiltonien d'Anderson.\newline
$\bullet $ \textbf{\emph{Mod\`{e}le sd}}\newline
Le mod\`{e}le d'\'{e}change $sd$ fut le premier mod\`{e}le introduit pour d\'{e}crire l'interaction entre les \'{e}lectrons de conduction $s$ du m\'{e}tal et les \'{e}lectrons $d$ d'une impuret\'{e} de transition caract\'{e}ris\'{e}e par son spin $S$. Les \'{e}lectrons de conduction n'int\'{e}ragissent avec l'impuret\'{e} que lorsqu'ils sont sur le site de celle-ci. Le mod\`{e}le d'\'{e}change $sd$ d\'{e}crit donc le couplage entre des \'{e}lectrons de conduction et un spin
localis\'{e} \cite{I4,t33}. L'hamiltonien s'\'{e}crit :
\begin{equation}
H_{sd}=-\sum \limits_{kk^{\prime }}J_{kk^{\prime }}S.\sum \limits_{\sigma
\sigma ^{\prime }}a_{k\sigma }^{\dag }\sigma _{\sigma \sigma ^{\prime
}}a_{k^{\prime }\sigma ^{\prime }},  \label{ow}
\end{equation}
o\`{u} $a_{k\sigma }^{\dag }$ est l'op\'{e}rateur d'annihilation d'un \'{e}lectron de conduction de spin $\sigma $ et de moment $k$, $a_{k^{\prime }\sigma ^{\prime }}$ est l'op\'{e}rateur de cr\'{e}ation d'un \'{e}lectron de conduction de spin $\sigma ^{\prime }$ et de moment $k^{\prime }$, $J_{kk^{\prime }}$ est le couplage d'\'{e}change entre le moment local de l'impuret\'{e} et les \'{e}lectrons de conduction, $S$ est le spin localis\'{e} et $\sigma $ le vecteur form\'{e} des trois matrices
de Pauli
\begin{equation}
\begin{array}{ccccc}
\sigma _{x}=\left(
\begin{array}{cc}
0 & 1 \\
1 & 0%
\end{array}%
\right) & , & \sigma _{y}=\left(
\begin{array}{cc}
0 & -i \\
i & 0%
\end{array}%
\right) & , & \sigma _{z}=\left(
\begin{array}{cc}
1 & 0 \\
0 & -1%
\end{array}%
\right) .%
\end{array}%
\end{equation}
Par exemple, si $J_{kk^{\prime }}\propto \delta _{kk^{\prime }}$, la composante $z$ de $\sum \limits_{\sigma \sigma ^{\prime }}a_{k\sigma }^{\dag
}\sigma _{\sigma \sigma ^{\prime }}a_{k\sigma ^{\prime }}$ a pour valeur
\begin{equation}
\begin{array}{ccccc}
\sum \limits_{\sigma \sigma ^{\prime }}a_{k\sigma }^{\dag }\sigma _{\sigma
\sigma ^{\prime }}^{z}a_{k\sigma ^{\prime }} & = & a_{k\uparrow }^{\dag
}a_{k\uparrow }-a_{k\downarrow }^{\dag }a_{k\downarrow } & = & n_{k\uparrow
}-n_{k\downarrow }.%
\end{array}%
\end{equation}
Ce terme d\'{e}crit donc la polarisation de spin. Notons que les param\`{e}tres du hamiltonien $sd$ sont reli\'{e}s \`{a} ceux du hamiltonien d'Anderson par les relations de Schrieffer-Wolff :
\begin{equation}
J_{kk^{\prime }}=\frac{V_{k}^{\ast }V_{k^{\prime }}}{2}\left( \frac{1}{%
U+\epsilon _{d}-\epsilon _{k^{\prime }}}+\frac{1}{\epsilon _{k}-\epsilon _{d}%
}\right) .
\end{equation}
\vspace{-1.5cm}
\subsection{Interaction magn\'{e}tique dipolaire}

La premi\`{e}re interaction cens\'{e}e jouer un r\^{o}le crucial dans les ph%
\'{e}nom\`{e}nes de magn\'{e}tisme est l'interaction magn\'{e}tique
dipolaire \cite{I1,I4}. L'\'{e}nergie des deux dip\^{o}les magn\'{e}tiques $%
\mu _{1}$ et $\mu _{2}$ s\'{e}par%
\'{e}s par le vecteur $r$ est donn\'{e}e par :
\begin{equation}
E=\frac{\mu _{0}}{4\pi r^{3}}\left( \mu _{1}.\mu _{2}-\frac{3}{r^{2}}\left(
\mu _{1}.r\right) \left( \mu _{2}.r\right) \right) ,
\end{equation}
Pour une estimation de cette \'{e}nergie, nous choisissons des valeurs
typiques avec $\mu _{1}=\mu _{2}=1\mu _{B}$ et $r=2\mathring{A}$. Nous trouvons :
\begin{equation}
E=\frac{\mu _{0}\mu _{B}^{2}}{2\pi r^{3}}=2.1.10^{-24}J.
\end{equation}
La temp\'{e}rature correspondant $\left( E=kT\right) $ est bien au-dessous de $1K$. L'interaction magn%
\'{e}tique dipolaire est trop faible pour provoquer le ferromagn\'{e}tisme \cite{I4}.
\vspace{-0.8cm}
\subsection{Interaction directe}
\begin{figure}[th]
\centering
\includegraphics[scale=0.45]{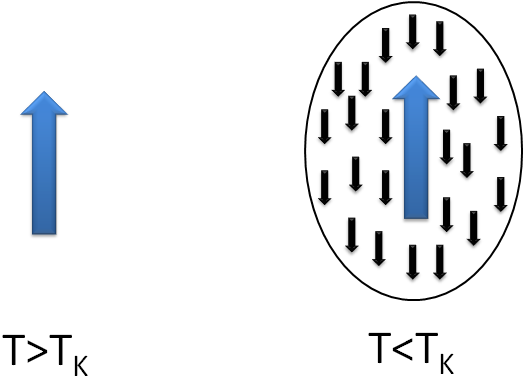}
\caption{Pour $T>T_{K}$, l'impuret\'{e} magn\'{e}tique est d\'{e}coupl%
\'{e}e des \'{e}lectrons de conduction. Pour $T<T_{K}$, les \'{e}lectrons de
conduction et l'impuret\'{e} magn\'{e}tique int\'{e}ragissent de mani\`{e}re coh%
\'{e}rente et forment le nuage de Kondo.}
\label{fig05}
\end{figure}

\vspace{-0.3cm}\hspace{-0.5cm}$\bullet $ \textbf{\emph{\'{E}change direct}}\newline
Pour un mod\`{e}le simple avec seulement deux \'{e}lectrons qui pr\'{e}%
sentent des vecteurs de position $r_{1}$ et $r_{2}$, la
fonction d'onde totale est compos\'{e}e du produit des fonctions d'onde de
deux \'{e}lectrons $\psi _{1}\left( r_{1}\right) $ et $%
\psi _{2}\left( r_{2}\right) $. Par cons\'{e}quent, le module carr\'{e} de la
fonction d'onde doit \^{e}tre invariant par \'{e}change des deux \'{e}%
lectrons. Puisque les \'{e}lectrons sont des fermions, le principe
d'exclusion de Pauli doit \^{e}tre respect\'{e}, ce qui conduit \`{a} une
fonction d'onde antisym\'{e}trique \cite{I4,I3}. Prenant en consid\'{e}ration le spin des
\'{e}lectrons, deux possibilit\'{e}s sont donn\'{e}es : une partie spatiale
sym\'{e}trique en combinaison avec une partie de spin antisym\'{e}trique ou
une partie spatiale antisym\'{e}trique en combinaison avec une partie de
spin sym\'{e}trique. Le premier cas repr\'{e}sente un \'{e}tat singulet $\chi _{S}$ avec $S_{total}=0$, le second cas pr\'{e}sente un \'{e}tat triplet $\chi _{T}$ avec $S_{total}=1$. Les fonctions d'onde totales correspondantes sont :
\begin{equation}
\begin{array}{ccc}
\psi _{S} & = & \frac{1}{\sqrt{2}}\left( \psi _{1}\left( r_{1}\right) \psi
_{2}\left( r_{2}\right) +\psi _{1}\left( r_{2}\right) \psi _{2}\left(
r_{1}\right) \right) .\chi _{S}, \\
\psi _{T} & = & \frac{1}{\sqrt{2}}\left( \psi _{1}\left( r_{1}\right) \psi
_{2}\left( r_{2}\right) -\psi _{1}\left( r_{2}\right) \psi _{2}\left(
r_{1}\right) \right) .\chi _{T}.%
\end{array}%
\end{equation}
Les \'{e}nergies des \'{e}tats singulet et triplet s'\'{e}l\`{e}vent \`{a} :
\begin{equation}
\begin{array}{ccc}
E_{S} & = & \int \psi _{S}^{\ast }H\psi _{S}dr_{1}dr_{2}, \\
E_{T} & = & \int \psi _{T}^{\ast }H\psi _{T}dr_{1}dr_{2},%
\end{array}%
\end{equation}
en tenant compte des parties normalis\'{e}es de spin des fonctions d'onde
singulet et triplet :
\begin{equation}
S^{2}_{total}=\left( S_{1}+S_{2}\right) ^{2}=S_{1}^{2}+S_{2}^{2}+2S_{1}.S_{2}.
\end{equation}
o\`{u} $S_{1}$ et $S_{2}$ sont les op\'{e}rateurs de spin des deux \'{e}lectrons. Ainsi, nous obtenons :
\begin{equation}
\begin{array}{ccc}
S_{1}.S_{2} & = & \frac{1}{2}S_{total}\left( S_{total}+1\right) -\frac{1}{2}%
S_{1}\left( S_{1}+1\right) -\frac{1}{2}S_{2}\left( S_{2}+1\right) \text{ \ }
\\
& = & \frac{1}{2}S_{total}\left( S_{total}+1\right) -\frac{3}{4}\qquad \text{%
avec}\qquad S_{1}=S_{2}=\frac{1}{2} \\
& = & \left \{
\begin{array}{c}
-\frac{3}{4}\qquad \text{pour\qquad }S_{total}=0\qquad \text{(singulet),} \\
+\frac{1}{4}\qquad \text{pour\qquad }S_{total}=1\qquad \text{(triplet). \ }%
\end{array}%
\right. \qquad%
\end{array}%
\end{equation}
L'hamiltonien effectif s'exprime alors comme suit :
\begin{equation}
H=\frac{1}{4}\left( E_{S}+3E_{T}\right) -\left( E_{S}-E_{T}\right)
S_{1}.S_{2}.\label{3-34}
\end{equation}
Le premier terme est constant et souvent inclus dans d'autres contributions
d'\'{e}nergie. Le second terme, qui d\'{e}pend du spin, joue un r\^{o}le important dans l'\'{e}tude des propri\'{e}t\'{e}s ferromagn\'{e}tiques \cite{I4}.\newline
En d\'{e}finissant la constante d'\'{e}change ou l'int\'{e}grale d'\'{e}%
change $J$ par:
\begin{equation}
J=\frac{E_{S}-E_{T}}{2}=\int \psi _{1}^{\ast }\left( r_{1}\right) \psi
_{2}^{\ast }\left( r_{2}\right) H\psi _{1}\left( r_{2}\right) \psi
_{2}\left( r_{1}\right) dV_{1}dV_{2},
\end{equation}
le terme d\'{e}pendant du spin dans l'hamiltonien effectif dans l'\'{e}quation (\ref{3-34}) devient alors :
\begin{equation}
H_{spin}=-2JS_{1}.S_{2}.
\end{equation}
Si l'int\'{e}grale d'\'{e}change $J$ est positif alors $E_{S}>E_{T}$. Dans ce cas, l'\'{e}%
tat de triplet qui satisfait $S_{total}=1$ est favoris\'{e} \'{e}nerg\'{e}tiquement. Si l'int\'{e}%
grale d'\'{e}change $J$ est n\'{e}gatif alors $E_{S}<E_{T}$, et c'est l'\'{e}tat de
singulet avec $%
S_{total}=0$ qui sera favoris\'{e} \'{e}nerg\'{e}tiquement dans ce cas.\newline
Contrairement \`{a} ce mod\`{e}le simple qui consid\`{e}re seulement deux \'{e}lectrons, la r\'{e}alit\'{e} est plus complexe. Les atomes dans les syst\`{e}mes magn\'{e}tiques contiennent un grand nombre d'\'{e}lectrons \cite{I1}. L'\'{e}quation de Schr%
\"{o}dinger de ces syst\`{e}mes \`{a} plusieurs corps ne peut \^{e}tre r\'{e}%
solue que par des approximations. Si l'on ne prend en compte que les interactions d'\'{e}change entre premier voisins, l'hamiltonien effectif peut \^{e}tre r\'{e}\'{e}crite sous la forme suivante :
\begin{equation}
H=-\sum_{ij}J_{ij}S_{i}.S_{j},
\end{equation}\newline
avec $J_{ij}$ \'{e}tant la constante d'\'{e}change entre spin $i$ et spin $j$. Le facteur $2$ est inclus dans la somme. Souvent, une bonne approximation est donn\'{e}e par :
\begin{equation}
J_{ij}=\left \{
\begin{array}{c}
J\qquad \text{pour les spins plus proches voisins,} \\
0\qquad \text{pour les autres.}\qquad \qquad \qquad \qquad\  \  \
\end{array}%
\right.
\end{equation}
G\'{e}n\'{e}ralement, $J$ est positive pour les \'{e}lectrons du m\^{e}me atome et n\'{e}gative si les deux \'{e}lectrons
appartiennent \`{a} diff\'{e}rents atomes.\newline
$\bullet $ \textbf{\emph{Effet Kondo}}\newline
L'interaction des \'{e}lectrons de conduction avec les impuret\'{e}s magn%
\'{e}tiques dans les m\'{e}taux a \'{e}t\'{e} trait\'{e}e pour la premi\`{e}re fois par Kondo \cite{t33,t29}. L'effet Kondo se produit lorsque le spin d'une
impuret\'{e} magn\'{e}tique se couple de fa\c{c}on antiferromagn\'{e}tique aux spins des \'{e}lectrons de conduction d'un m\'{e}tal pour former un
\'{e}tat singulet \cite{t41} comme illustr\'{e} sur la figure (\ref{fig05}).
\newline
L'interaction de Kondo est exprim\'{e}e sous la forme :
\begin{equation}
H_{K}=J_{K}\sum \limits_{i,\alpha }\delta \left( r_{\alpha }-R_{i}\right)
S_{i}.s_{\alpha },
\end{equation}
o\`{u} $s_{\alpha }$ repr\'{e}sente le spin d'un porteur itin\'{e}rant, $S_{i}$ le spin de
l'impuret\'{e} au site $i$ et $R_{i}$ et $r_{\alpha} $ repr\'{e}sentent respectivement les positions des impuret\'{e}s magn\'{e}tiques et celles des porteurs. La pr\'{e}sence de la distribution de Dirac $\delta $ indique que cette interaction est locale. $J_{K}$ est le
couplage local entre le spin des porteurs de charge et celui d'une impuret%
\'{e} occupant un site du r\'{e}seau h\^{o}te. L'hamiltonien total minimal s'\'{e}crit comme la somme suivante :
\begin{equation}
H=H_{0}+H_{K},
\end{equation}
o\`{u} l'hamiltonien $H_{0}$ d\'{e}crit les porteurs de
charge dans le mat\'{e}riau h\^{o}te. La
partie $H_{0}$ contient donc l'information relative \`{a} la structure de bande de
l'h\^{o}te en absence d'impurt\'{e}. \`{A} ce stade, il est important de noter que ce
mod\`{e}le est g\'{e}n\'{e}ral \cite{I3}.\newline
\`{A} haute temp\'{e}rature, l'impuret\'{e} magn\'{e}tique se comporte comme
un spin libre. Mais lorsque la temp\'{e}rature d\'{e}cro\^{\i}t en dessous
d'une temp\'{e}rature caract\'{e}ristique de l'alliage, appel\'{e}e temp\'{e}%
rature de Kondo $T_{K}$, l'interaction de Kondo devient de plus en plus importante.
L'impuret\'{e} perd alors progressivement son caract\`{e}re pour former
\`{a} tr\`{e}s basse temp\'{e}rature une impuret\'{e} statique compl\`{e}%
tement non-magn\'{e}tique. Cette transition lente et \'{e}largie de l'impuret%
\'{e} d'un \'{e}tat magn\'{e}tique vers un \'{e}tat non-magn\'{e}tique ;
s'exprime par une modification radicale des contributions de l'impuret\'{e}
\`{a} la plupart des propri\'{e}t\'{e}s \'{e}lectroniques de l'alliage, et en particulier par une modification importante de la diffusion
\'{e}lectronique \cite{t29,t41}.
\vspace{-0.5cm}
\subsection{Interaction indirecte}

\begin{figure}[!ht]
  \centering
\includegraphics[scale=0.65]{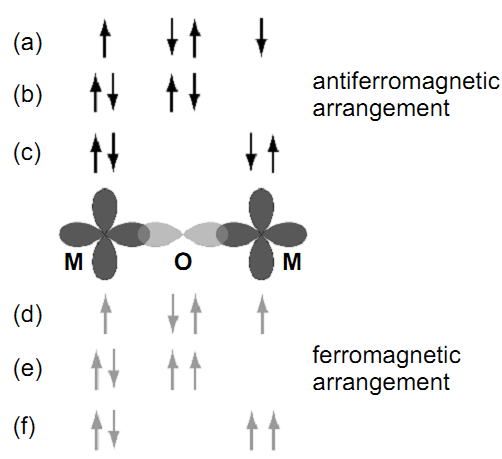}
\caption{\label{pic14}Pr\'{e}sence d'une interaction super\'{e}change en un oxyde magn%
\'{e}tique. Les fl\`{e}ches repr\'{e}sentent les spins des \'{e}lectrons
impliqu\'{e}s dans l'interaction entre le m\'{e}tal $(M)$ et un atome d'oxyg\`{e}%
ne $(O)$ \cite{I1}.}
\end{figure}

Selon le type des mat\'{e}riaux magn\'{e}tiques consid\'{e}r\'{e}s, nous distinguons diff\'{e}rentes classes des interactions d'\'{e}change indirect.\newline
$\bullet $ \textbf{\emph{Interaction super\'{e}change}}\newline
Ce type d'interaction d'\'{e}change indirect se produit dans les solides
ioniques. L'interaction d'\'{e}change entre les ions magn\'{e}tiques
non-voisins s'effectue via l'interm\'{e}diaire d'un ion non-magn\'{e}%
tique situ\'{e} entre les deux. La distance entre les ions magn\'{e}tiques doit \^{e}tre assez grande pour que l'\'{e}change direct puisse avoir lieu.\newline
Un exemple de solide ionique antiferromagn\'{e}tique est l'oxyde de mangan\`{e}se $MnO$. Dans ce mat\'{e}riau, chaque ion $%
Mn^{2+}$ pr\'{e}sente $5$ \'{e}lectrons dans sa couche $d$ avec des spins tous parall\`{e}les
en raison de la r\`{e}gle de Hund \cite{I4}. Les ions $O^{2-} $ par contre poss\`{e}dent des \'{e}lectrons qui occupent enti\`{e}rement les orbitales $p$. De ce fait, les spins sont tous align\'{e}s de fa\c{c}ons antiparall\`{e}les. Il existe deux possibilit\'{e}s pour
l'alignement des spins dans les atomes de $Mn$ voisins ; un
alignement parall\`{e}le qui conduit \`{a} un arrangement ferromagn\'{e}tique ou un alignement antiparall\`{e}le qui m\`{e}ne \`{a} un arrangement
antiferromagn\'{e}tique. Cette derni\`{e}re configuration est favoris\'{e}e \'{e}nerg\'{e}%
tiquement et induit une d\'{e}localisation des \'{e}lectrons impliqu\'{e}s
en raison d'une diminution de l'\'{e}nergie cin\'{e}tique, comme le montre la figure (\ref{pic14}).
\newline
Dans le cas antiferromagn\'{e}tique, les \'{e}lectrons avec leur \'{e}tat
fondamental donn\'{e} en (a) peuvent \^{e}tre \'{e}chang\'{e}s via les \'{e}%
tats excit\'{e}s indiqu\'{e}s en (b) et (c) menant \`{a} une d\'{e}%
localisation. Pour l'alignement ferromagn\'{e}tique avec l'\'{e}tat
fondamental correspondant pr\'{e}sent\'{e} en (d), le principe d'exclusion
de Pauli interdit les arrangements figurant dans (e) et (f). Ainsi, aucune d%
\'{e}localisation ne se produit \cite{I4}. Par cons\'{e}quent, le couplage
antiferromagn\'{e}tique entre deux atomes de $Mn$ est \'{e}nerg\'{e}tiquement
favoris\'{e}. Il est important que les \'{e}lectrons de l'atome d'oxyg\`{e}%
ne se trouvent dans la m\^{e}me orbitale ; \`{a} ce titre, l'atome doit
relier les deux atomes de $Mn$. Par ailleurs, les r\`{e}gles de Goodenough-Kanamori permettent de d\'{e}finir le type d'intercations entre les atomes si les configurations \'{e}lectroniques sont connues. Si les liaisons entre les ions de mangan\`{e}se et les ions d'oxyg\`{e}ne font des angles d'environ $180%
%TCIMACRO{\U{b0}}%
%BeginExpansion
{{}^\circ}%
%EndExpansion
$, l'interaction d'\'{e}change entre les ions magn\'{e}tiques sera antiferromagn\'{e}tique. Si les angles sont de $90%
%TCIMACRO{\U{b0}}%
%BeginExpansion
{{}^\circ}%
%EndExpansion
$, les interactions seront ferromagn\'{e}tiques.\newline
Le super\'{e}change n'est pas suffisant pour expliquer la pr\'{e}sence de la m\'{e}tallicit\'{e} dans les compos\'{e}s \`{a} valence mixte tels que les manganites. En $1950$, Jonker et Van Santen ont montr\'{e} par des mesures d'aimantation et de r\'{e}sistivit\'{e} le comportement ferromagn\'{e}tique et m\'{e}tallique de certains compos\'{e}s dop\'{e}s d\'{e}riv\'{e}s de $LaMnO_{3}$. En $1956$, ils corr\'{e}laient la structure cristalline, la structure magn\'{e}tique et les propri\'{e}t\'{e}s de transport \`{a} l'aide des interactions magn\'{e}tiques propos\'{e}es par Zener, dites de double \'{e}change.\newline
$\bullet $ \textbf{\emph{Interaction double-\'{e}change}}\newline
Le m\'{e}canisme de double-\'{e}change est un type d'\'{e}change magn\'{e}%
tique qui peut surgir entre les ions dans diff\'{e}rents \'{e}tats
d'oxydation. Cette interaction appara\^{\i}t entre les ions $3d$ ayant des \'{e}%
lectrons $d$ localis\'{e}s et d\'{e}localis\'{e}s. Ce mod\`{e}le ressemble
superficiellement \`{a} l'interaction super\'{e}change. Cependant, dans cette interaction, un alignement ferromagn\'{e}tique ou
antiferromagn\'{e}tique se produit entre deux atomes avec le m\^{e}me nombre
d'\'{e}lectrons de valence ; alors que l'interaction double-\'{e}change se produit uniquement lorsqu'un atome a un \'{e}lectron suppl%
\'{e}mentaire par rapport \`{a} l'autre \cite{I4}.\newline
En plus de l'interaction super\'{e}change et de double-\'{e}change, une autre interaction peut \^{e}tre pr\'{e}sente dans les mat\'{e}riaux antiferromagn\'{e}tiques pr\'{e}sentant un faible moment magn\'{e}tique, il s'agit de l'interaction antisym\'{e}trique.\newline
$\bullet $ \textbf{\emph{\'{E}change antisym\'{e}trique}}\newline
L'interaction antisym\'{e}trique connue aussi par interaction de Dzyaloshinskii-Moriya est, une correction relativiste du super\'{e}change et sa force est proportionnelle \`{a} la constante de couplage spin orbite. La th\'{e}orie Dzyaloshinskii-Moriya a \'{e}t\'{e} d\'{e}velopp\'{e}e pour expliquer le faible moment magn\'{e}tique pr\'{e}sent dans des antiferromagn\'{e}tiques telle que l'h\'{e}matite $\alpha Fe_{2}O_{3}$. L'interaction antisym\'{e}trique est proportionnelle au couplage spin orbite, est en g\'{e}n\'{e}ral faible par rapport au super\'{e}change magn\'{e}tique \cite{I4,t900}.
\begin{figure}[th]
\centering
\includegraphics[scale=0.28]{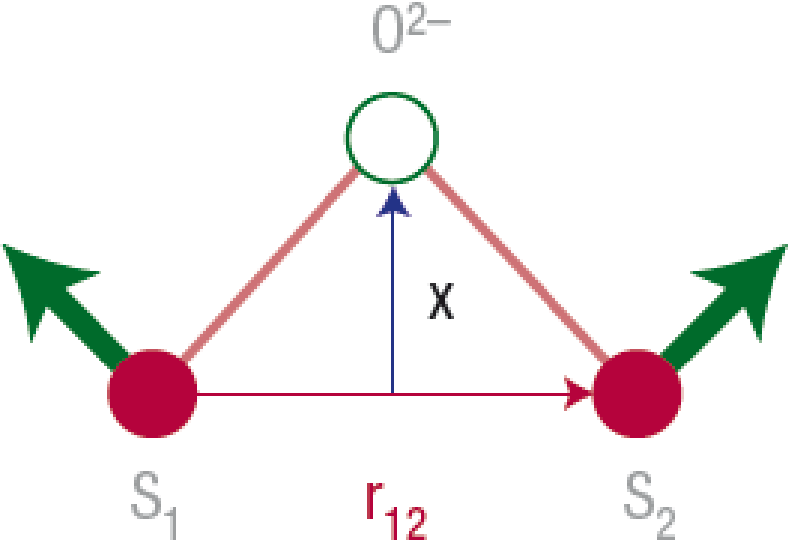}
\caption{Effet de l'interaction antisym\'{e}trique Dzyaloshinskii-Moriya.}
\label{pic24}
\end{figure}\newline
Soit deux atomes magn\'{e}tiques $M_{1}$ et $M_{2}$ portant chacun un spin s\'{e}par\'{e} par un atome diamagn\'{e}tique $O$ en g\'{e}n\'{e}ral un atome d'oxyg\`{e}ne. L'hamiltonien du syst\`{e}me peut alors s'\'{e}crire sous la forme :
\begin{equation}
H=-D.\left( S_{1}\times S_{2}\right) ,
\end{equation}
o\`{u} $D$ est le vecteur de Dzyaloshinskii qui est un vecteur constant parall\`{e}le \`{a} l'axe d'ordre $3$ et $S_{1}$ et $S_{2}$ sont deux spins voisins non \'{e}quivalents. Le vecteur Dzyaloshinskii $D$ est proportionnel au couplage spin orbit, il s'\'{e}crit comme suit :
\begin{equation}
D\propto \lambda x\times r_{12},
\end{equation}
o\`{u} $r_{12}$ est le vecteur unitaire le long de la ligne connectant les ions magn\'{e}tiques $1$ et $2$, $x$ repr\'{e}sente le d\'{e}placement de l'ion $O$ par rapport \`{a} cette ligne comme montr\'{e} sur la figure (\ref{pic24}) et $\lambda$ est le couplage spin orbit.\newline
L'interaction antisym\'{e}trique est reli\'{e}e \`{a} la sym\'{e}trie du syst\`{e}me et favorise l'alignement non colin\'{e}aire des moments magn\'{e}tiques \cite{I4}. \newline
$\bullet $ \textbf{\emph{Interaction RKKY}}\newline
L'interaction RKKY (Ruderman, Kittel, Kasuya, Yosida) se produit dans les m%
\'{e}taux avec des moments magn\'{e}tiques localis\'{e}s. C'est une
interaction d'\'{e}change indirect entre les spins de deux impuret\'{e}s magn%
\'{e}tiques, r\'{e}alis\'{e}e par l'interm\'{e}diaire des \'{e}lectrons de
conduction du cristal dans lequel les impuret\'{e}s sont introduites \cite{I4,t900}. L'interaction RKKY est une interaction tr\`{e}s forte et \`{a}
longue distance. L'hamiltonien de l'interaction RKKY est un
hamiltonien d'\'{e}change entre le spin $S_{1}$ d'un premier moment magn\'{e}tique
local et le spin $S_{2}$ d'un second moment magn\'{e}tique local. Il s'exprime comme suit :
\begin{equation}
H_{RKKY}=J_{RKKY}\left( r\right) \vec{S}_{1}\vec{S}_{2},
\end{equation}
o\`{u} $J_{RKKY}$ est la constante de couplage de l'interaction RKKY. L'interaction
RKKY est formul\'{e}e par :
\begin{equation}
J_{RKKY}\left( r\right) \propto F\left( 2k_{F}r\right) ,
\end{equation}
o\`{u} $k_{F}$ est le nombre d'onde de Fermi et la fonction $F\left( r\right)$ qui d\'{e}pend de la distance $r$ s\'{e}parant les deux moments magn\'{e}tiques locaux prend la forme suivante :%
\begin{equation}
F\left( r\right) =\frac{\sin r-r\cos r}{r^{4}}.
\end{equation}
Ce type de couplage d'\'{e}change \`{a} longue port\'{e}e et anisotrope se
traduit souvent par des arrangements compliqu\'{e}s de spin. Il poss\`{e}de
un comportement oscillant comme montr\'{e} dans la figure (\ref{fig07}).
\begin{figure}[th]
\centering
\includegraphics[scale=0.43]{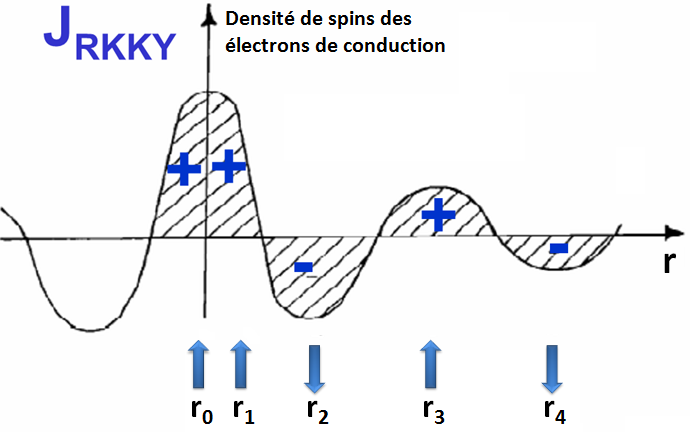}
\caption{Signe de la fonction de couplage $J_{RKKY}$\ et type de couplage en
fonction de la position atomique $r_{i}$ \cite{15}.}
\label{fig07}
\end{figure}\newline
Par ailleurs, l'interaction RKKY et l'interaction
Kondo ont un effet oppos\'{e}. Si l'interaction Kondo tend \`{a} faire
dispara\^{\i}tre le magn\'{e}tisme du syst\`{e}me, via un \'{e}crantage des
moments localis\'{e}s par les \'{e}lectrons de conduction, l'interaction
RKKY favorise l'\'{e}change entre les sites magn\'{e}tiques et tend \`{a} faire
appara\^{\i}tre un ordre magn\'{e}tique \cite{I4} comme illustr\'{e} dans la figure (\ref{fig08}).
\begin{figure}[h]
\centering
\includegraphics[scale=0.33]{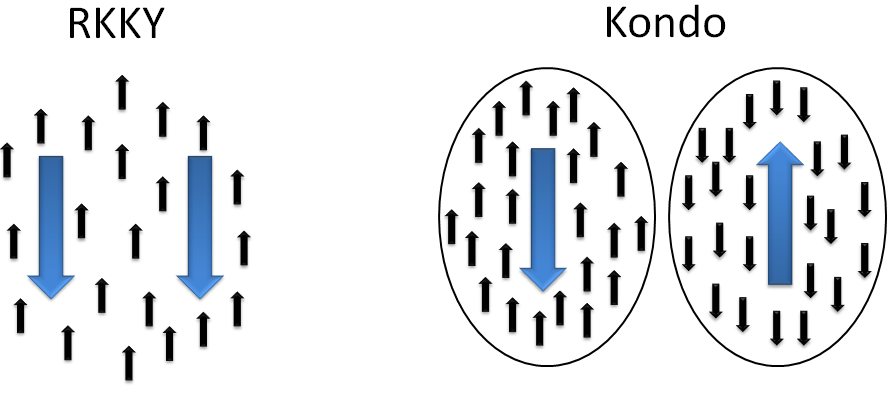}
\caption{\`{A} gauche: la polarisation des \'{e}lectrons de conduction
engendre une interaction effective magn\'{e}tique RKKY entre les spins
localis\'{e}s, ce qui peut conduire \`{a} un ordre magn\'{e}tique. \`{A}
droite, lorsque $J$\ domine, l'effet Kondo va former un singulet \`{a}
partir d'un spin localis\'{e} et d'un nuage d'\'{e}lectrons de conduction;
cet \'{e}tat ne poss\`{e}de donc pas d'ordre magn\'{e}tique.}
\label{fig08}
\end{figure}
\vspace{-1cm}
\section{Anisotropie magn\'{e}tique}

Si les interactions magn\'{e}tiques d\'{e}terminent les propri\'{e}t\'{e}s
magn\'{e}tiques des mat\'{e}riaux, l'anisotropie magn\'{e}tique les modifie et les contr\^{o}le. En effet, les mat\'{e}riaux magn\'{e}tiquement anisotropes ont des
directions pr\'{e}f\'{e}rentielles pour \^{e}tre magn\'{e}tis\'{e}s connues par directions d'aimantation facile ou axes faciles. En absence du champ magn\'{e}tique ext\'{e}rieur, les moments magn\'{e}tiques
d'un mat\'{e}riau anisotrope ont tendance \`{a} s'aligner le long de l'axe
facile \cite{I5,S2,I7}. Ces directions d'aimantation facile sont donn\'{e}es par les minima
de l'\'{e}nergie d'anisotropie magn\'{e}tique, qui est une somme de
plusieurs contributions. Dans les mat\'{e}riaux, l'\'{e}nergie magn%
\'{e}tocristalline et magn\'{e}tostatique sont les principales sources
d'anisotropie. Dans ce qui suit, nous allons d\'{e}crire quelque types d'anisotropie.
\vspace{-0.5cm}
\subsection{Anisotropie de forme}

L'anisotropie de forme appel\'{e}e aussi \'{e}nergie magn\'{e}tostatique provient du champ d\'{e}magn\'{e}tisant ; ce dernier d\'{e}pend de
l'orientation de l'aimantation par rapport \`{a} la forme du mat\'{e}riau. La pr\'{e}sence de ce champ d\'{e}magn\'{e}tisant rend anisotropes les propri\'{e}t\'{e}s magn\'{e}tiques du mat\'{e}riau selon sa forme. En effet, lorsque le mat\'{e}riau est sph\'{e}rique, l'\'{e}nergie magn\'{e}tostatique est nulle, mais elle prend des valeurs non n\'{e}gligeables si la sym\'{e}trie n'est plus sph\'{e}rique. Lorsque le mat\'{e}riau est un ellipso\"{\i}de allong\'{e}, la contribution de l'\'{e}nergie magn\'{e}tostatique peut \^{e}tre comparable ou d'un ordre de grandeur sup\'{e}rieur \`{a} l'\'{e}nergie magn\'{e}tocristalline. Par cons\'{e}quent, l'anisotropie de forme est une propri\'{e}t\'{e} extrins\`{e}que des mat\'{e}riaux non sph\'{e}riques \cite{I4}.
\vspace{-0.5cm}
\subsection{Anisotropie magn\'{e}tocristalline}

L'anisotropie magn\'{e}tocristalline est une propri\'{e}t\'{e} intrins\`{e}%
que du mat\'{e}riau, car elle d\'{e}coule directement de l'interaction d'%
\'{e}change et de la sym\'{e}trie du cristal. Cette anisotropie est caus\'{e}%
e par l'interaction spin-orbite. Ce couplage, qui est responsable de
l'orientation des spins en fonction de la sym\'{e}trie du r\'{e}seau, tend \`{a} aligner les moments magn\'{e}tiques le long des directions
cristallographiques pr\'{e}f\'{e}rentielles. Il en r\'{e}sultent certains
axes ou plans d'aimantation facile ou difficile. L'\'{e}nergie magn\'{e}%
tocristalline est g\'{e}n\'{e}ralement faible par rapport \`{a} l'\'{e}%
nergie d'\'{e}change. Cependant, la direction de l'aimantation est uniquement d%
\'{e}termin\'{e}e par l'anisotropie car l'interaction d'\'{e}change permet
d'aligner les moments magn\'{e}tiques parall\`{e}les ; peu importe la direction \cite{S4}.
\vspace{-0.5cm}
\subsection{Anisotropie de surface}

Les atomes de surface ont une sym\'{e}trie inf\'{e}rieure compar\'{e}e \`{a}
celle des atomes au sein du mat\'{e}riau. Leur influence sur l'\'{e}nergie
de la particule peut d\'{e}pendre de l'orientation de l'aimantation. Cela
donne lieu \`{a} l'anisotropie de surface qui peut aussi \^{e}tre d\'{e}%
pendante des impuret\'{e}s adsorb\'{e}es \`{a} la surface. L'ampleur de la contribution de la surface \`{a} l'\'{e}nergie de l'anisotropie magn\'{e}tique
augmente avec la diminution de la taille des mat\'{e}riaux, en particulier, elle devient importante seulement pour
des mat\'{e}riaux inf\'{e}rieures \`{a} $10nm$ \cite{I1,I4}.
\vspace{-0.5cm}
\subsection{Anisotropie magn\'{e}to\'{e}lastique}

L'\'{e}nergie d'anisotropie magn\'{e}to\'{e}lastique ou magn\'{e}tostriction
subsiste lors d'une d\'{e}formation m\'{e}canique du mat\'{e}riau. Cette d%
\'{e}formation change la direction de l'aimantation au sein du mat\'{e}riau
et induit donc une modification des propri\'{e}t\'{e}s magn\'{e}tiques.
L'existence d'une anisotropie magn\'{e}to\'{e}lastique est essentiellement due au couplage spin-orbite \cite{I5}.
\vspace{-0.5cm}
\section{Propri\'{e}t\'{e}s hyst\'{e}r\'{e}tiques}

\subsection{Structure des domaines magn\'{e}tiques}

Les mat\'{e}riaux magn\'{e}tiques poss\`{e}dent des r\'{e}gions uniform\'{e}%
ment aimant\'{e}es, qui pr\'{e}sentent d'une part des moments magn\'{e}tiques ordonn\'{e}s au sein des domaines magn\'{e}tiques connus aussi comme les domaines de Weiss, et d'autre part des moments magn\'{e}tiques d\'{e}sordonn\'{e}s dans
diff\'{e}rents domaines \cite{S1,I5,I7}.\newline
Ainsi, un mat\'{e}riau d\'{e}magn\'{e}tis\'{e} se compose de domaines
ordonn\'{e}s avec une aimantation totale nulle. Les interfaces entre ces
domaines sont appel\'{e}s parois des domaines magn\'{e}tiques. Dans ce qui suit, nous allons pr\'{e}senter les principes physiques indispensables
\`{a} la compr\'{e}hension du comportement des domaines magn\'{e}tiques et
de leurs parois dans les syst\`{e}mes macroscopiques.\newline
$\bullet $ \textbf{Domaines de Weiss}\newline
Les domaines de Weiss se forment dans des mat%
\'{e}riaux magn\'{e}tiquement ordonn\'{e}s tels les mat\'{e}riaux ferromagn%
\'{e}tiques, antiferromagn\'{e}tiques et ferrimagn\'{e}tiques. Ces domaines de Weiss sont des r\'{e}gions avec le m\^{e}me ordre magn%
\'{e}tique dans un mat\'{e}riau magn\'{e}tique. Ils sont s\'{e}par\'{e}s par
des parois de domaine \cite{I5,S2}. Les moments magn\'{e}tiques dans un domaine
s'alignent dans la m\^{e}me direction, et produisent une aimantation nette. Les aimantations des diff\'{e}rents domaines ont des directions diff\'{e}rentes.
\begin{figure}[!ht]
  \centering
\includegraphics[scale=0.4]{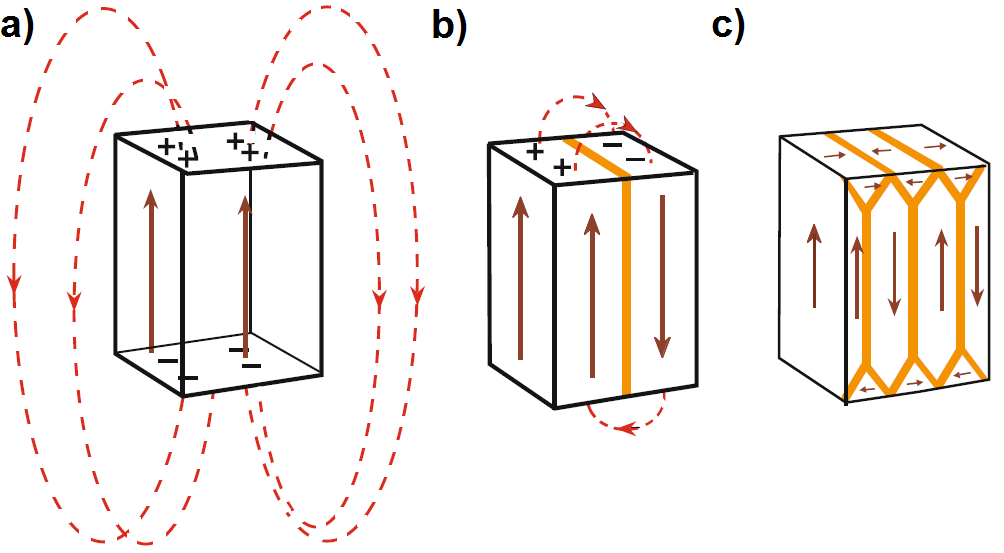}
\caption{\label{pic15}Diff\'{e}rentes structures de domaines pour un mat\'{e}riau
ferromagn\'{e}tique : (a) structure monodomaine, (b) structure poss\`{e}de
deux domaines et (c) structure de domaines de fermeture \cite{S2}.}
\end{figure}\newline
Une structure monodomaine, comme repr\'{e}sent\'{e}e sur la figure (\ref{pic15}-a), cr%
\'{e}e un champ magn\'{e}tique et minimise les \'{e}nergies d'\'{e}change et d'anisotropie mais l'\'{e}nergie dipolaire est alors importante car les p\^{o}les sont \'{e}loign\'{e}s les uns des autres. Pour r\'{e}duire l'\'{e}nergie dipolaire, le mat\'{e}riau peut se diviser en deux domaines avec une aimantation oppos\'{e}e (figure \ref{pic15}-b). Les lignes de champs forment une boucle d'un domaine \`{a} l'autre et son de direction oppos\'{e}e, r\'{e}duisant le champ \`{a} l'ext\'{e}rieur du mat\'{e}riau. Pour r\'{e}duire \`{a} nouveau le champ, chacun de ces domaines peut \`{a} nouveau se diviser, obtenant ainsi des domaines parall\`{e}les plus petits avec une aimantation dans diverses directions, ce qui supprime les p\^{o}les et donc le champ magn\'{e}tique \`{a} l'ext\'{e}rieur du mat\'{e}riau, comme illustr\'{e} sur la figure (\ref{pic15}-c). Les structures de domaine de fermeture peuvent \^{e}tre form\'{e}%
es pour minimiser l'\'{e}nergie dipolaire \cite{S2}. En consid\'{e}rant
le co\^{u}t \'{e}nerg\'{e}tique d\'{e}coulant de la formation des parois de
domaines magn\'{e}tiques, un \'{e}quilibre entre l'\'{e}nergie dipolaire et
le co\^{u}t \'{e}nerg\'{e}tique des parois de domaines est atteint.\newline
$\bullet $ \textbf{Paroi magn\'{e}tique}\newline
En magn\'{e}tisme, la paroi de domaine est l'interface s\'{e}parant les
domaines magn\'{e}tiques \`{a} l'int\'{e}rieur de laquelle l'aimantation va
tourner progressivement de la direction de l'aimantation d'un domaine \`{a}
celle de l'aimantation de son voisin. L'\'{e}nergie d'une paroi de domaine
est simplement la diff\'{e}rence entre les moments magn\'{e}tiques avant et
apr\`{e}s la cr\'{e}ation de la paroi de domaine. Cette valeur est g\'{e}n%
\'{e}ralement exprim\'{e}e en \'{e}nergie par unit\'{e} de surface de la
paroi \cite{S1,I2}.\newline
La largeur de la paroi de domaine varie en raison des deux \'{e}nergies oppos%
\'{e}es qui la cr\'{e}ent, notamment l'\'{e}nergie d'anisotropie magn\'{e}%
tocristalline et l'\'{e}nergie d'\'{e}change. Ces deux \'{e}nergies ont tendance \`{a}
\^{e}tre aussi faible que possible afin d'\^{e}tre dans un \'{e}tat \'{e}nerg%
\'{e}tique plus favorable. L'\'{e}nergie de l'anisotropie est minimale
lorsque les moments magn\'{e}tiques individuels sont align\'{e}s avec les
axes du r\'{e}seau cristallin r\'{e}duisant ainsi la largeur de la paroi de
domaine \cite{I4}. En revanche, l'\'{e}nergie d'\'{e}change est r\'{e}duite quand les
moments magn\'{e}tiques sont align\'{e}s parall\`{e}lement entre eux rendant
la paroi de domaine plus \'{e}paisse en raison de leur r\'{e}pulsion.%
\newline
Il existe deux types de parois de domaine selon l'angle que constituent deux aimantations de domaines voisins.
\begin{figure}[!ht]
  \centering
\includegraphics[scale=0.5]{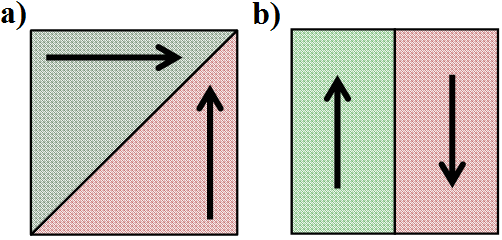}
\caption{\label{pic16}a) Paroi de domaine \`{a} $90%
%TCIMACRO{\U{b0}}%
%BeginExpansion
{{}^\circ}%
%EndExpansion
$ et b) paroi de domaine \`{a} $180%
%TCIMACRO{\U{b0}}%
%BeginExpansion
{{}^\circ}%
%EndExpansion
$ \cite{S2}.}
\end{figure}\newline
Une paroi de domaine \`{a} $90%
%TCIMACRO{\U{b0}}%
%BeginExpansion
{{}^\circ}%
%EndExpansion
$ qui s\'{e}pare deux domaines dans lesquels les aimantations forment un angle qui peut
\^{e}tre notablement diff\'{e}rent de $90%
%TCIMACRO{\U{b0}}%
%BeginExpansion
{{}^\circ}%
%EndExpansion
$, comme le montre la figure (\ref{pic16}-a). La direction
perpendiculaire d'une paroi de domaine correspond \`{a} la bissectrice entre
les directions de l'aimantation des domaines adjacents. Une paroi de domaine \`{a} $180%
%TCIMACRO{\U{b0}}%
%BeginExpansion
{{}^\circ}%
%EndExpansion
$ qui s\'{e}pare deux domaines dont les aimantations
sont antiparall\`{e}les, comme illustr\'{e} sur la figure (\ref{pic16}-b). Une inspection plus minutieuse de ce type de parois de domaines \`{a} r\'{e}v\`{e}le que nous pouvons distinguer deux classes selon la structure cristallographique du mat\'{e}riau \cite{I4} :\newline
$\bullet $ \textbf{Paroi de Bloch}\newline
Dans une paroi de Bloch, la rotation de l'aimantation se produit dans un
plan \'{e}tant parall\`{e}le \`{a} celui de la paroi de domaine. Par
ailleurs, l'aimantation peut tourner h\'{e}lico\"{\i}dalement tout en \'{e}tant
perpendiculaire au plan de la paroi, comme indiqu\'{e} sur la figure (\ref{pic17}).
\begin{figure}[!ht]
  \centering
\includegraphics[scale=0.66]{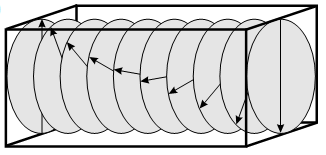}
\caption{\label{pic17}Rotation de l'aimantation dans une paroi de Bloch \cite{S2}.}
\end{figure}\newline
Les parois de Bloch apparaissent dans les mat\'{e}riaux magn\'{e}tiques dont la taille est consid\'{e}%
rablement plus grande que la largeur de la paroi de domaine. Dans ce cas, l'%
\'{e}nergie de d\'{e}magn\'{e}tisation n'a pas d'impact sur la structure
micromagn\'{e}tique de la paroi \cite{I5}.\newline
$\bullet $ \textbf{Paroi de N\'{e}el}\newline
Une paroi de N\'{e}el est une paroi dans laquelle la rotation du vecteur
d'aimantation se d\'{e}roule dans un plan perpendiculaire \`{a} celui de la
paroi de domaine, comme le montre la figure (\ref{pic18}).
\begin{figure}[!ht]
  \centering
\includegraphics[scale=0.66]{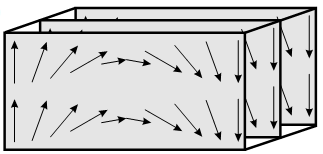}
\caption{\label{pic18}Rotation de l'aimantation dans une paroi de N\'{e}el \cite{S2}.}
\end{figure}\newline
Une paroi de N\'{e}el est une paroi dans laquelle l'aimantation tourne en restant parall\`{e}le \`{a} la surface du mat\'{e}riau, comme illustr\'{e} sur la figure (\ref{pic19}). Elle se produit souvent dans les syst\`{e}mes de couches minces ferromagn\'{e}tiques \cite{I4}.
\begin{figure}[!ht]
  \centering
\includegraphics[scale=0.32]{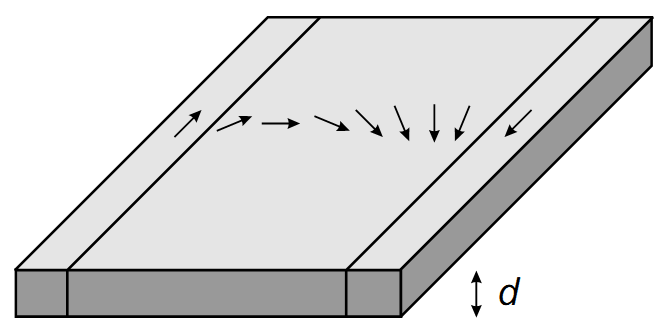}
\caption{\label{pic19}La paroi de N\'{e}el est \'{e}nerg\'{e}tiquement favorable dans
les syst\`{e}mes de couches minces pr\'{e}sentant seulement une petite \'{e}%
paisseur $d$ avec une aimantation dans le plan \cite{S2}.}
\end{figure}
\vspace{-0.8cm}
\subsection{Cycles d'hyst\'{e}r\'{e}sis}

\begin{figure}[!ht]
  \centering
\includegraphics[scale=0.45]{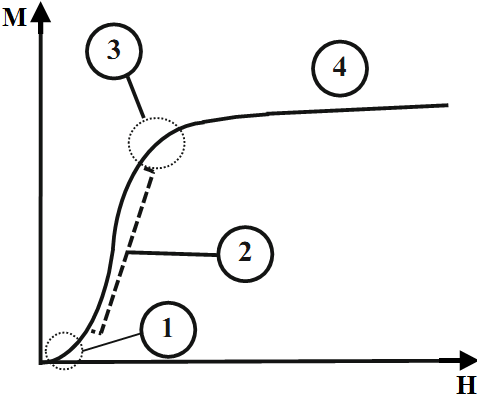}
\caption{\label{pic20}Courbe de premi\`{e}re aimantation.}
\end{figure}

Les cycles d'hyst\'{e}r\'{e}sis est l'une des caract\'{e}ristiques les plus distinctives des mat\'{e}riaux magn%
\'{e}tiques qui donnent la variation de l'aimantation en fonction du champ magn\'{e}tique
appliqu\'{e}. Le cycle d'hyst\'{e}r\'{e}sis est observ\'{e} pour les mat\'{e}%
riaux ferromagn\'{e}tiques et ferrimagn\'{e}tiques au-dessous de leur temp\'{e}rature critique. Il d\'{e}coule du r\'{e}arrangement des parois de domaines magn\'{e}tiques dans le mat\'{e}riau. Dans ce qui suit, nous \'{e}tudions en d\'{e}tail la courbe de premi\`{e}re
aimantation et le cycle d'hyst\'{e}r\'{e}sis.\newline
$\bullet $ \textbf{Courbe de premi\`{e}re aimantation}\newline
Lors de l'application d'un champ magn\'{e}tique externe \`{a} un flux
ferromagn\'{e}tique initialement d\'{e}magn\'{e}tis\'{e},
l'aimantation nette augmente. \`{A} noter que le taux d'augmentation \`{a} une
valeur de champ magn\'{e}tique donn\'{e}e d\'{e}pend de
nombreux facteurs. Parmi ces facteurs nous citons : l'orientation de champ externe par rapport \`{a}
l'orientation de chaque domaine individuel, l'amplitude de l'aimantation
spontan\'{e}e, la structure d\'{e}fectueuse du mat\'{e}riau ainsi que l'anisotropie et la g\'{e}om\'{e}trie du mat\'{e}riau \'{e}tudi\'{e}.\newline
La figure (\ref{pic20}) pr\'{e}sente la courbe de premi\`{e}re
aimantation d'un mat\'{e}riau polycristallin ferromagn\'{e}tique \cite{I7,I3}.
\newline
Il est clair que la courbe comporte quatre r\'{e}gions not\'{e}es $1$, $2$, $3$ et $4$.\newline
$\bullet $ La r\'{e}gion (1) est le d\'{e}but de la courbe o%
\`{u} le d\'{e}placement des parois de Bloch s\'{e}parant les domaines est r%
\'{e}versible. Cette partie de la courbe montre une augmentation de
l'aimantation lors de l'application d'un champ magn\'{e}tique \`{a} un mat\'{e}riau d\'{e}magn\'{e}tis\'{e}. Toutefois, apr\`{e}s la suppression du
champ appliqu\'{e}, l'aimantation macroscopique revient \`{a} sa valeur
initiale z\'{e}ro \cite{I3}.\newline
$\bullet $ Dans la r\'{e}gion (2) de la courbe de premi\`{e}re
aimantation, le d\'{e}placement des parois de Bloch s\'{e}parant les
domaines sont irr\'{e}versibles. L'aimantation nette augmente de telle sorte
que les parois de domaine vont occuper de nouvelles positions et ne peuvent plus retrouver leur position initiale. Dans cette r\'{e}%
gion, les d\'{e}placements des parois sont drastiques.\newline
$\bullet $ Avec cette nouvelle augmentation du champ magn\'{e}tique, la
courbe d'aimantation atteint la r\'{e}gion (3) qui commence
avant la fin de la r\'{e}gion de d\'{e}placement des parois de
domaine irr\'{e}versible, afin que les r\'{e}gions (2) et (3) se
chevauchent. La r\'{e}gion (3) est \`{a} la fois reconnue comme r\'{e}gion
de d\'{e}placement des parois de domaine irr\'{e}versible et de rotation des
aimantations des domaines. Dans cette r\'{e}gion, l'aimantation nette augmente avec le champ appliqu\'{e}.\newline
$\bullet $ Pour de fortes excitations magn\'{e}tiques, se produit une rotation des aimantations des domaines dans la direction du champ magn\'{e}tique appliqu\'{e}. Finalement, dans la r\'{e}gion (4), l'aimantation du mat\'{e}riau se rapproche pour atteindre la saturation \cite{I3}.%
\newline
$\bullet $ \textbf{Cycle d'hyst\'{e}r\'{e}sis}\newline
Dans les mat\'{e}riaux ferromagn\'{e}tiques, les courbes d'aimantation en terme du champ magn\'{e}tique appliqu\'{e} sont non lin\'{e}aires \`{a} cause du changement dans la structure de domaine magn\'{e}tique. Ces mat\'{e}riaux pr\'{e}sentent le cycle d'hyst\'{e}r\'{e}sis dans lequel l'aimantation $M$ et le champ magn\'{e}tique ext\'{e}rieur ont des comportements diff\'{e}rents \cite{I5,S2}.
\begin{figure}[!ht]
  \centering
\includegraphics[scale=0.55]{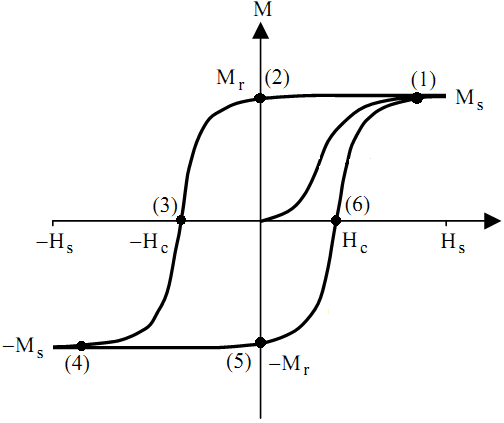}
\caption{\label{pic23}Repr\'{e}sentation d'un cycle d'hyst\'{e}r\'{e}sis traduisant le
retournement des parois s\'{e}parant les domaines magn\'{e}tiques.}
\end{figure}\newline
En effet, $M$ augmente avec le champ magn\'{e}tique ext\'{e}rieur et ne revient pas \`{a} z\'{e}ro apr\`{e}s l'annulation du champ comme montr\'{e} dans la figure (\ref{pic23}) en partant de l'aimantation \`{a} saturation, nous distinguons six points diff\'{e}rents :\newline
1) Au point (1), tous les domaines magn\'{e}tiques sont align\'{e}s dans le m%
\^{e}me sens que le champ magn\'{e}tique appliqu\'{e}.\newline
2) En se d\'{e}pla\c{c}ant du point (1) au point (2), le champ magn%
\'{e}tique se r\'{e}duit \`{a} z\'{e}ro. Le mat\'{e}riau garde une
certaine aimantation dite r\'{e}manente m\^{e}me si le champ magn\'{e}tique est
nul.\newline
3) Quand le champ est invers\'{e}, l'aimantation d\'{e}cro\^{\i}t jusqu'au point
de c{\oe}rcivit\'{e} (3), o\`{u} elle s'annule.\newline
4) En augmentant la valeur du champ magn\'{e}tique dans
le sens n\'{e}gatif, le mat\'{e}riau atteint \`{a} nouveau son aimantation
\`{a} saturation dans le sens inverse au point (4).\newline
5) En se d\'{e}pla\c{c}ant du point (4) au point (5), le champ magn%
\'{e}tique se r\'{e}duit \`{a} z\'{e}ro et le mat\'{e}riau demeure aimant\'{e} sans champ magn\'{e}tique ext\'{e}rieur.\newline
6) En r\'{e}augmentant le champ magn\'{e}tique dans le sens positif cette fois-ci, la
courbe passe par le point de c{\oe}rcivit\'{e} (6) o\`{u} l'aimantation est
nulle, pour enfin atteindre le point de saturation initial (1). On obtient
ainsi le cycle d'hyst\'{e}r\'{e}sis.\newline
Les cycles d'hyst\'{e}r\'{e}sis magn\'{e}tiques ne rel\`{e}vent que des
mat\'{e}riaux ferromagn\'{e}tiques et ferrimagn\'{e}tiques. D'autres ordres
magn\'{e}tiques, tels que les verres de spin, pr\'{e}sentent \'{e}galement
ce comportement dans la courbe d'\'{e}volution de l'aimantation en fonction du champ magn\'{e}tique ext\'{e}rieur \cite{I1}.
\vspace{-0.5cm}
\subsection{Param\`{e}tres caract\'{e}ristiques des cycles d'hyst\'{e}r\'{e}%
sis}

Le cycle d'hyst\'{e}r\'{e}sis est un moyen pour caract\'{e}riser les mat\'{e}%
riaux magn\'{e}tiques. Il nous procure des informations sur l'aimantation \`{a} saturation $M_{s}$ qui est une propri\'{e}t\'{e} magn\'{e}tique intrins\`{e}que des mat\'{e}riaux. Il nous fournit \'{e}galement deux propri\'{e}t\'{e}s extrins\`{e}ques qui sont
l'aimantation r\'{e}manente $M_{r}$ et le champ c{\oe}rcitif $H_{c}$. Ces deux propri\'{e}t\'{e}s d\'{e}pendent de plusieurs facteurs ext\'{e}rieurs tel que la forme du mat\'{e}riau, la rugosit\'{e} de surface, les d\'{e}fauts microscopiques
et les propri\'{e}t\'{e}s thermiques sans oublier la variation de l'intensit\'{e} du champ magn\'{e}tique appliqu\'{e}.\newline
Dans ce qui suit, nous d\'{e}finissons bri\`{e}vement les quatre param\`{e}tres caract\'{e}ristiques des cycles d'hyst\'{e}r\'{e}sis que nous retrouvons.\newline
$\bullet $ \textbf{Champ \`{a} saturation}\newline
Le champ \`{a} saturation est la valeur minimale du champ magn\'{e}tique appliqu\'{e} pour lequel l'aimantation du mat\'{e}riau atteint sa valeur \`{a} saturation \cite{S2}.\newline
$\bullet $ \textbf{Aimantation \`{a} saturation}\newline
L'aimantation \`{a} saturation $M_{s}$ correspond \`{a} la valeur maximale de
l'aimantation du mat\'{e}riau, o\`{u} tous les moments magn\'{e}tiques sont
parall\`{e}les au champ magn\'{e}tique appliqu\'{e} \cite{S2}.\newline
$\bullet $ \textbf{Aimantation r\'{e}manente}\newline
L'aimantation r\'{e}manente $M_{r}$ est l'aimantation r\'{e}siduelle du mat\'{e}riau
qui est obtenue en r\'{e}duisant le champ de fa\c{c}on monotone \`{a} z\'{e}%
ro apr\`{e}s la saturation. Dans ce cas, une partie des moments magn\'{e}%
tiques reste orient\'{e}e dans la direction du champ appliqu\'{e} du fait du blocage des parois de Bloch \cite{S2}.\newline
$\bullet $ \textbf{Champ c{\oe}rcitif}\newline
Le champ c{\oe}rcitif est le champ qui doit \^{e}tre appliqu\'{e} dans le
sens oppos\'{e} au champ \`{a} saturation, pour r\'{e}duire l'aimantation
\`{a} z\'{e}ro afin de d\'{e}magn\'{e}tiser le mat\'{e}riau.\newline
Ces propri\'{e}t\'{e}s auxquelles nous nous r\'{e}f\'{e}rons dans toutes les applications magn\'{e}tiques sont tr\`{e}s importantes \cite{I7}.
\vspace{-0.5cm}
\subsection{Mat\'{e}riaux magn\'{e}tiques durs et doux}

\begin{figure}[!ht]
  \centering
\includegraphics[scale=0.45]{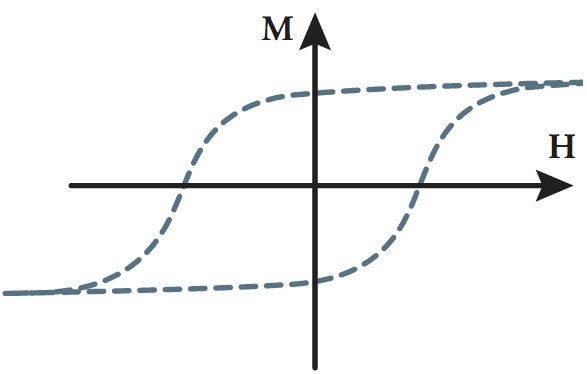}
\caption{\label{pic21}Cycle d'hyst\'{e}r\'{e}sis d'un mat\'{e}riau ferromagn\'{e}tique
dur.}
\end{figure}

\begin{figure}[!ht]
  \centering
\includegraphics[scale=0.6]{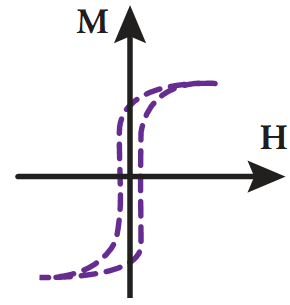}
\caption{\label{pic22}Cycle d'hyst\'{e}r\'{e}sis d'un mat\'{e}riau ferromagn\'{e}tique
doux.}
\end{figure}

Les mat\'{e}riaux ferromagn\'{e}tiques et ferrimagn\'{e}tiques peuvent \'{e}%
galement \^{e}tre class\'{e}s en deux groupes selon la largeur de leur cycle d'hyst\'{e}r\'{e}sis et la facilit\'{e} avec lesquelles ils peuvent \^{e}tre magn\'{e}tis\'{e}s.
En effet, quand nous tra\c{c}ons le cycle d'hyst\'{e}r\'{e}sis d'un mat\'{e}%
riau ferromagn\'{e}tique ou ferrimagn\'{e}tique, une certaine quantit\'{e} d'%
\'{e}nergie est perdue sous forme de chaleur pendant le processus du
mouvement des parois de domaines magn\'{e}tiques \cite{S1,I3}. Ces pertes par hyst\'{e}r\'{e}sis sont donn\'{e}es par :
\begin{equation}
W=\int H.dM,
\end{equation}
o\`{u} $H$ est le champ magn\'{e}tique appliqu\'{e} et $M$ est l'aimantation du syst%
\`{e}me. Nous constatons qu'elles sont proportionnelles \`{a} la
surface du cycle d'hyst\'{e}r\'{e}sis. Il s'ensuit que nous pouvons distinguer la surface du cycle d'hyst\'{e}r%
\'{e}sis, les mat\'{e}riaux magn\'{e}tiquement doux qui sont caract\'{e}ris\'{e}s par des petites surfaces de leur cycle d'hyst\'{e}r\'{e}sis et les mat\'{e}riaux magn\'{e}tiquement dur qui ont de grandes surfaces \cite{I4}.\newline
$\bullet $ \textbf{Mat\'{e}riaux durs}\newline
Les mat\'{e}riaux magn\'{e}tiques durs sont des mat\'{e}riaux ferromagn\'{e}%
tiques ou ferrimagn\'{e}tiques ayant un cycle d'hyst%
\'{e}r\'{e}sis large et des valeurs \'{e}lev\'{e}es de champ c{\oe}rcitif et d'aimantation r\'{e}manente comme indiqu\'{e} dans la figure (\ref{pic21}).
\newline
Ces mat\'{e}riaux sont difficiles \`{a} magn\'{e}tiser et \`{a} d\'{e}magn%
\'{e}tiser \cite{S1}. \`{A} ce titre, ils sont utilis\'{e}s comme aimants permanents.
Les dispositifs qui utilisent des aimants permanents sont de deux types
principaux :
\newline
$\bullet $ Dans le premier type, les propri\'{e}t\'{e}s de l'aimant
permanent sont utilis\'{e}es pour g\'{e}n\'{e}rer un champ magn\'{e}tique, g%
\'{e}n\'{e}ralement dans un entrefer entre les p\^{o}les de l'aimant. Nous retrouvons ce type de mat\'{e}riau dans les haut-parleurs, les aimants de r\'{e}frig%
\'{e}rateur, les instruments galvanom\'{e}triques, les dynamos, les g\'{e}n%
\'{e}rateurs et les moteurs \'{e}lectriques\ldots.
\newline
$\bullet $ Dans le deuxi\`{e}me type, le but est de g\'{e}n\'{e}rer une
force entre l'aimant et une sorte d'armature mobile. Ce type de mat\'{e}riau est appliqu\'{e} dans toutes sortes de dispositifs de levage et de serrage.\newline
Il existe plusieurs types de mat\'{e}riaux magn\'{e}tiques durs utilis\'{e}s comme aimants
permanents, parmi lesquels nous trouvons : les mat\'{e}riaux m\'{e}talliques
de type Alnico, les ferrites dures ou les mat\'{e}riaux interm\'{e}%
talliques\ldots. Ces derniers sont de nos jours tr\`{e}s pr\'{e}sents sur le march%
\'{e} \cite{S3}.\newline
$\bullet $ \textbf{Mat\'{e}riaux doux}\newline
Les mat\'{e}riaux ferromagn\'{e}tiques ou ferrimagn\'{e}tiques doux ont \'{e}%
t\'{e} d\'{e}velopp\'{e}s avec des applications techniques, afin de
permettre aux changements d'aimantation de se produire facilement dans les
champs faibles \cite{S1,S3}. Ils sont caract\'{e}ris\'{e}s par un cycle d'hyst\'{e}r\'{e}%
sis tr\`{e}s \'{e}troit de petite surface, comme le montre la figure (\ref{pic22}).
\newline
Ces mat\'{e}riaux ont donc de petites pertes par hyst\'{e}r\'{e}sis. Ils ont
\'{e}galement une aimantation r\'{e}manente importante et un champ c{\oe}rcitif faible. Ces mat\'{e}riaux contiennent g\'{e}n\'{e}ralement moins
d'impuret\'{e}s et les parois de domaines magn\'{e}tiques peuvent facilement
\^{e}tre d\'{e}plac\'{e}es avec une petite quantit\'{e} d'\'{e}nergie. Par
ailleurs, les mat\'{e}riaux doux sont faciles \`{a} magn\'{e}tiser. Ils sont
utilis\'{e}s dans les bobines des transformateurs, les g\'{e}n\'{e}rateurs
et les moteurs. Dans ces applications, l'aimantation doit \^{e}tre invers%
\'{e}e plusieurs fois par seconde et il est important que l'\'{e}nergie
dissip\'{e}e par cycle soit r\'{e}duite au minimum. Les mat\'{e}riaux
ferromagn\'{e}tiques ou ferrimagn\'{e}tiques doux sont compos\'{e}s des \'{e}%
l\'{e}ments principaux : fer, cobalt, nickel, magn\'{e}sium, molybd\`{e}ne
et silicium \cite{I3}.
\newline
Dans ce chapitre, nous avons soulign\'{e} le r\^{o}le important qui jouent les interactions d'\'{e}change directes ou indirectes dans l'apparition de l'ordre magn\'{e}tique dans les mat\'{e}riaux. Nous avons \'{e}galement donner des classes de mat\'{e}riaux magn\'{e}tiques selon la nature des interactions entre les constituants \'{e}l\'{e}mentaires ou selon la forme de la boucle d'hyst\'{e}r\'{e}sis. Ainsi, avec tout les outils pr\'{e}sent\'{e}s dans les trois premiers chapitres, nous allons consacrer le dernier chapitre \`{a} nos contributions dans le domaine.
\def\cleardoublepage{\clearpage}
\newpage
\strut
\newpage
\setcounter{chapter}{3}
\chapter{Contributions \`{a} l'\'{e}tude Monte Carlo des propri\'{e}t\'{e}s magn\'{e}tiques des nanomat\'{e}riaux type graphyne et graphone}
\graphicspath{{Chapitre4/figures/}}
Ce chapitre est consacr\'{e} \`{a} la pr\'{e}sentation des r\'{e}sultats de
nos contributions relatives \`{a} l'\'{e}tude Monte Carlo des propri\'{e}t\'{e}s magn\'{e}tiques des nanomat\'{e}riaux type graphyne et graphone. Par d\'{e}%
faut d'espace et pour ne pas alourdir le manuscrit, nous allons pr\'{e}senter les r\'{e}sum\'{e}s de tous nos travaux de recherche dans le domaine, mais nous n'en exposerons que ceux d\'{e}j\`{a} publi\'{e}s dans des journaux internationaux. \`{A} ce titre, nous avons rassembl\'{e} nos travaux de recherche dans trois sections principales. Dans la premi\`{e}re section, nous pr\'{e}sentons les contributions traitant des propri\'{e}t\'{e}s magn\'{e}tiques et hyst\'{e}r\'{e}tiques des mat\'{e}riaux ferromagn\'{e}tiques types graphone. Dans la deuxi\`{e}me section se trouvent deux de nos travaux de recherche o\`{u} nous avons examin\'{e} les propri\'{e}t\'{e}s thermodynamiques et hyst%
\'{e}r\'{e}tiques des mat\'{e}riaux c{\oe}ur-coquille type nanoruban de graph\`{e}ne et type nanoparticule de graphyne. Finalement, la troisi\`{e}me section exhibe trois de nos publications o\`{u} nous avons \'{e}tudi\'{e} dans les deux premi\`{e}res l'effet des d\'{e}fauts sur les propri\'{e}t\'{e}s magn\'{e}tiques et hyst\'{e}r\'{e}tiques des nanorubans des mat\'{e}riaux type graph\`{e}ne. Tandis que la troisi\`{e}me s'int\'{e}resse \`{a} l'\'{e}tude de l'effet de surface sur le comportement de compensation d'un nanocube avec la morphologie surface-volume.

\section{Mat\'{e}riaux ferromagn\'{e}tiques type graphone}

Depuis son isolation, le graph\`{e}ne suscite un immense int\'{e}r\^{e}t
dans la recherche scientifique et technologique, en raison de ses propri\'{e}%
t\'{e}s exotiques. La d\'{e}couverte de ce nouveau mat\'{e}riau a \'{e}t\'{e}
une force motrice de la communaut\'{e} scientifique pour synth\'{e}tiser et
caract\'{e}riser de nouveaux mat\'{e}riaux ayant des morphologies similaires
du fait de leurs propri\'{e}t\'{e}s uniques \`{a} l'\'{e}chelle nanom\'{e}%
trique. Bien que le graph\`{e}ne ne soit pas magn\'{e}tique, il est consid%
\'{e}r\'{e} parmi les mat\'{e}riaux des plus prometteurs en spintronique. En
effet, les nanostructures \`{a} base de graph\`{e}ne, tel que le graphone, montrent des remarquables propri\'{e}t%
\'{e}s magn\'{e}tiques aussi bien en pratique qu'en th\'{e}orie. Le graphone
est un semiconducteur ferromagn\'{e}tique avec un petit gap indirect. Ce
nouveau d\'{e}riv\'{e} du graph\`{e}ne est issu de son hydrog\'{e}nation
partielle. Il a \'{e}t\'{e} propos\'{e} th\'{e}oriquement en $2012$ et synth%
\'{e}tis\'{e} en $2014$. En outre, signalons que le magn\'{e}tisme, dans ce d%
\'{e}riv\'{e} de graph\`{e}ne, provient des \'{e}lectrons localis\'{e}s sur
les atomes de carbone non-hydrog\'{e}n\'{e}s. Le graphone est donc un
candidat prometteur pour les nouvelles applications dans les domaines de la
nano\'{e}lectronique et la spintronique.\newline
Nous avons tenu compte de l`importance actuelle des mat\'{e}riaux \`{a}
base de graph\`{e}ne, notamment le graphone, et du r\^{o}le qu`ils seront amen%
\'{e}s \`{a} prendre dans le futur du fait de leurs propri\'{e}t\'{e}s
exotiques et de leurs nombreuses applications dans le domaine de la
nanotechnologie et de la spintronique. \`{A} ce titre, nous avons essay\'{e}
modestement d'enrichir la liste des contributions dans ce domaine des nanomat%
\'{e}riaux par deux travaux de recherche cons\'{e}cutifs.

\subsection{Magnetic phase transitions in pure zigzag graphone nanoribbons%
\newline
(J. Phys. Lett. A 379 (2015) 753-760)}

$\bullet $ \textbf{R\'{e}sum\'{e} de la publication 1 :}

Dans cette publication, nous avons \'{e}tudi\'{e} les propri\'{e}t\'{e}s magn%
\'{e}tiques et hyst\'{e}r\'{e}tiques d'un nanoruban de graphone pur de type
zigzag en utilisant la simulation Monte Carlo et le calcul de champ moyen.
Dans ce travail, nous avons pr\'{e}sent\'{e} une \'{e}tude d\'{e}taill\'{e}e
permettant de comprendre le comportement magn\'{e}tique et hyst\'{e}r\'{e}%
tique d'un nanoruban de graphone pur avec des bords de type zigzag.
Signalons que le magn\'{e}tisme, dans cette d\'{e}riv\'{e}e de graph\`{e}ne,
provient des \'{e}lectrons localis\'{e}s sur les atomes de carbone qui ne
sont pas hydrog\'{e}n\'{e}s.

%$\bullet $ \textbf{Contenu de la publication 1 :}
%\includepdf[pages={1-8}]{article1.pdf}
\subsection{Edge effect on magnetic phases of doped zigzag graphone
nanoribbons\newline
(J. Magn. Magn. Mater. 374 (2015) 394-401)}

$\bullet $ \textbf{R\'{e}sum\'{e} de la publication 2 :}

Dans ce papier, nous nous sommes bas\'{e}s sur deux outils tr\`{e}s
importants : la m\'{e}thode Monte Carlo et la th\'{e}orie de champ moyen
afin d'investiguer les propri\'{e}t\'{e}s magn\'{e}tiques des nanorubans de
graphone. Nous avons pr\'{e}sent\'{e} une alternative qui permet de contr%
\^{o}ler le magn\'{e}tisme dans des nanorubans de graphone de type zigzag en
introduisant des impuret\'{e}s magn\'{e}tiques dans diff\'{e}rentes
positions. Nous avons aussi examin\'{e} l'effet des positions et du nombre
des atomes magn\'{e}tiques substitu\'{e}s sur les transitions de phase des
nanorubans de graphone de type zigzag dans trois cas diff\'{e}rents : mono-,
bi- et tri-dop\'{e}.

%$\bullet $ \textbf{Contenu de la publication 2 :}
%\includepdf[pages={1-8}]{article2.pdf}

\section{Mat\'{e}riaux c{\oe}ur-coquille type nanoruban de graph\`{e}ne et type nanoparticule de graphyne}

Au cours des derni\`{e}res ann\'{e}es, le ph\'{e}nom\`{e}ne de
ferrimagn\'{e}tisme a \'{e}t\'{e} l'un des sujets les plus \'{e}tudi\'{e}s en m%
\'{e}canique statistique et physique de la mati\`{e}re condens\'{e}e, en
raison de ses propri\'{e}t\'{e}s physiques int\'{e}ressantes. En outre,
l'existence \'{e}ventuelle d'une temp\'{e}rature de compensation sous
certaines conditions rend les mat\'{e}riaux ferrimagn\'{e}tiques tr\`{e}s
prometteurs pour des applications technologiques importantes dans le domaine
de l'enregistrement thermo-optique. D'autre part, les r\'{e}sultats int\'{e}%
ressants obtenus \`{a} partir des \'{e}tudes th\'{e}oriques des syst\`{e}mes
magn\'{e}tiques nanostructur\'{e}s ont ouvert un nouveau champ \`{a} la
recherche des ph\'{e}nom\`{e}nes magn\'{e}tiques critiques \`{a} l'\'{e}%
chelle nanom\'{e}trique. En effet, un int\'{e}r\^{e}t tout particulier est
port\'{e} aux nanostructures avec la morphologie c{\oe}ur-coquille. Ce genre de syst%
\`{e}me rev\^{e}t des comportements magn\'{e}tiques importants et
inhabituels. Dans cette section, nous pr\'{e}sentons les r\'{e}sum\'{e}s de
nos deux contributions ainsi que les contenus de ces deux articles dont le r\'{e}%
sultat met en evidence les propri\'{e}t\'{e}s magn\'{e}tiques et hyst\'{e}r%
\'{e}tiques int\'{e}ressantes de mat\'{e}riaux avec la morphologie c{\oe}ur-coquille.

\subsection{Monte Carlo study of magnetic behavior of core-shell nanoribbon%
\newline
(J. Magn. Magn. Mater. 374 (2015) 639-646)}

$\bullet $ \textbf{R\'{e}sum\'{e} de la publication 3 :}

Dans cette publication, nous avons consid\'{e}r\'{e} les propri\'{e}t\'{e}s
magn\'{e}tiques et hyst\'{e}r\'{e}tiques d'un nanoruban ferrimagn\'{e}tique
avec la morphologie c{\oe}ur-coquille. Ce genre de syst\`{e}me pr\'{e}sente des
comportements magn\'{e}tiques importants et inhabituels. Ainsi, nous avons
analys\'{e} l'\'{e}tat fondamental ainsi que les diagrammes de phase de ce
syst\`{e}me. Nous avons montr\'{e} l'existence d'une temp\'{e}rature de
compensation de tr\`{e}s haute importance dans le stockage d'information et
plus particuli\`{e}rement dans l'enregistrement thermo-optique. Nous avons m%
\^{e}me sp\'{e}cifi\'{e} l'effet des valeurs du couplage du c{\oe}ur et de la coquille ainsi que le couplage interm\'{e}diaire sur la temp\'{e}rature de
compensation, tout en donnant explicitement les conditions permettant son
apparition dans notre syst\`{e}me.

%\textbf{Contenu de la publication 3 :}
%\includepdf[pages={1-8}]{article3.pdf}

\subsection{Graphyne core/shell nanoparticles : Monte Carlo study of thermal and magnetic properties \newline (Submitted, (2016))}

$\bullet $ \textbf{R\'{e}sum\'{e} de la publication 4 :}

Dans ce papier, notre objectif principal consistait \`{a} l'\'{e}tude des
propri\'{e}t\'{e}s magn\'{e}tiques et hyst\'{e}r\'{e}tiques d'une nanoparticule c{\oe}ur-coquille
avec une structure crystallographique semblable au graphyne. Pour ce faire,
nous avons utilis\'{e} la simulation Monte Carlo. Nous avons examin\'{e} les
effets des param\`{e}tres de l'hamiltonien sur les propri\'{e}t\'{e}s magn%
\'{e}tiques et thermodynamiques du syst\`{e}me, \`{a} savoir : l'aimantation
totale, la susceptibilit\'{e}, les cycles d'hyst\'{e}r\'{e}sis et la temp%
\'{e}rature de compensation. Par ailleurs, nous avons examin\'{e} l'\'{e}tat
fondamental, ainsi que les diagrammes de phase de ce syst\`{e}me. \`{A} ce
titre, nous avons not\'{e} l'apparition de deux points de compensation,
ainsi que l'existence de deux nouveaux type du compensation. Ceux-ci n'ont
pas \'{e}t\'{e} classifi\'{e}s dans la nomenclature du N\'{e}el plus les
types $Q$, $P$ et $N$. L'\'{e}tude des propri\'{e}t\'{e}s hyst\'{e}r\'{e}%
tiques a r\'{e}v\'{e}l\'{e} que la nanoparticule c{\oe}ur-coquille de type graphyne pr\'{e}sente des
cycles simples et triples avec diverses formes.

%\textbf{Contenu de la publication 4 :}
%\includepdf[pages={1-10}]{article4.pdf}

\section{Effet des d\'{e}fauts et de surface sur les propri\'{e}t\'{e}s magn\'{e}tiques des nanomat\'{e}riaux}

Les nanomat\'{e}riaux font l'objet d'un int\'{e}r\^{e}t grandissant gr\^{a}ce \`{a} leurs propri\'{e}t\'{e}s singuli\`{e}res, vis-\`{a}-vis des mat\'{e}riaux  massifs correspondants, qui les rendent tr\`{e}s attractifs pour de nombreuses applications en opto\'{e}lectronique, \'{e}lectronique, magn\'{e}tisme etc. Le  d\'{e}veloppement de ces applications n\'{e}cessite une parfaite connaissance de leurs structures. Par ailleurs, comme dans tous les mat\'{e}riaux, les d\'{e}fauts de structures sont in\'{e}vitables et affectent leurs propri\'{e}t\'{e}s. Les d\'{e}fauts connues, concernant les mat\'{e}riaux structur\'{e}s en nids d'abeille peuvent se pr\'{e}senter sous la forme, soit de site vacant, soit des d\'{e}fauts de type Stone-Wales correspondent \`{a} la transformation de quatre hexagones en deux pentagones et deux heptagones par rotation \`{a} $90%
%TCIMACRO{\U{b0}}%
%BeginExpansion
{{}^\circ}%
%EndExpansion
$ d'une liaison. Tous ces d\'{e}fauts peuvent coexister sur diff\'{e}rents plans et migrer sous l'effet de la temp\'{e}rature.

Outre l'effet de d\'{e}faut structurels, la morphologie des mat\'{e}riaux affecte aussi leurs propri\'{e}t\'{e}s magn\'{e}tiques. En effet, les nanomat\'{e}riaux dans
leurs diff\'{e}rentes morphologies, pr\'{e}sentent de nombreuses propri\'{e}t%
\'{e}s magn\'{e}tiques int\'{e}ressantes, dont l'existence d'une temp\'{e}%
rature de compensation, caract\'{e}ristique de ces nanostructures. Par ailleurs, plusieurs mat\'{e}riaux avec la morphologie
surface-volume, montrent que les propri\'{e}t\'{e}s magn\'{e}tiques des
surfaces diff\`{e}rent consid\'{e}rablement de celles du volume auquel elles
sont coupl\'{e}es, conduisant ainsi \`{a} des nouvelles propri\'{e}t\'{e}s.

Dans cette section, et faisant suite aux d\'{e}veloppements r\'{e}cents en physique des nanomat\'{e}riaux, nous \'{e}talons trois articles de recherche qui s'int\'{e}ressent particuli\`{e}rement \`{a} l'\'{e}tude de l'effet des d\'{e}fauts et de surface sur les propri\'{e}t\'{e}s magn\'{e}tiques des nanomat\'{e}riaux.

Le premier pr\'{e}sente l'\'{e}tude des d\'{e}faut Stone-Wales sur les propri\'{e}t\'{e}s magn\'{e}tiques et hyst\'{e}r\'{e}tique des mat\'{e}riaux mixtes \`{a} base mol\'{e}culaire qui ont provoqu\'{e}
r\'{e}cemment un consid\'{e}rable int\'{e}r\^{e}t dans leur synth\`{e}se et
l'\'{e}tude de leurs propri\'{e}t\'{e}s magn\'{e}tiques et qui offrent la possibilit\'{e} de cr\'{e}er des nouvelles
architectures de r\'{e}seaux. Compar\'{e}s \`{a} ceux du m\'{e}tal classique et m\'{e}tal-oxyde, ces types
de mat\'{e}riaux ont l'avantage d'\^{e}tre obtenus \`{a} travers une s\'{e}%
lection de sources appropri\'{e}e de spin, comme des ions de m\'{e}taux de
transition et des radicaux organiques. Le domaine
\'{e}mergent du magn\'{e}tisme mol\'{e}culaire, constitue une nouvelle
branche de la science des mat\'{e}riaux, qui traite des propri\'{e}t\'{e}s
magn\'{e}tiques des mol\'{e}cules ou des assemblages de mol\'{e}cules. L'int%
\'{e}r\^{e}t croissant pour la compr\'{e}hension de l'origine de l'ordre magn%
\'{e}tique dans ces mati\`{e}res, est d'obtenir de nouveaux mat\'{e}riaux magn%
\'{e}tiques mol\'{e}culaires avec une temp\'{e}rature de transition tr\`{e}s
\'{e}lev\'{e}e. Ainsi, plusieurs familles des mat\'{e}riaux \`{a} base
mol\'{e}culaire ont \'{e}t\'{e} examin\'{e}es au cours des deux derni\`{e}%
res d\'{e}cennies. Parmi ces mat\'{e}riaux, se trouvent des syst\`{e}mes
mixtes de type $AFe^{II}Fe^{III}(C_{2}O_{4})_{3}$, qui ont fait l'objet de
nombreuses recherches, depuis leur premi\`{e}re synth\`{e}se par diffraction
des rayons $X$. Les mat\'{e}riaux \`{a} base mol\'{e}culaire de type $%
AFe^{II}Fe^{III}(C_{2}O_{4})_{3}$ $(A=N(n-C_{n}H_{2n+1})_{4},n=3-5)$ ont une
structure en nid d'abeille en couches. Dans chaque couche, les atomes magn%
\'{e}tiques $Fe^{II}$ et $Fe^{III}$ sont coupl\'{e}s avec des ions oxalates $%
C_{2}O_{4}^{2-}$ pour former une structure hexagonale \`{a} deux dimensions ; tandis que les cations $A^{+}$ sont positionn\'{e}s entre les couches. Les
mat\'{e}riaux \`{a} base mol\'{e}culaire de type $AFe^{II}Fe^{III}(C_{2}O_{4})_{3}$
peuvent \^{e}tre d\'{e}crits par des syst\`{e}mes de spin d'Ising mixtes.
Ils ont engendr\'{e} un immense int\'{e}r\^{e}t, car ils pr\'{e}sentent de
nombreux ph\'{e}nom\`{e}nes qui ne peuvent pas \^{e}tre observ\'{e}s dans
leurs homologues avec un seul type de spin.

Le deuxi\`{e}me article traite l'effet des sites vacants sur les propri\'{e}t\'{e}s thermodynamiques du nanoruban $FeS_{2}$, qui a fourni des opportunit\'{e}s uniques pour visualiser les interactions du
fer avec les chalcog\`{e}nes, qui pr\'{e}sentent un immense int\'{e}r\^{e}t.
Ce nouveau mat\'{e}riau est un m\'{e}tal ferromagn\'{e}tique en nid
d'abeille o\`{u} les moments magn\'{e}tiques sont situ\'{e}s \`{a}
l'orbitale des atomes de fer. L'analyse des propri\'{e}t\'{e}s magn\'{e}%
tiques a r\'{e}v\'{e}l\'{e} que $FeS_{2}$ est un bon candidat pour les
applications de la spintronique.

Le troisi\`{e}me article concerne l'\'{e}tude de l'effet de surface sur un nanocube avec la morphologie surface-volume.

\subsection{Stone-Wales defected molecular-based $%
AFe^{II}Fe^{III}(C_{2}O_{4})_{3}$ nanoribons : Thermal and magnetic
properties \newline (In preparation)}

$\bullet $ \textbf{R\'{e}sum\'{e} de la publication 5 :}
\begin{sloppypar}
Dans cette publication, les propri\'{e}t\'{e}s magn\'{e}tiques et hyst\'{e}r%
\'{e}tiques du nanoruban de type $AFe^{II}Fe^{III}(C_{2}O_{4})_{3}$ avec des d\'{e}%
fauts de Stone-Wales ont \'{e}t\'{e} examin\'{e}es \`{a} l'aide de la
simulation Monte Carlo. Plus pr\'{e}cis\'{e}ment, nous nous sommes interess%
\'{e}s \`{a} l'effet du nombre et de la position des d\'{e}fauts de
Stone-Wales sur les quantit\'{e}s thermodynamiques qui d\'{e}crivent ce
syst\`{e}me. Parmi les principaux r\'{e}sultats obtenus lors de notre
analyse, nous avons montr\'{e} l'apparition de la temp\'{e}rature de
compensation qui pr\'{e}sente une importance essentielle dans les
applications technologiques. Nous avons \'{e}galement trouv\'{e} que la temp%
\'{e}rature de compensation et la temp\'{e}rature critique d\'{e}pendent du
nombre de d\'{e}fauts de Stone-Wales et des effets de bords. L'\'{e}tude des
propri\'{e}t\'{e}s hyst\'{e}r\'{e}tiques a r\'{e}v\'{e}l\'{e} l'apparition
des cycles d'hyst\'{e}r\'{e}sis pr\'{e}sents sous diff\'{e}rentes formes et avec un nombre \'{e}lev\'{e} de pas.
\end{sloppypar}

\subsection{Monte Carlo study of edge effect on magnetic and hysteretic
behaviors of sulfur vacancy defected zigzag $FeS_{2}$ nanoribbon \newline (In preparation)}

$\bullet $ \textbf{R\'{e}sum\'{e} de la publication 6 :}

Dans cette contribution, dont nous ne donnerons que le r\'{e}sum\'{e}, nous
avons utilis\'{e} la m\'{e}thode Monte Carlo pour examiner en d\'{e}tail
l'effet des sites vacants de soufre sur les propri\'{e}t\'{e}s magn\'{e}%
tiques et hyst\'{e}r\'{e}tiques des nanorubans de $FeS_{2}$. Nous avons
recherch\'{e} en particulier la d\'{e}pendance en temp\'{e}rature de
l'aimantation et de la susceptibilit\'{e}. Nous avons aussi concentr\'{e}
notre attention sur l'effet du nombre et de la position des sites vacants de
soufre sur le comportement thermodynamique et les caract\'{e}ristiques hyst%
\'{e}r\'{e}tiques du syst\`{e}me. Puis nous avons d\'{e}termin\'{e} la temp%
\'{e}rature critique de toutes les configurations des nanorubans de $FeS_{2}$%
\ avec les sites vacants de soufre. Un certain nombre de comportements caract%
\'{e}ristiques ont \'{e}t\'{e} trouv\'{e}s. Plus pr\'{e}cis\'{e}ment, nous
avons montr\'{e} l'apparition des boucles d'hyst\'{e}r\'{e}sis carr\'{e}es.

\subsection{Surface effect on compensation and hysteretic behavior in surface/bulk nanocube \newline (Submitted, (2016))}

$\bullet $ \textbf{R\'{e}sum\'{e} de la publication 7 :}

Dans ce papier, nous nous sommes attach\'{e}s \`{a} l'\'{e}tude de l'effet
de la surface sur les propri\'{e}t\'{e}s magn\'{e}tiques et hyst\'{e}r\'{e}%
tiques d'un nanocube avec la morphologie surface-volume en utilisant la
simulation Monte Carlo. Nous avons rapport\'{e} les effets des champs magn%
\'{e}tiques et cristallins ainsi que les couplages interm\'{e}diaires et le
couplage de volume, la temp\'{e}rature et la taille sur le diagramme de phase,
l'aimantation, la susceptibilit\'{e}, les cycles d'hyst\'{e}r\'{e}sis, la
temp\'{e}rature critique et la temp\'{e}rature de compensation du mod\`{e}%
le. Cette \'{e}tude a d\'{e}montr\'{e} un certain nombre de comportements
caract\'{e}ristiques, tels que l'existence de comportement de type $Q$ et $N$
dans la classification de N\'{e}el et aussi l'apparition des boucles d'hyst%
\'{e}r\'{e}sis simples et triples avec un nombre \'{e}lev\'{e} de pas.

%\textbf{Contenu de la publication 7 :}
%\includepdf[pages={1-12}]{article5.pdf}

%\include{Chapitre4/chapitre4}
\backmatter \pagestyle{fancy} \fancyhf{}
\renewcommand{\chaptermark}[1]{\markboth{Conclusion g\'en\'erale et perspectives}{}}
\fancyhead[LE,RO]{Conclusion g\'en\'erale et perspectives}
\fancyfoot[RO]{\thepage}
\fancyfoot[LE]{\thepage}
\renewcommand{\headrulewidth}{0.5pt}
\renewcommand{\footrulewidth}{0pt}

\chapter{Conclusion g\'en\'erale et perspectives}\vspace{0.5cm}
De nos jours, les mat\'{e}riaux magn\'{e}tiques sont devenus omnipr\'{e}sents et indispensables dans notre vie quotidienne en raison de leur capacit\'{e}
\`{a} guider le flux magn\'{e}tique et \`{a} m\'{e}moriser l'information.
Pour les optimiser et les fabriquer, il est n\'{e}cessaire de comprendre leur
structure et leur comportement. Int\'{e}ress\'{e}s par ce domaine de recherche pertinent, nous avons consacr\'{e} ce m\'{e}moire de th\`{e}se \`{a} l'\'{e}tude des propri\'{e}t\'{e}s magn\'{e}tiques des nanomat\'{e}riaux type graphyne et graphone par la m\'{e}thode Monte Carlo. Pour ce fait, nous avons pr\'{e}sent\'{e} les approximations de la physique statistique telles que : l'approximation du champ moyen, l'approximation du champ effectif et la m\'{e}thode de la matrice de transfert. Ces m\'{e}thodes sont cruciales dans la d\'{e}termination des propri\'{e}t\'{e}s critiques des syst\`{e}mes consid\'{e}r\'{e}s.

Dans ce manuscrit, qui comporte quatre chapitres en plus d'une introduction g\'{e}%
n\'{e}rale, nous avons consacr\'{e} tout un chapitre \`{a} nos travaux de recherche et nos r\'{e}%
cents r\'{e}sultats obtenus dans le domaine. Plus pr\'{e}cis\'{e}ment, ce volet pr\'{e}sente nos \'{e}tudes par la m\'{e}thode Monte Carlo et la th\'{e}orie du champ moyen des diff\'{e}rentes propri\'{e}t\'{e}s magn\'{e}tiques et hyst\'{e}r\'{e}tiques des nouveaux mat\'{e}riaux vu leur int\'{e}r\^{e}t et leur importance dans le domaine
de la mati\`{e}re condens\'{e}e et de la nanotechnologie. Cette conclusion g\'{e}n\'{e}rale pr\'{e}sente un
bilan condens\'{e} des id\'{e}es cl\'{e}s d\'{e}velopp\'{e}es dans ce manuscrit.

R\'{e}cemment, les chercheurs de la mati\`{e}re condens\'{e}e se sont int%
\'{e}ress\'{e}s aux mat\'{e}riaux \`{a} base de graph\`{e}ne en raison de leur propri\'{e}t%
\'{e}s physiques fascinantes. Dans ce cadre, nous avons contribu\'{e} avec une \'{e}tude des propri\'{e}t\'{e}s magn\'{e}%
tiques et hyst\'{e}r\'{e}tiques des nanorubans de graphone. Nous avons utilis\'{e} la simulation Monte Carlo et la th\'{e}orie du champ
moyen, qui sont consid\'{e}r\'{e}s comme des outils cruciaux dans l'approche des ph\'{e}nom\`{e}nes magn\'{e}tiques et critiques. Ces m\'{e}thodes
permettent d'examiner de mani\`{e}re simple et rigoureuse les propri\'{e}t%
\'{e}s magn\'{e}tiques et hyst\'{e}r\'{e}tiques du syst\`{e}me.

La substitution d'un atome de carbone par un atome magn\'{e}tique
portant un spin $3/2$\ dans le nanoruban de graphone m\`{e}ne \`{a} diff\'{e}%
rentes propri\'{e}t\'{e}s thermodynamiques. En utilisant \`{a} la fois la m%
\'{e}thode Monte Carlo et la th\'{e}orie du champ moyen, nous avons calcul%
\'{e} l'aimantation, la susceptibilit\'{e} et la temp\'{e}rature de
transition des nanorubans de graphone de type zigzag dans trois cas diff\'{e}%
rents : mono-dop\'{e}, bi- dop\'{e} et tri-dop\'{e}. Nous avons montr\'{e}
une alternative qui permet de contr\^{o}ler le magn\'{e}tisme dans des
nanorubans de graphone de type zigzag, en fonction de la position et du
nombre des impuret\'{e}s magn\'{e}tiques dans le nanoruban.

Encourag\'{e}s par l'importance des nanomat\'{e}riaux dans les
applications de la nanotechnologie, nous avons construit un nouveau mod\`{e}%
le de nanoruban ferrimagn\'{e}tique avec la morphologie c{\oe}ur-coquille. Dans ce
cadre, nous avons analys\'{e} l'\'{e}tat fondamental ainsi que les
diagrammes de phase de ce syst\`{e}me en utilisant la simulation Monte
Carlo. Nous avons \'{e}tudi\'{e} les propri\'{e}t\'{e}s magn\'{e}tiques et
hyst\'{e}r\'{e}tiques du c{\oe}ur-coquille nanoruban qui ont r\'{e}v\'{e}l\'{e} l'occurence de la temp\'{e}rature
de compensation. Cette derni\`{e}re est d'une importance capitale pour le stockage
de l'information et en particulier dans l'enregistrement thermo-optique.

Nous avons \'{e}galement examin\'{e} les propri\'{e}t\'{e}s magn\'{e}tiques et hyst\'{e}r\'{e}tiques de nanoparticule c{\oe}ur-coquille type graphyne. Plus particuli\`{e}rement, nous nous sommes int\'{e}ress\'{e}s \`{a} l'effet des param\`{e}tres de l'hamiltonien sur le comportement de compensation. Nous avons montr\'{e} l'apparition de deux points de compensation, ainsi que l'existence de deux nouveaux type de l'aimantation, qui n'ont pas \'{e}t\'{e} classifi\'{e}s dans la nomenclature du N\'{e}el plus les types $Q$, $P$ et $N$.

Dans cette pr\'{e}sente \'{e}tude, un int\'{e}r\^{e}t particulier a \'{e}t\'{e} apport\'{e} \`{a} l'effet de la morphologie des nanomat\'{e}riaux sur leur propri\'{e}t\'{e}s thermodynamiques. Dans ce contexte, nous avons \'{e}tudi\'{e} l'effet de surface sur un nanocube avec la morphologie surface-volume. Nous avons examin\'{e} l'\'{e}tat fondamental, ainsi que les diagrammes de phase de ce syst\`{e}me. Cette \'{e}tude a r\'{e}v\'{e}l\'{e} \`{a} un certain nombre de
comportements caract\'{e}ristiques, tels que l'existence de comportement de type $Q$ et $N$ dans la classification de N\'{e}el et aussi l'apparition des boucles d'hyst\'{e}r\'{e}sis simples et triples avec un nombre \'{e}lev\'{e} de pas.

Par ailleurs, nous avons pr\'{e}sent\'{e} dans ce m\'{e}moire
de th\`{e}se des facettes de l'\'{e}volution de toutes les \'{e}tudes pr\'{e}%
cit\'{e}es en mettant l'accent sur quelques d\'{e}veloppements r\'{e}cents
et certains r\'{e}sultats inattendus.

De tout ce qui pr\'{e}lude, il est clair que gr\^{a}ce aux d\'{e}%
veloppements rapides qu'ont connus les nouveaux mat\'{e}riaux, les mat\'{e}%
riaux \`{a} base de graph\`{e}ne et les nanoparticules ferrimagn\'{e}tiques demeurent des sujets de recherche d'actualit\'{e} majeurs. Plusieurs obstacles sont encore \`{a} franchir, diverses cl\'{e}%
s demeurent perdues et de nombreuses questions sans r\'{e}ponses \'{e}%
mergent dont les \'{e}tudes pourront faire l'objet de d\'{e}veloppements ult%
\'{e}rieurs. Ceci nous a motiv\'{e} \`{a} lancer de futures perspectives. En effet, nous envisageons d'aller plus loin dans ce
domaine, d'essayer d'autres m\'{e}thodes et codes utilis\'{e}s par la
communaut\'{e} scientifique. Il serait \'{e}galement int%
\'{e}ressant d'examiner diff\'{e}rentes propri\'{e}t\'{e}s et caract\'{e}%
ristiques des nanomat\'{e}riaux qui sont primordiales dans les domaines industriels et commercials des nanotechnologies.
\appendix
\setcounter{figure}{0} \setcounter{table}{0}
\setcounter{footnote}{0} \setcounter{equation}{0} \pagestyle{fancy}
\fancyhf{}
\renewcommand{\chaptermark}[1]{\markboth{\MakeUppercase{#1 }}{}}
\renewcommand{\sectionmark}[1]{\markright{\thesection~ #1}}
\fancyhead[RO]{\bfseries\rightmark}
\fancyhead[LE]{\bfseries\leftmark} \fancyfoot[RO]{\thepage}
\fancyfoot[LE]{\thepage}
\renewcommand{\headrulewidth}{0.5pt}
\renewcommand{\footrulewidth}{0pt}
\def\thechapter {\arabic{chapter}}
\def\thesection {\arabic{section}}
\def\thesubsection {\thesection\arabic{subsection}}
\def\thesubsubsection {\thesubsection \arabic{subsubsection}.}

\makeatletter
\renewcommand\thefigure{A.\arabic{figure}}
\renewcommand\thetable{A.\arabic{table}}
\makeatother
\def\cleardoublepage{\clearpage}
\newpage
\strut

\addtocontents{toc}{\vspace{-0.8cm}}
\clearpage
\newpage
\markboth{\uppercase}{\uppercase}
\renewcommand{\headrulewidth}{0pt}
\fancyhf{}

\newcommand{\RAPAi}{Discipline}
\newcommand{\RAPAj}{\textbf{Physique}}
\newcommand{\RAPBi}{Sp\'{e}cialit\'{e}}
\newcommand{\RAPBj}{\textbf{Physique Math\'{e}matique}}
\newcommand{\RAPCi}{Laboratoire}
\newcommand{\RAPCj}{\textbf{Physique des Hautes Energies,}}
\newcommand{\RAPCk}{\textbf{Mod\'{e}lisation et Simulation}}
\newcommand{\RAPDi}{Responsable du laboratoire}
\newcommand{\RAPDj}{\textbf{El Hassan SAIDI}}
\newcommand{\PRESi}{P\'{e}riode d'accr\'{e}ditation}
\newcommand{\PRESj}{\textbf{2013-2016}}

\newcommand{\affiliation}{
\begin{tabular}{l@{\protect\hspace{0.5cm}}l@{\protect\hspace{0.5cm}}l}
%\textbf{\large{\textbf{Composition du Examinateurs :}}}
\vspace{-0.25cm}\RAPAi :&\RAPAj\\
\vspace{-0.25cm}\RAPBi :&\RAPBj\\
\vspace{-0.25cm}\RAPCi :&\RAPCj\\
\vspace{-0.25cm}&\RAPCk\\
\vspace{-0.25cm}\RAPDi :&\RAPDj\\
\vspace{-0.25cm}\PRESi :&\PRESj\\
\end{tabular}
}
\clearpage
\newpage
\markboth{\uppercase}{\uppercase}
\renewcommand{\headrulewidth}{0pt}
\fancyhf{}

\newcommand{\RAi}{\large{Pr\'{e}nom, Nom :}}
\newcommand{\RAj}{\large{\textbf{Sanae ZRIOUEL}}}
\newcommand{\RBi}{\large{R\'{e}sum\'{e} :}}

\newcommand{\affect}{
\begin{tabular}{l@{\protect\hspace{0.5cm}}l@{\protect\hspace{-1cm}}l}
%\textbf{\large{\textbf{Composition du Examinateurs :}}}
\hspace{-0.25cm}\vspace{0.2cm}\RAi &\RAj\\
\hspace{-0.25cm}\vspace{0.2cm}\RBi

\end{tabular}
}
\vspace*{-3.6cm}
\begin{tabular}{l@{\protect\hspace{4cm}}c@{\protect\hspace{4cm}}r}
\hspace{-0.85cm}\includegraphics[scale=1.3]{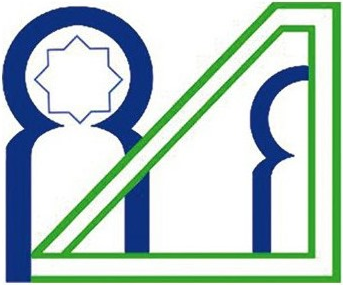} &
\hspace{7.5cm}\includegraphics[scale=1]{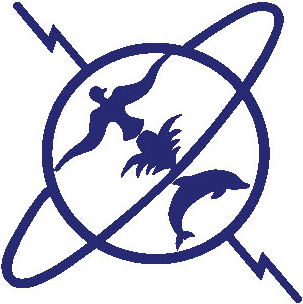}
\end{tabular}
\begin{center}
%\hbox{\raisebox{0.4em}{\vrule depth 0pt height 1pt width 17cm}}\setlength{\baselineskip}{13pt}~\\
\vspace{-0.2cm}\hbox{\raisebox{0.4em}{\vrule depth 0pt height 1pt width 17cm}}\setlength{\baselineskip}{13pt}~\\
{\Large{\textbf{}}}\\[\baselineskip]
      \PRESENTATION\\[\baselineskip]
\vspace{-0.15cm}\hspace{0.7cm}{\large{\textbf{\textbf{UNIVERSIT\'{E} MOHAMMED V-RABAT}}}}\\[\baselineskip]
    \vspace{-0.5cm}\hspace{0.7cm}  \large{\textbf{FACULT\'{E} DES SCIENCES}} \\[\baselineskip]
    \vspace{-0.8cm}\hspace{0.7cm}\large{\textbf{RABAT}} {~~ \vspace{2.7cm }~~}\\[\baselineskip]
    \vspace{-2.5cm} \hspace{12.325cm}\large{\textbf{N$%
%TCIMACRO{\U{b0}}%
%BeginExpansion
{{}^\circ}%
%EndExpansion
$ d'ordre : $2845$}}{~~ \vspace{0.cm }~~}\\[\baselineskip]
      \vspace{-0.3cm}\Large{\textbf{\bf{DOCTORAT}}} \\[\baselineskip]
     \vspace{-0.6cm}\large{\bf{R\'{e}sum\'{e} de la Th\`{e}se}} \\[\baselineskip]
\vspace{-0.3cm}
       \affiliation\vspace{0.5cm}\\[\baselineskip]\vspace{-0.7cm}
 %\hbox{\raisebox{0.2em}{\vrule depth 0pt height 3.5pt width 17cm}}
            %\setlength{\baselineskip}{4pt}
           \hbox{\raisebox{0.4em}{\vrule depth 0pt height 1pt width 17cm}}\setlength{\baselineskip}{10pt}~\\
            \vspace*{-20pt}
            \vspace{0.5cm}{\large{Titre de la th\`{e}se : \textbf{Contributions \`{a} l'\'{e}tude Monte Carlo des propri\'{e}t\'{e}s magn\'{e}tiques des nanomat\'{e}riaux type graphyne et graphone}}}~\\[\baselineskip]
        \hbox{\raisebox{0.4em}{\vrule depth 0pt height 1pt width 17cm}}~\\\vspace{0.5cm}
\end{center}
\vspace{-1cm}\affect\newline
Attir\'{e}s par l'importance de nouveaux mat\'{e}riaux dans le domaine des
nanotechnologies, cette th\`{e}se d\'{e}veloppe cet axe de recherche tout en
approfondissant les r\'{e}sultats. Nous avons commenc\'{e} par
introduire les m\'{e}thodes de simulation et de calcul les plus sophistiqu%
\'{e}es, telles que : la m\'{e}thode Monte Carlo, la th\'{e}orie du champ
moyen, la th\'{e}orie du champ effectif et la m\'{e}thode de la matrice de
transfert. Par la suite, nous avons \'{e}tudi\'{e} les propri\'{e}t\'{e}s
magn\'{e}tiques et hyst\'{e}r\'{e}tiques des mat\'{e}riaux. Ensuite, nous
avons d\'{e}taill\'{e} certaines de nos contributions correspondantes aux mat%
\'{e}riaux \`{a} base de graph\`{e}ne et des nanomat\'{e}riaux ferrimagn\'{e}%
tiques avec diff\'{e}rentes morphologies. Nous avons d\'{e}battu de l'effet
des d\'{e}fauts sur les propri\'{e}t\'{e}s thermodynamiques de ces nouveaux
mat\'{e}riaux. Une attention particuli\`{e}re a \'{e}t\'{e} port\'{e}e aux
param\`{e}tres physiques qui influencent la temp\'{e}rature de compensation.
Celle-ci ayant une tr\`{e}s grande importance dans le stockage d'information
et plus particuli\`{e}rement dans l'enregistrement thermo-optique. Avec tous
ces \'{e}l\'{e}ments, nous nous sommes ouverts aux d\'{e}veloppements les
plus r\'{e}cents de la physique de nouveaux mat\'{e}riaux. Enfin, nous avons termin\'{e} par la conclusion et des perspectives.\vspace{0.31cm}
\begin{center}

 %\hbox{\raisebox{0.2em}{\vrule depth 0pt height 3.5pt width 17cm}}
            %\setlength{\baselineskip}{4pt}
           \hbox{\raisebox{0.4em}{\vrule depth 0pt height 1pt width 17cm}}\setlength{\baselineskip}{13pt}~\\
            \vspace*{-26pt}
            \vspace{0.5cm}{\Large{\textbf{Mots-clefs :} \large{Monte Carlo, graphone, graphyne, nanoruban, c{\oe}ur-coquille, temp\'{e}rature de compensation, propri\'{e}t\'{e}s thermodynamiques et magn\'{e}tiques}}}~\\[\baselineskip]\vspace*{-10pt}
        \hbox{\raisebox{0.4em}{\vrule depth 0pt height 1pt width 17cm}}
\end{center}
\end{document}

%% file: Remerciement/remerciement.tex
\chapter*{Remerciement}
%===================================================================
\vspace{-1cm} Les travaux pr\'{e}sent\'{e}s dans cette th\`{e}se ont \'{e}t\'{e} r\'{e}alis%
\'{e}s au sein du Laboratoire de Physique des Hautes Energies, Mod\'{e}%
lisation et Simulation (LPHE-MS) du d\'{e}partement de physique de la Facult%
\'{e} des Sciences de Rabat, sous la direction de Monsieur El
Hassan SAIDI, Professeur de l'Enseignement Sup\'{e}rieur \`{a} la Facult\'{e} des Sciences de Rabat et la co-direction de Mme Lalla Btissam DRISSI, Professeure Habilit\'{e} \`{a} la
Facult\'{e} des Sciences de Rabat.

Toute ma gratitude et mes sinc\`{e}res remerciements \`{a} mon directeur de
th\`{e}se Monsieur El Hassan SAIDI, Professeur de l'Enseignement Sup\'{e}rieur \`{a} la Facult\'{e} des sciences de Rabat et ma Co-directrice Madame Lalla Btissam DRISSI, Professeure Habilit\'{e} \`{a} la Facult\'{e} des Sciences de Rabat, pour
m'avoir guid\'{e}e, encourag\'{e}e, conseill\'{e}e et pour le formidable
encadrement qu'ils m'ont accord\'{e} tout au long de ce travail. Leurs
disponibilit\'{e}, leurs exp\'{e}rience et leurs sens de transmission des
connaissances scientifiques m'ont permis de mener \`{a} bien ce travail. Je
ne peux que leurs remercier, non seulement pour leurs comp\'{e}tences
scientifiques, mais aussi pour leurs qualit\'{e}s humaines, leurs conseils
judicieux et leurs attentions au d\'{e}tail.

J'adresse mes remerciements les plus profonds \`{a} Madame Najia KOMIHA, Professeure
de l'Enseignement Sup\'{e}rieur \`{a} la Facult\'{e} des Sciences de Rabat, qui m'a fait l'honneur d'\^{e}tre la pr\'{e}sidente de
mon Jury de th\`{e}se. Je la remercie aussi pour son enseignement durant la pr\'{e}paration du Master avec une tr%
\`{e}s grande comp\'{e}tence. Son aide et l'enseignement de haut niveau que
j'ai re\c{c}u aupr\`{e}s de lui durant ma formation de Master m'ont \'{e}t%
\'{e} d'un grand int\'{e}r\^{e}t.

Je tiens \'{e}galement \`{a} remercier Monsieur Lahoucine BAHMAD, Professeur de
l'Enseignement Sup\'{e}rieur \`{a} la Facult\'{e} des Sciences de Rabat, pour avoir accept\'{e} d'\^{e}tre rapporteur de ma th\`{e}se. Veuillez recevoir, Monsieur, l'expression de mon respect et de ma profonde
gratitude.

Je suis tr\`{e}s honor\'{e} que Monsieur Najem HASSANAIN, Professeur \`{a} la Facult%
\'{e} des Sciences de Rabat, ait accept\'{e} d'\^{e}tre Examinateur de ce travail. Qu'il trouve ici, l'expression de ma
profonde consid\'{e}ration.

Un grand merci \`{a} Monsieur Mohammed DAOUD, Professeur de l'Enseignement
Sup\'{e}rieur \`{a} la Facult\'{e} des Sciences, Ain Chock Casablanca, pour sa participation \`{a} mon jury de th\`{e}se en qualit\'{e} de rapporteur de mon travail et pour toutes remarques int\'{e}ressantes qu'il m'a faites. Veuillez agr\'{e}er, Monsieur, l'assurance
de mon profond respect.

Je remercie \'{e}norm\'{e}ment tous les membres du jury pour leur
disponibilit\'{e} le jour de ma soutenance, pour le privil\`{e}ge qu'ils
m'ont accord\'{e} par leur pr\'{e}sence et pour l'int\'{e}r\^{e}t qu'ils ont
attribu\'{e} \`{a} mon m\'{e}moire de th\`{e}se. Je tiens \`{a} vous
remercier encore une fois d'avoir accept\'{e} de vous d\'{e}placer pour
partager avec moi l'un des moments les plus m\'{e}morables de toute ma vie.
Ces moments qui risquent de cr\'{e}er un \'{e}v\'{e}nement inoubliable pour
le reste de mes jours.

Je voudrais tout particuli\`{e}rement exprimer ma reconnaissance \`{a} Monsieur Rachid AHL LAAMARA, Professeur Assistant au CRMEF, Meknes, qui m'a suivi tout  au long de cette th\`{e}se. Je tiens \`{a} l'assurer de ma profonde reconnaissance, pour son aide, sa disponibilit\'{e}, ses nombreux
conseils et son soutien sans faille. Soyez assur\'{e}, Monsieur, de ma
profonde gratitude.

Je voudrais adresser un remerciement particulier \`{a} tous le corps
professoral du Master, pour leur soutien et pour leur patience. A Prof M.
Ait Ben Haddou, Prof L. Bahmad, Prof A. Belhaj, Prof M. Bennai, Prof F.
Bentayeb, Prof M. Daoud, Prof A. El Kenz, Prof N. E. Fahssi, Dr. A. Hamama,
Dr J. Khan, Prof. N. Komiha, Prof T. Lhallabi, Prof E.H. Saidi et Prof M. B.
Sedra, qui ont contribu\'{e} \`{a} nous transmettre leur savoir pour assurer
notre formation. Merci \`{a} vous tous chers professeurs pour m'avoir fait go%
\^{u}ter au charme de la physique th\'{e}orique et pour m'avoir fait d\'{e}%
guster le d\'{e}lice des diff\'{e}rents axes de la physique des hautes \'{e}%
nergies, mod\'{e}lisation et simulation.

Si j'en suis arriv\'{e}e l\`{a} aujourd'hui, c'est aussi parce que j'ai rencontr\'{e} sur mon chemin des personnes qui m'ont apport\'{e} le meilleur d'elles-m\^{e}mes et m'ont hiss\'{e} vers l'excellence. Je remercie l'ensemble de l'\'{e}quipe de recherche du LPHE-MS pour les
nombreuses et toujours fructueuses discussions. Travailler avec eux a \'{e}t%
\'{e} un r\'{e}el plaisir que du bonheur.

Ne pouvant malheureusement pas citer toutes les personnes que j'ai rencontr%
\'{e} durant mon parcours et qui ont pu contribu\'{e} d'une fa\c{c}on ou d'une
autre, de pr\`{e}s ou de loin, \`{a} l'aboutissement de cette th\`{e}se, je
leur dis \`{a} toutes merci d'avoir \'{e}t\'{e} l\`{a} \`{a} cette instant pr%
\'{e}cis o\`{u} je les ai rencontr\'{e}es et o\`{u} ils m'ont apport\'{e}e
cette aide qui a surement contribu\'{e}e \`{a} aller au bout de cette th\`{e}%
se.

Cette th\`{e}se s'est r\'{e}dig\'{e}e en une bonne partie \`{a} la maison
avec l'encouragement continu, le soutien financier et moral constant et la
bienveillance sensationnelle de ma tendre tr\`{e}s ch\`{e}re m\`{e}re Fatima
et mon adorable aimable p\`{e}re Said. Merci \`{a} vous deux pour tout ce
que vous avez fait pour moi durant tout le parcours de ma vie et gr\^{a}ce
auquel je suis l\`{a} pr\'{e}sente aujourd'hui. Mille merci \`{a} toi
Hamada, Karima et Oussama pour toutes les petites aussi bien que les tr\`{e}%
s grandes choses que vous m'avez pr\'{e}sent\'{e}es et que je n'oublierai
jamais, Merci ma petite famille, vous \'{e}tiez tous pr\'{e}sents chaque
moments o\`{u} j'avais besoin de vous, merci pour votre d\'{e}vouement
illimit\'{e} et votre amour z\'{e}l\'{e}.

Merci \`{a} tous les membres de ma grande famille et surtout \`{a} mon
exquise grand-m\`{e}re et ma tante Fatima pour chaque douce parole prononc%
\'{e}e et chaque belle pri\`{e}re provenant de leur coeur purs et limpides, traversant
mon \^{a}me comme une fl\`{e}che, me redonnant un grand et nouvel espoir pour demain
et repoussant tout sentiment r\'{e}pugnant.

J'exprime \'{e}galement mes sinc\`{e}res remerciements \`{a} Madame et Monsieur AZIZI, pour leur gentillesse et leur encouragement durant mes ann\'{e}es d'\'{e}tudes.

Je ne peux cl\^{o}turer cette partie de remerciements sans pr\'{e}senter mes
vifs remerciements au CNRST qui m'a attribu\'{e} la bourse d'excellence
durant mes trois ann\'{e}es d'\'{e}tude doctorale. Merci pour ce soutien qui
m'a \'{e}t\'{e} tr\`{e}s lucratif.
\newpage
\thispagestyle{empty} \ \
 \ \

 \ \

\ \

\ \

\ \

\ \

\ \

\ \

\ \

\ \

\begin{center}
\textit{\`{A} la m\'{e}moire de mes grands-parents}

\textit{Puisse Dieu les accueillir dans son infinie
Mis\'{e}ricorde}

\textit{\`{A} mes tr\`{e}s chers parents}

\textit{\`{A} ma tr\`{e}s ch\`{e}re s\oe ur Karima}

\textit{\`{A} mes tr\`{e}s chers fr\`{e}res Hamada et Oussama}

\textit{\`{A} tous ceux qui m'ont aid\'{e}e}

\textit{\`{A} tous ceux qui m'ont soutenue de pr\`{e}s ou de loin}

\textit{\`{A} tous ceux qui ont cru en moi}

\textit{\`{A} tous ceux qui m'ont faits confiance}

\textit{\`{A} tous ceux qui ont support\'{e} mes questions incessantes}

\textit{Je vous d\'{e}die ce Manuscrit avec toute la gr\^{a}ce du monde}
 \end{center}

%% file: Resume/Resume.tex
\chapter*{R\'{e}sum\'{e}}
%===================================================================

Attir\'{e}s par l'importance de nouveaux mat\'{e}riaux dans le domaine des
nanotechnologies, cette th\`{e}se d\'{e}veloppe cet axe de recherche tout en
approfondissant les r\'{e}sultats. Nous avons commenc\'{e} par
introduire les m\'{e}thodes de simulation et de calcul les plus sophistiqu%
\'{e}es, telles que : la m\'{e}thode Monte Carlo, la th\'{e}orie du champ
moyen, la th\'{e}orie du champ effectif et la m\'{e}thode de la matrice de
transfert. Par la suite, nous avons \'{e}tudi\'{e} les propri\'{e}t\'{e}s
magn\'{e}tiques et hyst\'{e}r\'{e}tiques des mat\'{e}riaux. Ensuite, nous
avons d\'{e}taill\'{e} certaines de nos contributions correspondantes aux mat%
\'{e}riaux \`{a} base de graph\`{e}ne et des nanomat\'{e}riaux ferrimagn\'{e}%
tiques avec diff\'{e}rentes morphologies. Nous avons d\'{e}battu de l'effet
des d\'{e}fauts sur les propri\'{e}t\'{e}s thermodynamiques de ces nouveaux
mat\'{e}riaux. Une attention particuli\`{e}re a \'{e}t\'{e} port\'{e}e aux
param\`{e}tres physiques qui influencent la temp\'{e}rature de compensation.
Celle-ci ayant une tr\`{e}s grande importance dans le stockage d'information
et plus particuli\`{e}rement dans l'enregistrement thermo-optique. Avec tous
ces \'{e}l\'{e}ments, nous nous sommes ouverts aux d\'{e}veloppements les
plus r\'{e}cents de la physique de nouveaux mat\'{e}riaux. Enfin, nous avons termin\'{e} par la conclusion et des perspectives.\vspace{0.31cm}

 %\hbox{\raisebox{0.2em}{\vrule depth 0pt height 3.5pt width 17cm}}
            %\setlength{\baselineskip}{4pt}

            \vspace{2cm}{\Large{\textbf{Mots-clefs :} \large{Monte Carlo, graphone, graphyne, nanoruban, c{\oe}ur-coquille, temp\'{e}rature de compensation, propri\'{e}t\'{e}s thermodynamiques et magn\'{e}tiques}}}